\documentclass[11pt,oneside]{article}
\DeclareUnicodeCharacter{2605}{\ensuremath{\star}}
\usepackage{epsfig,lscape}
\usepackage{rotating}
\usepackage{graphicx} 
\usepackage{amsmath}
\usepackage{amsthm}
\usepackage{amssymb}
\usepackage{subcaption}
\usepackage{bbm}
\usepackage{dsfont}
\usepackage{float}
\usepackage{hyperref}
\usepackage{enumitem}
\usepackage{algpseudocode}
\usepackage[authoryear]{natbib} 
\usepackage[resetlabels]{multibib}
\newcites{Supp}{Supplement References}

\usepackage[ruled]{algorithm}
\usepackage{lipsum} 
\usepackage{titling} 

\providecommand{\keywords}[1]{\bigskip\noindent\textbf{Keywords: }#1}

\usepackage{color}
\usepackage[dvipsnames]{xcolor}

\usepackage{float}

\usepackage[dvipsnames]{xcolor}
\newcommand{\zt}{\textcolor{blue}}

\DeclareMathOperator*{\argmin}{argmin}
\def\E{\mathrm E}
\def\var{\mathrm{Var}}

\def\Softmax{\mathrm{Softmax}}
\def\TV{\mathrm{TV}}
\def\log{\mathrm{log}}
\def\ESS{\mathrm{ESS}}

\def\Unif{\mathrm{Unif}}
\def\exp{\mathrm{exp}}

\def\vech{\mathrm{vech}}
\def\vec{\mathrm{vec}}

\def\tr{\mathrm{tr}}
\def\T{ {\mathrm{\scriptscriptstyle T}} }

\makeatother

\newtheorem{proposition}{Proposition}

\allowdisplaybreaks

\begin{document}

\setlength{\abovedisplayskip}{5pt}
\setlength{\belowdisplayskip}{5pt}


\begin{titlepage}

\begin{center}
{\Large Preconditioned Discrete-HAMS: A Second-order Irreversible Discrete Sampler }

\vspace{.1in}Yuze Zhou\footnotemark[1] \& Zhiqiang Tan \footnotemark[1]

\vspace{.1in}
\today
\end{center}

\footnotetext[1]{Department of Statistics, Rutgers University. Address: 110 Frelinghuysen Road,
Piscataway, NJ 08854. E-mails: yz909@scarletmail.rutgers.edu, ztan@stat.rutgers.edu.}

\begin{abstract}
Gradient-based Markov Chain Monte Carlo methods have recently received much attention for sampling discrete distributions, with notable examples such as
Norm Constrained Gradient (NCG), Auxiliary Variable Gradient (AVG), and Discrete Hamiltonian Assisted Metropolis Sampling (DHAMS).
In this work, we propose the Preconditioned Discrete-HAMS (PDHAMS) algorithm, which extends DHAMS by incorporating a second-order, quadratic approximation of the potential function, and uses Gaussian integral trick to avoid directly sampling a pairwise Markov random field. 
The PDHAMS sampler not only satisfies generalized detailed balance, hence enabling irreversible sampling, but also is a rejection-free property for a target distribution with a quadratic potential function.
In various numerical experiments, PDHAMS algorithms consistently yield superior performance compared with other methods.
\end{abstract}


\end{titlepage}
\section{Introduction}

Markov Chain Monte Carlo (MCMC) is one of the most widely used techniques for sampling from discrete distributions with intractable probability mass functions $\pi(s)$ \citep{robert2013MCrule}, where $\pi(s) \propto \exp(f(s))$ is defined over a discrete domain $\mathcal{S} \subset \mathbb{R}^{d}$
with a negative potential function $f(s)$. The support $\mathcal{S}$ is typically assumed to be a discrete $d$-dimensional lattice of the form $\mathcal{S} = \{a_{1}, a_{2}, \cdots, a_{k}\}^{d}$.

Recently, gradient-based samplers for discrete distributions have received considerable attention. These methods leverage a first-order Taylor expansion of $f(s)$ at the current state $s_t$:
\begin{align}
f(s) \approx f(s_t) + \nabla f(s_t)^{\T}(s - s_t), \label{eqn:first-order-approx}
\end{align}
as an approximation to the potential function. This linearization facilitates the construction of tractable proposal distributions, enabling more efficient sampling. Notable approaches include Norm Constrained Gradient (NCG) \citep{Zhang2012MRF, Rhodes2022GradientMC}, which is a direct analogue to continuous sampler MALA \citep{Besag1994mala, Roberts1996mala} and Auxiliary Variable Gradient (AVG) \citep{Rhodes2022GradientMC}, which is analogous to MALA via an auxiliary variable approach. Another recent approach is Discrete-HAMS (DHAMS) \citep{Zhou2025Dhams},
which introduces a momentum variable to form a Hamiltonian and mimics Hamiltonian Assisted Metropolis Sampling (HAMS) \citep{Song2023hams}.

The first-order approximation inherently assumes independence across dimensions, modeling the target distribution as a product of independent marginals guided by the gradient. This simplification overlooks interactions among different dimensions in the state $s$. A natural extension is to include second-order terms in the approximation. However, doing so results in a quadratic potential function, effectively yielding a pairwise Markov random field (MRF), from which direct sampling is generally intractable due to the dependency structure.

To overcome this challenge, \cite{martens2010gitml} exploited the Gaussian integral trick, which originates in statistical physics \citep{Hertz1991GIT}. This technique introduces Gaussian auxiliary variables to facilitate sampling from MRFs using a Gibbs sampling scheme, where interaction terms are decoupled during the discrete update step. Building on this idea, \cite{Rhodes2022GradientMC} proposed the Preconditioned Auxiliary Variable Gradient (PAVG) method. In PAVG, a Gaussian auxiliary variable is introduced following a second-order expansion of $f(s)$, and new proposals for $s$ are generated in a Metropolis-within-Gibbs scheme. The proposed state $s^*$ is then accepted or rejected. This method is further studied and generalized in \cite{sun2023anyscale} by including a scaling parameter applied to the potential.

In this work, we propose the Preconditioned Discrete-HAMS (PDHAMS) sampler by carefully incorporating the second-order approximation of $f(s)$ into discrete-HAMS
for preconditioning.
Remarkably, we show that with the second-order approximation of $f(s)$, the main steps underlying Discrete-HAMS can be effectively extended, including 
an auxiliary-variable proposal scheme, negation and gradient correction for updating the momentum
variable, and over-relaxation for updating the state variable.
We conduct several numerical experiments with discrete Gaussian, quadratic mixture, and clock Potts distributions, 
and find that PDHAMS algorithms consistently
yield superior results over existing first-order algorithms NCG, AVG, DHAMS and second-order algorithm PAVG, as measured by the total variation (TV) distances
over iterations (if feasible to compute) and effective sample size (ESS) estimated from multiple chains.

Our PDHAMS algorithm is a discrete sampler which, \textit{for the first time}, achieve two key properties simultaneously. 
First, PDHAMS satisfies a generalized detailed balance condition, where the reverse transition is related to the forward transition by negating the momentum.
Due to the momentum negation,
generalized detailed balance enables irreversible exploration of the target distribution, similarly to irreversible continuous samplers such as 
Underdamped Langevin Sampling \citep{Bussi2207UDL} and HAMS \citep{Song2023hams}.
Second, PDHAMS becomes rejection-free when the target distribution has a quadratic potential, hence representing a pairwise MRF. In contrast, DHAMS is an irreversible discrete sampler and is rejection-free when the target distribution has a linear potential. The PAVG algorithm is rejection-free when the target distribution has a quadratic potential
but achieves standard detailed balance (hence reversible sampling).

From the development of PDHAMS, we also identify several interesting connections for PAVG and PDHAMS.
First, we show that PAVG admits two distinct constructions of the auxiliary variable (mean or variance version), both of which lead to the same algorithm. Similarly, we present three distinct constructions (mean, variance, or momentum version) for the auxiliary variable in PDHAMS, which are also algorithmically equivalent.
The momentum version is convenient in analytical discussion, whereas the variance version is preferred in numerical implementation. 
Second, we demonstrate that PDHAMS reduces to PAVG in a specific boundary case, and PAVG, in turn, recovers the Gaussian integral trick when the target distribution is a pairwise MRF. Third, we also establish connections of PAVG and PDHAMS with the continuous samplers modified-MALA and HAMS respectively \citep{Song2023hams}.

\paragraph{Notation.}
For a (vector-valued) state variable $s$, we denote as $s_i$ the $i$-th coordinate of $s$.
When a sequence of draws is discussed, $s_{t}$ refers to the $t$-th draw, while $s_{t,i}$ denotes the $i$-th coordinate of $s_{t}$. 
For matrix operations, we use $\tr$ for the trace, $\vec$ for the vectorization of a matrix, and $\vech$ for the vectorization of the upper triangular part of a symmetric matrix. For a positive definite matrix $A$ with eigen-decomposition $A = U\Lambda U^{\T}$, the square root is defined as $A ^{1/2} = U\Lambda ^{1/2}U$.
We denote as $\Unif([a,b))$ a uniform distribution with support $[a,b)$. We use $\mathcal{N}(\mu, V)$ to represent a Gaussian distribution with mean $\mu$ and variance matrix $V$, and $\mathcal{N}(\cdot|\mu, V)$ to represent the corresponding density function. We use $\Softmax$ to denote
the probabilities in a discrete distribution with negative potential $g(\cdot)$ over the discrete set $\mathcal{A} = \{a_1, \ldots, a_K\}$ such that
\begin{align}
   \Softmax( g(a_k))  & = \frac{\exp(g(a_k))}{\sum\limits_{j=1}^{K} \exp(g(a_j))}, \quad k= 1, \ldots, K.\nonumber
\end{align}
where the dependency on $\mathcal{A}$ is suppressed.

\section{Related Methods}

\subsection{Gaussian Integral Trick}\label{sec:Gaussian_integral}

The Gaussian integral trick \citep{Hertz1991GIT}, also known as the Hubbard--Stratonovich transform \citep{Hubbard1959GIT}, was introduced to deal with quadratic 
potential functions in statistical physics. The method was used by \cite{martens2010gitml} and \cite{Zhang2012MRF} to facilitate sampling from a binary Markov random field with a quadratic potential,
\begin{align}
    \pi(s) \propto \exp(\frac{1}{2}s^{\T} W s + b^{\T} s), \nonumber
\end{align}
where $s$ is a vector of binary components with support $\mathcal{S} = \{0,1\}^{d}$ and $W$ is a symmetric matrix of dimension $d \times d$. The method can also be applied to a non-binary support where $\mathcal{S} = \{a_{1}, a_{2}, \cdots, a_{k}\}^{d}$.

The Gaussian integral trick introduces a continuous auxiliary variable $z$ which is conditionally Gaussian given $s$,
\begin{align}
    \pi(z|s) \sim \mathcal{N}(z;A(W+D)s, A(W+D)A^{\T}).
    \label{eqn:Gaussint2}
\end{align}
The additional matrix $D$ above is diagonal such that $W+D$ is positive definite, denoted as $(W+D) \succ 0$. Then the joint distribution and the conditional distribution of $s$ on $z$ can be shown to be
\begin{align}
    \pi(s, z) \propto \exp(-\frac{1}{2}z^{\T}(A^{-1})^{\T}(W+D)^{-1}A^{-1}z+z^{\T}A^{-1}s + b^{\T}s-\frac{1}{2}s^{\T}Ds), \nonumber
\end{align}
\begin{align}
    \pi(s|z) &  \propto \exp(z^{\T}A^{-1}s + b^{\T}s-\frac{1}{2}s^{\T}Ds) \nonumber \\
    & \propto \prod\limits_{i=1}^{d}\Softmax\left( - \frac{1}{2}d_{i}s_{i}^{2}+(((A^{-1})^{\T}z)_{i}+b_{i})s_{i}\right).     \label{eqn:Gaussint3}
\end{align}
As long as $D$ is diagonal, the conditional distribution $\pi(s| z)$ factorizes across dimensions. This property eliminates the need of joint updates in multiple dimensions and significantly improves the efficiency of the sampling algorithm. Leveraging the Gaussian Integral Trick, a straightforward Gibbs auxiliary sampler in \cite{martens2010gitml} can be implemented as follows.
\begin{itemize}
\item[(i)] Draw $z_{t}$ from $\pi(z|s_t)$ in \eqref{eqn:Gaussint2},

\item[(ii)] Draw $s_{t+1}$ from $\pi(s|z_t)$ in \eqref{eqn:Gaussint3}.
\end{itemize}
While different choices of $A$ are allowed in the preceding algorithm, 
the choice of $A = (W+D)^{-1/2}$ is recommended by \cite{Zhang2012MRF} such that $z$ given $s$ has an identity variance matrix and hence 
the components of $z$ are conditionally independent.

\subsection{Preconditioned Auxiliary Variable Gradient}\label{sec:PAVG}

For a discrete distribution whose potential function is not quadratic, the Gaussian integral trick can be combined with other sampling techniques
such as auxiliary variable or informed proposals \citep{Titsias2018auxavg,Zanella2020lbp}. For the first-order versions of these methods, with current state $s_{t}$,
the negative potential function (or log of the probability function) $f(s)$ is approximated using a first-order Taylor expansion at $s_t$ such that $f(s) \approx f(s_{t}) + \nabla f(s_{t})^{\T}(s-s_{t})$ \citep{Grathwohl2021gwg, Rhodes2022GradientMC, Zhang2022dmala}.
As an extension, second-order methods are proposed in \cite{Rhodes2022GradientMC} and \cite{sun2023anyscale} by using a second-order approximation,
\begin{align}
    f(s) \approx  \nabla f(s_{t})^{\T}(s-s_{t})+\frac{1}{2}(s-s_{t})^{\T}W(s-s_{t}).
    \label{eqn:precond_second_approx}
\end{align}
When $W = \nabla^{2}f(s_{t})$ is the Hessian matrix, \eqref{eqn:precond_second_approx} corresponds to the second-order Taylor expansion at $s_t$. However, querying the Hessian at every step of sampling is computationally expensive for complex or high-dimensional distributions. Thus, following \cite{Rhodes2022GradientMC}, we employ a global matrix $W$ for all states $s_t$. From the quadratic approximation \eqref{eqn:precond_second_approx}, the target distribution is approximated at $s_t$ as
\begin{equation}
\tilde{\pi}(s;s_{t}) \propto \exp(\nabla f(s_{t})^{\T}(s-s_{t})+\frac{1}{2}(s-s_{t})^{\T}W(s-s_{t})).
\label{eqn:PAVG_approx}
\end{equation}

In the following, we discuss the second-order method in \cite{Rhodes2022GradientMC}, called Preconditioned Auxiliary Variable Gradient (PAVG),
and present two equivalent versions using seemingly different auxiliary variable schemes. As in Section \ref{sec:Gaussian_integral}, let $D$ be a diagonal matrix such that $(W+D) \succ 0$ and
further decompose $W+D = LL^{\T}$ for some matrix $L$.

\textit{Preconditioning by mean.}\; The original verision of PAVG
in \cite{Rhodes2022GradientMC} introduces auxiliary variable $z$ which is conditionally Gaussian given $s$ such that
\begin{align}
    \pi(z|s) \sim \mathcal{N}(z; L^{\T}s, I).
    \label{eqn:PAVG_mean_z}
\end{align}
Together with the second-order approximation \eqref{eqn:PAVG_approx} to the target distribution $\pi(s)$, the conditional distribution of $s$ given $z$ can be approximated as
    \begin{align}
        \tilde{\pi}(s|z; s_{t}) & \propto \tilde{\pi}(s;s_{t})\pi(z|s) \nonumber \\
        & \propto \exp(\nabla f(s_{t})^{\T}(s-s_{t})+\frac{1}{2}(s-s_{t})^{\T}W(s-s_{t})) \exp(-\frac{1}{2}\|z- L^{\T}s\|_{2}^{2}) \nonumber \\
        & \propto \exp(-\frac{1}{2}s^{\T}Ds +(\nabla f(s_{t})-Ws_{t} +Lz)^{\T}s) \nonumber \\
        & \propto \prod\limits_{i=1}^{d}\Softmax(-\frac{1}{2}d_{i}s_{i}^{2} + [\nabla f(s_{t})_{i} -(Ws_{t})_{i} + (Lz)_{i}]s_{i}). \label{eqn:PAVG_mean_s}
    \end{align}
Based on the auxiliary variable technique in \cite{Titsias2018auxavg}, PAVG proceeds as follows with current state $s_t$.
\begin{itemize}
\item[(i)] Sample $z_{t} = L^{\T}s_{t}+Z$, where $Z \sim \mathcal{N}(0, I)$ is drawn independently of $s_t$.
\item[(ii)] Propose $s^{*} \sim Q(s|z_{t}; s_{t}) \propto \tilde{\pi}(s|z_{t}; s_{t})$ in \eqref{eqn:PAVG_mean_s}.
\item[(iii)] Accept $s_{t+1} = s^{*}$ with the following Metropolis-within-Gibbs probability,
\begin{equation}
\min\{1, \exp(f(s^{*})-f(s_{t}))\frac{\mathcal{N}(z_{t}| L^{\T}s^{*}, I)Q(s_{t}|z_{t};s^{*})}{\mathcal{N}(z_{t} | L^{\T}s_{t}, I)Q(s^{*}|z_{t};s_{t})}\},
\label{eqn:PAVG_mean_acc}
\end{equation}
or otherwise reject at $s_{t+1} = s_{t}$.
\end{itemize}
From the above description, the PAVG sampler can be seen to
first generate a proposal by applying the Gaussian integral trick with $A=L^{-1}$ to the approximate distribution \eqref{eqn:PAVG_approx}
and then perform acceptance or rejection, which is needed for $\pi(s)$ with a non-quadratic potential.
By the product structure of the proposal distribution in \eqref{eqn:PAVG_mean_s},
the components of $s^*$ can be drawn independently, hence amenable to parallel sampling.

\textit{Preconditioning by variance.}\;
The auxiliary variable $z$ in \eqref{eqn:PAVG_mean_z} is Gaussian with identity variance matrix but mean $L^\T s$, parameterized depending on $L$ and hence $(W,D)$.
Alternatively, by taking $A=(W+D)^{-1}$ in the Gaussian integral trick, we observe
that an auxiliary variable $z$ can also be constructed with mean just $s$
but variance matrix $(W+D)^{-1}$ depending on $(W,D)$,
\begin{equation}
    \pi(z|s) \sim \mathcal{N}(z; s, (W+D)^{-1}),
    \label{eqn:PAVG_var_z}
\end{equation}
This scheme seems to align with the usual concept of preconditioning to capture variance structures.
Together with the approximation \eqref{eqn:PAVG_approx} to the target distribution $\pi(s)$, the conditional distribution of $s$ on $z$ can be approximated as
    \begin{align}
        \tilde{\pi}(s|z; s_{t}) & \propto \tilde{\pi}(s;s_{t})\pi(z|s) \nonumber  \\
        & \propto \prod\limits_{i=1}^{d}\Softmax(-\frac{1}{2}d_{i}s_{i}^{2} + [\nabla f(s_{t})_{i} -(Ws_{t})_{i} + (W+D)z)_{i}]s_{i}).
        \label{eqn:PAVG_var_s}
    \end{align}
With current state $s_t$, this leads to the following sampler.
\begin{itemize}
\item[(i)] Sample $z_{t} = s_{t}+(L^{\T})^{-1}Z$, where $Z \sim \mathcal{N}(0, I)$ is drawn independently of $s_t$.
\item[(ii)] Propose $s^{*} \sim Q(s|z_{t}; s_{t})  \propto \tilde{\pi}(s|z_{t}; s_{t})$ in \eqref{eqn:PAVG_var_s}.
\item[(iii)] Accept $s_{t+1} = s^{*}$ with the following probability,
\begin{equation}
\min\{1, \exp(f(s^{*})-f(s_{t}))\frac{\mathcal{N}(z_{t}| s^{*}, (W+D)^{-1})Q(s_{t}|z_{t};s^{*})}{\mathcal{N}(z_{t} |s_{t}, (W+D)^{-1})Q(s^{*}|z_{t};s_{t})}\},
\label{eqn:PAVG_var_acc}
\end{equation}
or otherwise set $s_{t+1} = s_{t}$.
\end{itemize}

We make several important remarks. First, the two versions of PVAG, preconditioning by mean or by variance,
are equivalent to each other, in producing identical transitions from $s_t$ to $s_{t+1}$ (including acceptance-rejection).
Intuitively, this equivalence can be explained by the fact that $z_t$ in the mean version is $z_t$ in the variance version left-multiplied by $L^{\T}$.
See a formal proof in Supplement Section~\ref{prop_pavg_id}.
Second, the PAVG sampler does not depend on the choice of $L$, as shown by
\eqref{eqn:PAVG_var_z} and \eqref{eqn:PAVG_var_s} in the variance version, although not obvious from the mean version.
Third, PAVG is rejection-free when the negative potential $f(s)$ is quadratic,
and its second-order coefficient matrix is exactly $W$ in \eqref{eqn:PAVG_approx} (Supplement Section \ref{sec:PAVG_reject_free}).
In this special case, the PAVG algorithm reduces to the Gaussian integral trick in Section \ref{sec:Gaussian_integral}. 
Finally, if $s$ is a continuous variable, then $z_t$ can be integrated out and the resulting 
marginalized sampler reduces to modified MALA in \cite{Song2023hams} for certain choices of $(W,D)$. See Supplement Section ~\ref{sec:PAVG_cont}.
For conciseness in later sections, we always refer to the PAVG algorithm as derived from preconditioning by variance.

\section{Preconditioned Discrete-HAMS}\label{sec:PDHAMS}
\subsection{Auxiliary Variable Scheme}\label{sec:auxiliary_PHAMS}

As in the Discrete-HAMS sampler, we introduce a momentum variable $u \in \mathbb{R}^{d}$ to augment the target distribution for the Preconditioned version. This variable follows a standard normal distribution, $u \sim \mathcal{N}(0, I)$. Consequently, the augmented target probability function takes the Hamiltonian form of
\begin{equation}
    \pi(s, u) \propto \exp(f(s) -\frac{1}{2}\|u\|_{2}^{2}). \label{eqn:HAMS_hamiltonian-u}
\end{equation}
Similarly as in PVAG, we use the second-order approximation  \eqref{eqn:precond_second_approx} and \eqref{eqn:PAVG_approx} with current state $s_t$, so that the joint distribution $\pi(s, u)$ is approximated as
\begin{equation}
     \tilde{\pi}(s, u; s_{t}) \propto \exp(\nabla f(s_{t})^{\T}(s-s_{t})+\frac{1}{2}(s-s_{t})^{\T}W(s-s_{t}) -\frac{1}{2}\|u\|_{2}^{2}).
     \label{eqn:pHAMS_su}
\end{equation}

We consider three different ways of introducing an auxiliary variable $z$ as a linear combination of $s$ and $u$:
the first two ways resemble preconditioning by mean or by variance in PAVG, and the third way involves preconditioning on the momentum $u$.
For consistency, we use the same matrices $D$ and $L$ satisfying $W+D= LL^\T \succ 0$ as in Section~\ref{sec:PAVG}.

\textit{Preconditioning by mean.}\; In the first approach, we construct the auxiliary variable $z$ as
\begin{equation}
    z = L^{\T}s+u.
    \label{eqn:precondHAMS_mean_z}
\end{equation}
Together with the approximation \eqref{eqn:pHAMS_su} to the joint distribution $\pi(x,u)$, the conditional distribution of $(s, u)$ given $z$ can be approximated as
\begin{align}
        \tilde{\pi}(s, u|z;s_{t}) &\propto \tilde{\pi}(s, u; s_{t})\pi(z|s, u) \nonumber \\
        & \propto \exp(\nabla f(s_{t})^{\T}(s-s_{t})+\frac{1}{2}(s-s_{t})^{\T}W(s-s_{t}) -\frac{1}{2}\|u\|_{2}^{2}) \mathds{1}\{z = L^{\T} s + u\}.    \label{eqn:PHAMS_mean_cond_onz}
\end{align}
With current state and momentum $(s_t, u_t)$ and $z_t = L^{\T}s_t +u_t$, we generate a proposal $(s^*, u^*)$ by sampling from
the conditional distribution $\tilde{\pi}(s, u|z_t= L^{\T}s_t +u_t; s_{t}) $, defined as \eqref{eqn:PHAMS_mean_cond_onz} with $z=z_t$.
By setting $u = z_t - L^\T s$, the sampling can be implemented in two steps as follows.
\begin{itemize}
    \item[(i)] Propose \begin{align}
    s^{*} &\sim Q(s|z_t= L^{\T}s_t +u_t; s_{t}) \nonumber \\
    & \propto  \prod\limits_{i=1}^{d}\Softmax(-\frac{1}{2}d_{i}s_{i}^{2} + [\nabla f(s_{t})_{i} -(Ws_{t})_{i} + (Lz_{t})_{i}]s_{i}) \nonumber \\
    & \propto  \prod\limits_{i=1}^{d}\Softmax(-\frac{1}{2}d_{i}s_{i}^{2} + [\nabla f(s_{t})_{i} +(Ds_{t})_{i} +(Lu_{t})_{i}]s_{i}). \label{eqn:phams_mean_s}
\end{align}
\item [(ii)] Compute \begin{align}
u^{*} = z_{t}-L^{\T}s^{*}= u_{t} + L^{\T}(s_{t}-s^{*}). \label{eqn:phams_mean_u}
\end{align}
\end{itemize}
The construction of auxiliary variable $z$ ensures that the quadratic term associated with $W$ is canceled out in \eqref{eqn:phams_mean_s}, and the proposal distribution for state variable $s$ can also be factorized in a product form. We then either accept the proposal and set $(s_{t+1}, u_{t+1})=(s^*, u^*)$ with following Metropolis--Hastings probability
\begin{align}
\min\{1, \exp(f(s^{*})-\frac{1}{2}\|u^{*}\|_{2}^{2}-f(s_{t})+\frac{1}{2}\|u_{t}\|_{2}^{2})\frac{Q(s_{t}, u_{t}|s^{*}, u^{*})}{Q(s^{*}, u^{*}|s_{t}, u_{t})})\},
\label{eqn:phams_mean_acc}
\end{align}
or otherwise reject the proposal and set $(s_{t+1}, u_{t+1}) = (s_t, u_t)$. The forward proposal $Q(s^*, u^*|s_t, u_t)$ is evaluated as $Q(s^*|z_t =  L^{\T}s_t +u_t; s_t)$ and the backward proposal $Q(s_t, u_t|s^*, u^*)$ is evaluated as $Q(s_t|z_t=L^{\T}s^* +u^*; s^*)$.

\textit{Preconditioning by variance.}\; In the second approach for constructing auxiliary variable $z$, instead of applying a linear transformation to the state variable $s$, we perform the transformation on the momentum $u$ such that
\begin{equation}
z = s + (L^{\T})^{-1}u.
    \label{eqn:phams_var_z}
\end{equation}
Together with the approximation \eqref{eqn:pHAMS_su} to the joint distribution $\pi(x,u)$, the conditional distribution of $(s, u)$ given $z$ can be approximated as
    \begin{align}
        \tilde{\pi}(s, u|z;s_{t}) &\propto \tilde{\pi}(s, u; s_{t})\pi(z|s, u) \nonumber \\
        & \propto \exp(\nabla f(s_{t})^{\T}(s-s_{t})+\frac{1}{2}(s-s_{t})^{\T}W(s-s_{t}) -\frac{1}{2}\|u\|_{2}^{2}) \mathds{1}\{z = s + (L^{\T})^{-1}u\}. \label{eqn:phams_var_cond_onz}
    \end{align}
With current state and momentum $(s_t, u_t)$ and $z_t = s_t + (L^{\T})^{-1}u_t$, we generate a proposal $(s^*, u^*)$ by sampling from
$\tilde{\pi}(s, u|z_t= s_t + (L^{\T})^{-1}u_t; s_{t})$ in two steps as follows.
\begin{itemize}
\item[(i)] Propose
\begin{align}
    s^{*} &\sim Q(s|z_t= L^{\T}s_t +u_t; s_{t}) \nonumber \\
    & \propto  \prod\limits_{i=1}^{d}\Softmax(-\frac{1}{2}d_{i}s_{i}^{2} + [\nabla f(s_{t})_{i} -(Ws_{t})_{i} + ((W+D)z_{t})_{i}]s_{i}) \nonumber \\
    & \propto  \prod\limits_{i=1}^{d}\Softmax(-\frac{1}{2}d_{i}s_{i}^{2} + [\nabla f(s_{t})_{i} +(Ds_{t})_{i} + (Lu_{t})_{i}]s_{i}).\label{eqn:phams_var_s}
\end{align}
\item [(ii)] Compute
\begin{align}
u^{*} = L^{\T}(z_{t}-s^{*}) = u_{t} + L^{\T}(s_{t}-s^{*}).
\label{eqn:phams_var_u}
\end{align}
\end{itemize}
Then we either accept the proposal and set $(s_{t+1}, u_{t+1}) = (s^*, u^*)$ with Metropolis--Hastings probability as in \eqref{eqn:phams_mean_acc},
or reject the proposal and set $(s_{t+1}, u_{t+1}) = (s_t, u_t)$. It is easy to verify that the proposal of the state $s^*$ in the variance approach \eqref{eqn:phams_var_s} and that in the mean approach \eqref{eqn:phams_mean_s} are the same; so are the proposal of the momentum $u^*$ in \eqref{eqn:phams_var_u} and that in \eqref{eqn:phams_mean_u}. The pair $(s^*,u^*)$ is also accepted with the same probability as in \eqref{eqn:phams_mean_acc}.

\textit{Preconditioning on momentum.}\; In this approach, we first transform the momentum variable as $u = L^{\T} v$ and adjust our target distribution accordingly:
\begin{equation} \pi(s, v) \propto \exp\left( f(s) - \frac{1}{2} v^{\T} (W+D) v \right).  \label{eqn:phams_mom1}\end{equation}
Using the transformed momentum, we introduce the auxiliary variable $z$ as
$ z = s + v $ and approximate the conditional distribution of $(s, v)$ given $z$ as
\begin{align}
        \tilde{\pi}(s, v|z;s_{t}) &\propto \tilde{\pi}(s, v; s_{t})\pi(z|s, v) \nonumber \\
        & \propto \exp(\nabla f(s_{t})^{\T}(s-s_{t})+\frac{1}{2}(s-s_{t})^{\T}W(s-s_{t}) -\frac{1}{2}v^{\T}(W+D)v) \mathds{1}\{z = s + v\}. \label{eqn:phams_mom_condonz}
\end{align}
With current state and transformed momentum $(s_t, v_t)$ and $z_t = s_t + v_{t}$, we generate a proposal $(s^*, v^*)$
by sampling from $\tilde{\pi}(s, v|z_t= s_t +v_{t}; s_{t}) $ in two steps as follows.
\begin{itemize}
\item[(i)] Propose
\begin{align}
    s^{*} &\sim Q(s|z_t = s_{t}+ v_{t};s_t) \nonumber \\
    & \propto  \prod\limits_{i=1}^{d}\Softmax(-\frac{1}{2}d_{i}s_{i}^{2} + [\nabla f(s_{t})_{i} -(Ws_{t})_{i} + ((W+D)z_{t})_{i}]s_{i}) \nonumber \\
    & \propto  \prod\limits_{i=1}^{d}\Softmax(-\frac{1}{2}d_{i}s_{i}^{2} + [\nabla f(s_{t})_{i} +(Ds_{t})_{i} +((W+D)v_{t})_{i}]s_{i}). \label{eqn:phams_mom_s}
\end{align}
\item[(ii)] Compute
\begin{align}
    v^{*} = v_{t}+s_t-s^{*}.
     \label{eqn:phams_mom_u}
\end{align}
\end{itemize}
Then we either accept the proposal and set $(s_{t+1}, v_{t+1}) = (s^*, v^*)$ with following Metropolis--Hastings probability,
\begin{equation}
\min\{1, \exp(f(s^{*})-\frac{1}{2}v^{*\T}(W+D)v^{*}-f(s_{t})+\frac{1}{2}v_{t}^{\T}(W+D)v_{t})\frac{Q(s_{t}, v_{t}|s^{*}, v^{*})}{Q(s^{*}, v^{*}|s_{t}, v_{t})})\},
\label{eqn:phams_mom_acc}
\end{equation}
or reject the proposal and set $(s_{t+1}, v_{t+1}) = (s_t, v_t)$. The forward proposal $Q(s^*, v^*|s_t, v_t)$ is evaluated as $Q(s^*|z_t = s_t+ v_t; s_t)$ and the backward proposal $Q(s_t, v_t|s^*, v^*)$ is evaluated as $Q(s_t|z_t = s^*+ v^*; s^*)$.

We make several remarks on the auxiliary variable schemes. The transformed momentum $v$ can be defined in all three scheme as $v = L^\T u$, and hence $u_t$ and $v_t$ are related via $v_t = L^\T u_t$.
First, all three schemes are equivalent to each other, in producing
identical transitions of the state and momentum from $(s_t,u_t)$ to $(s_{t+1}, u_{t+1})$, provided that the same $L$ is used.
See Supplement Section \ref{prop_precondHAMS_id} for details. Second, as directly shown in the momentum approach \eqref{eqn:phams_mom1}--\eqref{eqn:phams_mom_acc},
the choice of $L$ does not affect the generation of the state and transformed momentum $(s_t,v_t)$, as long as $W+D = LL^{\T}$.
Hence all three schemes produce
identical transitions of the state and transformed momentum from $(s_t,v_t)$ to $(s_{t+1}, v_{t+1})$ even for different choices of $L$ for fixed $(W,D)$.
Finally, similarly to PAVG, the three auxiliary variable schemes are rejection-free when the negative potential $f(s)$ is quadratic, and its second-order coefficient matrix is exactly $W$ in \eqref{eqn:PAVG_approx} (Supplement Section \ref{sec:PHAMS_reject_free}).

\subsection{Vanilla Preconditioned Discrete-HAMS}\label{sec:Vanilla_PHAMS}

In Section~\ref{sec:auxiliary_PHAMS}, we establish the equivalence among the three approaches for constructing auxiliary variables. For convenience of exposition, in the subsequent formulation of Vanilla Preconditioned Discrete-HAMS, we focus on the momentum approach in \eqref{eqn:phams_mom1}--\eqref{eqn:phams_mom_acc}.
The variance approach, however, is preferred in numerical implementation; see Section \ref{sec:L_calibration}.
This section is organized into three parts, an auto-regression step to deal with non-ergodicity, a momentum negation step to introduce generalized reversibility and the last step for gradient correction on momentum.

\subsubsection{The Auto-Regression Step}\label{sec:auto-regression}
The auxiliary variable scheme in \eqref{eqn:phams_mom1}--\eqref{eqn:phams_mom_acc} suffers from a degeneracy issue.
By construction of \eqref{eqn:phams_var_u}, the constraint $s^*+v^* = s_t+v_t$ is satisfied, and hence
$s_{t+1} + v_{t+1} =s_t + v_t $ always holds no matter whether the proposal $(s^*,v^*)$ is accepted or not.
In other words, $s_t + v_t$ remains constant throughout the sampling process. Since $s_t$ only takes a finite number of possible values on the lattice $\mathcal{S}$,
it follows that $v_t$ is also restricted to a finite set of possible values, depending on the initial value $v_0$. This behavior
indicates that the sequence of $v_t$ fails to be ergodic with respect to a Gaussian distribution.
Following Discrete-HAMS,
we introduce an auto-regression step by incorporating an intermediate momentum $v_{t+1/2}$ before proposing the next state and momentum, which is defined as
\begin{equation}
v_{t+1/2} = \epsilon v_{t} + \sqrt{1 - \epsilon^{2}}\, (L^{\T})^{-1} Z, \quad Z \sim \mathcal{N}(0, I).
\label{eqn:auto_regression1}
\end{equation}
The auto-regression step \eqref{eqn:auto_regression1} is always accepted, being reversible to the momentum distribution $v \sim \mathcal{N}(0, (W+D)^{-1})$ as well as the augmented target distribution $\pi(s,v)$. After generating $v_{t+1/2}$, we proceed with the auxiliary variable scheme described in \eqref{eqn:phams_mom_s}--\eqref{eqn:phams_mom_u} to propose $(s^{*}, v^{*})$ from the current state and momentum $(s_t, v_{t+1/2})$ such that
\begin{align}
 s^* & \sim Q( s | z_t = s_t+ v_{t+1/2} ;s_t),  \nonumber \\
        v^{*} &= v_{t+1/2} +s_{t} -s^{*}.  \nonumber
\end{align}
Then we either accept $(s_{t+1}, v_{t+1}) = (s^{*}, v^{*})$ with the Metropolis--Hastings
probability
\begin{equation}
\min\{1, \frac{\pi(s^{*}, v^{*}) Q(s_{t}, v_{t+1/2}|s^{*}, v^{*} )}{\pi(s_{t}, v_{t+1/2}) Q(s^{*}, v^{*}|s_{t}, v_{t+1/2} )}\},
\label{eqn:auto-regression-acc}
\end{equation}
or reject the proposal and set $(s_{t+1}, v_{t+1}) =(s_{t}, v_{t+1/2})$,
where the forward proposal probability $Q (s^*,v^*|s_t, v_{t+1/2} )$ is evaluated as $Q( s^* |z_t= s_t+ v_{t+1/2};s_t) $; and backward proposal $Q(s_{t}, v_{t+1/2}|s^*, v^*)$ is evaluated as $Q(s_{t}|z_t = s^*+  v^*;s^*)$.

\subsubsection{Momentum Negation}\label{sec:negation}
To introduce generalized reversibility as in Discrete-HAMS, we modify the proposal after auto-regression in \eqref{eqn:auto_regression1}--\eqref{eqn:auto-regression-acc} by further reversing the intermediate momentum $v_{t+1/2}$ in \eqref{eqn:auto_regression1} to $-v_{t+1/2}$. This change leads to the following proposal with momentum negation.
\begin{itemize}
\item[(i)] Propose
\begin{align}
    s^{*} &\sim Q(s|z_t = s_{t} -v_{t+1/2};s_t) \nonumber \\
    & \propto  \prod\limits_{i=1}^{d}\Softmax(-\frac{1}{2}d_{i}s_{i}^{2} + [\nabla f(s_{t})_{i} -(Ws_{t})_{i} + ((W+D)s_{t}-(W+D)v_{t+1/2})_{i}]s_{i}). \label{eqn:phams_negation-s}
\end{align}
\item[(ii)] Compute
\begin{align}
    v^{*} = -v_{t+1/2}+s_t-s^{*}.
     \label{eqn:phams_negation-v}
\end{align}
\end{itemize}
To account for momentum negation, we either accept the proposal and set $(s_{t+1}, v_{t+1}) =(s^*, v^*)$ with a generalized Metropolis--Hastings probability,
\begin{equation}
\min\{1, \frac{\pi(s^{*}, -v^{*}) Q(s_{t}, -v_{t+1/2}|s^{*}, -v^{*} )}{\pi(s_{t}, v_{t+1/2}) Q(s^{*}, v^{*}|s_{t}, v_{t+1/2} )}\}, \nonumber
\end{equation}
or rejected the proposal and set $(s_{t+1}, v_{t+1}) = (s_t, -v_{t+1/2})$. The forward proposal $Q(s^*, v^*|s_t, v_{t+1/2})$ is evaluated as $Q(s^*|z_t=s_t-v_{t+1/2}; s_t)$ and the backward proposal $Q(s_t, -v_{t+1/2}|s^*, -v^*)$ is evaluated as $Q(s_t|z_t =s^*+v^*; s^*)$.

\subsubsection{Gradient Correction on Momentum}\label{sec:grad_correction}

A potential limitation of the negation proposal \eqref{eqn:phams_negation-s}--\eqref{eqn:phams_negation-v} is that, the gradient at the new state $\nabla f(s^{*})$ is ignored when updating the new momentum $v^{*}$. Such gradient information from the new proposal state $s^*$ (not the current state $s_t$) has been introduced for Hamiltonian-based continuous samplers, such as HMC \citep{duane1987HMC} through the leap-frog scheme
and HAMS \citep{Song2023hams} through a direct gradient correction, also used in Discrete-HAMS for discrete samplers. Motivated by this consideration, we keep the update of $s^*$ \eqref{eqn:phams_negation-s} and modify \eqref{eqn:phams_negation-v} as follows,
\begin{align}
        v^{*} &= -v_{t+1/2}+s_{t}-s^{*} + \phi (\nabla f(s^{*})-\nabla f(s_{t})+W(s_t-s^*)),
    \label{eqn:p-mod1}
\end{align}
where $\phi \geq 0$ is a tuning parameter. The update \eqref{eqn:p-mod1} can be rearranged to
\begin{align}
        -v_{t+1/2} &= v^{*} +s^{*}-s_{t} + \phi (\nabla f(s^{*})-\nabla f(s_{t})+W(s_t-s^*)).
    \label{eqn:p-mod1-rev}
\end{align}
Importantly, the $v$-updates $(s_t, v_{t+1/2}, s^*) \mapsto v^*$ in \eqref{eqn:p-mod1} and $(s^*, -v^*, s_t) \mapsto - v_{t+1/2}$ in \eqref{eqn:p-mod1-rev} satisfy the \textit{same} mapping deterministically. To distinguish the joint proposal of $(s,v)$ using \eqref{eqn:phams_negation-s} and \eqref{eqn:p-mod1} from that using \eqref{eqn:phams_negation-s} and \eqref{eqn:phams_negation-v}, we denote the new proposal with gradient correction as $Q_{\phi}(s^{*}, v^{*}|s_{t}, v_{t+1/2})$.
Then we either accept $(s_{t+1},v_{u+1}) = (s^*,v^*)$ with the following probability
\begin{equation}
   \min\{1, \frac{\pi(s^{*}, -v^{*}) Q_{\phi}(s_{t}, -v_{t+1/2}|s^{*}, -v^{*})}{\pi(s_{t}, v_{t+1/2}) Q_{\phi} (s^{*}, v^{*}|s_{t}, v_{t+1/2})}\},
    \label{eqn:p-mod-acc}
\end{equation}
or reject the proposal and set $(s_{t+1}, v_{t+1}) = (s_t, -v_{t+1/2})$. The forward transition probability $Q_{\phi}(s^{*}, v^{*}|s_{t}, v_{t+1/2})$ is evaluated as $Q(s^*|z_t = s_t -v_{t+1/2};s_t)$ and the backward transition $Q_{\phi}(s_t, v_{t+1/2}|s^{*}, v^*)$ is evaluated as $Q(s_t|z_t=s^*+ v^*;s^*)$. 
The resulting transition can be shown to satisfy generalized detailed balance, similarly as in \cite{Zhou2025Dhams} and \cite{Song2023hams}.

\begin{proposition}\label{prop:1}
Let $K_{\phi}( s_{t+1}, v_{t+1} |s_{t}, v_{t+1/2})$ be the transition kernel defined by proposal \eqref{eqn:phams_negation-s} and \eqref{eqn:p-mod1} and acceptance probability \eqref{eqn:p-mod-acc}. Then the generalized detailed balance \eqref{eqn:PHAMS-gdbc} holds between $(s_{t}, v_{t+1/2})$ and $(s_{t+1}, v_{t+1})$ such that
\begin{equation}
    \pi(s_{t}, v_{t+1/2})K_{\phi}(s_{t+1}, v_{t+1}|s_{t}, v_{t+1/2}) = \pi(s_{t+1}, -v_{t+1})K_{\phi}(s_{t}, -v_{t+1/2}|s_{t+1}, -v_{t+1}).
    \label{eqn:PHAMS-gdbc}
\end{equation}
Additionally, the augmented target distribution $\pi(s, v)$ is a stationary distribution of the defined Markov chain.
\end{proposition}

We summarize all the operations from Sections \ref{sec:auxiliary_PHAMS}--\ref{sec:Vanilla_PHAMS} in Algorithm~\ref{algo:Vanilla-PHAMS}, referred to as Vanilla Preconditioned Discrete-HAMS. Additionally, we make two important remarks. First, Vanilla Preconditioned Discrete-HAMS reduces to PAVG when $\epsilon=0$ and $\phi=0$, hence recovering the latter as a boundary case (Supplement Section \ref{sec:pham_to_pavg}). Second, similarly to PAVG, Vanilla Preconditioned Discrete-HAMS would preserve the rejection-free property
when the negative potential $f(s)$ is quadratic, and its second-order coefficient matrix is exactly $W$ in \eqref{eqn:PAVG_approx}
(Supplement Section \ref{sec:PHAMS_reject_free}). Finally, when $s$ is continuous, Vanilla Preconditioned Discrete-HAMS reduces to the continuous HAMS sampler \citep{Song2023hams} for certain choices of $(W,D)$ (Supplement Section ~\ref{sec:V-PDHAMS_cont}). 

\begin{algorithm}[tbp]
With current state and momentum $(s_{t}, v_{t})$
\begin{itemize}
\item Generate $v_{t+1/2} = \epsilon v_{t} + \sqrt{1 - \epsilon^{2}}\, (L^{\T})^{-1} Z, \quad Z \sim \mathcal{N}(0, I)$.
\item Generate $z_{t} = s_{t} - v_{t+1/2}$.
\item Propose $s^*$ by \eqref{eqn:phams_negation-s}, and compute $v^{*}$ by \eqref{eqn:p-mod1}.
\item Accept $(s_{t+1}, v_{t+1}) = (s^{*}, v^{*})$ with probability \eqref{eqn:p-mod-acc}, or otherwise set $(s_{t+1}, v_{t+1}) = (s_{t}, -v_{t+1/2})$.
\end{itemize}
\caption{Vanilla Preconditioned Discrete-HAMS (V-PDHAMS)}
\label{algo:Vanilla-PHAMS}
\end{algorithm}

\subsection{Over-relaxed Preconditioned Discrete-HAMS}

All the operations in Section~\ref{sec:Vanilla_PHAMS}, auto-regression, negation, and gradient correction, are
concerned about updating the momentum variable.
In this section, we incorporate the discrete over-relaxation technique of \cite{Zhou2025Dhams} for updating
the state variable, as motivated by the use of Gaussian over-relaxation \citep{Adler1981overrelaxation, Neal1998overrelax} in the continuous HAMS sampler.
We refer to \cite{Zhou2025Dhams} for details in the development of the discrete over-relaxation technique.

In Vanilla Preconditioned Discrete-HAMS, the proposal distribution \eqref{eqn:phams_negation-s} is in a product form, which allows us to apply coordinate-wise the discrete over-relaxation technique of \cite{Zhou2025Dhams}, which is summarized in Algorithm~\ref{algo:over-relaxation}. In the over-relaxed algorithm, we draw a proposal $s^*$ component by component with the reference distribution $Q (s |z_t = s_t+ v_{t+1/2};s_t)$ in \eqref{eqn:phams_negation-s},
i.e., apply Algorithm \ref{algo:over-relaxation} with
\begin{align}
 \begin{split}
  x_0 & = \text{$i$th component of $s_t$},  \\
  p(x) &= \text{$i$th component in $Q (s |z_t= s_t+ v_{t+1/2};s_t)$}, \\
  x_1 &= \text{$i$th component of $s^*$}
\end{split}
\label{eqn:over-relaxation}
\end{align}
The resulting proposal probability with over-relaxation, denoted as $\tilde Q(s |z_t =s_t-v_{t+1/2};s_t)$, is also a product of individual components, where each component distribution is determined as the conditional distribution $ p(x_1|x_0)$ in Algorithm \ref{algo:over-relaxation}.
\begin{algorithm}[tbp]
 \begin{itemize}
       \item Given a discrete variable $x_0$ and a discrete reference distribution $p(\cdot)$ with CDF $F(\cdot)$.
       \item Sample  $w_{0} \sim Unif([F(x_{0}^{-}), F(x_{0})))$ and $\tilde{w} \sim Unif([0,1))$ independently.
       \item Compute $w_{1} = ( -w_{0}+ \beta \tilde{w} )\%1$.
       \item Output $x_1$ such that $F(x_1^- ) \leq w_1 < F(x_1)$.
\end{itemize}
\caption{Discrete Over-relaxation}
\label{algo:over-relaxation}
\end{algorithm}

The rest of the over-relaxed algorithm is similar as in the vanilla algorithm. We still generate the momentum proposal $u^*$ by \eqref{eqn:p-mod1}
with gradient correction. Denote the joint proposal of $(s,v)$ as $\tilde{Q}_\phi (s^{*}, v^{*}|s_{t}, v_{t+1/2})$.
Then we either accept $(s_{t+1},u_{u+1}) = (s^*,u^*)$ with the following probability
\begin{equation}
 \min\{1, \frac{\pi(s^{*}, -v^{*}) \tilde Q_{\phi} (s_{t}, -v_{t+1/2}|s^{*}, -v^{*})}{\pi(s_{t}, v_{t+1/2}) \tilde Q_{\phi} (s^{*}, v^{*}|s_{t}, v_{t+1/2})}\},
    \label{eqn:over-phams-acc}
\end{equation}
or reject the proposal and set $(s_{t+1},v_{t+1}) = (s_t, -v_{t+1/2})$,
where the forward transition probability $\tilde Q_{\phi}(s^{*}, v^{*}|s_{t}, v_{t+1/2})$ can be evaluated as $\tilde Q( s^* | z_t = s_t- v_{t+1/2};s_t) $,
and backward transition probability $\tilde Q_{\phi}(s_{t}, -v_{t+1/2}|s^{*}, -v^{*})$ can be evaluated as $\tilde Q (s_t |z_t = s^* + v^*; s^*)$. 
The resulting transition can be shown to satisfy generalized detailed balance by a similar proof as in \cite{Zhou2025Dhams} and \cite{Song2023hams}.

\begin{proposition}\label{prop:2-new}
Let $\tilde K_\phi (s_{t+1}, v_{t+1}|s_{t}, v_{t+1/2})$ be the over-relaxed transition kernel from $(s_t,  v_{t+1/2})$ to $(s_{t+1},v_{t+1})$ defined by
the proposal scheme \eqref{eqn:over-relaxation} and \eqref{eqn:p-mod1} and acceptance probability \eqref{eqn:over-phams-acc}.
Then the following generalized detailed balance holds:
\begin{align*}
    \pi(s_{t}, v_{t+1/2}) \tilde K_\phi (s_{t+1}, v_{t+1}|s_{t}, v_{t+1/2}) = \pi(s_{t+1}, -v_{t+1}) \tilde K_\phi (s_{t}, -v_{t+1/2}|s_{t+1}, -v_{t+1}).
\end{align*}
Moreover, the augmented target distribution $\pi(s, v)$ is a stationary distribution of the resulting Markov chain.
\end{proposition}

We summarize, in Algorithm \ref{algo:overrelaxePDHAMS}, the Over-relaxed Preconditioned Discrete-HAMS method by incorporating over-relaxation as well as the operations from Section~\ref{sec:Vanilla_PHAMS}. We make two remarks about this new sampler. First,  
when $\beta = \pm1$ is used in Algorithm \ref{algo:over-relaxation}, $s^*$ is drawn independently from the reference distribution as noted in \citet{Zhou2025Dhams}, and then Over-relaxed Preconditioned Discrete-HAMS (Algorithm \ref{algo:overrelaxePDHAMS}) reduces to the Vanilla Preconditioned Discrete-HAMS (Algorithm \ref{algo:Vanilla-PHAMS}).
Second, the Over-relaxed Preconditioned Discrete-HAMS can be shown to maintain the rejection-free property when the negative potential $f(s)$ is quadratic, and its second-order coefficient matrix is exactly $W$ in \eqref{eqn:PAVG_approx},
provided that the generalized acceptance probability \eqref{eqn:over-phams-acc} is used. See Supplement Section~\ref{sec:O-PHAMS_reject_free} for details.

In the subsequent sections, V-PDHAMS refers to Vanilla Preconditioned Discrete-HAMS (Algorithm \ref{algo:Vanilla-PHAMS}) and O-PDHAMS refers to Over-relaxed Preconditioned Discrete-HAMS (Algorithm \ref{algo:overrelaxePDHAMS}).
\begin{algorithm}[tbp]
With current state and momentum $(s_{t}, v_{t})$
\begin{itemize}
\item Generate $v_{t+1/2}$ according to \eqref{eqn:auto_regression1}.
\item  Calculate the original proposal distribution $Q(s|s_{t},  v_{t+1/2})$ from \eqref{eqn:phams_negation-s} as the reference distribution, and \\
Propose $s^*$  from the over-relaxed proposal probability $\tilde{Q}(s|s_{t} ,v_{t+1/2})$ via component-wise application of Algorithm \ref{algo:over-relaxation}.
\item Compute $v^{*}$ by \eqref{eqn:p-mod1}.
\item Accept $(s_{t+1}, v_{t+1}) = (s^{*}, v^{*})$ with probability \eqref{eqn:over-phams-acc}, or otherwise set $(s_{t+1}, v_{t+1}) =(s_{t}, -v_{t+1/2})$.
\end{itemize}
\caption{Over-relaxed Preconditioned Discrete-HAMS (O-PDHAMS)}
\label{algo:overrelaxePDHAMS}
\end{algorithm}

\section{Parameter Calibration}

In both algorithms PAVG (Section~\ref{sec:PAVG}) and Preconditioned Discrete-HAMS (Section~\ref{sec:PDHAMS}), we need to consider the choice of matrices $W$, $D$ and $L$ to realize efficient sampling. In this section, we discuss the calibration process in detail.

\subsection{Calibration of \textit{W}} \label{sec:W_calibration}
 A natural choice for $W$ arises when the target distribution has a quadratic potential. In this case, selecting $W$ as the matrix associated with the second-order term leads to a rejection-free algorithm. For non-quadratic potentials, however, choosing a suitable $W$ becomes more challenging. The intuition behind \eqref{eqn:precond_second_approx} is to employ a global matrix $W$ that approximates the interaction between different components of the state without exactly querying the Hessian. This idea is reminiscent of the empirical average Hessians used in Quasi-Newton methods for optimization \citep{quasi-newton}, which motivates us to estimate $W$ from samples collected during a burn-in phase using other non-preconditioned samplers, or using a preconditioned sampler initialized with a rough guess of $W$. However, unlike in Quasi-Newton algorithms such as BFGS \citep{goldfarb1970family}, the matrix $W$ used here is not required to be negative semi-definite.

Suppose we collect a sample with $T$ draws $\{s_t\}_{t=1}^T$ during a burn-in phase. \cite{sun2023anyscale} suggests that the matrix $W$ can be estimated by minimizing discrepancies in the first-order derivatives as follows:
\begin{align}
    \hat{W} &= \argmin\limits_{W =W^{\T}} \sum\limits_{t=1}^{T} \|\nabla f(s_{t+1}) -\nabla f(s_{t})- W(s_{t+1}-s_{t})\|_{2}^{2}.
    \label{eqn:W_calibration1}
\end{align}
Alternatively, $W$ can be estimated by minimizing discrepancies in the function values $f(s)$ by using second-order expansions, following \cite{Rhodes2022GradientMC} for PAVG:
\begin{align}
    \hat{W} &= \argmin\limits_{W =W^{\T}} \sum\limits_{t=1}^{T} [f(s_{t+1})-f(s_{t}) -\nabla f(s_{t})^{\T}(s_{t+1}-s_{t}) -\frac{1}{2}(s_{t+1}-s_{t})^{\T}W(s_{t+1}-s_{t}) ]^{2}.
    \label{eqn:W_calibration2}
\end{align}
Details of these two optimization problems are discussed in Supplement Section~\ref{sec:W_estimate1} and ~\ref{sec:W_estimate2}.

We make several remarks. First, when $f(s)$ is quadratic in $s$,
calibration method~\eqref{eqn:W_calibration1} or~\eqref{eqn:W_calibration2} can be verified to recover the exact second-order coefficient matrix in $f(s)$.
Hence PAVG or Preconditioned Discrete-HAMS using the above calibration of $W$ is rejection-free for a quadratic potential,
whereas AVG or Discrete-HAMS without preconditioning is rejection-free for a linear potential.
Second, in Supplement Section~\ref{sec:scaling} , we show that
when using estimated $W$ from calibration method~\eqref{eqn:W_calibration1} or~\eqref{eqn:W_calibration2},
PAVG or Preconditioned Discrete-HAMS is invariant to rescaling of the state variable $s$. 
Finally, a comparison of the two calibration method~\eqref{eqn:W_calibration1} or~\eqref{eqn:W_calibration2} in terms of which method should be used 
is provided in Supplement Section ~\ref{sec:W_choice}.

\subsection{Calibration of \textit{D}}\label{sec:D_calibration}
After choosing $W$, we need to determine a suitable $D$ to ensure positive definite property $W+D \succ 0$. Given that $W$ is symmetric, this choice depends on the smallest eigenvalue of $W$, denoted as $\lambda_{\min}(W)$.
Following \cite{martens2010gitml}, we employ a $\lambda$-shift strategy\zt{:} setting $D = \lambda I$ with
\begin{align}
\lambda = \delta + \min\{0, \lambda_{\min}(W)\}. \nonumber
\end{align}
From \eqref{eqn:PAVG_mean_s} and \eqref{eqn:phams_mom_s}, we observe that the diagonal matrix $D$ mimics an inverse variance matrix for the proposal distribution of state variable $s$, and a larger value $\delta$ leads to smaller ''variance''. Therefore $1/\delta$ acts as a stepsize parameter controlling the dispersion of the proposal distribution and the magnitude of changes between consecutive draws.

\subsection{Calibration of \textit{L}}\label{sec:L_calibration}

Although PAVG or Preconditioned Discrete-HAMS (concerning the state variable) does not depend on the choice of $L$, selecting an appropriate $L$ can
reduce computational cost and improve sampling process. Given $(W,D)$, the only constraint on $L$ is $W+D = LL^\T$, hence allowing for various choices of $L$. A common choice is the square root $L = (W+D)^{1/2}$ \citep{sun2023anyscale}. We propose an alternative approach for choosing $L$ based on the condition number of $W+D$.

When the condition number of $W+D$ is relatively low (below 100 in our experiments), we choose $L$ as the lower-triangular matrix obtained from the Cholesky decomposition, $W+D = L L^\T$.
On the other hand, if the condition number is high, we set $L = U\Lambda^{1/2}$ from the eigen decomposition, $W+D = U \Lambda U^{\T}$ for some orthonormal matrix $U$ and a diagonal matrix $\Lambda$. The corresponding matrix $L$ from Cholesky decomposition is much sparser, which accelerates subsequent computations in the sampling process for both PAVG and Preconditioned Discrete-HAMS if implemented in the variance approach. However, when $W+D$ is ill-conditioned, Cholesky decomposition becomes less robust. In such cases, we set $L = U\Lambda^{1/2}$ to ensure better numerical stability \citep{golub2013matrix}.

\section{Numerical Studies}

We present simulation studies comparing various gradient-based discrete samplers, including 
both first-order methods NCG, AVG, V-DHAMS and O-DHAMS (which are vanilla and over-relaxed version of DHAMS) and second-order methods PAVG, V-PDHAMS and O-PDHAMS as well as a Metropolis sampler (with the implementation details as described in \cite{Zhou2025Dhams}, Supplement Section C). The tuning procedures for second-order methods are detailed in Supplement Section~\ref{sec:precond_tuning}.

To compare the performance of different samplers, we use the following metrics: total variation distance (TV-distance) and effective sample size (ESS).
The TV-distance is an important metric for evaluating convergence of Markov chains \citep{liu2001montetv}.
For each chain, we compute the TV-distance between the empirical distribution $\hat{\pi}$ and the target distribution $\pi$,
which for discrete distributions with support $\mathcal{S}$ is defined as
\begin{align}
    \TV(\pi, \hat{\pi}) = \frac{1}{2} \sum\limits_{s \in \mathcal{S}} |\pi(s) - \hat{\pi}(s)|. \nonumber
\end{align}
Similarly as in \cite{ma2019tv} and \cite{Jiang2020tv}, we report the mean and standard deviation of the TV-distances across multiple chains from repeated independent runs.

The ESS estimator we use is the \textit{multivariate batch mean ESS estimator} from \cite{vats2019ESS}, which incorporates within-chain and between-chain variances from multiple independent chains. Suppose that we are interested in a variable $x$ and have $m$ chains, each of length $T$. Let $x_{i,t}$ be the variable of interest from the $t$-th draw of the $i$-th chain. The ESS is computed by:
\begin{equation}
    \ESS (x)= T\frac{W}{B}, \; W = \frac{1}{m(T-1)}\sum\limits_{i,t} (x_{i,t}-\bar{x}_{i,\cdot})^{2},\; B = \frac{T}{m-1}\sum\limits_{i}(\bar{x}_{i,\cdot}-\bar{x})^{2},
    \nonumber
\end{equation}
where $\bar{x}_{i,\cdot} = \frac{1}{T}\sum\limits_{t}x_{i,t}$ is the mean of the $i$-th chain and $\bar{x} = \frac{1}{m}\sum\limits_{i} \bar{x}_{i,\cdot}$ is the overall mean across all chains.

\subsection{Discrete Gaussian}\label{sec:preconditioned_exp_gauss}

\begin{figure}[tbp]
    \centering
    \includegraphics[width=0.5\linewidth]{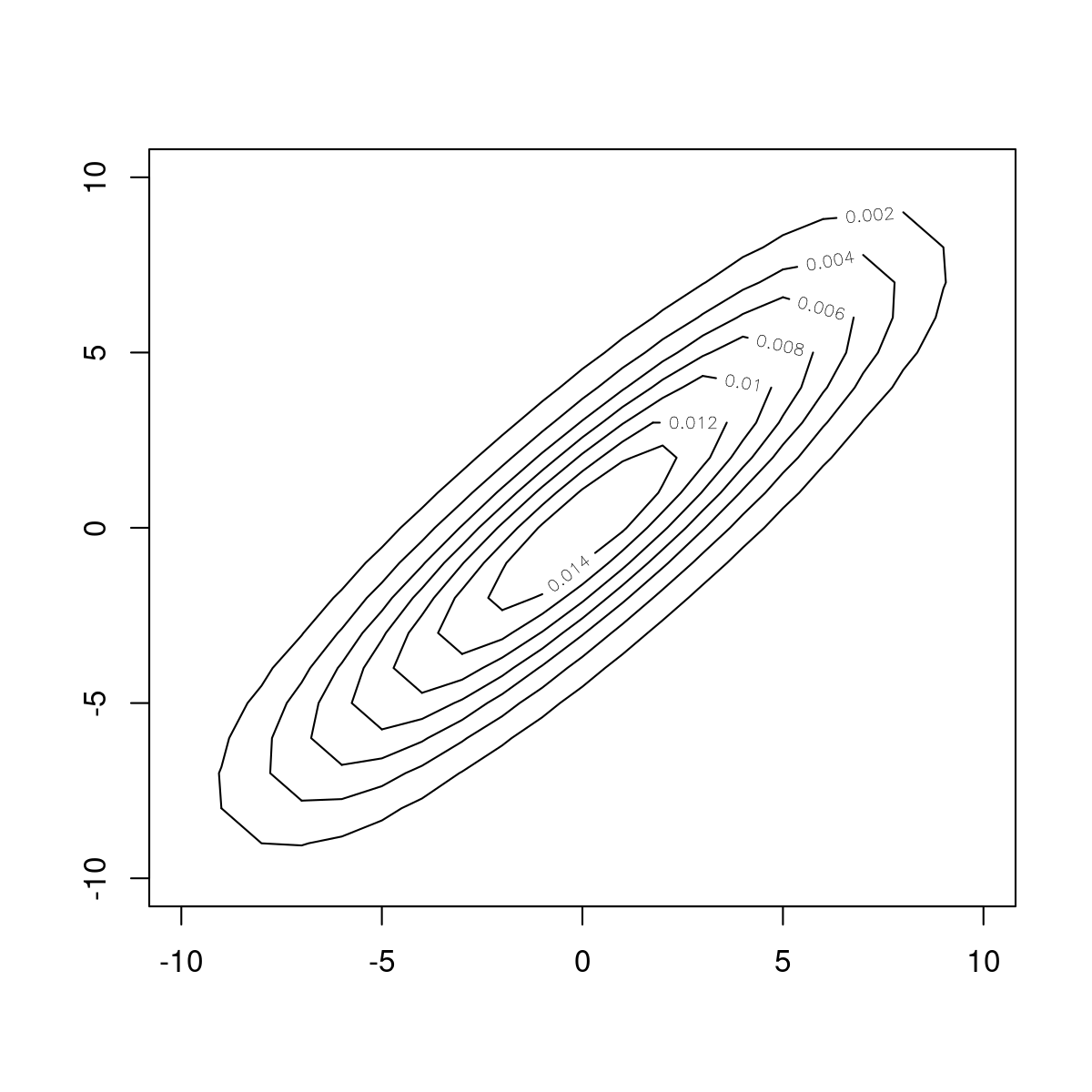} 
    \caption{Contour plot of discrete Gaussian distribution}
    \label{fig:heatmap} 
\end{figure}

We use the same discrete Gaussian distribution as in \cite{Zhou2025Dhams}. The distribution is restricted to a $d$-dimensional lattice $\mathcal{S} \in \mathbb{R}^{d}$, $\mathcal{S} = \{-k, -(k-1), \cdots, -1,0, 1, \cdots, (k-1), k\}^{d}$, and the negative potential function $f(s)$ is of a quadratic form
\begin{align}
   f(s) &= -\frac{1}{2}s^{\T}\Sigma^{-1}s, \nonumber \\
   \Sigma &= \sigma^{2}[\rho\textbf{1}\textbf{1}^{\T}+(1-\rho)I].
    \label{eq:disgau_f}
\end{align}
where $\Sigma$ is a positive definite and has an equi-correlation structure. In our example, $d=8$, $k=10$, $\sigma=5$ and $\rho = 0.9$. The marginal distribution of the first two coordinates is visualized using a contour plot in Figure \ref{fig:heatmap}. Given the equi-correlation structure of $\Sigma$ in \eqref{eq:disgau_f}, the marginal distribution of any two coordinates is also visualized by the same contour, and the joint distribution is invariant under permutations
of the indices. We choose $W = -\Sigma$ and the quadratic form of $f(s)$ in \eqref{eq:disgau_f} ensures that all preconditioned samplers studied are rejection-free. For our discrete Gaussian distribution, the exact distributions of up to any four-dimensional marginals can be calculated by direct enumeration, $\pi(s_{j_{1}}, s_{j_{2}}, s_{j_{3}}, s_{j_{4}})$, at manageable computational costs.

For parameter tuning with each sampler, we search over 50 different parameters. For each parameter, we run 50 independent chains, each of length 5,500 and burn-in 500.
As in \cite{sun2023discs}, we select the best parameter by the highest ESS of the negative potential function $f(s)$. After completing parameter tuning, we conduct 100 independent chains for each sampler using the optimal parameter setting. For each chain, the initial 500 draws were discarded as burn-in, and the subsequent 15,000 draws were retained for analysis.

\begin{figure}[tbp]
    \centering
    \begin{subfigure}{0.45\textwidth}
        \centering
        \includegraphics[width=\linewidth]{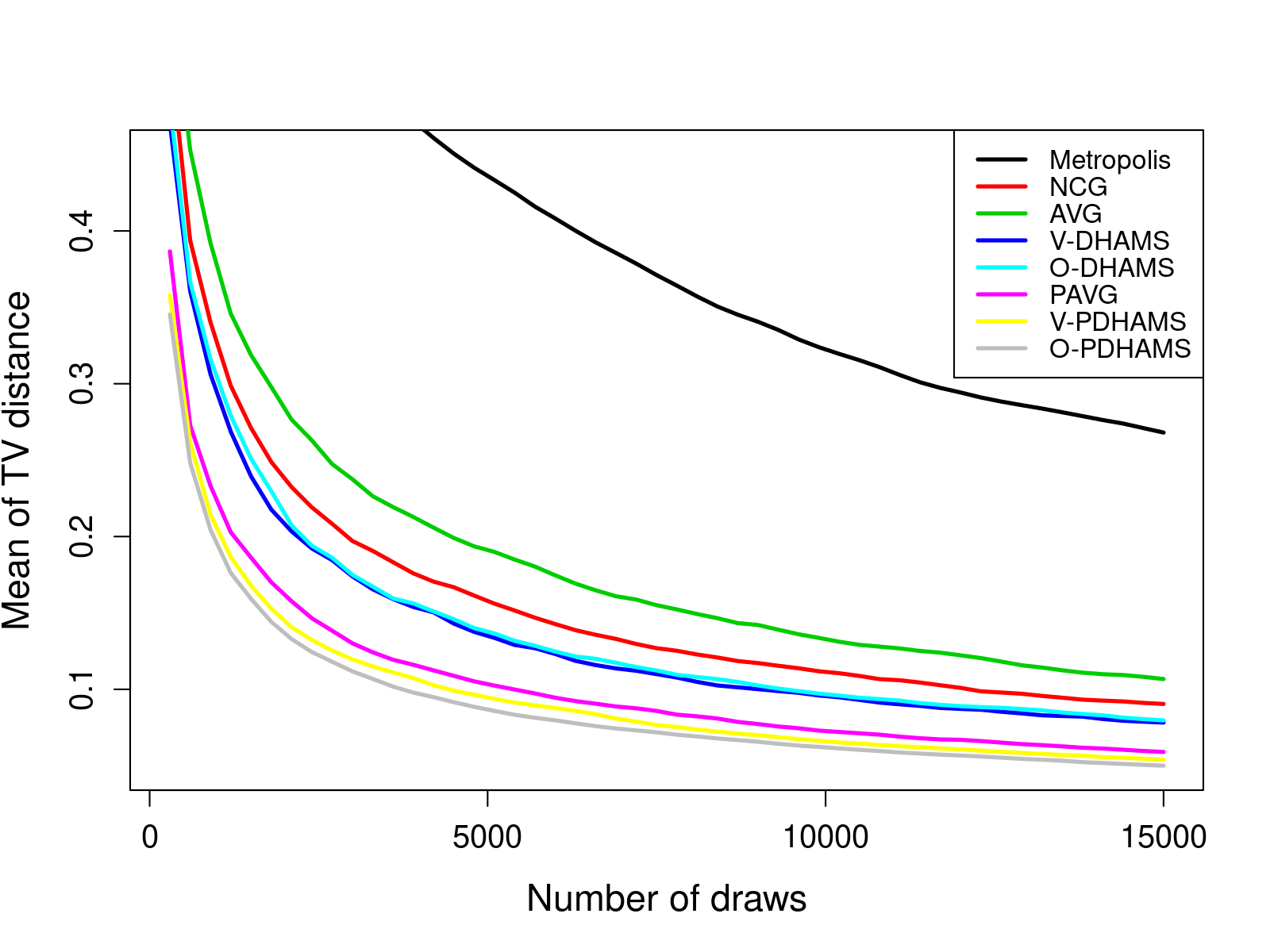}
        \caption{Average mean of TV-distances for all two-dimensional marginal distributions}
        \label{fig:precond_tv2}
    \end{subfigure}
     \begin{subfigure}{0.45\textwidth}
        \centering
        \includegraphics[width=\linewidth]{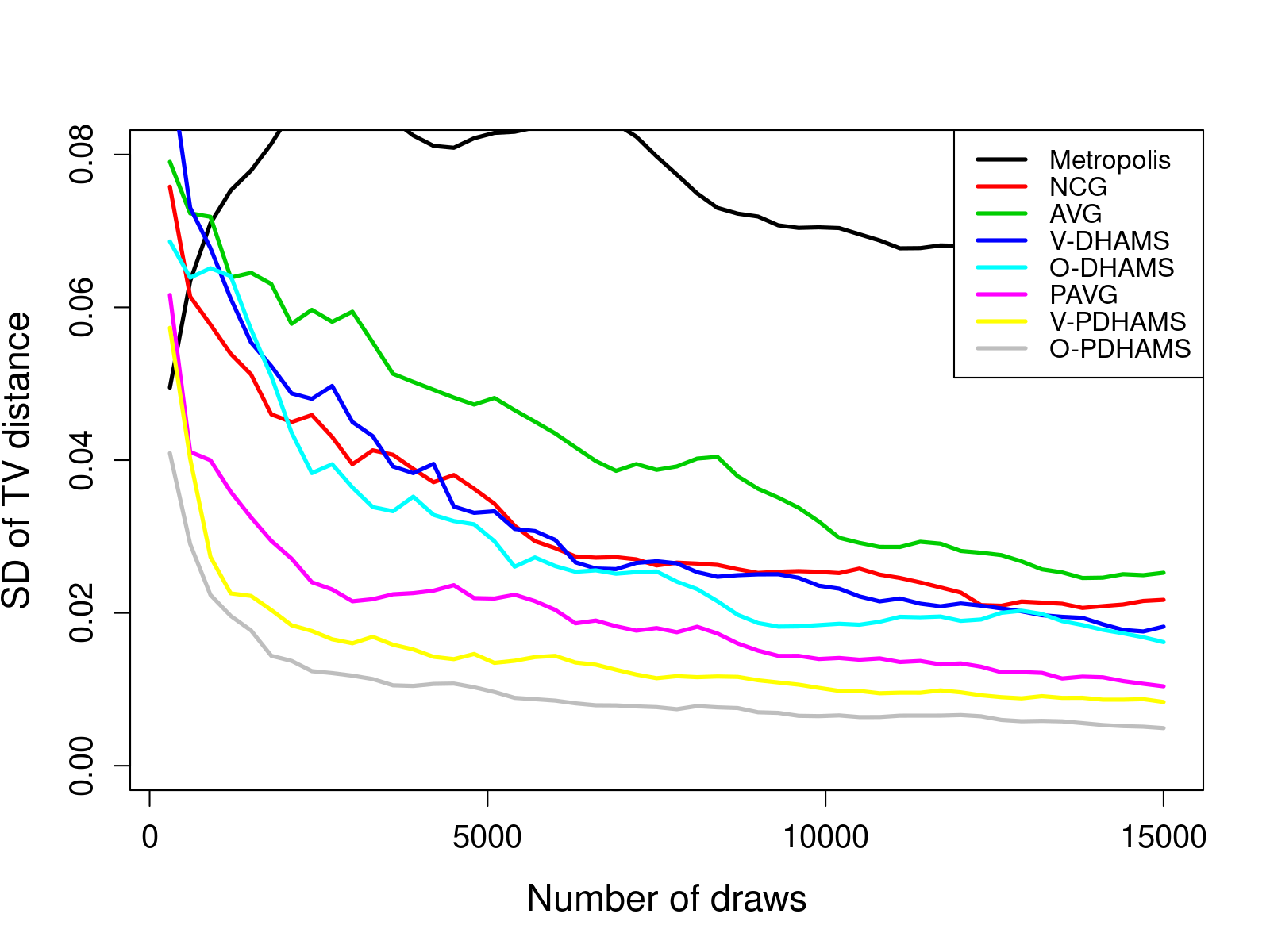}
        \caption{Average standard deviation of TV-distances for all two-dimensional marginal distributions}
        \label{fig:precond_tv2sd}
    \end{subfigure}
    \begin{subfigure}{0.45\textwidth}
        \centering
        \includegraphics[width=\linewidth]{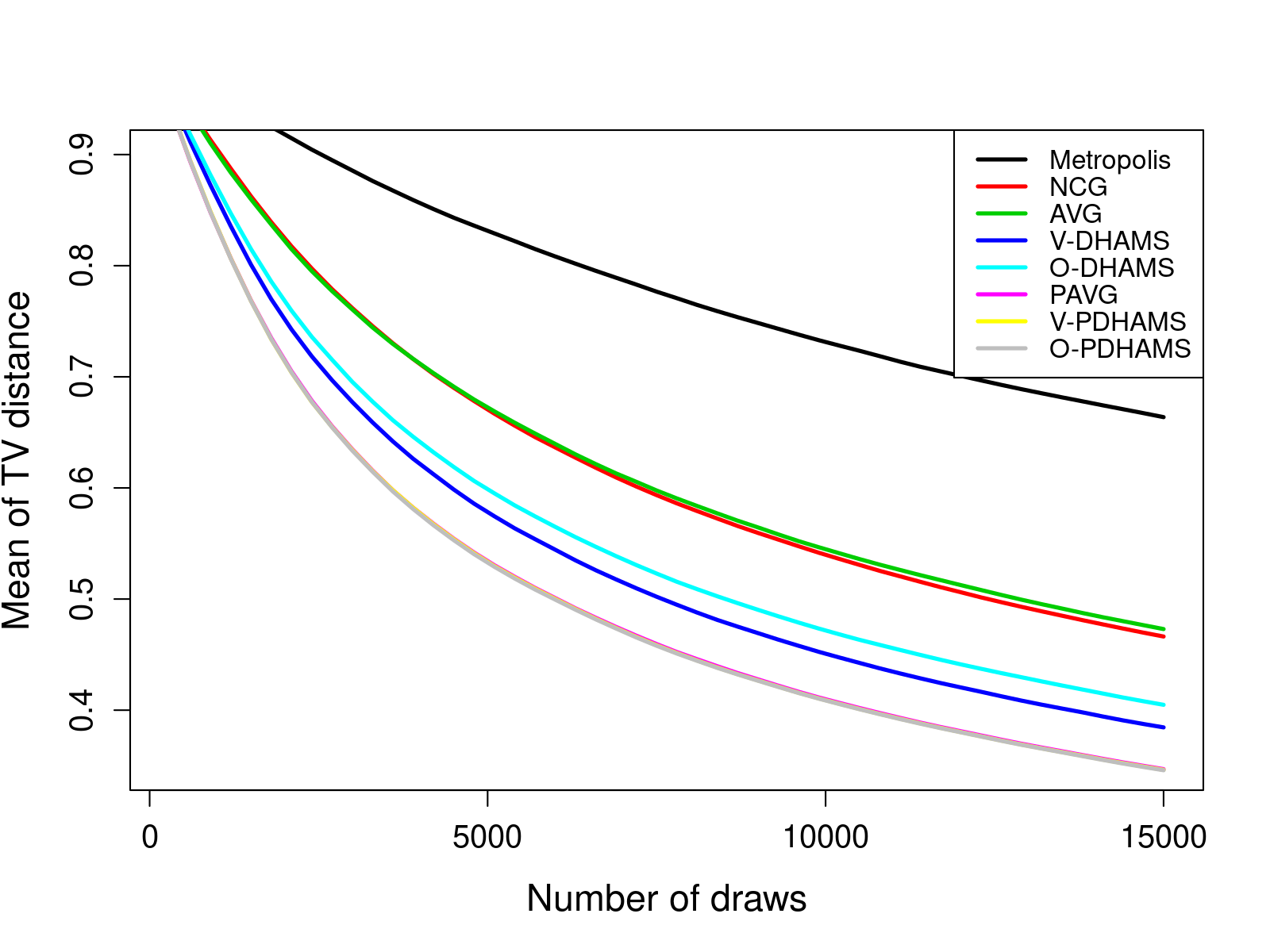}
        \caption{Average mean of TV-distances for all four-dimensional marginal distributions}
        \label{fig:precond_tv4}
    \end{subfigure}
     \begin{subfigure}{0.45\textwidth}
        \centering
        \includegraphics[width=\linewidth]{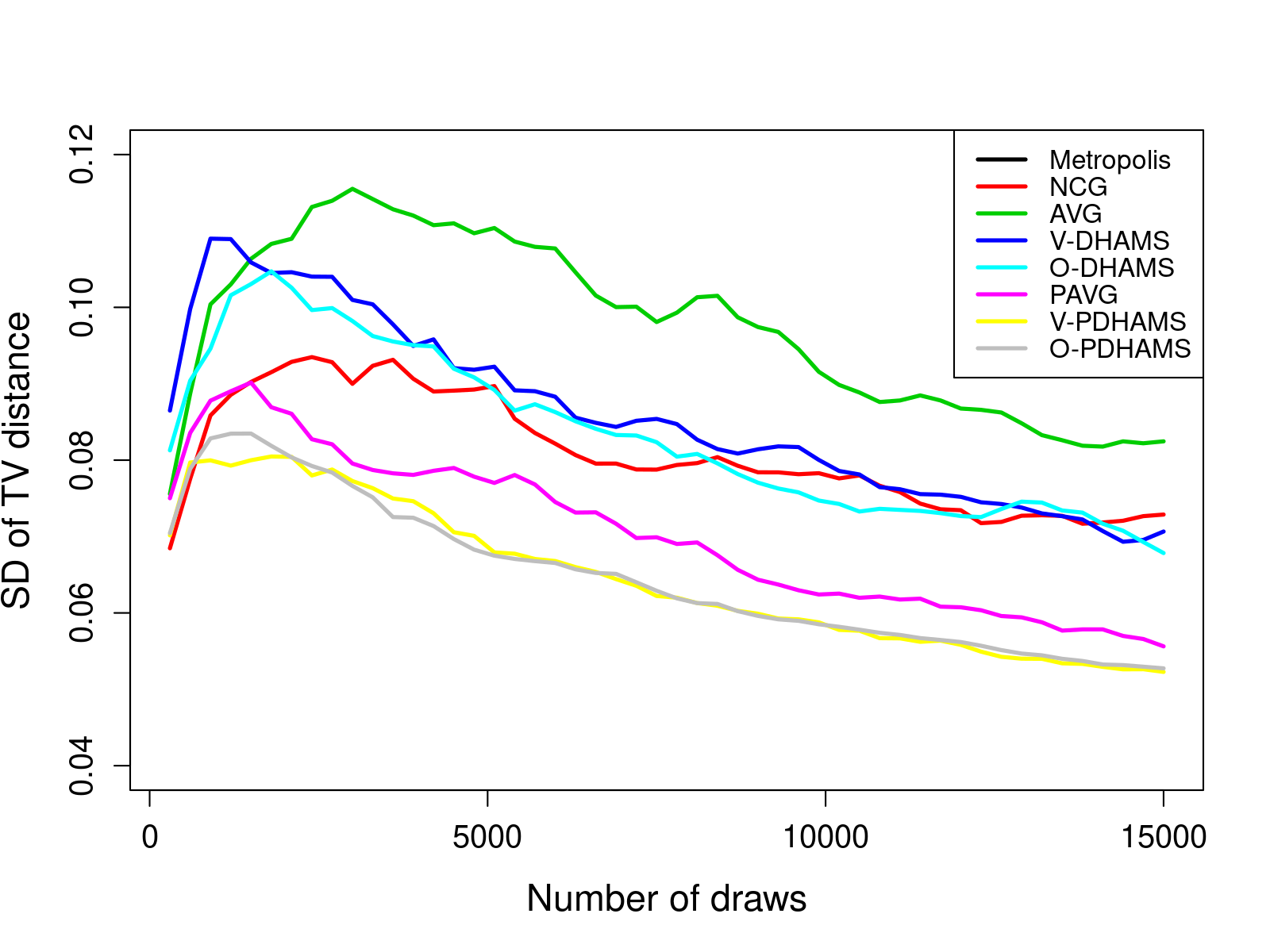}
        \caption{Average standard deviation of TV-distances for all four-dimensional marginal distributions}
        \label{fig:precond_tv4sd}
    \end{subfigure}\caption{TV-distance results for discrete Gaussian distribution}
    \label{fig:precond_tvs}
\end{figure}

In our analysis, we focus on the following marginal distributions $\pi(s_i, s_j)$ and  $\pi(s_{j_{1}}, s_{j_{2}}, s_{j_{3}}, s_{j_{4}})$, and further average the means and standard deviations of the TV-distances (each over 100 chains) across all index pairs $(s_i, s_j)$ for the bivariate marginals, and across all index combinations  $(s_{j_{1}}, s_{j_{2}}, s_{j_{3}}, s_{j_{4}})$ for the four-dimensional marginals as in \cite{Zhou2025Dhams}. The results are presented in Figure~\ref{fig:precond_tvs}, plotted against the number of draws.

\begin{figure}[tbp]
    \begin{subfigure}{0.45\textwidth}
        \centering
        \includegraphics[width=\linewidth]{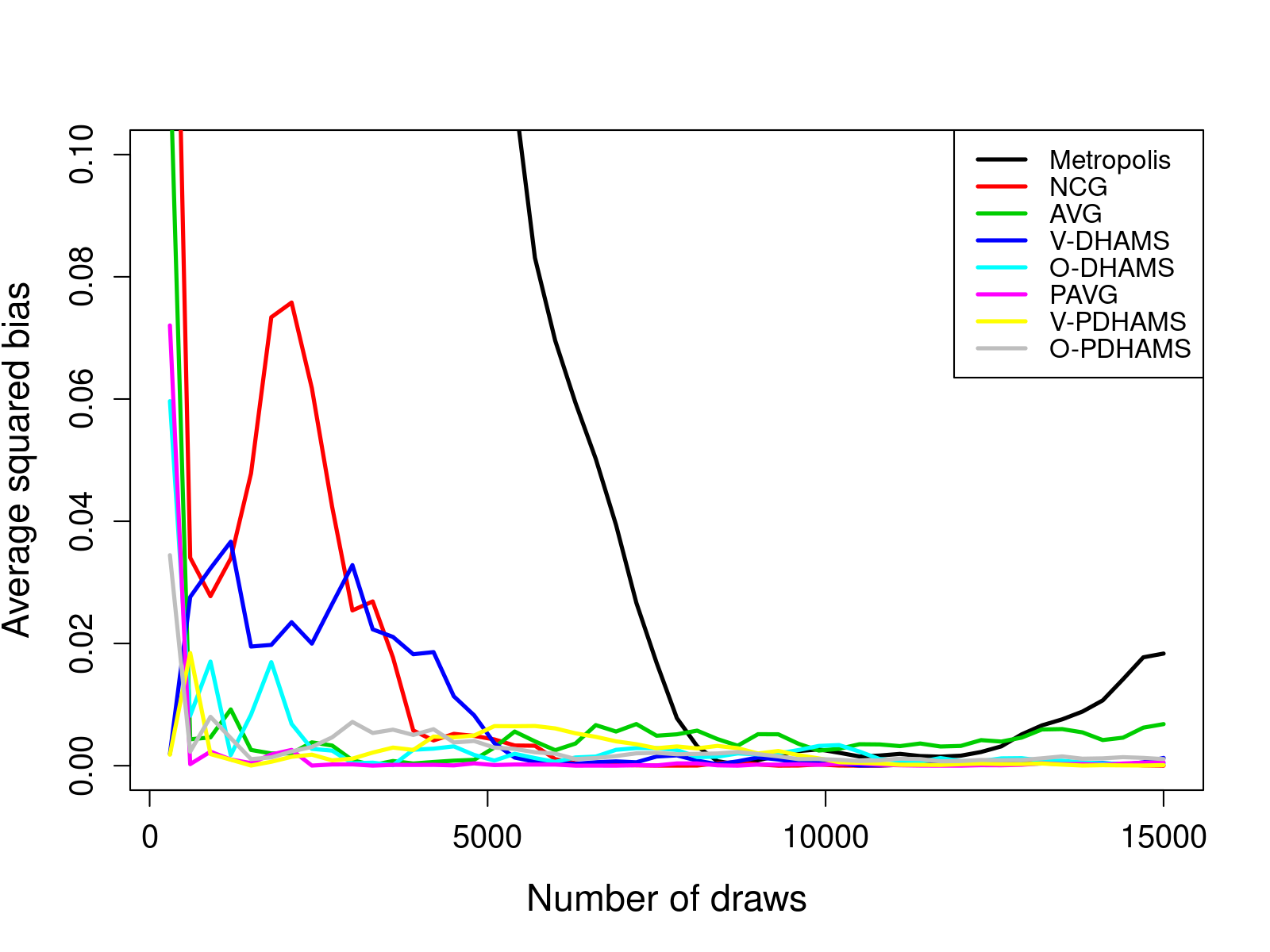}
        \caption{Average bias for $E[s_{j}]$}
        \label{fig:precond_biass1}
    \end{subfigure}
    \hfill
    \begin{subfigure}{0.45\textwidth}
        \centering
        \includegraphics[width=\linewidth]{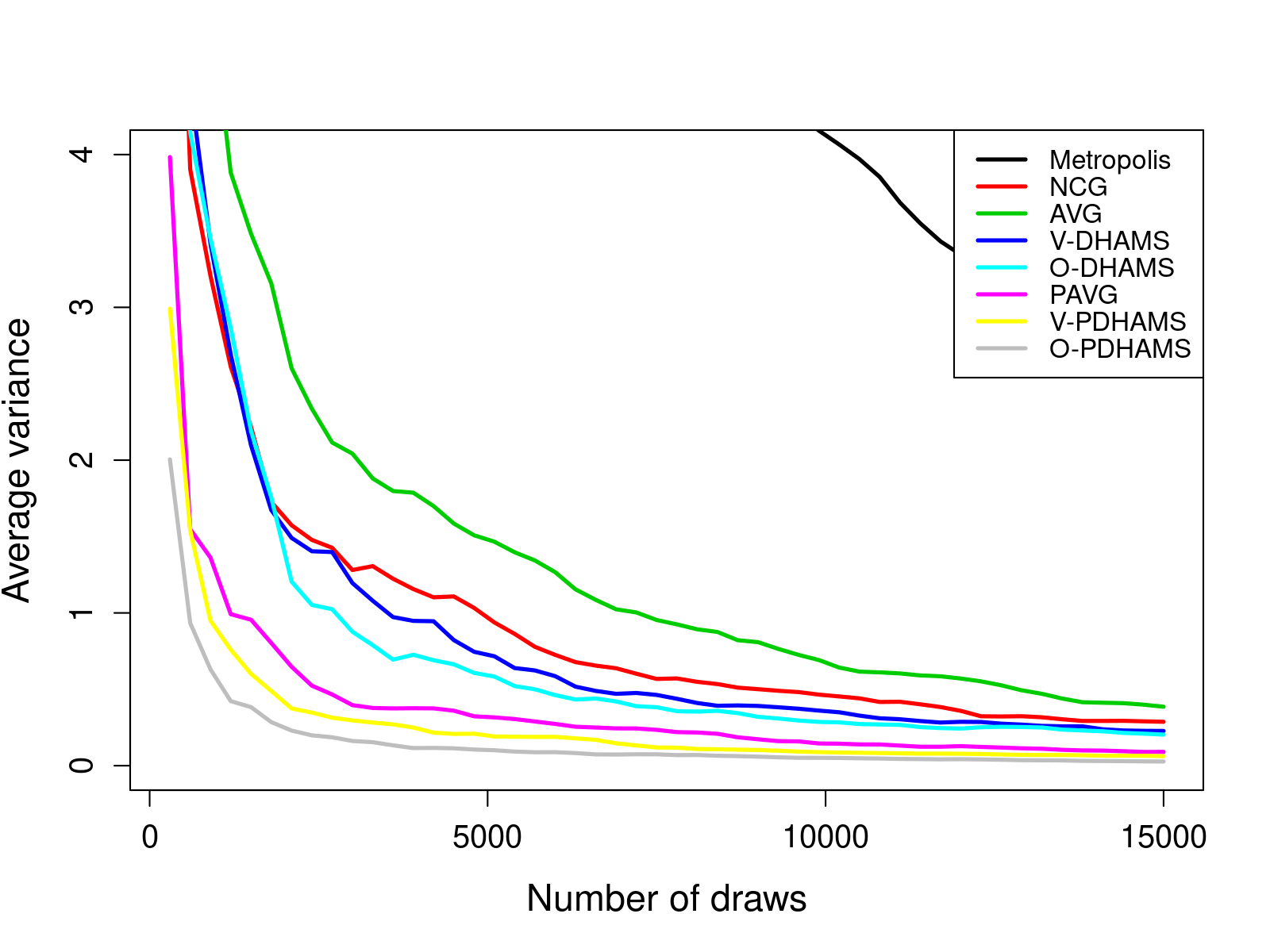}
        \caption{Average variance for $E[s_{j}]$}
        \label{fig:precond_vars1}
    \end{subfigure}

    \begin{subfigure}{0.45\textwidth}
        \centering
        \includegraphics[width=\linewidth]{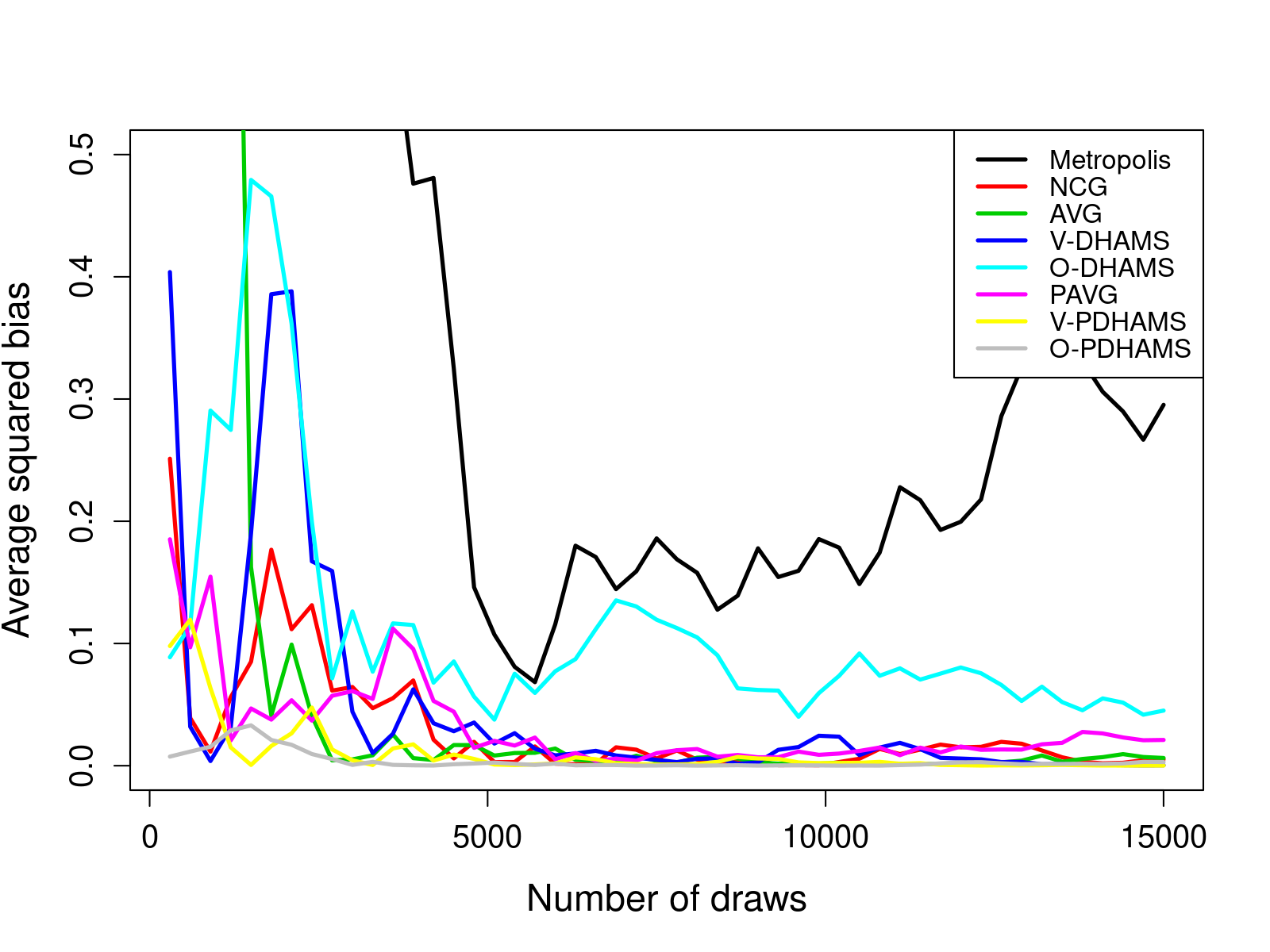}
        \caption{Average bias for $E[s_{j}^{2}]$}
        \label{fig:precond_biass12}
    \end{subfigure}
    \hfill
    \begin{subfigure}{0.45\textwidth}
        \centering
        \includegraphics[width=\linewidth]{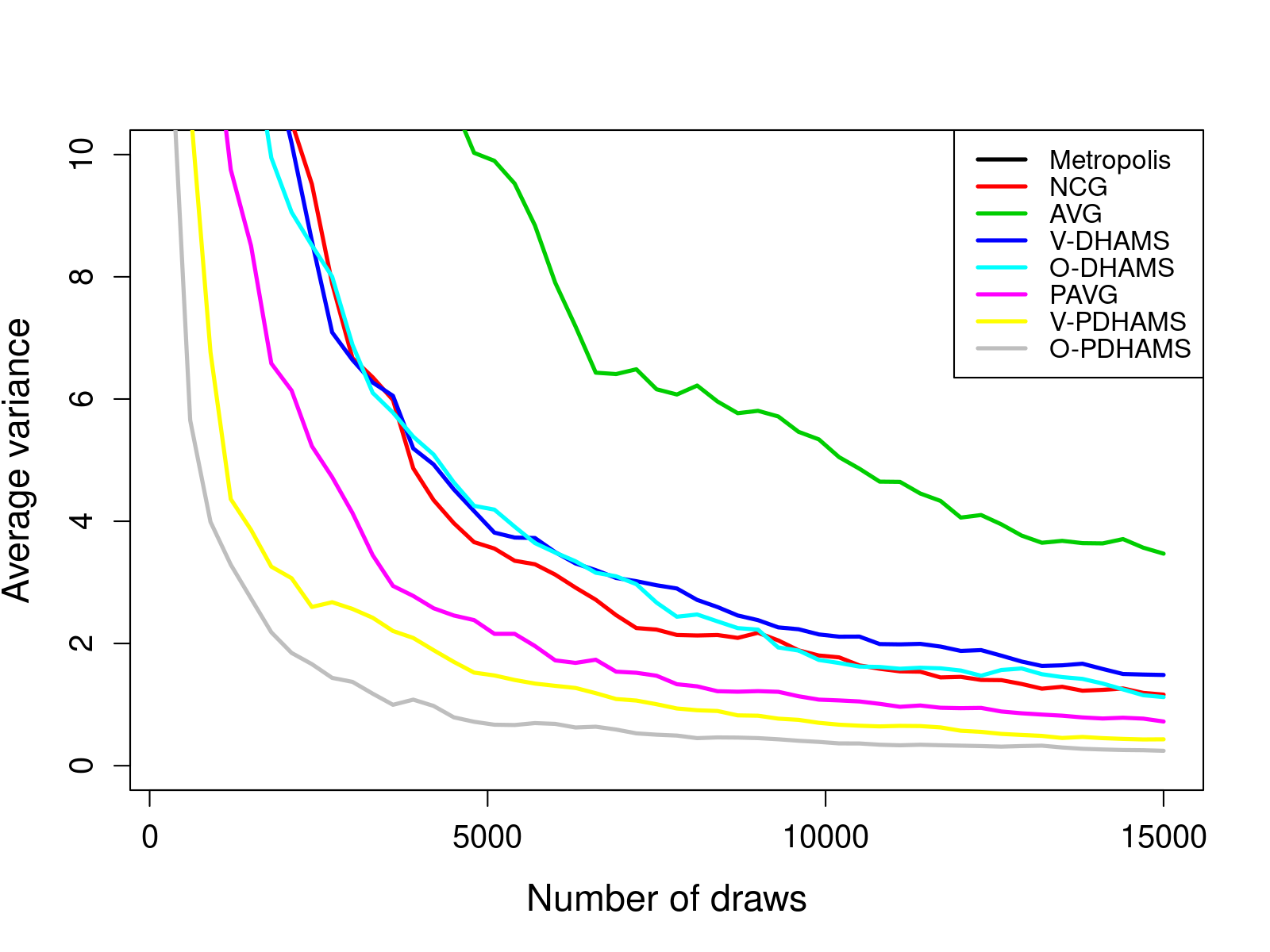}
        \caption{Average variance for $E[s_{j}^{2}]$}
        \label{fig:precond_vars12}
    \end{subfigure}

    \begin{subfigure}{0.45\textwidth}
        \centering
        \includegraphics[width=\linewidth]{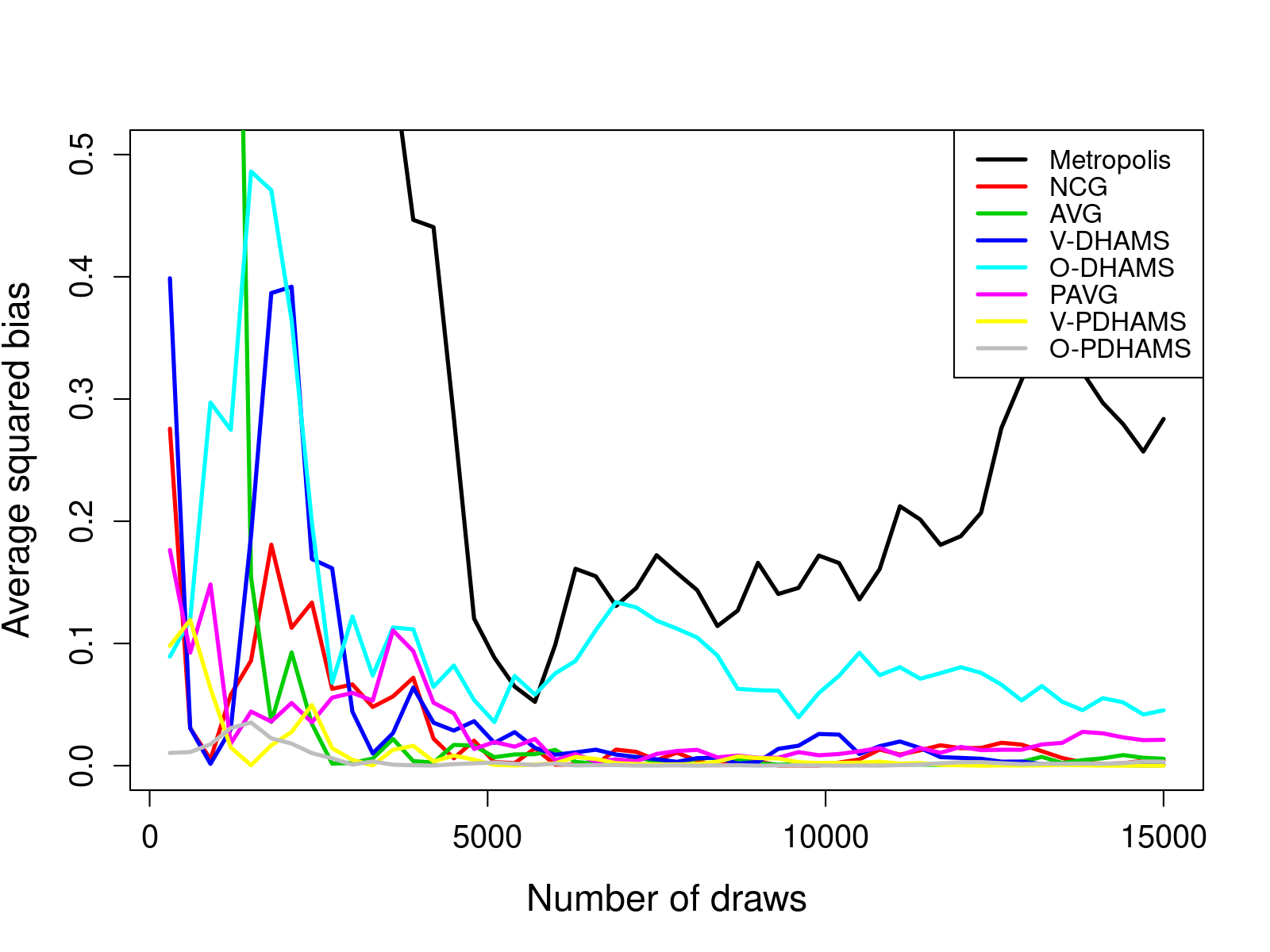}
        \caption{Average bias for $E[s_{i}s_{j}]$}
        \label{fig:precond_biass1s2}
    \end{subfigure}
    \hfill
    \begin{subfigure}{0.45\textwidth}
        \centering
        \includegraphics[width=\linewidth]{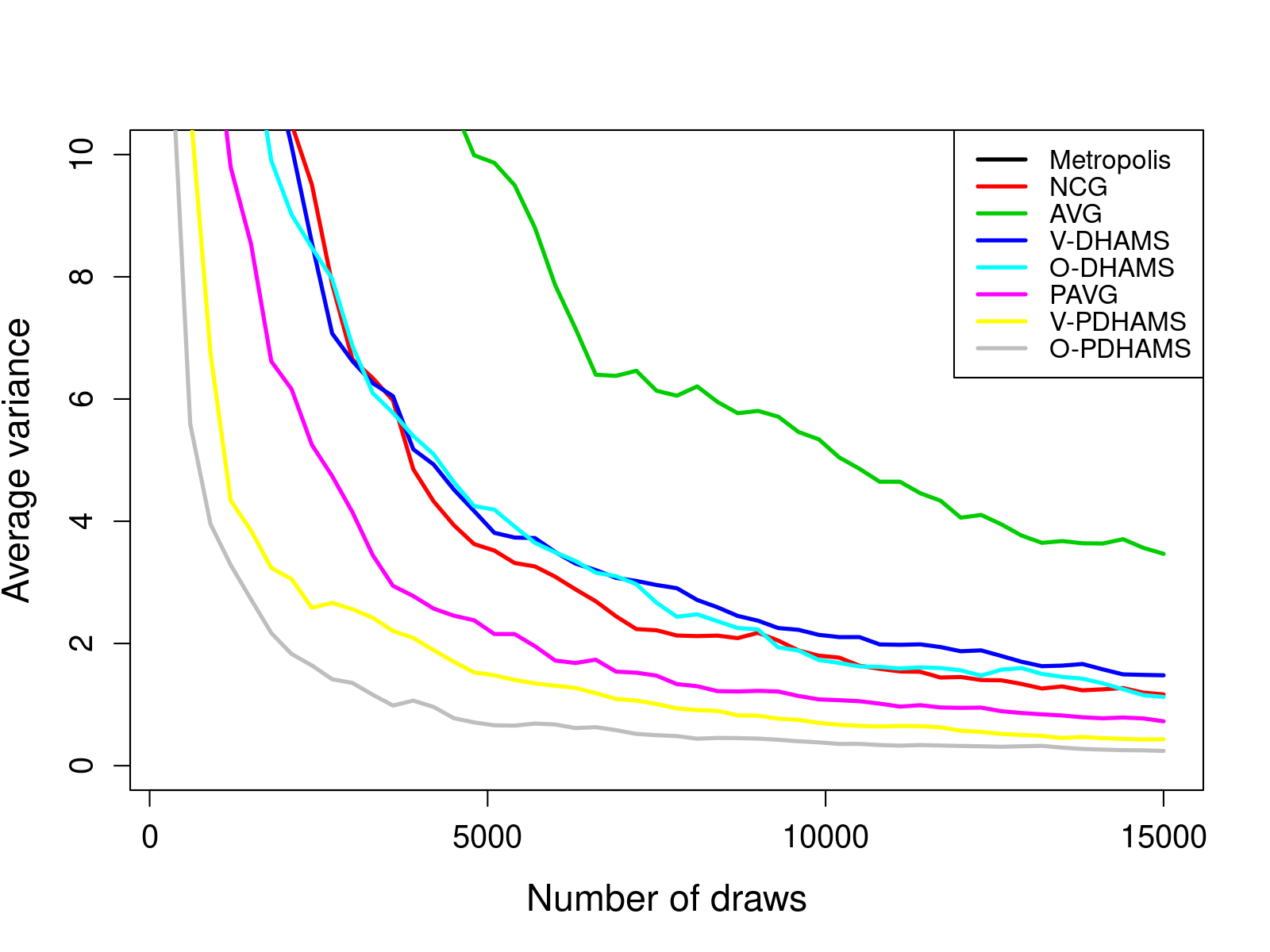}
        \caption{Average variance for $E[s_{i}s_{j}]$}
        \label{fig:precond_vars1s2}
    \end{subfigure}

    \caption{Estimation results for discrete Gaussian distribution}
    \label{fig:estimations_disgau_precond}
\end{figure}

We are also interested in the estimation of the three quantities, $\E[s_i]$, $\E[s_i^2]$ and $\E[s_is_j]$. For each chain, we compute estimates of these quantities, and then, using the 100 estimates obtained from the 100 repeated chains for each sampler, we evaluate the squared bias and variance. The squared bias and variance for $\E[s_i]$, $\E[s_i^2]$ are averaged over all dimensions, while those for $\E[s_is_j]$ are averaged over all index pairs. The resulting average metrics are plotted against the number of draws in Figure \ref{fig:estimations_disgau_precond}.

Similarly as in \cite{girolami2011ESS_marginal}, we report the minimum, median, and maximum of ESS across all coordinates. In addition, the ESS for the negative potential function $f(s)$ is also reported. The results are presented in Table \ref{tab:ess_gaussian_precond}.
\begin{table}[tbp]
\centering
\begin{tabular}{|c|c|c|c|c|}
\hline
    Sampler & ESS Minimum & ESS Median & ESS Maximum & ESS Energy\\
    \hline
    Metropolis & 4.67 & 4.72 & 4.81 & 180.50 \\
    NCG & 58.50 & 58.97 & 59.55 & 3388.48\\
       AVG & 43.02 & 43.67 & 43.94 & 2254.74\\
   V-DHAMS & 73.87 & 75.09 & 76.14 & 3841.09\\
    O-DHAMS & 82.25 & 82.73 & 83.78 & 3167.07\\
    PAVG & 186.12 & 189.35 & 192.09& 9795.28\\
    V-PDHAMS & 269.08 & 275.84 & 283.65 & 9996.29\\
    O-PDHAMS & 615.94 & 630.43 & 636.29 & 4485.22\\

    \hline
\end{tabular}
\caption{ESS table for discrete Gaussian distribution}
\label{tab:ess_gaussian_precond}
\end{table}

The preconditioned second-order samplers outperform first-order samplers , consistently achieving lower TV-distances. Among the three preconditioned samplers, O-PDHAMS achieves the lowest standard deviations in TV-distances. In the estimation tasks, preconditioned samplers exhibit significantly lower variances, especially O-PDHAMS. From Table~\ref{tab:ess_gaussian_precond}, O-PDHAMS achieves the highest marginal ESS and V-PDHAMS achieves the highest ESS of $f(s)$. The overall performance in simulating the discrete Gaussian distribution follows the order:
\[
\text{O-PDHAMS} > \text{V-PDHAMS} > \text{PAVG} > \text{O-DHAMS} > \text{V-DHAMS} > \text{NCG} > \text{AVG} > \text{Metropolis}.
\]
Additional results (including selected parameter settings, average acceptance rates, and trace, frequency and auto-correlation plots from individual runs) are presented in Supplement Section~\ref{sec:precond_gaussian_results}.

\subsection{Quadratic Mixture} \label{sec:poly_mix}
\begin{figure}[tbp]
    \centering
    \includegraphics[width=0.5\linewidth]{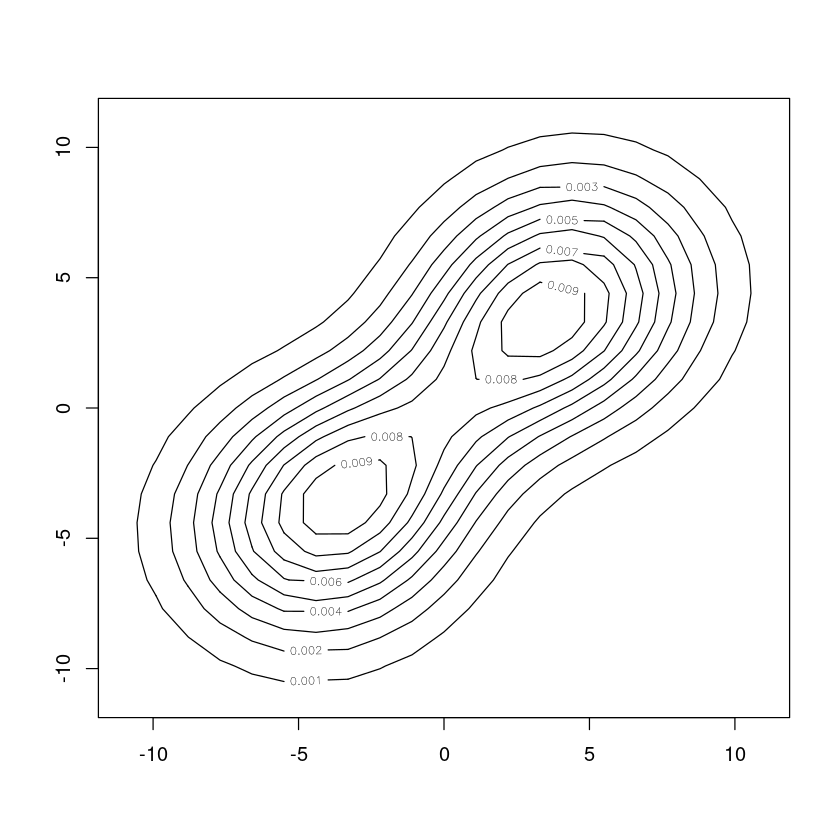} 
    \caption{Contour plot of quadratic mixture distribution}
    \label{fig:poly_less_heatmap} 
\end{figure}

We consider a quadratic mixture distribution similarly as in \cite{Rhodes2022GradientMC} and \cite{Zhou2025Dhams}, but with more mixture components to allow more effective preconditioning.
The negative potential function $f(s)$ is defined as
\begin{align}
    f(s) \propto \log\left(\sum\limits_{m=1}^{M} \exp\left(-\frac{1}{2}(s - \mu_m)^{\T}\Sigma_m^{-1}(s - \mu_m)\right)\right).
    \label{eq:poly_less_mix_f}
\end{align}
The distribution is defined over a $d$-dimensional lattice $\mathcal{S} \subset \mathbb{R}^d$, given by $\mathcal{S} = \{-k, -(k-1), \ldots, -1, 0, 1, \ldots, k-1, k\}^d$. In our experiment, we set $d = 10$, $k = 10$, and $M = 9$. Each component has mean $\mu_m = (-5.625 + 1.125m)\mathbf{1}$ and covariance $\Sigma_m = \sigma_m^2I$, $\sigma_m^2 = 2.10 + 0.15|m-5|$. The means are equally spaced across components, and components with larger absolute mean values are assigned larger variances. The resulting distribution \eqref{eq:poly_less_mix_f} is symmetric along each coordinate axis and invariant under permutations of coordinates. For visualization, we present the contour plot over any two selected dimensions in Figure~\ref{fig:poly_less_heatmap}. The preconditioning matrix $W$ is calculated using the second calibration method \eqref{eqn:W_calibration2}.

 For each sampler, 50 different parameters are searched for tuning for each sampler. We run 100 independent chains, each of length 7,500 after 500 burn-in draws for each parameter. The best parameter is then selected by the highest ESS of negative potential function $f(s)$.
After tuning, we conduct 100 independent chains for each sampler using the optimal parameter setting. For each chain, the initial 500 draws were discarded as burn-in, and the subsequent 24,000 draws were retained for analysis.

\begin{figure}[tbp]
    \centering
    \begin{subfigure}{0.45\textwidth}
        \centering
        \includegraphics[width=\linewidth]{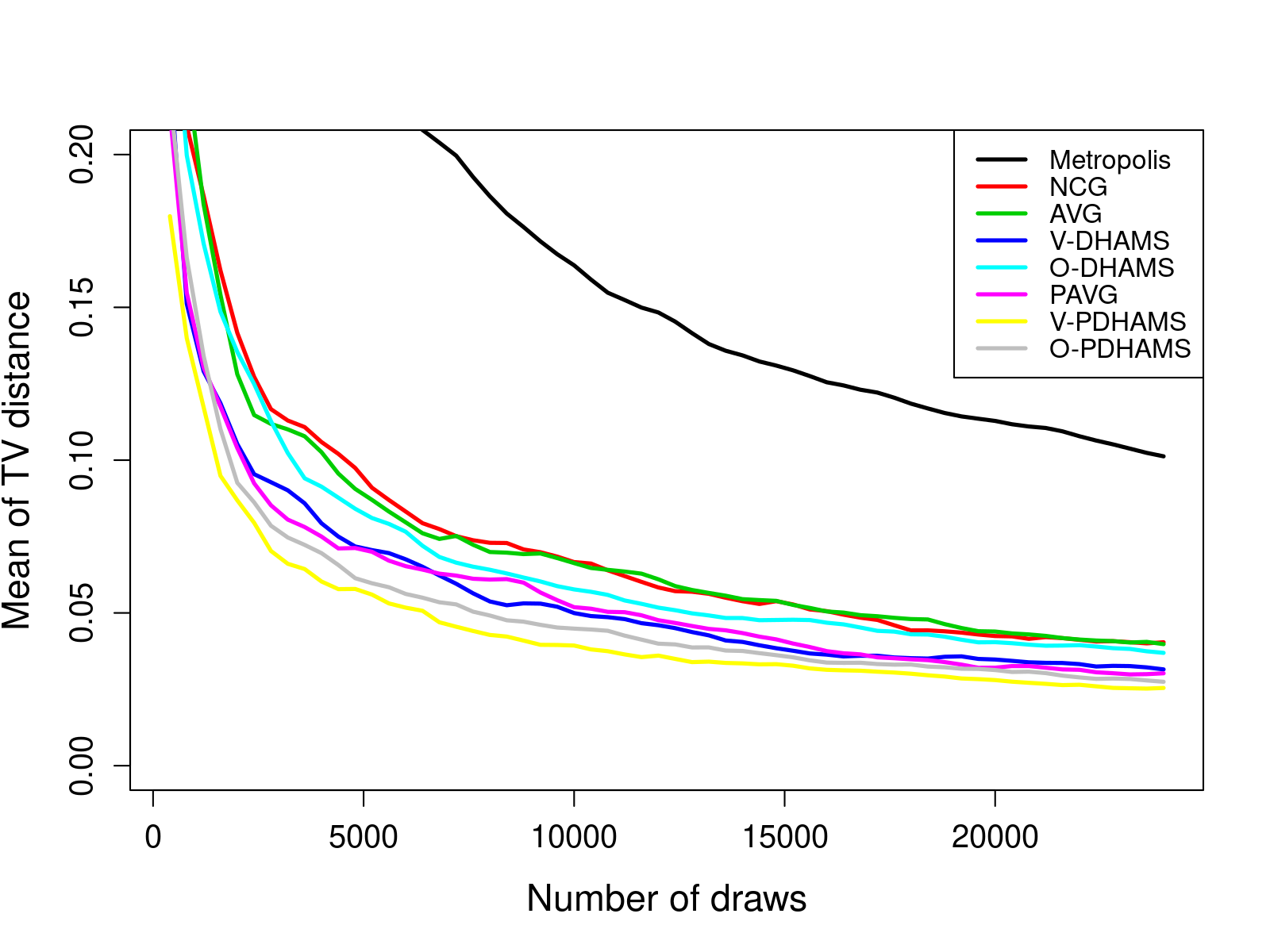}
        \caption{Average mean of TV-distance for all univariate marginal distributions}
        \label{fig:poly_less_tv2}
    \end{subfigure}
     \begin{subfigure}{0.45\textwidth}
        \centering
        \includegraphics[width=\linewidth]{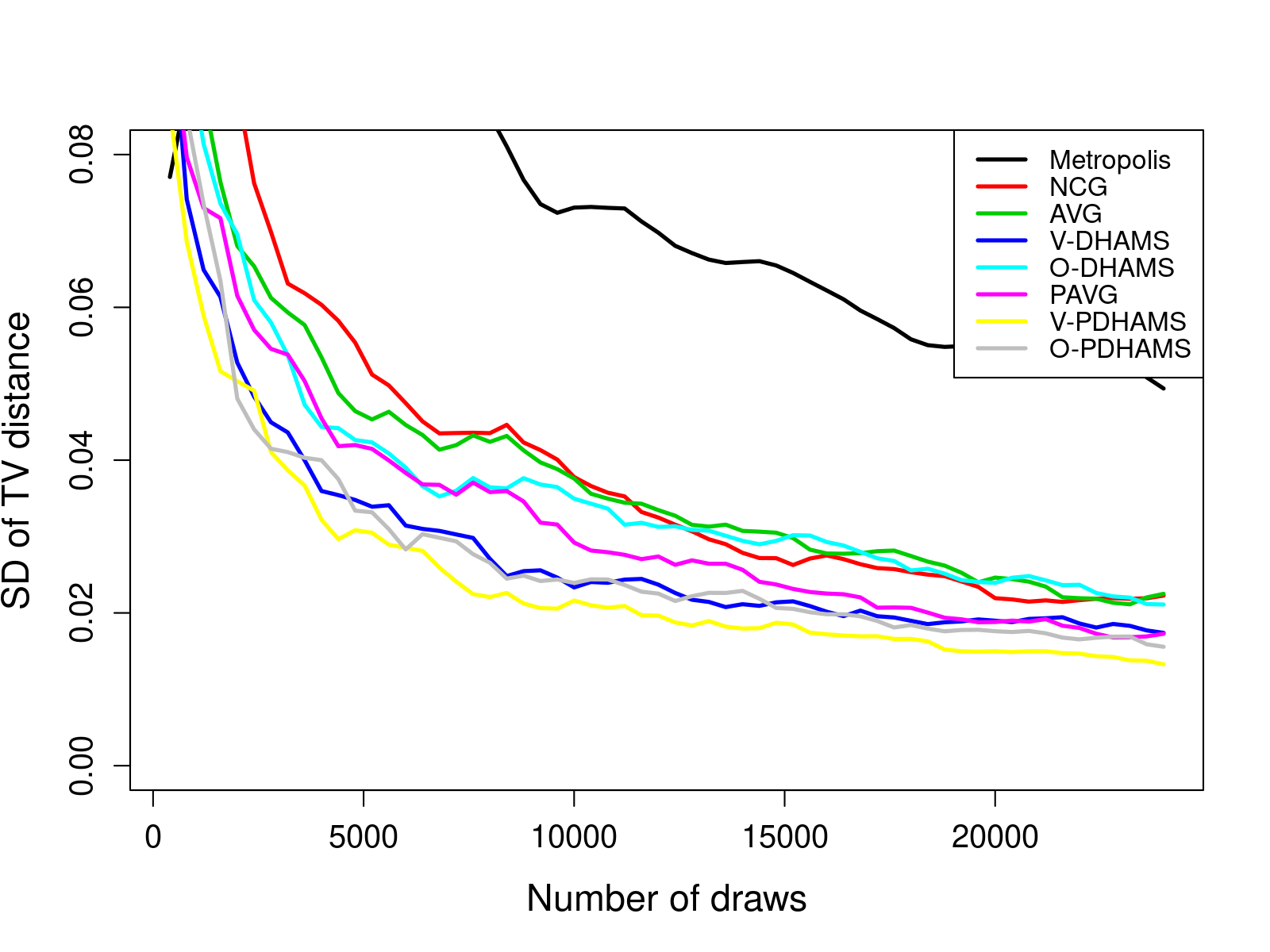}
        \caption{Average standard deviation of TV-distance for all univariate marginal distributions}
        \label{fig:poly_less_tv2sd}
    \end{subfigure}
    \begin{subfigure}{0.45\textwidth}
        \centering
        \includegraphics[width=\linewidth]{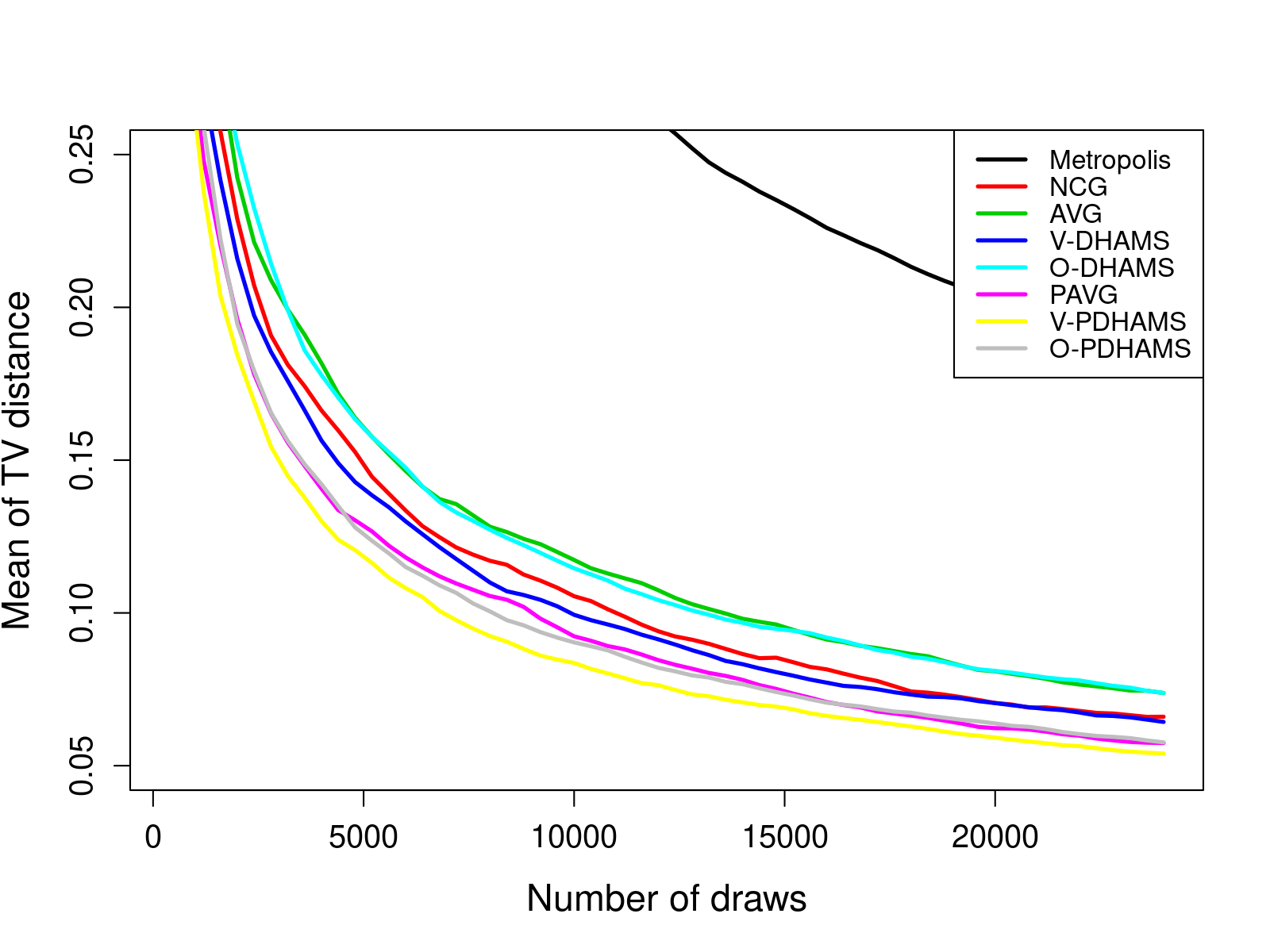}
        \caption{Average mean of TV-distance for all bivariate marginal distributions}
        \label{fig:poly_less_tv4}
    \end{subfigure}
     \begin{subfigure}{0.45\textwidth}
        \centering
        \includegraphics[width=\linewidth]{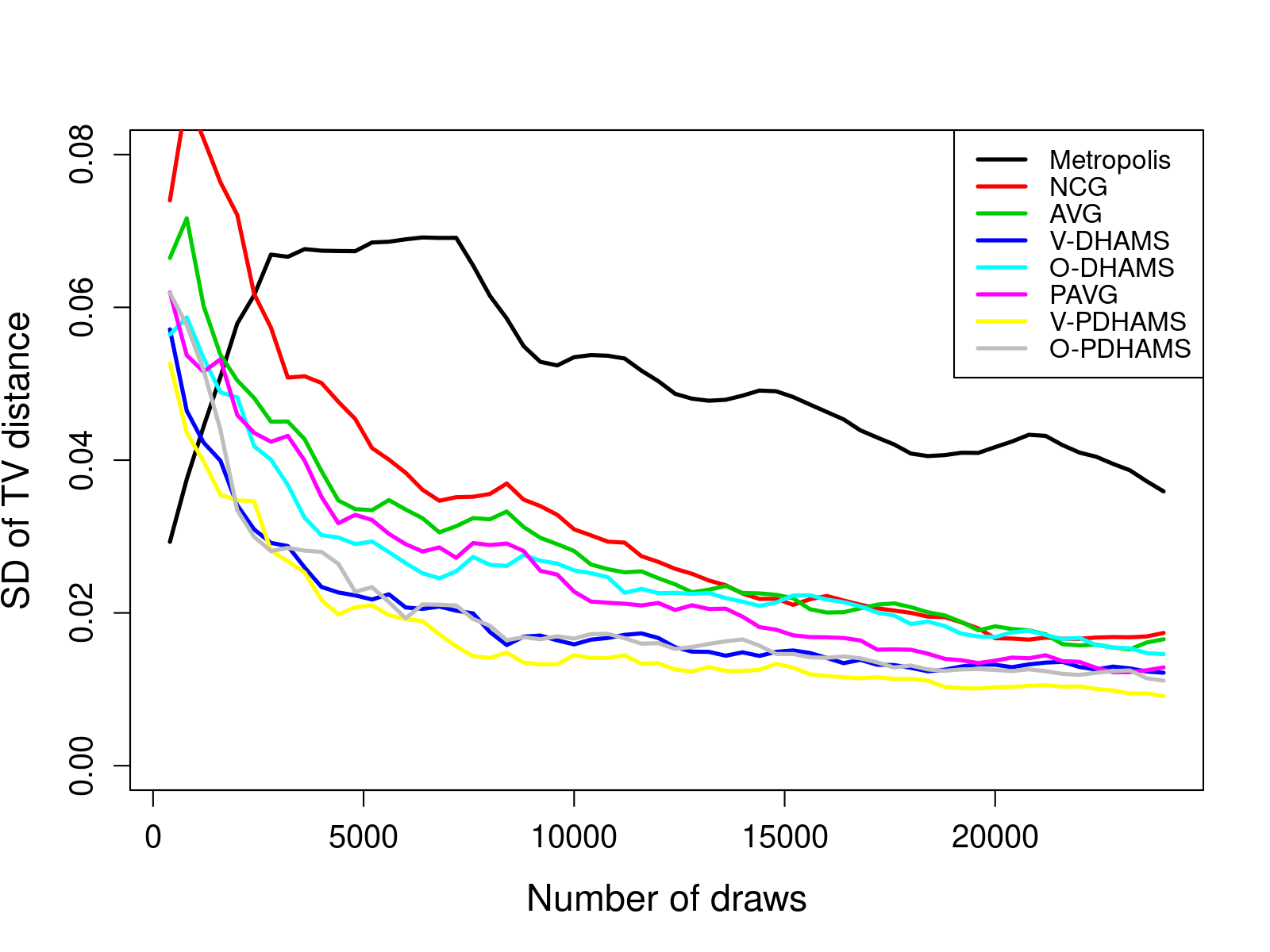}
        \caption{Average standard deviation of TV-distance for all bivariate marginal distributions}
        \label{fig:poly_less_tv4sd}
    \end{subfigure}\caption{TV-distances results for quadratic mixture distribution}
    \label{fig:poly_less_tvs}
\end{figure}

In the quadratic mixture experiment, we report results of TV-distances for univariate and bivariate marginal distributions $\pi(s_i)$ and $\pi(s_{i_1}, s_{i_2})$. We average the means and standard deviations of the TV-distances over 100 chains across all dimensions $s_i$ for the univariate marginals, and across all pairs $(s_{i_1}, s_{i_2})$ for bivariate marginals. The aggregated results plotted against the number of draws are presented in Figure~\ref{fig:poly_less_tvs}.

\begin{figure}[tbp]
    \begin{subfigure}{0.45\textwidth}
        \centering
        \includegraphics[width=\linewidth]{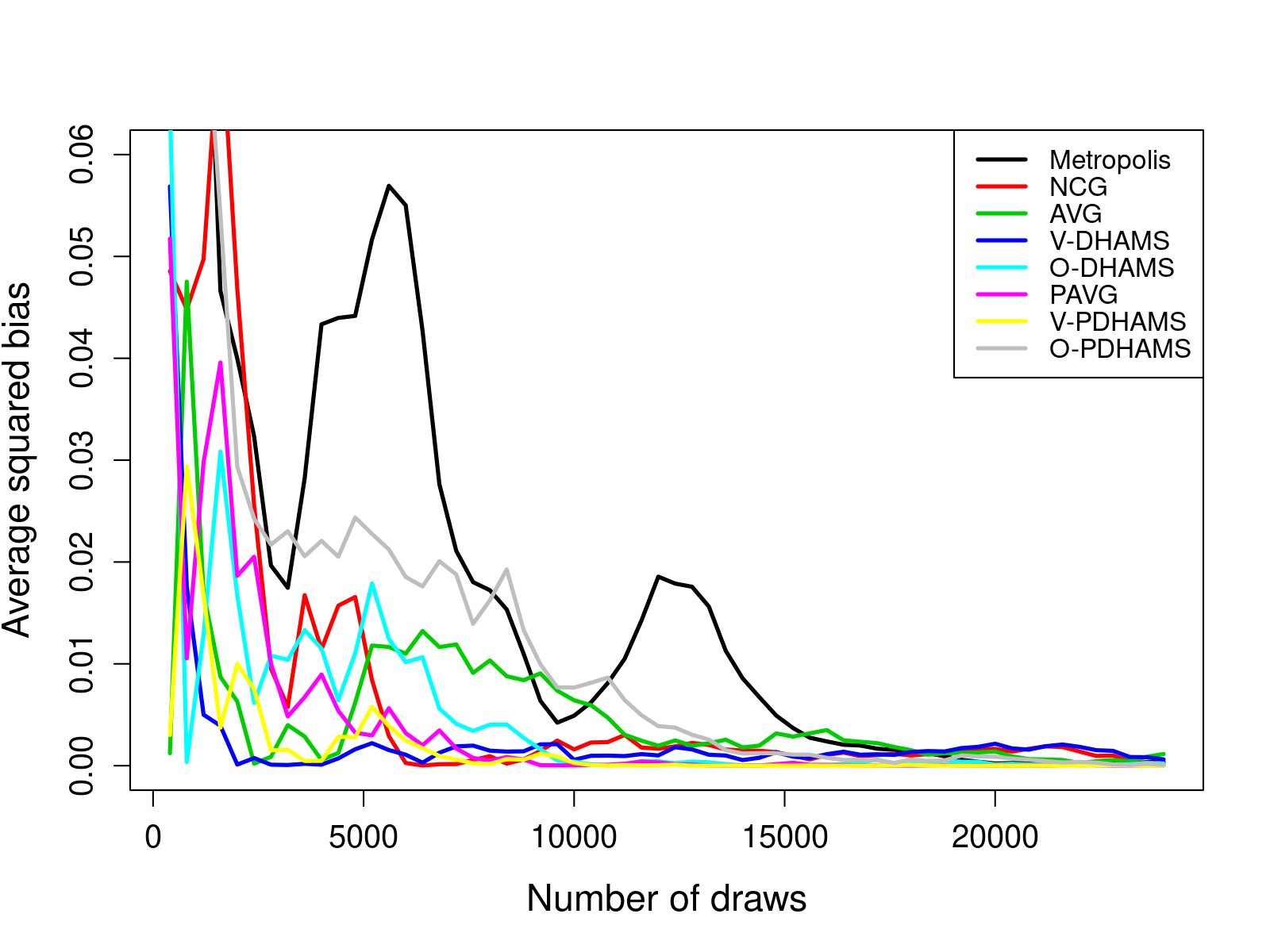}
        \caption{Average squared bias of $E[s_{i}]$}
        \label{fig:poly_less_biass1}
    \end{subfigure}
    \hfill
    \begin{subfigure}{0.45\textwidth}
        \centering
        \includegraphics[width=\linewidth]{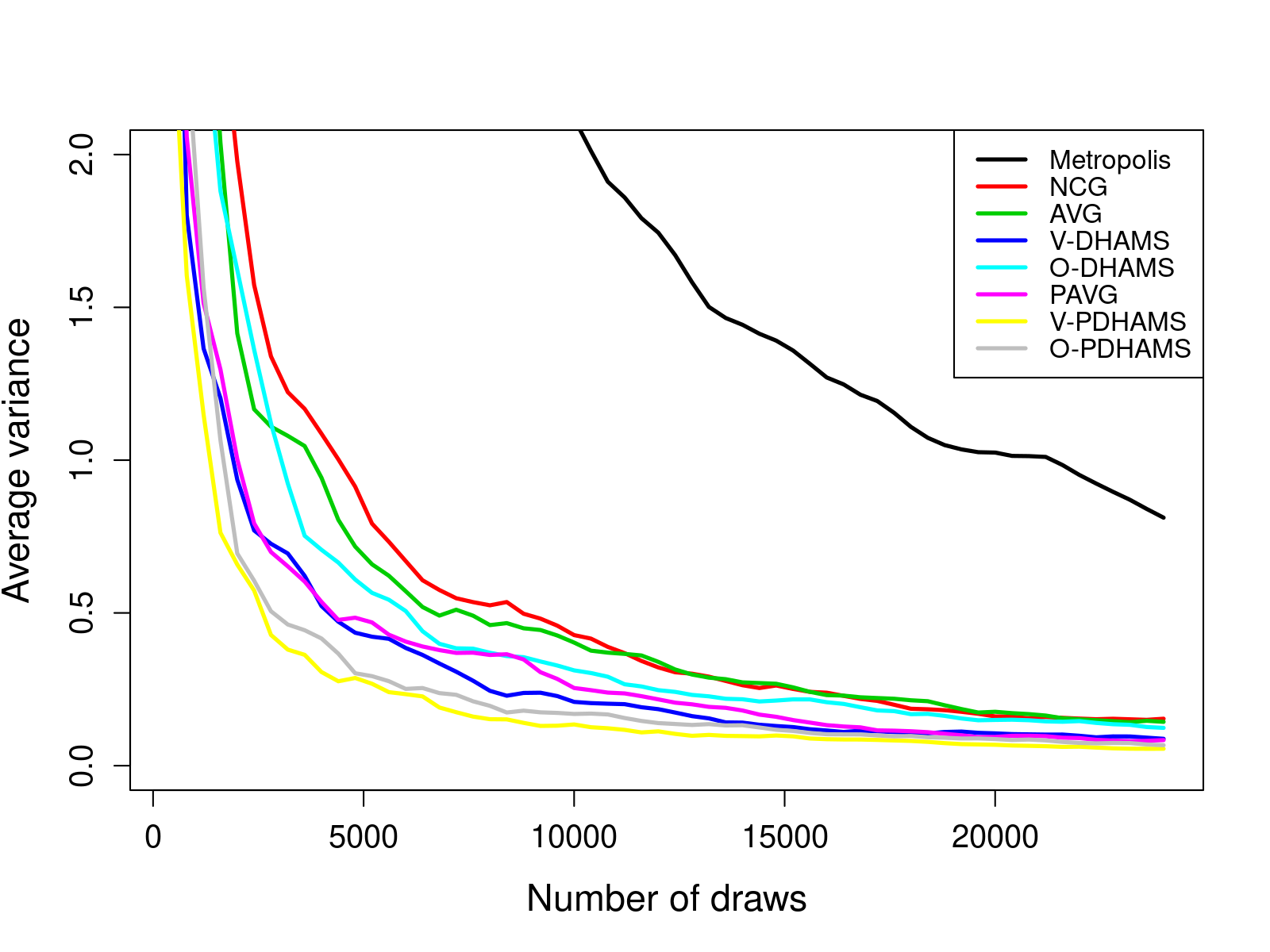}
        \caption{Average variance of $E[s_{i}]$}
        \label{fig:poly_less_vars1}
    \end{subfigure}

    \par\medskip

    \begin{subfigure}{0.45\textwidth}
        \centering
        \includegraphics[width=\linewidth]{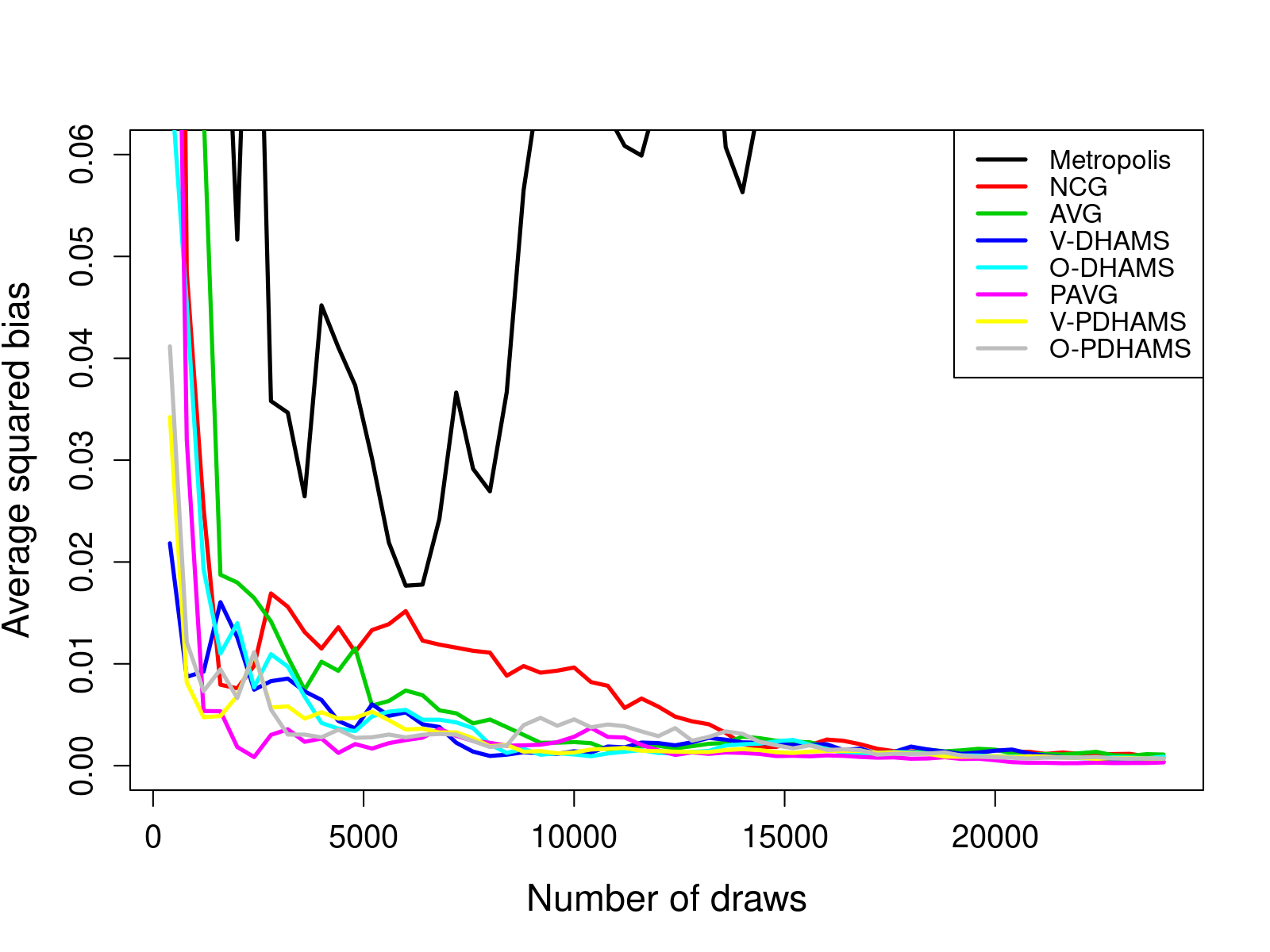}
        \caption{Average squared bias of $E[s_{i}^{2}]$}
        \label{fig:poly_less_biass12}
    \end{subfigure}
    \hfill
    \begin{subfigure}{0.45\textwidth}
        \centering
        \includegraphics[width=\linewidth]{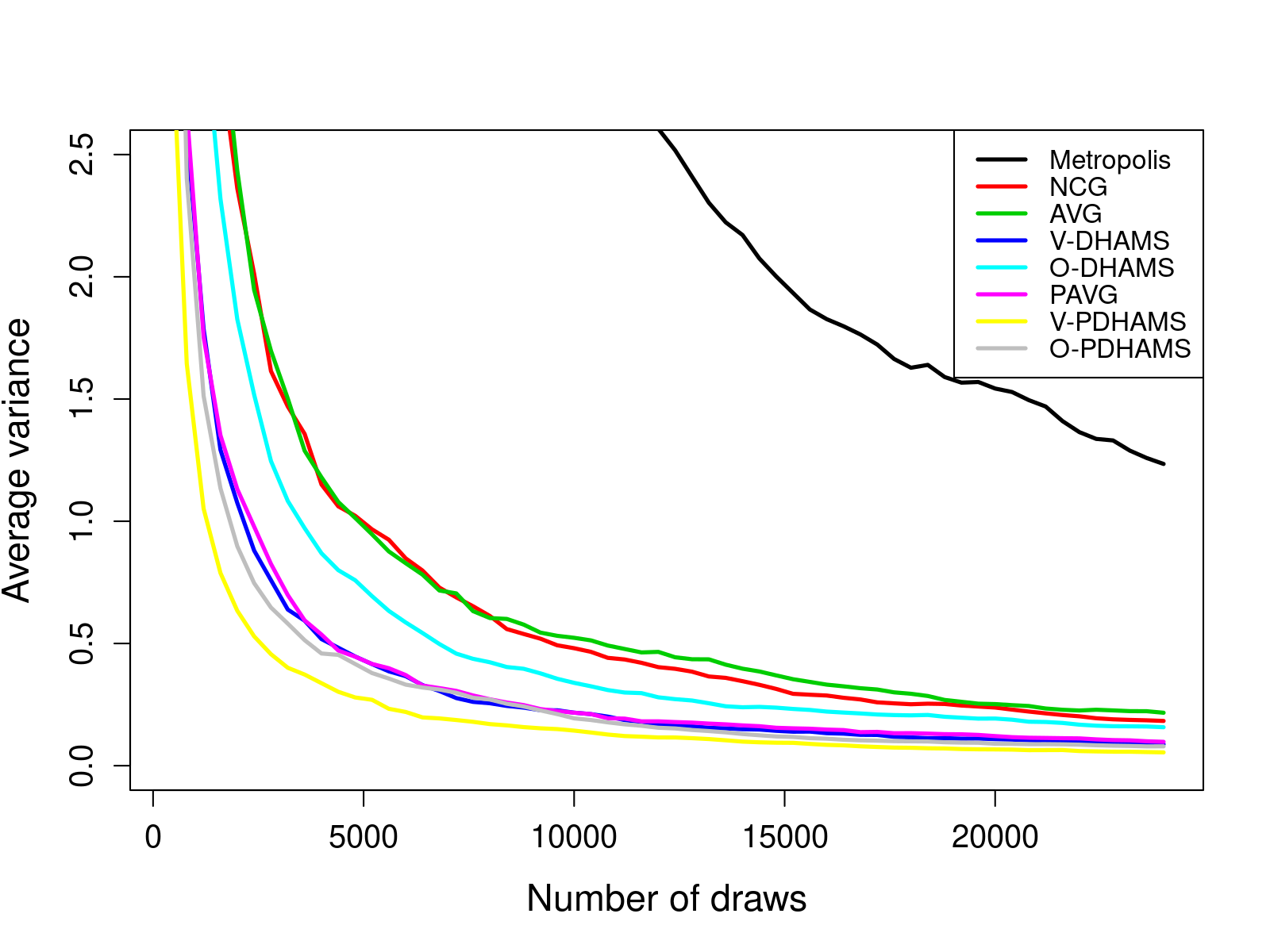}
        \caption{Average variance of $E[s_{i}^{2}]$}
        \label{fig:poly_less_vars12}
    \end{subfigure}

    \par\medskip

    \begin{subfigure}{0.45\textwidth}
        \centering
        \includegraphics[width=\linewidth]{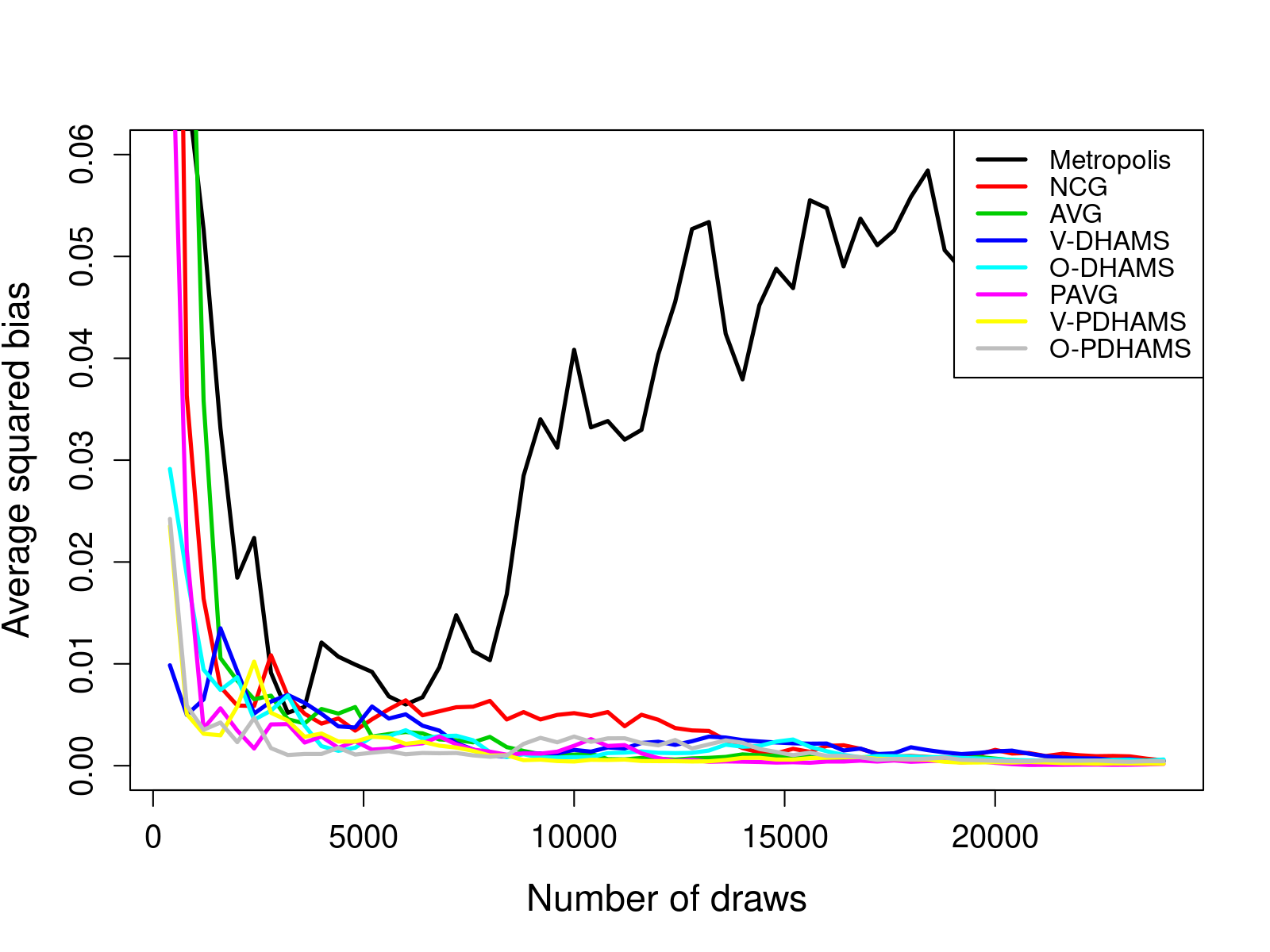}
        \caption{Average squared bias of $E[s_{i_1}s_{i_2}]$}
        \label{fig:poly_less_biass1s2}
    \end{subfigure}
    \hfill
    \begin{subfigure}{0.45\textwidth}
        \centering
        \includegraphics[width=\linewidth]{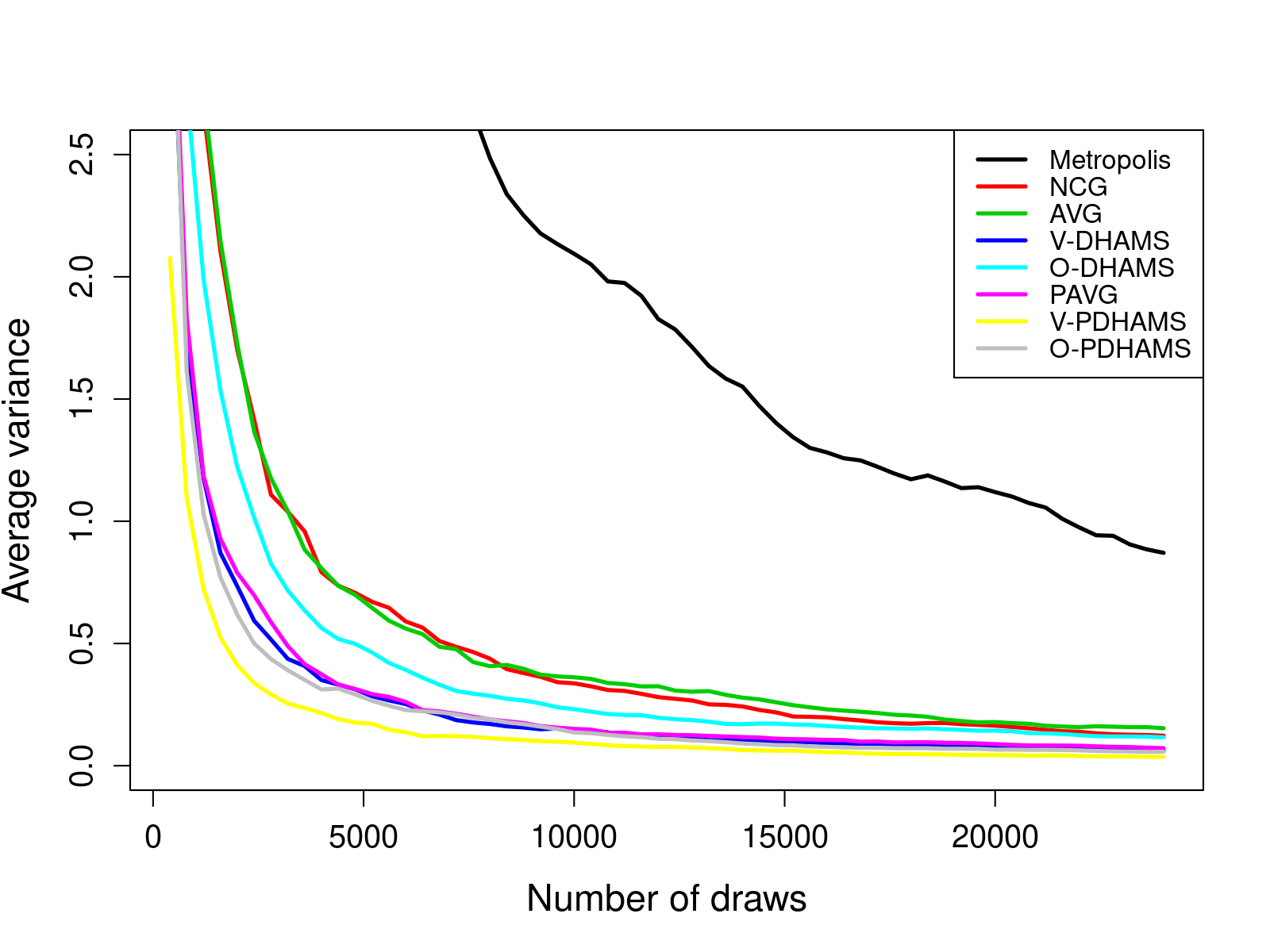}
        \caption{Average variance of $E[s_{i_1}s_{i_2}]$}
        \label{fig:poly_less_vars1s2}
    \end{subfigure}

    \caption{Estimation results for quadratic mixture distribution}
    \label{fig:estimations_poly_less}
\end{figure}

We also report the results of estimating $\E[s_i]$, $\E[s_i^2]$ and $\E[s_{i_1}s_{i_2}]$. The squared bias and variance for $\E[s_i]$, $\E[s_i^2]$ are averaged over all dimensions, and those for $\E[s_{i_1}s_{i_2}]$ are averaged over all index pairs. The averaged results are plotted against the number of draws in Figure \ref{fig:estimations_poly_less}.
The minimum, median, and maximum of ESS across all coordinates, as well as the ESS for the negative potential function $f(s)$ are reported in Table \ref{tab:ess_poly_less}.

The preconditioned samplers have exhibited lower TV-distances and higher ESS compared to first-order methods. V-PDHAMS and O-PDHAMS performs better than PAVG by yielding lower standard deviations in TV-distances and lower variance in estimation problems. The overall performance ranking of all samplers is summarized below:
\[
\text{O-PDHAMS} \approx \text{V-PDHAMS} > \text{PAVG} >\text{V-DHAMS} > \text{O-DHAMS} > \text{NCG} \approx \text{AVG} > \text{Metropolis}.
\]
Additional results (including parameter settings, average acceptance rates, and trace, frequency and auto-correlation plots from individual runs) are presented in Supplement Section \ref{sec:precond_mixture_results}.
\begin{table}[tbp]
\centering
\begin{tabular}{|c|c|c|c|c|}
\hline
    Sampler & ESS Minimum & ESS Median & ESS Maximum & ESS Energy\\
    \hline
    Metropolis & 20.13 & 21.27 & 22.27 & 629.39 \\
    NCG & 115.00 & 118.63 & 121.03 & 5483.64 \\
       AVG & 123.52 & 125.00 & 133.20 & 5363.65 \\
   V-DHAMS & 203.21 & 205.50 & 215.08 & 8148.31 \\
    O-DHAMS & 140.44 & 147.48 & 152.62 & 6432.11 \\
    PAVG & 213.33 & 215.64 & 224.95 & 12894.46 \\
    V-PDHAMS & 316.00 & 329.66 & 338.91 & 10184.64 \\
    O-PDHAMS & 266.45 & 271.63 & 284.62 & 12830.68 \\
    \hline
\end{tabular}
\caption{ESS table for quadratic mixture distribution}
\label{tab:ess_poly_less}
\end{table}

\subsection{Clock Potts Model}
\begin{figure}[tbp]
    \centering
    \includegraphics[width=0.4\linewidth]{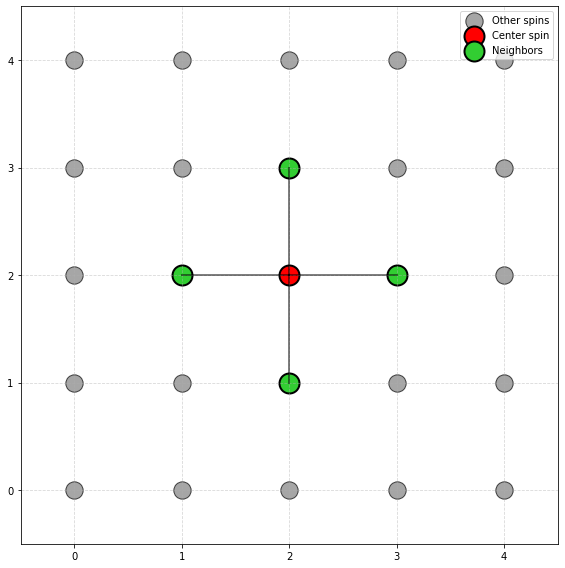}
    \caption{Lattice in clock Potts model}
    \label{fig:potts_lattice}
\end{figure}

The clock Potts model \citep{Potts1952, Villain1975, Suzuki1971} is a classical model in statistical mechanics that describes interacting spins on a crystalline lattice. Each spin $s_i$ resides on a two-dimensional square lattice of size $L \times L$ and takes one of $q$ discrete values from the set $\{0, 1, \dots, q-1\}$. The corresponding negative potential $f(s)$ is defined as
\begin{align}
    f(s) &\propto J \sum_{\langle i, j \rangle} \cos(\theta_i - \theta_j), \nonumber \\
    \theta_i &= \frac{2\pi s_i}{q}, \nonumber
\end{align}
where $\langle i, j \rangle$ denotes that the spins $s_i$ and $s_j$ are the nearest neighbors on the lattice, $J$ is the coupling constant and $T$ is the temperature. The neighbor structure of spins on the lattice is demonstrated in Figure ~\ref{fig:potts_lattice}. In our experiments, we set the lattice size to $L = 20$ with periodic boundaries and choose $q = 7$. The temperature is normalized such that the coupling strength is set to $J=1$ (ferromagnetic) or $J=-1$ (anti-ferromagnetic). For $J=1$, this corresponds to a temperature near the critical point of the 7-state clock model \citep{Li2020ClockModel}. The preconditioning matrix $W$ is calculated using the first calibration method \eqref{eqn:W_calibration1}. Hereafter, we refer to the ferromagnetic and anti-ferromagnetic models as FM and AFM, respectively.

For each sampler, 50 different parameters are searched for tuning for each sampler. We run 50 independent chains, each of length 20,000 after 10,000 burn-in draws for each parameter. The best parameter is then selected by the highest ESS of negative potential function $f(s)$.
After tuning, we conduct 50 independent chains for each sampler using the optimal parameter setting. For each chain, the initial 10,000 draws were discarded as burn-in, and the subsequent 60,000 draws were retained for analysis.

\begin{table}[tbp]
\centering
\begin{tabular}{|l|cc|cc|cc|cc|}
\hline
{Sampler} & \multicolumn{2}{c|}{ESS Minimum} & \multicolumn{2}{c|}{ESS Median} & \multicolumn{2}{c|}{ESS Maximum} & \multicolumn{2}{c|} {ESS Energy} \\
 & FM & AFM & FM & AFM & FM & AFM & FM & AFM\\
 \hline
    Metropolis & 0.20 & 0.21 & 0.44 & 0.42 & 0.86& 0.77 & 0.36 & 0.51 \\
    NCG & 0.29 & 0.52 & 0.62 & 0.94 & 1.19 & 2.06 & 23.85 & 4.61 \\ 
       AVG & 0.13 &  0.21 & 0.30 & 0.44 & 0.60 & 0.97 & 10.65 & 1.12 \\ 
   V-DHAMS & 0.55 & 0.51 & 1.00 & 0.99 & 1.82 & 1.96 & 33.04 & 9.46 \\ 
    O-DHAMS & 0.62 & 0.49 & 1.08 & 1.02 & 2.13 & 2.09 & 27.21 & 6.80 \\ 
    PAVG & 0.38 & 0.46 & 0.72 & 1.02 & 1.29 & 2.96 & 35.50 & 17.85 \\ 
    V-PDHAMS & 0.58 &  0.59 & 1.17 &  1.46 & 2.10 & 3.33 & 40.74 & 20.45 \\
    O-PDHAMS & 0.54 & 0.69 & 1.18 & 1.41 & 2.11& 3.20 & 37. 00 & 22.35 \\
    \hline
\end{tabular}
\caption{ESS table for clock Potts model}
\label{tab:ess_poly_potts}
\end{table}

\begin{figure}[tbp]
\begin{subfigure}[b]{0.32\textwidth}
        \centering
        \includegraphics[width=0.8\linewidth]{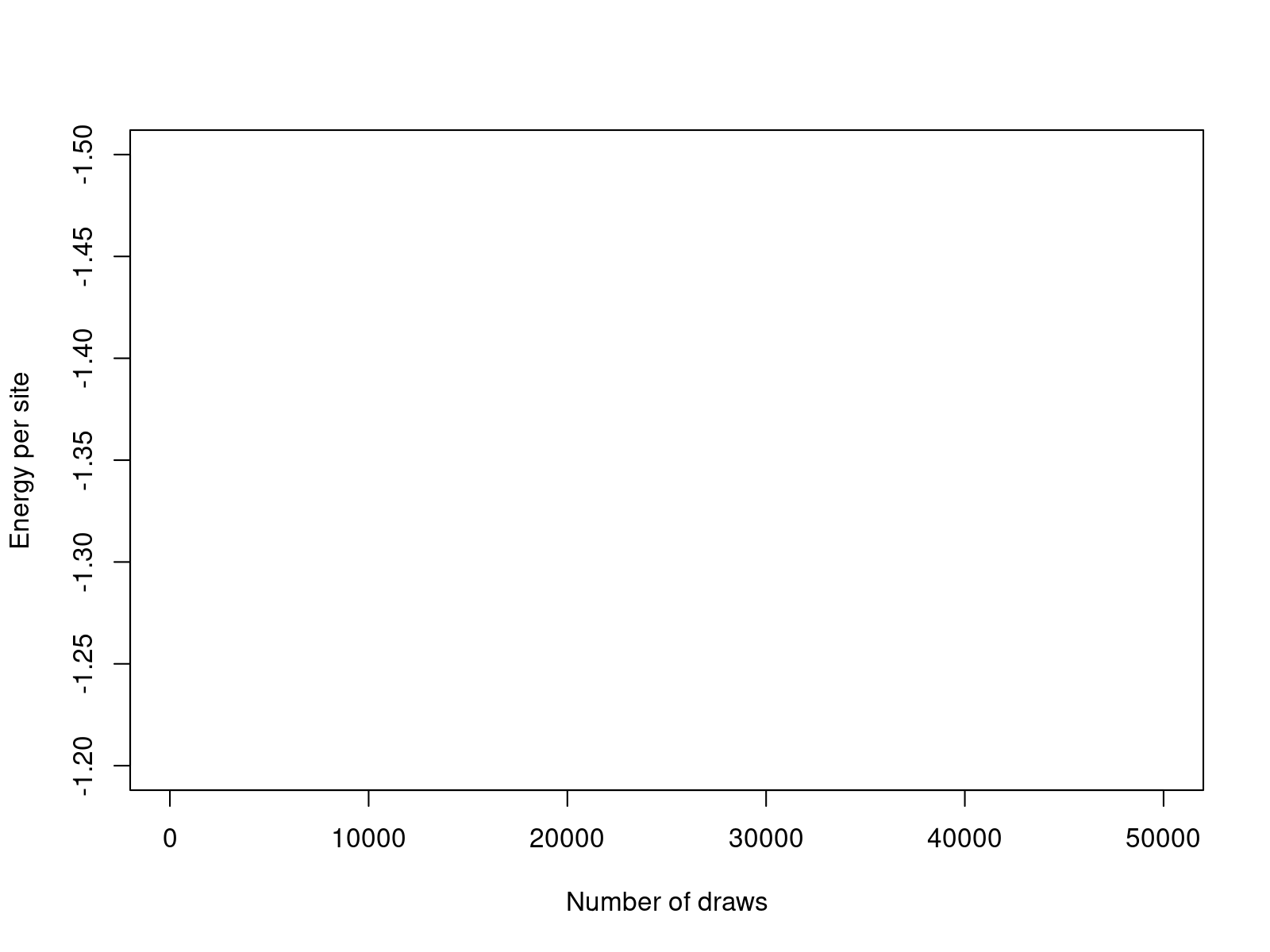}
        \caption{Metropolis}
        \label{fig:trace_precond_Metropolis_potts_ferro}
    \end{subfigure}
     \begin{subfigure}[b]{0.32\textwidth}
        \centering
        \includegraphics[width=0.8\linewidth]{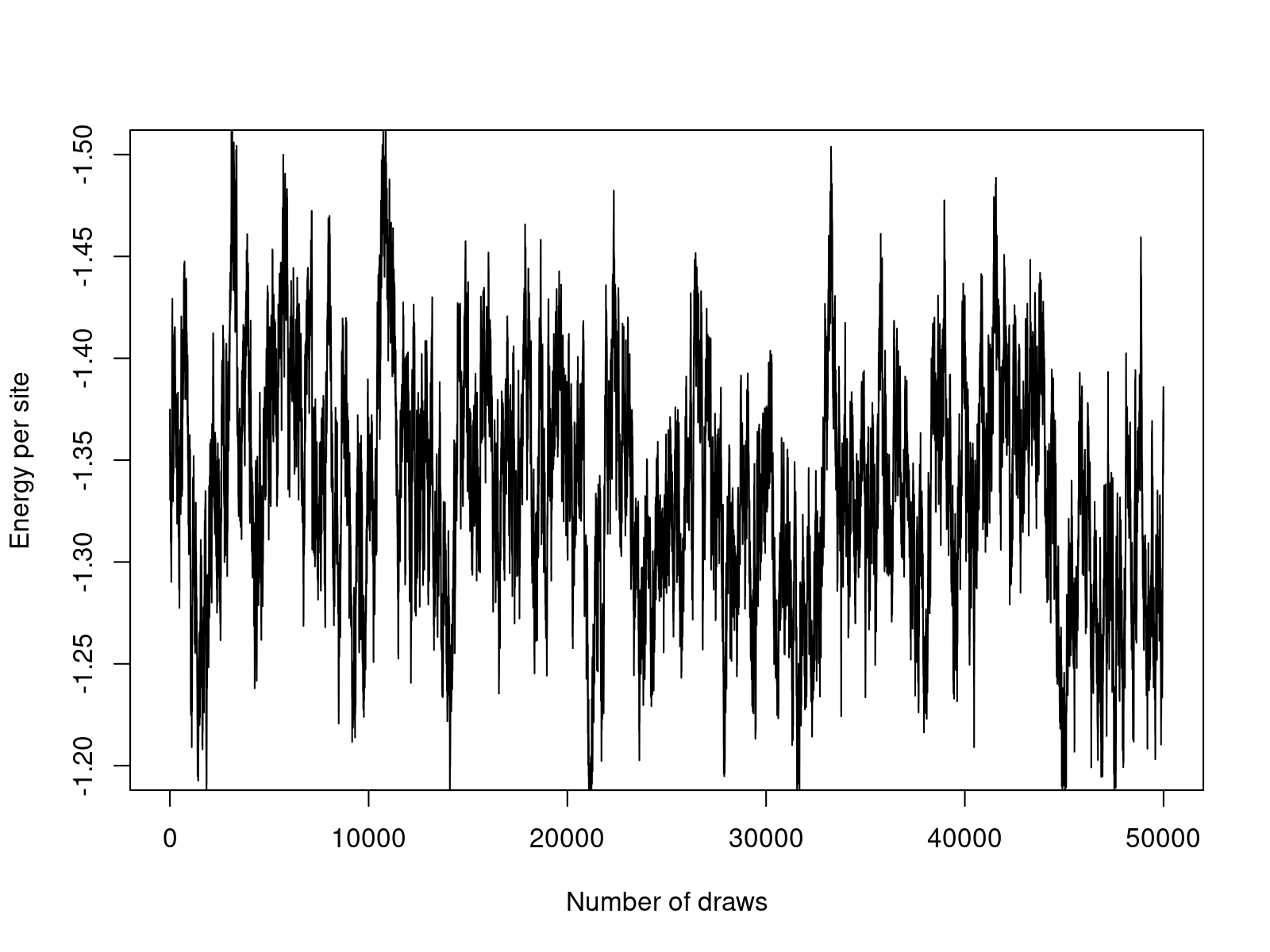}
        \caption{NCG}
        \label{fig:trace_precond_NCG_potts_ferro}
    \end{subfigure}
     \begin{subfigure}[b]{0.32\textwidth}
        \centering
        \includegraphics[width=0.8\linewidth]{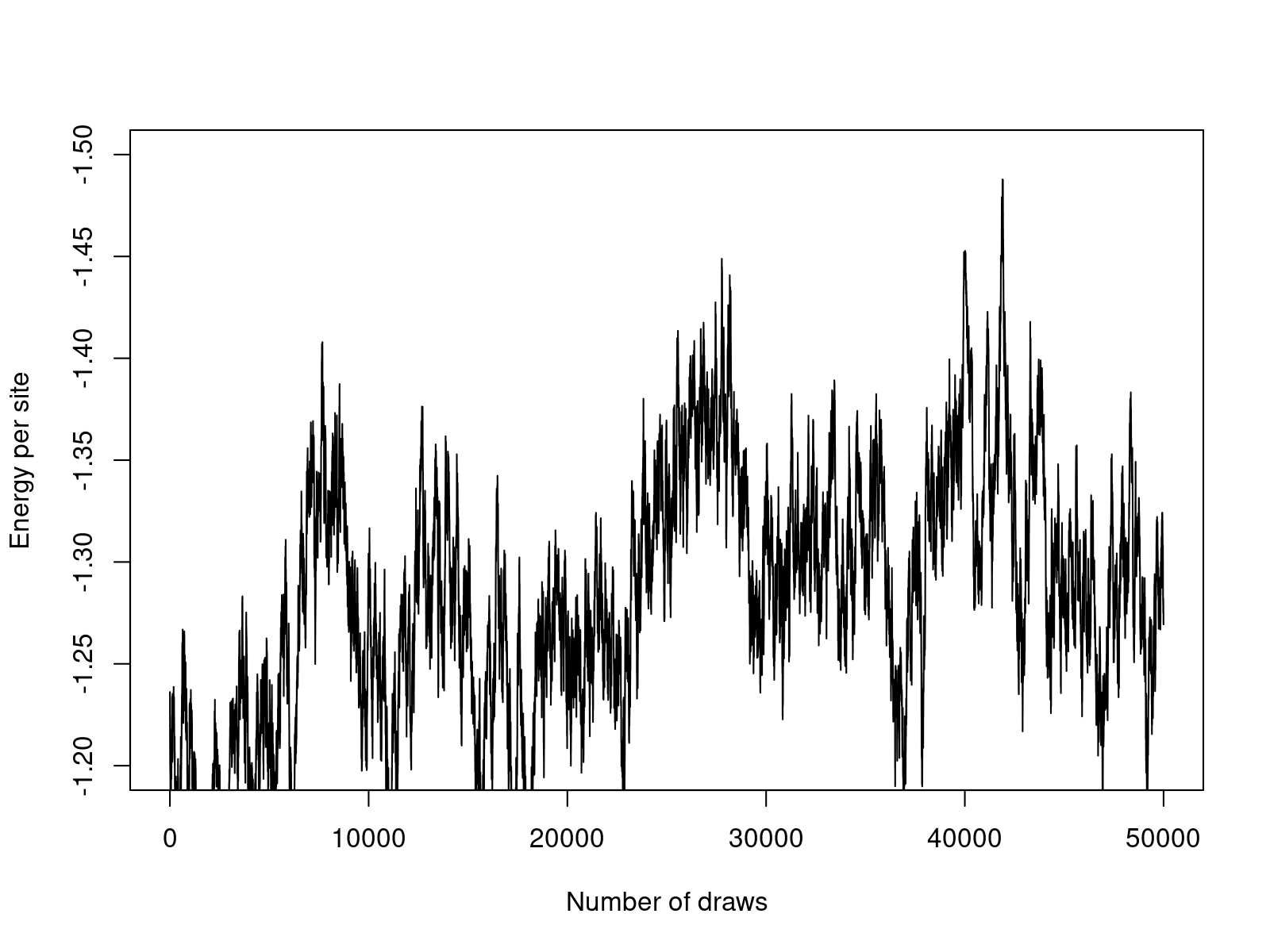}
        \caption{AVG}
        \label{fig:trace_precond_avg_potts_ferro}
    \end{subfigure}
     \begin{subfigure}[b]{0.32\textwidth}
        \centering
        \includegraphics[width=0.8\linewidth]{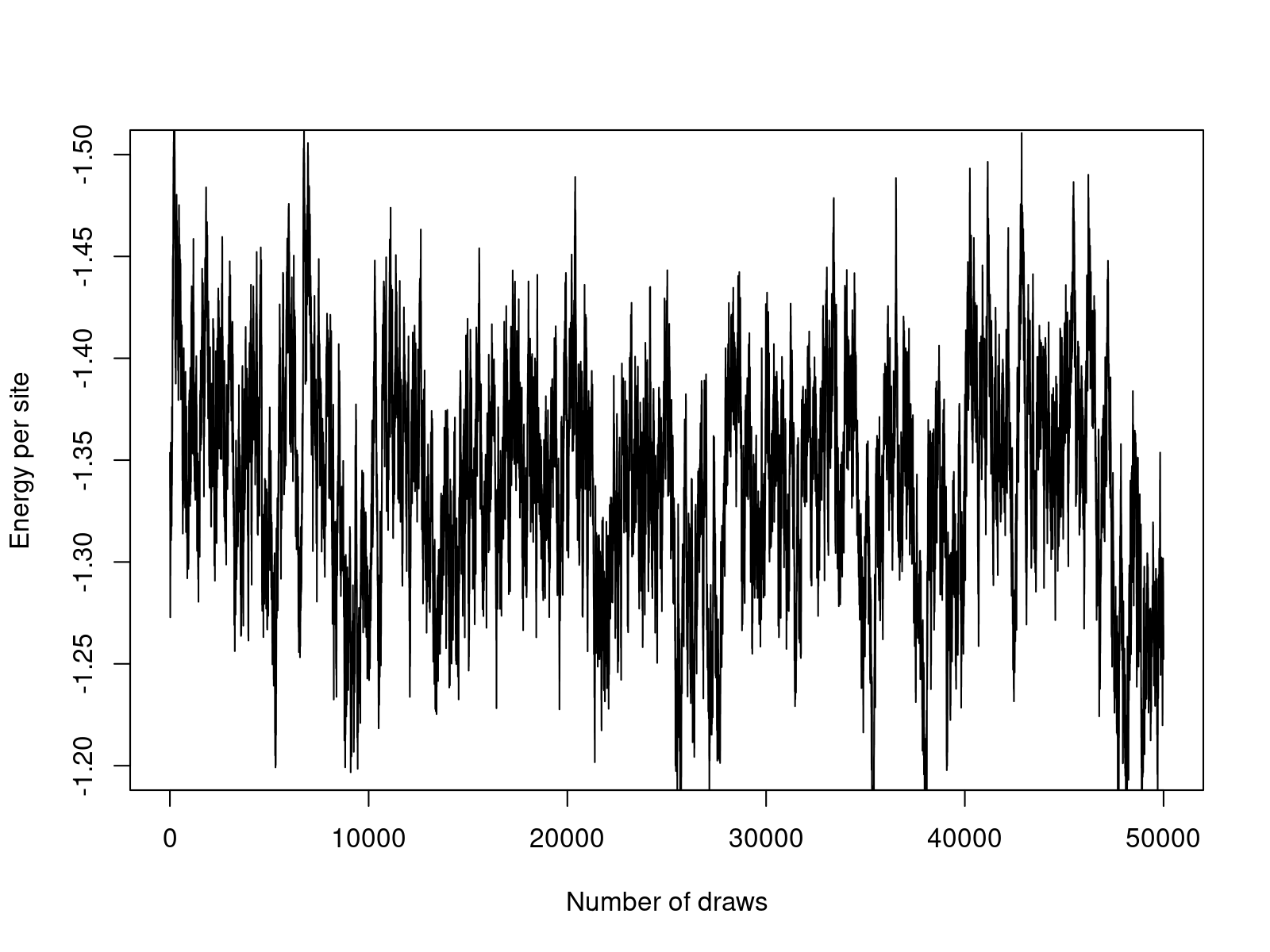}
        \caption{V-DHAMS}
        \label{fig:trace_precond_Hams_potts_ferro}
    \end{subfigure}
     \begin{subfigure}[b]{0.32\textwidth}
        \centering
        \includegraphics[width=0.8\linewidth]{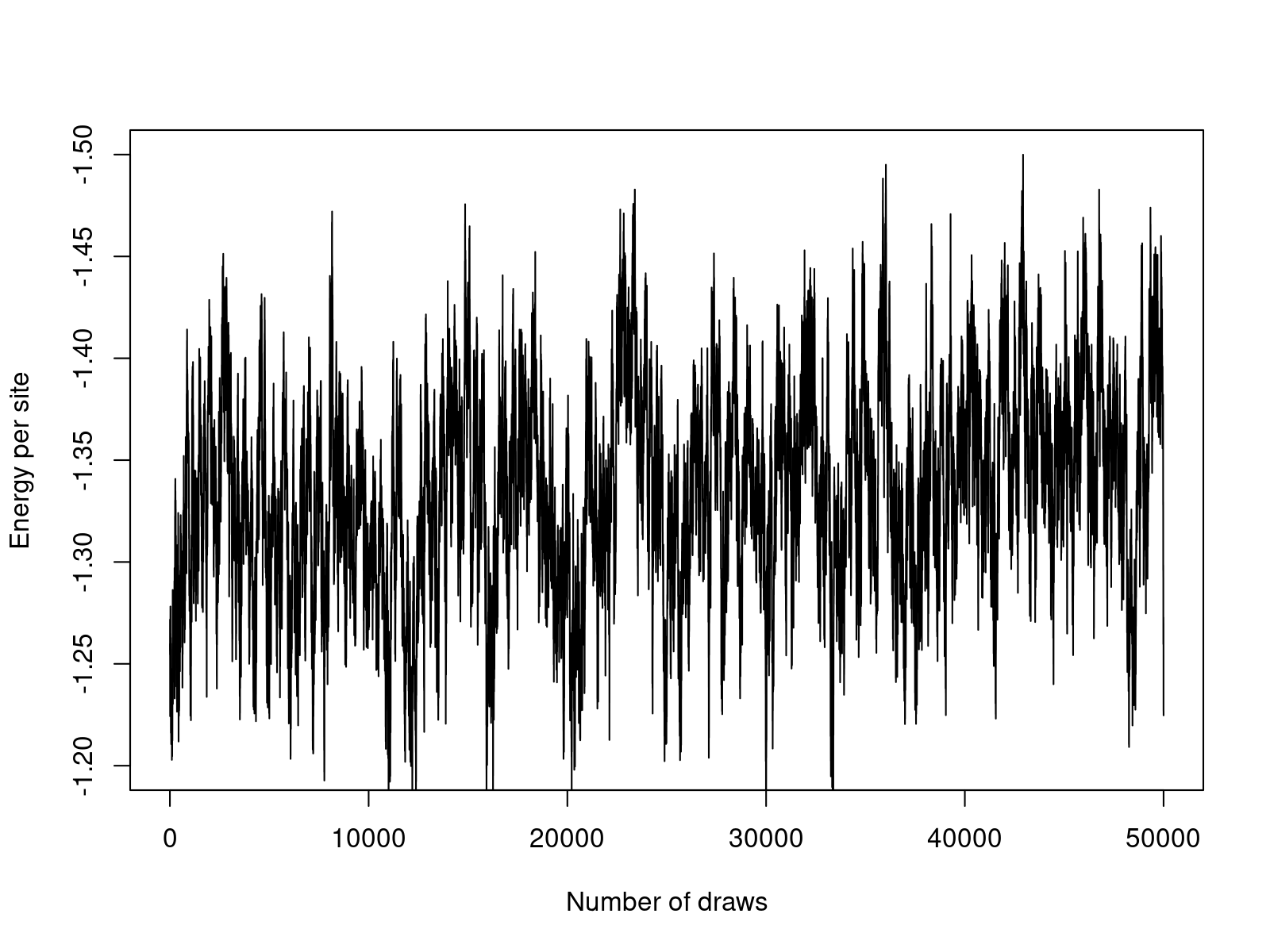}
        \caption{O-DHAMS}
        \label{fig:trace_precond_overhams_potts_ferro}
    \end{subfigure}
    \begin{subfigure}[b]{0.32\textwidth}
        \centering
        \includegraphics[width=0.8\linewidth]{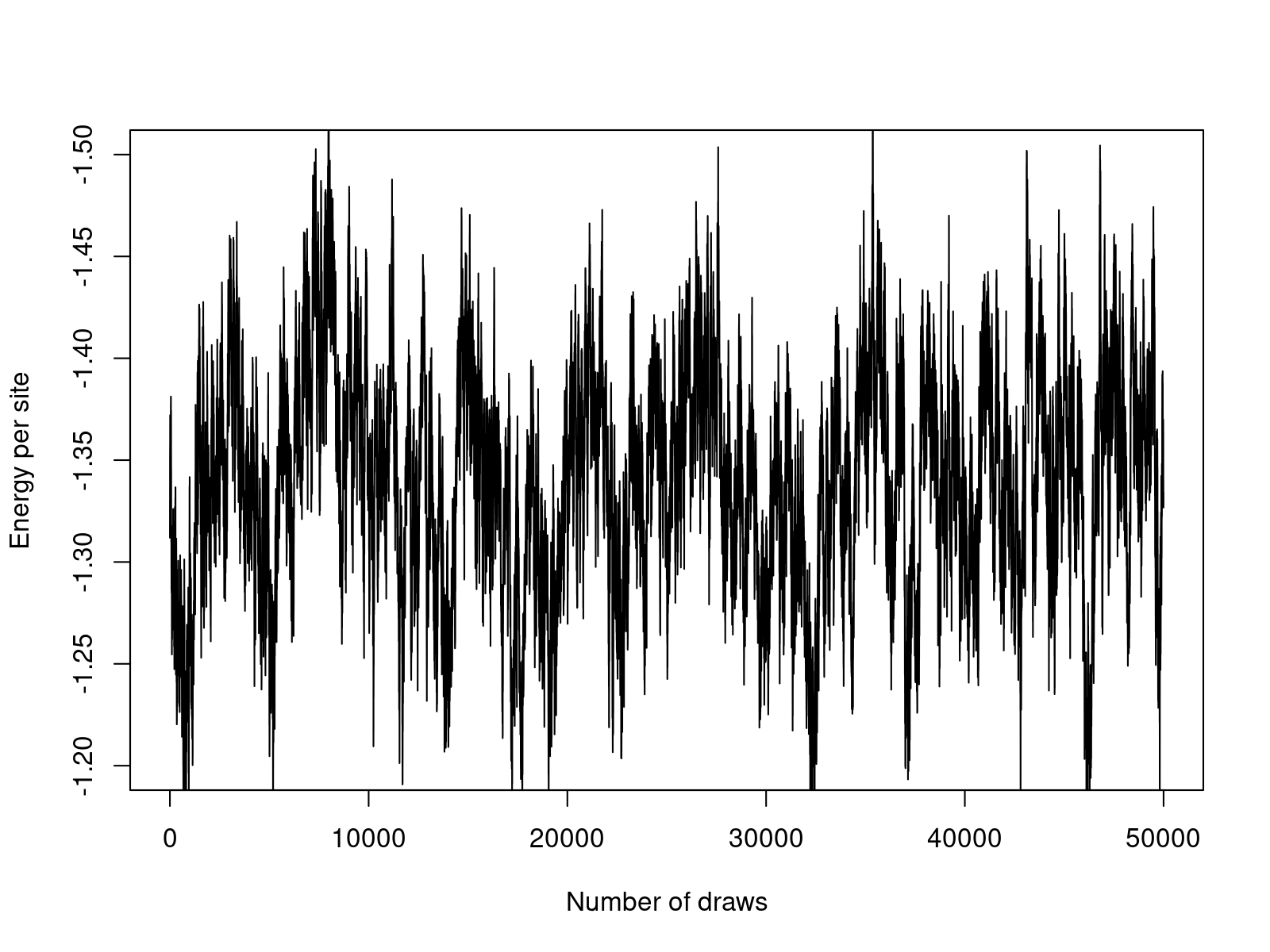}
        \caption{PAVG}
        \label{fig:trace_precond_pavg_potts_ferro}
    \end{subfigure}
    \begin{subfigure}[b]{0.32\textwidth}
        \centering
        \includegraphics[width=0.8\linewidth]{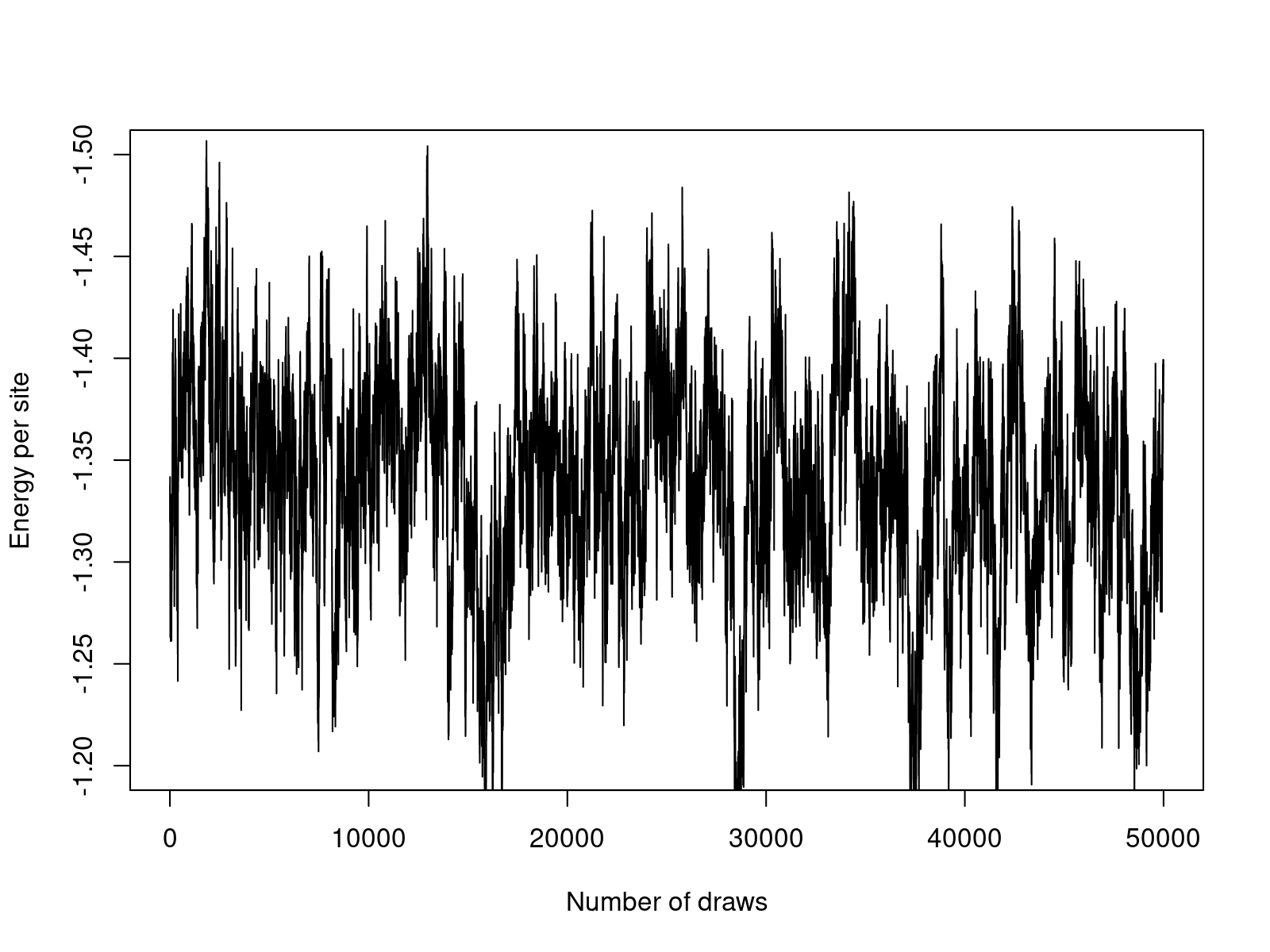}
        \caption{V-PDHAMS}
        \label{fig:trace_precond_vpdhams_potts_ferro}
    \end{subfigure}
    \begin{subfigure}[b]{0.32\textwidth}
        \centering
        \includegraphics[width=0.8\linewidth]{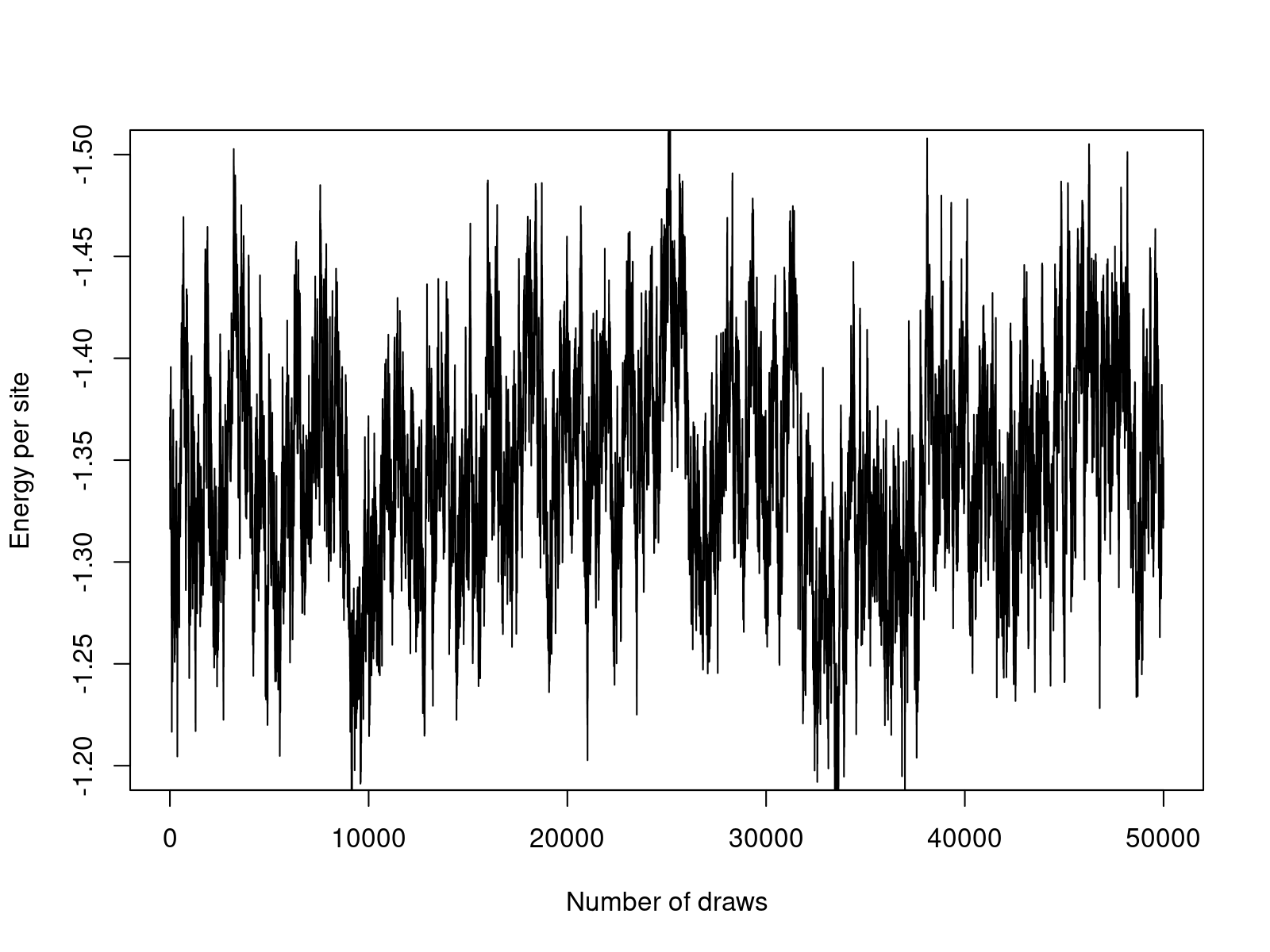}
        \caption{O-PDHAMS}
        \label{fig:trace_precond_opdhams_potts_ferro}
    \end{subfigure}
\caption{Trace plots of energy per site for clock Potts model (Ferromagnetic model)}
\label{fig:trace_precond_plots_potts_ferro}
\end{figure}

\begin{figure}[tbp]
\begin{subfigure}[b]{0.32\textwidth}
        \centering
        \includegraphics[width=0.8\linewidth]{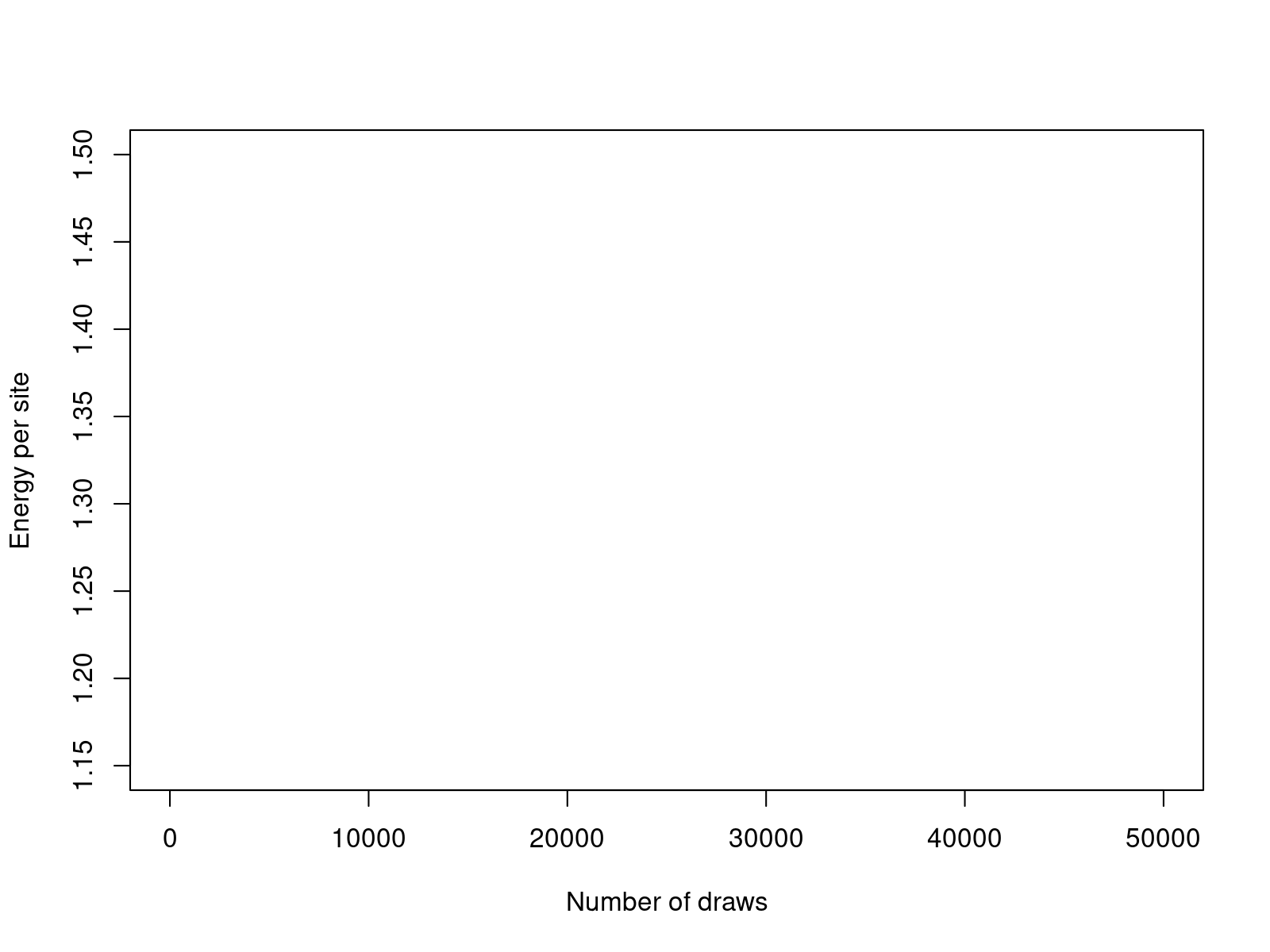}
        \caption{Metropolis}
        \label{fig:trace_precond_Metropolis_potts}
    \end{subfigure}
     \begin{subfigure}[b]{0.32\textwidth}
        \centering
        \includegraphics[width=0.8\linewidth]{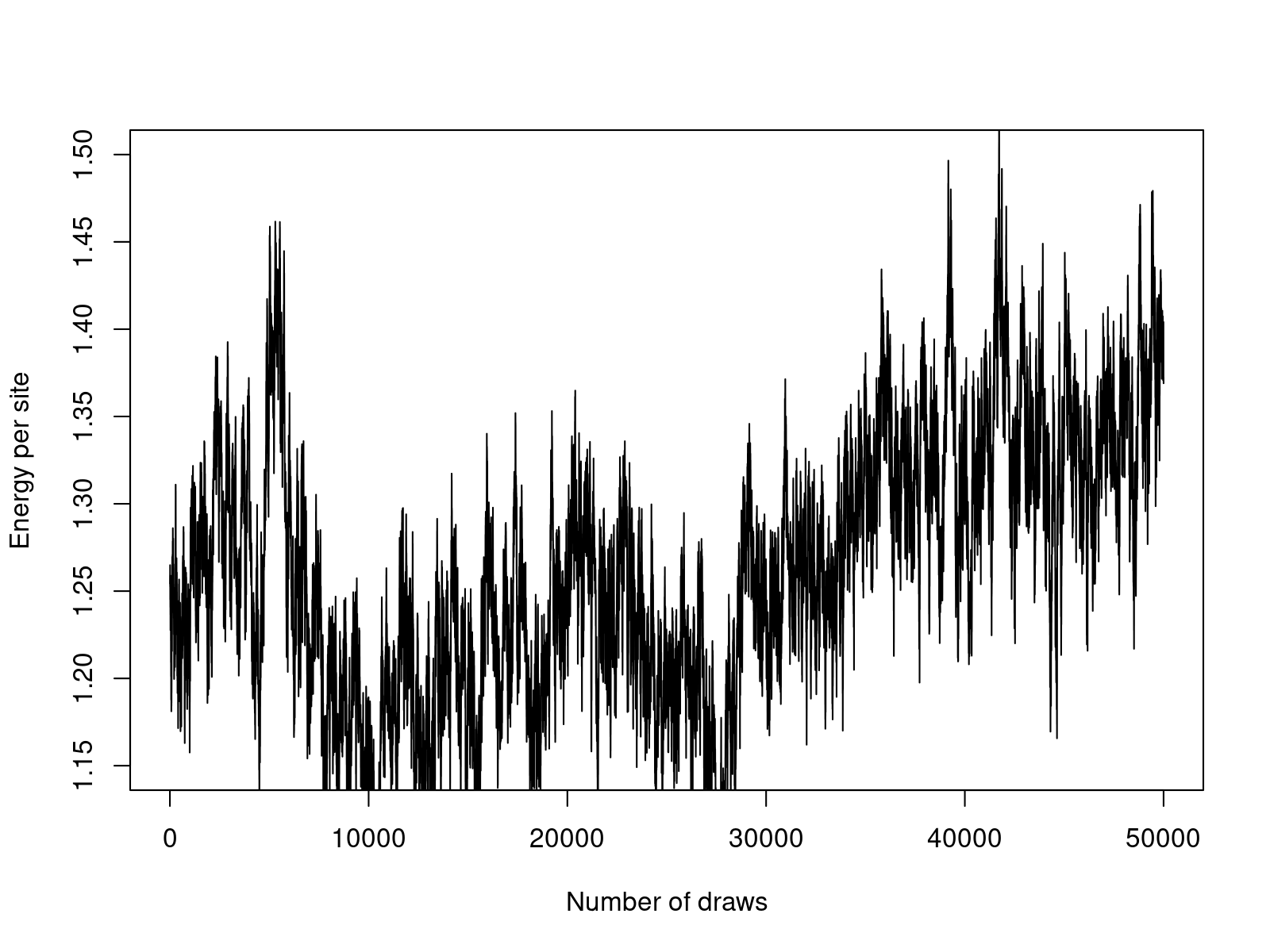}
        \caption{NCG}
        \label{fig:trace_precond_NCG_potts}
    \end{subfigure}
     \begin{subfigure}[b]{0.32\textwidth}
        \centering
        \includegraphics[width=0.8\linewidth]{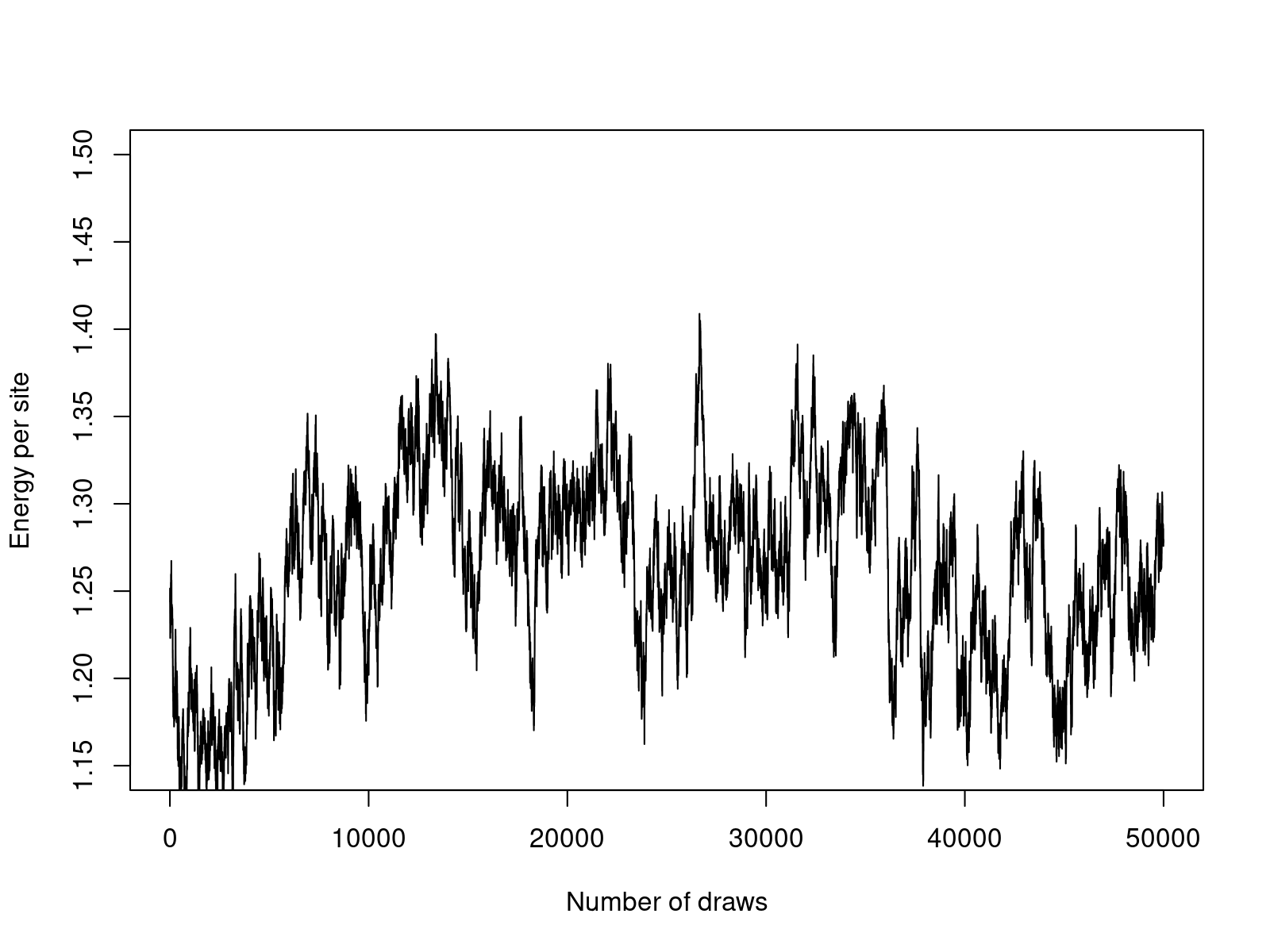}
        \caption{AVG}
        \label{fig:trace_precond_avg_potts}
    \end{subfigure}
     \begin{subfigure}[b]{0.32\textwidth}
        \centering
        \includegraphics[width=0.8\linewidth]{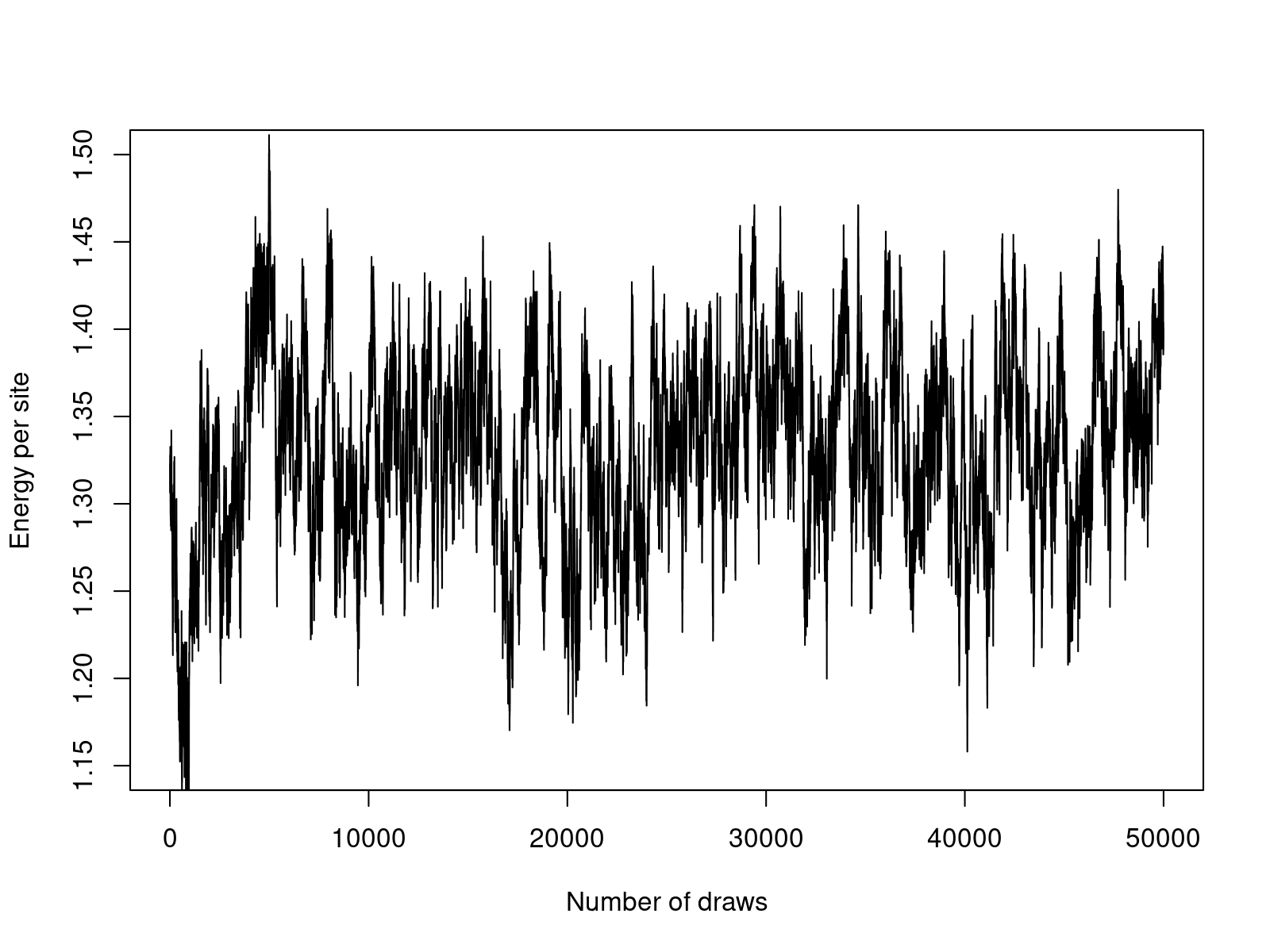}
        \caption{V-DHAMS}
        \label{fig:trace_precond_Hams_potts}
    \end{subfigure}
     \begin{subfigure}[b]{0.32\textwidth}
        \centering
        \includegraphics[width=0.8\linewidth]{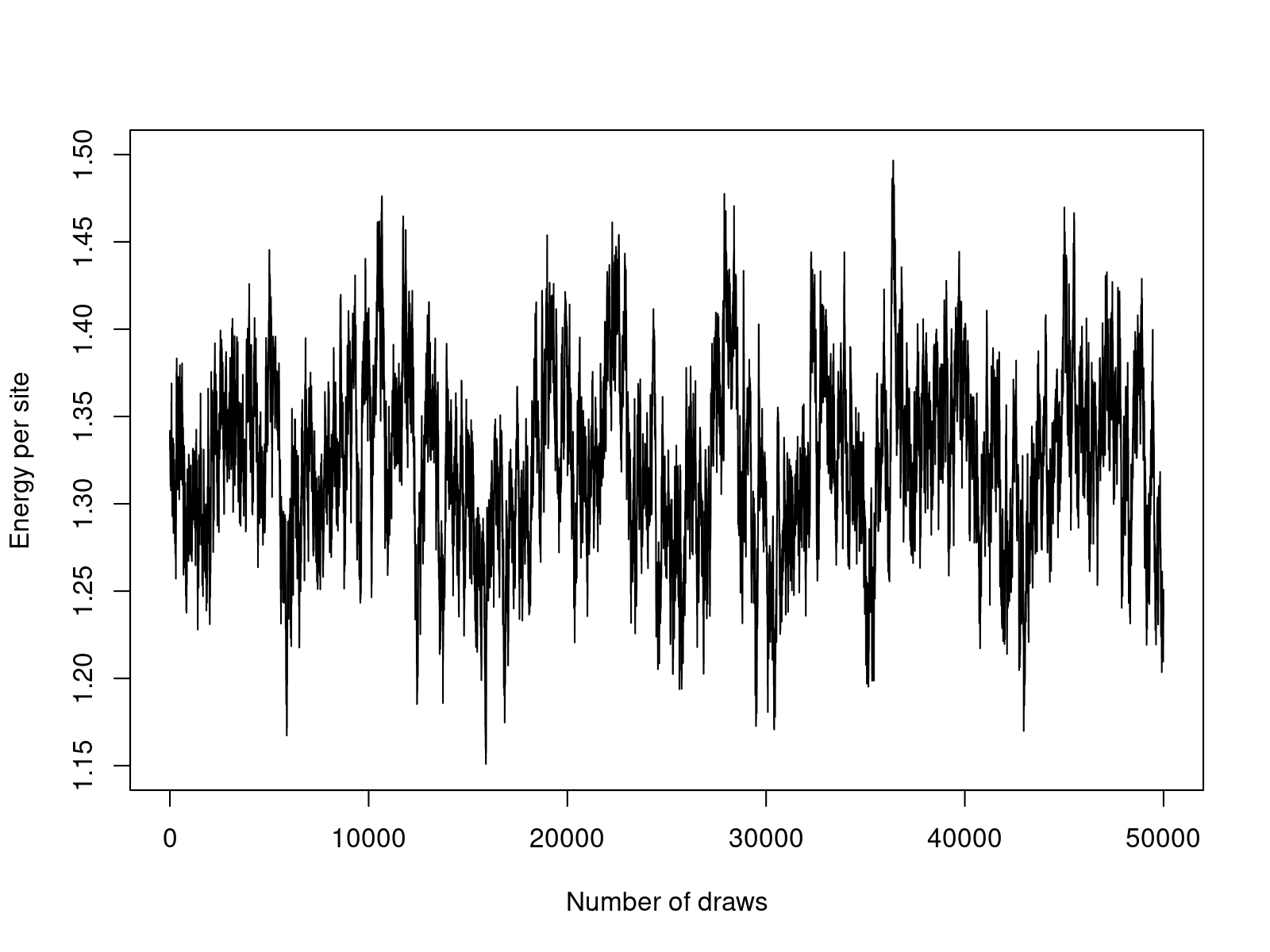}
        \caption{O-DHAMS}
        \label{fig:trace_precond_overhams_potts}
    \end{subfigure}
    \begin{subfigure}[b]{0.32\textwidth}
        \centering
        \includegraphics[width=0.8\linewidth]{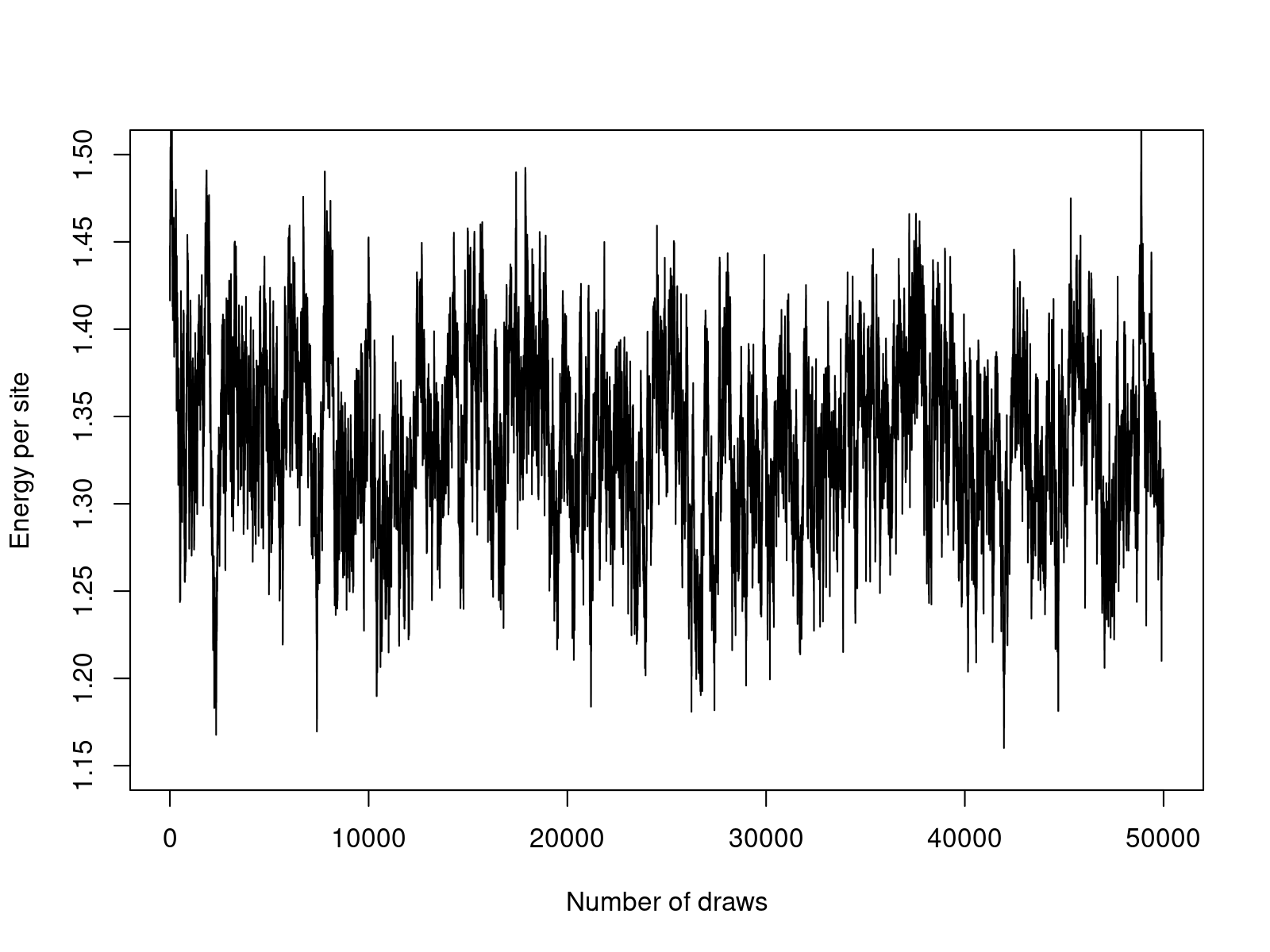}
        \caption{PAVG}
        \label{fig:trace_precond_pavg_potts}
    \end{subfigure}
    \begin{subfigure}[b]{0.32\textwidth}
        \centering
        \includegraphics[width=0.8\linewidth]{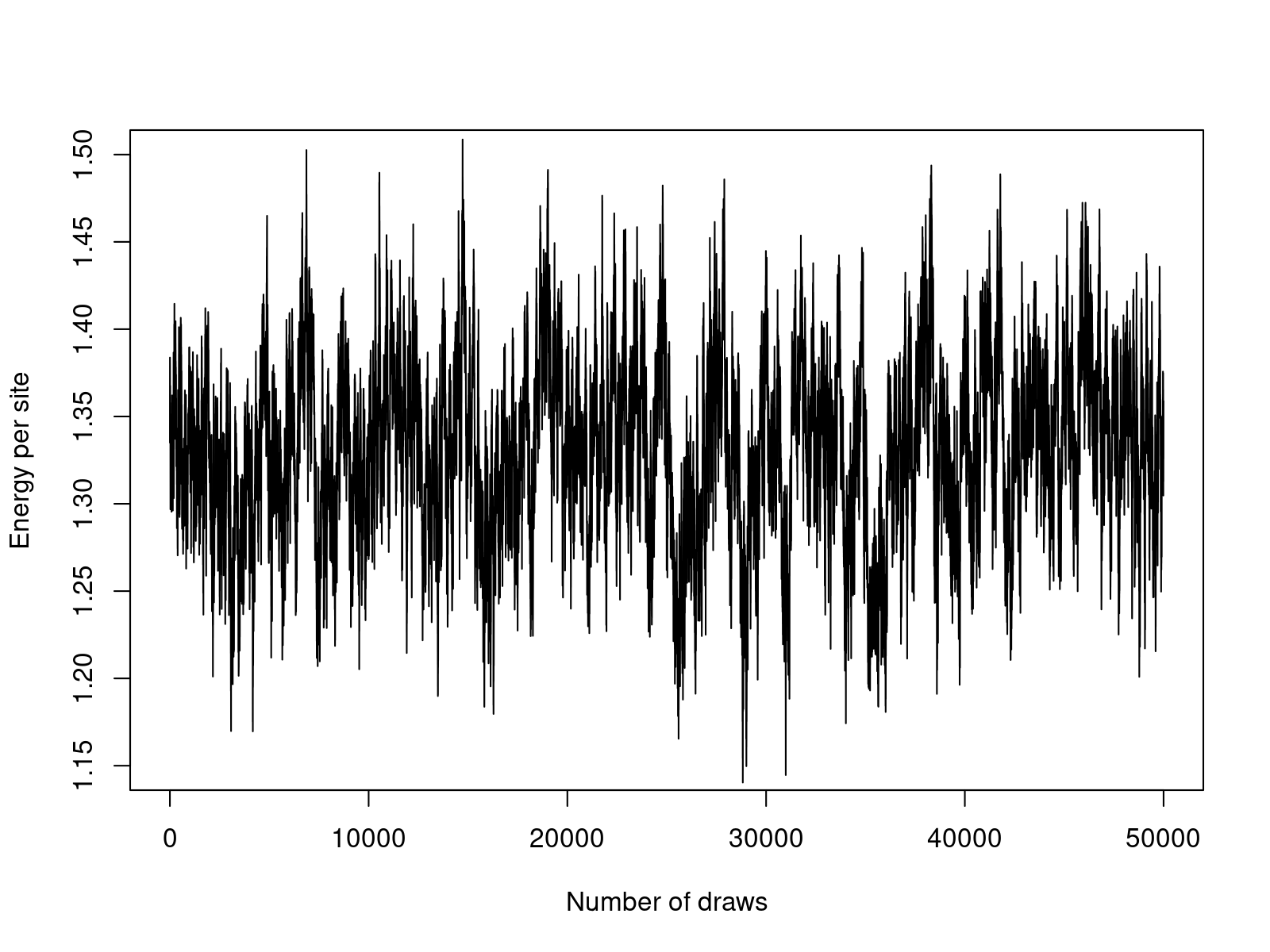}
        \caption{V-PDHAMS}
        \label{fig:trace_precond_vpdhams_potts}
    \end{subfigure}
    \begin{subfigure}[b]{0.32\textwidth}
        \centering
        \includegraphics[width=0.8\linewidth]{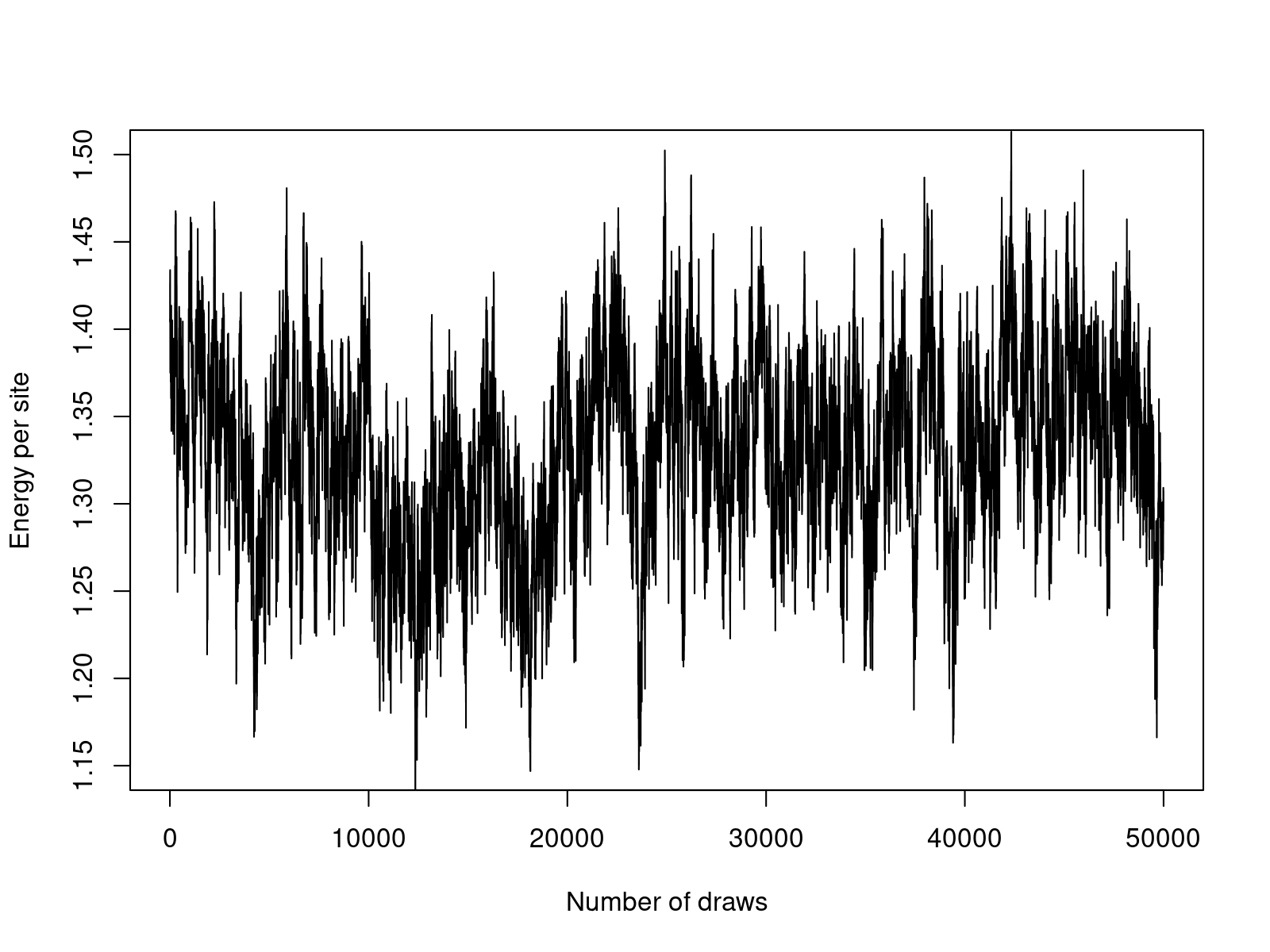}
        \caption{O-PDHAMS}
        \label{fig:trace_precond_opdhams_potts}
    \end{subfigure}
\caption{Trace plots of energy per site for clock Potts model (Anti-ferromagnetic model)}
\label{fig:trace_precond_plots_potts}
\end{figure}

Computation of the exact distribution of the clock Potts model is intractable due to the large lattice size $L$ and large number of different spin values $q$, making comparison via TV-distance or estimation accuracy of quantities impractical. We only report the minimum, median, and maximum of ESS across all spins, as well as the ESS for the negative potential function $f(s)$ in Table \ref{tab:ess_poly_potts}. We also present the trace plots of energy per site, ($f(s)/(-JL^2)$) from a single chain in Figure~\ref{fig:trace_precond_plots_potts_ferro} for the ferromagnetic model and in Figure~\ref{fig:trace_precond_plots_potts} for the anti-ferromagnetic model.
Additional results (including parameter settings, acceptance rates, and auto-correlation plots from individual runs) are presented in Supplement Section \ref{sec:precond_potts_results}.

The second-order samplers exhibit significantly higher ESS compared to first-order methods, particularly on the Anti-ferromagnetic model. Moreover, both PDHAMS samplers outperform PAVG. The trace plots of the second-order samplers demonstrate reduced autocorrelation and more efficient exploration, indicating improved mixing. Overall, the ranking of performance in simulating the clock Potts model is as follows,
\[
\text{O-PDHAMS} \approx \text{V-PDHAMS} > \text{PAVG} > \text{V-DHAMS} > \text{O-DHAMS} > \text{NCG} > \text{AVG} > \text{Metropolis}.
\]

\section{Conclusion}

In this work, we extend the DHAMS sampler \citep{Zhou2025Dhams} by incorporating a second-order, quadratic approximation of the potential function,
as used in \cite{Rhodes2022GradientMC} and \cite{sun2023anyscale}. The resulting PDHAMS sampler preserves the generalized detailed balance property of DHAMS while extending the rejection-free property to arbitrary distributions with quadratic potentials. Our development also establishes various novel connections between discrete and continuous sampling methods. Our numerical experiments demonstrate PDHAMS's superior performance compared with first-order samplers and second-order PAVG. An important direction for future research involves developing efficient calibration methods for the preconditioning matrix $W$, particularly through online adaptation schemes.

\bibliographystyle{apalike}
\bibliography{hams, precond}

\clearpage
\appendix

\begin{center}
{\Large Supplementary Material for}\\
{\Large ``Preconditioned Discrete-HAMS: A Second-order Irreversible Discrete Sampler''} \\
{{Yuze Zhou} \text{and} {Zhiqiang Tan}}

\end{center}

\section{Continuous Analogues}

\subsection{Continuous Analogue for PAVG}\label{sec:PAVG_cont}
Recall the PAVG sampler in \eqref{eqn:PAVG_var_s}--\eqref{eqn:PAVG_var_acc}
\begin{align}
    z_{t} & = s_{t}+(L^{\T})^{-1}Z, \quad Z \sim \mathcal{N}(0, I), \nonumber\\
   s^{*} & \sim Q(s|z_{t}, s_{t})  \propto \exp(-\frac{1}{2}s^{\T}Ds +(\nabla f(s_{t})-Ws_{t} +(W+D)z_t)^{\T}s). \nonumber
\end{align}
When $s$ is a continuous variable, the auxiliary variable $z_{t}$ can be integrated out similarly as in a marginalized sampler \citep{Titsias2018auxavg}.
The marginalized proposal is
\begin{align}
        s^{*} \sim Q(s|s_{t}) &= \int Q(s|z_{t},s_{t})\pi(z_{t}|s_{t}) dz_{t} \nonumber\\
        & \propto \int \exp(-\frac{1}{2}s^{\T}Ds +(\nabla f(s_{t})-Ws_{t} +(W+D)z_t)^{\T}s) \mathcal{N}(z_{t};s_{t}, (W+D)^{-1})dz_{t} \nonumber \\
        & \propto \exp(-\frac{1}{2}(s-s_t -D^{-1}\nabla f(s_t))^{\T}(2D^{-1}+D^{-1}WD^{-1})^{-1}(s-s_t - D^{-1}\nabla f(s_t))). \nonumber
\end{align}
The new proposal can be equivalently written as
\begin{align}
    s^{*} = s_{t}+ D^{-1}\nabla f(s_t) + Z, \quad Z \sim \mathcal{N}(0, 2D^{-1}+D^{-1}WD^{-1}). \label{eq:PAVG_cont}
\end{align}
It can be directly verified that if $W$ is negative definite and $s_t \sim \mathcal{N}(0, -W^{-1})$, then the proposal from \eqref{eq:PAVG_cont}
gives $s^{*} \sim \mathcal{N}(0, -W^{-1})$, following the same distribution. If we accept $s_{t+1} = s^*$ with the Metropolis--Hastings probability and otherwise set $s_{t+1} = s_t$, then the resulting algorithm is rejection-free when the target distribution is $\mathcal{N}(0, -W^{-1})$. Furthermore, when $W = -I$ and $D = \frac{1}{1 - \sqrt{1 - \epsilon^2}} I$, the proposal in \eqref{eq:PAVG_cont} simplifies to
\begin{align*}
    s^{*} = s_{t} + \frac{\epsilon^{2}}{1 + \sqrt{1 - \epsilon^2}} \nabla f(s_{t}) + \epsilon Z, \quad Z \sim \mathcal{N}(0, I),
\end{align*}
which is the proposal used in modified MALA \citep{Song2023hams}.

\subsection{Continuous Analogue for V-PDHAMS}\label{sec:V-PDHAMS_cont}
Recall the V-PDHAMS proposal via the momentum approach for constructing the auxiliary variable in \eqref{eqn:phams_mom_s}--\eqref{eqn:phams_mom_u}:
\begin{align*}
    s^* & \sim \exp(-\frac{1}{2}s^\T Ds + [\nabla f(s_t) +Ds_t +(W+D)v_t]^{\T}s), \\
    v^* &= v_t +s_t -s^*.
\end{align*}
When $s$ is a continuous variable, the proposal above can be written as
\begin{align*}
    s^* &= s_t +D^{-1}\nabla f(s_t) +D^{-1}(W+D)v_t +Z, \quad Z \sim \mathcal{N}(0, D^{-1}), \\
    v^* &= v_t +s_t -s^*.
\end{align*}
By relating the momentum variable $v$ in \eqref{eqn:phams_mom1} into standard Gaussian $u$ such that $u = L^{\T} v$, and replacing $u^* =  L^{\T} v^*$,  $u_t = L^{\T} v_t$, the proposal in term of $(s, u)$ is
\begin{align*}
    s^* &= s_t +D^{-1} \nabla f(s_t) +D^{-1}(W+D)v_t + Z,  \\
    u^* &= u_t - L^\T D^{-1} \nabla f(s_t) - L^\T D^{-1} L u_t - L^\T Z,
\end{align*}
where $Z \sim \mathcal{N}(0, D^{-1})$ is drawn independently. In a matrix form, the proposal above can be rewritten as
\begin{align}
\begin{pmatrix}
s^{*} \\
u^{*}
\end{pmatrix}
=
\begin{pmatrix}
s_{t} \\
u_{t}
\end{pmatrix}
- A
\begin{pmatrix}
- \nabla f(s_{t}) \\
u_{t}
\end{pmatrix}
+
\begin{pmatrix}
Z \\
-L^{\T}Z
\end{pmatrix},
\label{eq:vpdhams-cont-proposal}
\end{align}
where
\begin{align}
A =
\begin{pmatrix}
D^{-1} & -D^{-1}L \\
-L^{\T}D^{-1} & L^{\T}D^{-1}L
\end{pmatrix}.
\label{eq:vpdhams-matrix}
\end{align}

We make two important remarks about the proposals generated from \eqref{eq:vpdhams-cont-proposal} and \eqref{eq:vpdhams-matrix}. First, if $W$ is negative definite, and $s_t \sim \mathcal{N}(0, -W^{-1})$ and is independent of $u_t \sim \mathcal{N}(0,I)$, then $s^* \sim \mathcal{N}(0, -W^{-1})$ and is also independent of $u^* \sim \mathcal{N}(0,I)$. If the proposal \eqref{eq:vpdhams-cont-proposal} is accepted with the Metropolis--Hastings probability, then the corresponding sampling algorithm is rejection-free for the following Gaussian target distribution:
\begin{align*}
    \pi(s, u) \sim \mathcal{N}\left(
    \begin{pmatrix}
        0\\
        0
    \end{pmatrix},
    \begin{pmatrix}
        -W^{-1} & 0 \\
        0 & I
    \end{pmatrix}
    \right).
\end{align*}
Furthermore, the rejection-free property is preserved if momentum negation and gradient correction \eqref{eqn:p-mod1} is applied and the new proposal is accepted using generalized Metropolis--Hastings probability. Second, if $W = -I$, then matrix $A$ in \eqref{eq:vpdhams-matrix} can be verified to satisfy the following:
\begin{align*}
    2A-A^2
    &= \begin{pmatrix}
        D^{-1} & -D^{-1}L \\
        -L^{\T}D^{-1} &  L^{\T}D^{-1}L
    \end{pmatrix} \\
    &= \var\left(
    \begin{pmatrix}
        Z \\
        -L^\T Z
    \end{pmatrix}
    \right),
\end{align*}
which indicates that the proposal in \eqref{eq:vpdhams-cont-proposal} becomes a HAMS proposal before momentum negation and gradient correction. Specifically, if we choose $D = \frac{1+\delta^2}{\delta^2}I$ for some $\delta >0$ and $L = (W+D)^{1/2}$, then matrix $A$ in \eqref{eq:vpdhams-matrix} is
\begin{align*}
    \begin{pmatrix}
        \frac{\delta^2}{1+\delta^2} I & -\frac{\delta}{1+\delta^2} I \\
         -\frac{\delta}{1+\delta^2} I & \frac{1}{1+\delta^2} I
    \end{pmatrix} ,
\end{align*}
which is the matrix corresponding to HAMS-A before applying Gaussian over-relaxation and momentum negation.

\section{Properties of PAVG}
\subsection{Proof of Equivalency of Different Auxiliary Variable Schemes} \label{prop_pavg_id}
To show the equivalence of the mean approach and the variance approach, we first express the auxiliary variable $z_t$ in terms of the current state $s_t$ and the Gaussian noise $Z \sim \mathcal{N}(0, I)$ in both the proposal and the acceptance probability.

\textit{Preconditioning by mean.}\; In the first approach  \eqref{eqn:PAVG_mean_z}--\eqref{eqn:PAVG_mean_acc}, the auxiliary variable can be written as $z_t = L^{\T}s_t + Z$. Substituting this into the forward proposal, backward proposal, and conditional distributions, we obtain the following:
\begin{align}
    Q(s^{*}|z_{t}; s_{t}) & \propto \exp(-\frac{1}{2}s^{*\T}Ds^{*} +(\nabla f(s_{t})-Ws_{t} +Lz_{t})^{\T}s^{*}) \nonumber \\
    & \propto \exp(-\frac{1}{2}s^{*\T}Ds^{*} +(\nabla f(s_{t})-Ws_{t} +(W+D)s_{t}+LZ)^{\T}s^{*}), \nonumber
\end{align}
\begin{align}
    \pi(z_{t}|s^{*}) & \propto \mathcal{N}(z_t|L^{\T}s^*, I) \nonumber\\
    &\propto \exp(-\frac{1}{2}\|z_{t}-L^{\T}s^{*}\|_{2}^{2}) \nonumber \\
    & \propto \exp(-\frac{1}{2}\|L^{\T}(s_{t}-s^{*})\|_{2}^{2} -\frac{1}{2}\|Z\|_{2}^{2} -\frac{1}{2}(s_{t}-s^{*})^{\T}LZ), \nonumber
\end{align}
\begin{align}
    Q(s_{t}|z_{t}; s^{*}) & \propto \exp(-\frac{1}{2}s_{t}^{\T}Ds_{t} +(\nabla f(s^{*})-Ws^{*} +Lz_{t})^{\T}s_{t}) \nonumber \\
    & \propto \exp(-\frac{1}{2}s^{\T}_{t}Ds_{t} +(\nabla f(s^{*})-Ws^{*} +(W+D)s_{t}+LZ)^{\T}s_{t}). \nonumber
\end{align}

\textit{Preconditioning by variance.} In the second approach  \eqref{eqn:PAVG_var_z}--\eqref{eqn:PAVG_var_acc}, the auxiliary variable is $z_{t} = s_{t} + (L^{\T})^{-1}Z$,
and the corresponding terms are
\begin{align}
    Q(s^{*}|z_{t}; s_{t}) & \propto \exp(-\frac{1}{2}s^{*\T}Ds^{*} +(\nabla f(s_{t})-Ws_{t} +(W+D)z_{t})^{\T}s^{*}) \nonumber \\
    & \propto \exp(-\frac{1}{2}s^{*\T}Ds^{*} +(\nabla f(s_{t})-Ws_{t} +(W+D)s_{t}+LZ)^{\T}s^{*}), \nonumber
\end{align}
\begin{align}
    \pi(z_{t}|s^{*}) & \propto \mathcal{N}(z_t|s^*,  (W+D)^{-1} ) \nonumber \\
    &\propto \exp(-\frac{1}{2}\|L^{\T}(z_{t}-s^{*})\|_{2}^{2}) \nonumber \\
    & \propto \exp(-\frac{1}{2}\|L^{\T}(s_{t}-s^{*})\|_{2}^{2} -\frac{1}{2}\|Z\|_{2}^{2} -\frac{1}{2}(s_{t}-s^{*})^{\T}LZ), \nonumber
\end{align}
\begin{align}
    Q(s_{t}|z_{t}; s^{*}) & \propto \exp(-\frac{1}{2}s_{t}^{\T}Ds_{t} +(\nabla f(s^{*})-Ws^{*} +(W+D)z_{t})^{\T}s_{t}) \nonumber \\
    & \propto \exp(-\frac{1}{2}s^{\T}_{t}Ds_{t} +(\nabla f(s^{*})-Ws^{*} +(W+D)s_{t}+LZ)^{\T}s_{t}). \nonumber
\end{align}

All relevant terms above are identical between the two approaches in terms of $s_t$ and $Z$, which establishes their equivalence. As mentioned in the main paper, $z_t$ in the first approach \eqref{eqn:PAVG_mean_z} corresponds to 
$L^{\T}z_{t} = L^{\T}s_t + Z$ in the second approach \eqref{eqn:PAVG_var_z}.

\subsection{Exact Formulas of Acceptance Probability for PAVG}

Continuing from the previous subsection, we present the exact formula for calculating the acceptance probability \eqref{eqn:PAVG_var_acc} for PAVG.
We denote $C_1$ and $C_2$ as the normalization constants corresponding to the forward transition  $Q(s|z_t; s_t)$ and backward transition $Q(s|z_t; s^*)$:
\begin{align}
    C_1 &= \sum\limits_{s \in \mathcal{S}}\exp(-\frac{1}{2}s^{\T}Ds +(\nabla f(s_t)-Ws_t +(W+D)z_{t})^{\T}s) \nonumber \\
    &= \prod\limits_{i=1}^{d}[\sum\limits_{s_i}exp(-\frac{1}{2}d_is_i^2+(\nabla f(s_t)_i-(Ws_t)_i+((W+D)z_t)_i)s_i)],
    \label{eqn:C1} \\
    C_2 &= \sum\limits_{s \in \mathcal{S}}\exp(-\frac{1}{2}s^{\T}Ds +(\nabla f(s^{*})-Ws^{*} +(W+D)z_{t})^{\T}s) \nonumber\\
      &= \prod\limits_{i=1}^{d}[\sum\limits_{s_i}exp(-\frac{1}{2}d_is_i^2+(\nabla f(s^*)_i-(Ws^*)_i+((W+D)z_t)_i)s_i)] .\label{eqn:C2}
\end{align}
After substituting \eqref{eqn:C1} and \eqref{eqn:C2}, the ratio inside acceptance probability \eqref{eqn:PAVG_var_acc} is
\begin{align}
    & \exp(f(s^{*})-f(s_{t}))\frac{\mathcal{N}(z_{t}| s^{*}, ((W+D))^{-1})Q(s_{t}|z_{t},s^{*})}{\mathcal{N}(z_{t} |s_{t}, ((W+D))^{-1})Q(s^{*}|z_{t},s_{t})} \nonumber \\
    &= \frac{C_1 exp(f(s^*)-\frac{1}{2}\|L^{\T}(z_t-s^*)\|_2^2-\frac{1}{2}s^{\T}_{t}Ds_{t} +(\nabla f(s^{*})-Ws^{*} +(W+D)s_{t}+LZ)^{\T}s_{t})}{C_2 \exp(f(s_t) - \frac{1}{2}\|L^{\T}(z_t-s_t)\|_2^2-\frac{1}{2}s^{*\T}Ds^{*} +(\nabla f(s_{t})-Ws_{t} +(W+D)s_{t}+LZ)^{\T}s^{*})} \nonumber\\
    &= \frac{C_1 exp(f(s^*)-\frac{1}{2}\|L^{\T}(z_t-s^*)\|_2^2-\frac{1}{2}s^{\T}_{t}Ds_{t} +(\nabla f(s^{*})-Ws^{*} +(W+D)z_t)^{\T}s_{t})}{C_2 \exp(f(s_t) - \frac{1}{2}\|L^{\T}(z_t-s_t)\|_2^2-\frac{1}{2}s^{*\T}Ds^{*} +(\nabla f(s_{t})-Ws_{t} +(W+D)z_t)^{\T}s^{*})}.\label{eqn:PAVG_exact_acc}
\end{align}

\subsection{Proof of Rejection-free Property}\label{sec:PAVG_reject_free}

We aim to prove that PAVG is rejection-free when the target distribution has a quadratic potential of the form:
\begin{align}
    \pi(s) \propto \exp\left(\frac{1}{2} s^{\T} W s + b^{\T} s \right),
    \label{eqn:quadratic_energy}
\end{align}
where $W$ is the same matrix as in the approximation \eqref{eqn:PAVG_approx}. This target distribution implies $f(s) = \frac{1}{2} s^{\T} W s + b^{\T} s$, with its gradient given by
\begin{align}
    \nabla f(s) = Ws + b. \label{eqn:quadratic_grad}
\end{align}
With the gradient \eqref{eqn:quadratic_grad}, we notice that the approximate distribution in \eqref{eqn:PAVG_approx} is the target distribution itself:
\begin{align}
     &  \nabla f(s_{t})^{\T}(s-s_{t})+\frac{1}{2}(s-s_{t})^{\T}W(s-s_{t}) \nonumber\\
    & \propto  (W s_{t} + b)^{\T} (s - s_{t}) + \frac{1}{2} (s - s_{t})^{\T} W (s - s_{t}) \nonumber \\
    & \propto \frac{1}{2} s^{\T} W s + b^{\T} s. \nonumber
\end{align}

Following the variance approach described in \eqref{eqn:PAVG_var_z}-\eqref{eqn:PAVG_var_acc}, we express each step for the proposal as
\begin{equation}
    z_{t} = s_{t}+(L^{\T})^{-1}Z \quad Z \sim \mathcal{N}(0, I),
    \label{eqn:proof_pavg_rej1}
\end{equation}
    \begin{align}
        s^* \sim Q(s |z_{t}, s_{t}) & \propto \exp(-\frac{1}{2}s^{\T}Ds + [\nabla f(s_{t})-Ws_{t} +(W+D)z_{t}]^{\T}s) \nonumber \\
       & \propto \exp(-\frac{1}{2}s^{\T}Ds + [b +(W+D)z_{t}]^{\T}s).     \label{eqn:proof_pavg_rej2}
    \end{align}
We notice that the proposal in \eqref{eqn:proof_pavg_rej2} does not involve any terms related to $s_t$.
From the expressions \eqref{eqn:C1} and \eqref{eqn:C2} for the normalization constants $C_1$ and $C_2$, we have
\begin{align}
    C_1& = C_2= \sum\limits_{s \in \mathcal{S}}\exp(-\frac{1}{2}s^{\T}Ds + [b +(W+D)z_{t}]^{\T}s). \nonumber
\end{align}
We introduce an additional constant $C$:
\begin{align}
    C &= \sum\limits_{s \in \mathcal{S}}\exp(-\frac{1}{2}s^{\T}Ds +b^{\T}s) \int \exp(-\frac{1}{2}(z_{t}-s)^{\T}(W+D)(z_{t}-s))dz_{t}. \nonumber
\end{align}
Then in the denominator of the ratio inside the acceptance probability, we have
    \begin{align}
        &\exp(f(s_{t}))\mathcal{N}(z_{t} | L^{\T}s_{t}, I)Q(s^{*}|z_{t},s_{t}) \nonumber\\
        &=\frac{1}{C_1 C}\exp(\frac{1}{2}s_{t}^{\T}Ws_{t}+b^{\T}s_{t} - \frac{1}{2}(z_{t}-s_{t})^{\T}(W+D)(z_{t}-s_{t})-\frac{1}{2}s^{*\T}Ds^{*} + [b +(W+D)z_{t}]^{\T}s^{*}) \nonumber\\
        & = \frac{1}{C_1 C} \exp(-\frac{1}{2}s^{*\T}Ds^{*}-\frac{1}{2}s_{t}^{\T}Ds_{t}-\frac{1}{2}z_{t}^{\T}(W+D)z_{t} + b^{\T}(s_{t}+s^{*})+z_{t}^{\T}(W+D)(s_{t}+s^{*})).    \label{eqn:pavg_rej_denominator}
\end{align}
By swapping $s_{t}$ and $s^{*}$ in \eqref{eqn:pavg_rej_denominator}, we see that the numerator of the ratio inside the acceptance probability is
\begin{align}
& \exp(f(s^{*}))\mathcal{N}(z_{t} | L^{\T}s^{*}, I)Q(s_{t}|z_{t},s^{*}) \nonumber\\
&= \frac{1}{C_1 C} \exp(-\frac{1}{2}s^{*\T}Ds^{*}-\frac{1}{2}s_{t}^{\T}Ds_{t}-\frac{1}{2}z_{t}^{\T}(W+D)z_{t} + b^{\T}(s_{t}+s^{*})+z_{t}^{\T}(W+D)(s_{t}+s^{*})).
    \label{eqn:pavg_rej_numerator}
\end{align}
Comparing \eqref{eqn:pavg_rej_denominator} and \eqref{eqn:pavg_rej_numerator}, it is evident that the numerator and the denominator are identical, resulting in an acceptance probability of $1$, thereby proving rejection-free property.

We notice that \eqref{eqn:proof_pavg_rej1} corresponds to the step for sampling $z_t$ in \eqref{eqn:Gaussint2}, and \eqref{eqn:proof_pavg_rej2} corresponds to the step for sampling $s_{t+1}$ in \eqref{eqn:Gaussint3} for the Gaussian integral trick in Section~\ref{sec:Gaussian_integral} with $A = (W+D)^{-1}$. From the rejection-free property, the proposal is always accepted such that $s_{t+1} =s^*$, and PAVG reduces to the Gaussian integral trick in the presence of quadratic potentials.

\section{Properties of V-PDHAMS}

\subsection{Proof of Equivalency between Different Auxiliary Variable Schemes} \label{prop_precondHAMS_id}
It is easy to observe that the proposal $Q(s, u | s_t, u_t)$ is identical in the first two approaches: preconditioning by mean \eqref{eqn:precondHAMS_mean_z}--\eqref{eqn:phams_mean_acc} and preconditioning by variance \eqref{eqn:phams_var_z}--\eqref{eqn:phams_var_u}. We only need to establish their equivalence with the momentum approach \eqref{eqn:phams_mom1}-\eqref{eqn:phams_mom_acc}.

We first transform the momentum variable $v$ in \eqref{eqn:phams_mom1} into standard Gaussian $u$ such that $u = L^{\T} v$. The corresponding target distribution becomes
$\pi(s, u) \propto \exp \left( f(s) - \frac{1}{2} \|u\|_{2}^{2} \right)$ and the corresponding auxiliary variable is given by:
\begin{equation}
    z = s + v = s + (L^{\T})^{-1} u.  \nonumber
\end{equation}
The auxiliary variable is constructed in the same way as \eqref{eqn:phams_var_z}. Replacing $v_{t} = (L^{\T})^{-1}u_{t}$ and  $v^{*} = (L^{\T})^{-1}u^{*}$, we have the new proposal in terms of $(s, u)$ as
\begin{align}
    s^{*} & \sim Q(s|z_{t} =s_t + v_t; s_{t})  \nonumber\\
    & \propto  \prod\limits_{i=1}^{d}\Softmax(-\frac{1}{2}d_{i}s_{i}^{2} + [\nabla f(s_{t})_{i} -(Ws_{t})_{i} + (W+D)z_{t})_{i}]s_{i}) \nonumber\\
    & \propto  \prod\limits_{i=1}^{d}\Softmax(-\frac{1}{2}d_{i}s_{i}^{2} + [\nabla f(s_{t})_{i} -(Ws_{t})_{i} +
    ((W+D)s_{t}+(W+D)v_{t})_{i}]s_{i}) \nonumber\\
     & \propto  \prod\limits_{i=1}^{d}\Softmax(-\frac{1}{2}d_{i}s_{i}^{2} + [\nabla f(s_{t})_{i} +(Ds_{t})_{i} +(Lu_{t})_{i}]s_{i}),
\label{eqn:precond_HAMS_pf1}
\end{align}
and update on the momentum $u^*$ is
\begin{align}
        v^{*} &= z_{t} - s^{*}, \nonumber\\
        (L^{\T})^{-1}u^{*} &= s_{t}+ (L^{\T})^{-1}u_{t}- s^{*},\nonumber \\
        u^{*} &= L^{\T}(s_{t}-s^{*})+u_{t}. \label{eqn:precond_HAMS_pf2}
    \end{align}
The two steps above \eqref{eqn:precond_HAMS_pf1} and \eqref{eqn:precond_HAMS_pf2} are exactly the same as the proposals from the variance approach \eqref{eqn:phams_var_s} and \eqref{eqn:phams_var_u}. With the same proposals, the associated Metropolis--Hastings probabilities are also the same. Hence the three auxiliary variable schemes are equivalent.

\subsection{Exact Formulas of Acceptance Probability for V-PDHAMS}
Continuing from the previous subsection, we denote $C_1$ and $C_2$ as the normalization constants corresponding to the forward transition  $Q(s|z_t = s_t- v_{t+1/2};s_t)$ and backward transition $Q(s|z_t= s^* +v^*;s^*)$:
\begin{align}
    C_1 &= \sum\limits_{s \in \mathcal{S}}\exp(-\frac{1}{2}s^{\T}Ds +(\nabla f(s_t)-Ws_t +(W+D)(s_t-v_{t+1/2}))^{\T}s) \nonumber \\
     &= \prod\limits_{i=1}^{d}[\sum\limits_{s_i}exp(-\frac{1}{2}d_is_i^2+(\nabla f(s_t)_i+(Ds_t)_i-((W+D)v_{t+1/2})_i)s_i)],
     \label{eqn:HAMSC1} \\
    C_2& = \sum\limits_{s \in \mathcal{S}}\exp(-\frac{1}{2}s^{\T}Ds +(\nabla f(s^{*})-Ws^{*} +(W+D)(s^*+v^*))^{\T}s) \nonumber \\
     &= \prod\limits_{i=1}^{d}[\sum\limits_{s_i}exp(-\frac{1}{2}d_is_i^2+(\nabla f(s^*)_i+(Ds^*)_i+((W+D)v^*)_i)s_i)].
    \label{eqn:HAMSC2}
\end{align}
After substituting \eqref{eqn:HAMSC1} and \eqref{eqn:HAMSC2} into the acceptance probability in \eqref{eqn:p-mod-acc}, we have
\begin{align}
    &\frac{\pi(s^{*}, -v^{*}) Q_{\phi}(s_{t}, -v_{t+1/2}|s^{*}, -v^{*})}{\pi(s_{t}, v_{t+1/2}) Q_{\phi} (s^{*}, v^{*}|s_{t}, v_{t+1/2})} \nonumber\\
    &=\frac{C_1\exp(f(s^*)-\frac{1}{2}\|L^{\T}v^*\|_2^2-\frac{1}{2}s_t^{\T}Ds_t +(\nabla f(s^{*})-Ws^{*} +(W+D)(s^*+v^*))^{\T}s_t)}{C_2 \exp(f(s_t)-\frac{1}{2}\|L^{\T}v_{t+1/2}\|_2^2-\frac{1}{2}s^{*\T}Ds^* +(\nabla f(s_t)-Ws_t +(W+D)(s_t-v_{t+1/2}))^{\T}s^*)}. \label{eqn:HAMS_exact_acc}
\end{align}
We notice that the formula \eqref{eqn:HAMS_exact_acc} does not contain any term regarding $\phi$, which is implicitly incorporated into $v^*$.

\subsection{Proof of Rejection-free Property} \label{sec:PHAMS_reject_free}
When the target distribution has a quadratic potential, the target distribution is \eqref{eqn:quadratic_energy} and the corresponding gradient of $f(s)$ is \eqref{eqn:quadratic_grad}. The gradient correction term in \eqref{eqn:p-mod1} vanishes, and we have:
\begin{align}
    z_t = s_{t}-v_{t+1/2} = s^* +v^* .\label{eqn:prop_hams_free_1}
\end{align}
Then we substitute the equation \eqref{eqn:prop_hams_free_1} into both the forward and backward proposals in \eqref{eqn:p-mod-acc}:
    \begin{align}
        Q_{\phi}(s^{*}, v^{*}|s_{t}, v_{t+1/2}) & \propto \exp(-\frac{1}{2}s^{*\T}Ds^{*} + (\nabla f(s_{t})-Ws_{t}+(W+D)(s_{t}-v_{t+1/2}))^{\T}s^{*}) \nonumber \\
        & \propto \exp(-\frac{1}{2}s^{*\T}Ds^{*} + (b+(W+D)z_{t})^{\T}s^{*}),
            \label{eqn:prop6_forward}
    \end{align}
    \begin{align}
        Q_{\phi}(s_{t}, -v_{t+1/2}|s^{*}, -v^{*}) & \propto \exp(-\frac{1}{2}s_{t}^{\T}Ds_{t} + (\nabla f(s^{*})-Ws^{*}+(W+D)(s^{*}+v^{*}))^{\T}s_{t}) \nonumber \\
        & \propto \exp(-\frac{1}{2}s_{t}^{\T}Ds_{t} + (b+(W+D)z_{t})^{\T}s_{t}).
            \label{eqn:prop6_backward}
    \end{align}
The forward proposal probability \eqref{eqn:prop6_forward} and the backward proposal probability \eqref{eqn:prop6_backward} share the same potential function, which implies that the normalization constant $C_1$ \eqref{eqn:HAMSC1} and $C_2$ \eqref{eqn:HAMSC2} are actually equivalent:
\begin{equation}
    C_1 = C_2 =\sum\limits_{s \in \mathcal{S}} \exp\left( -\frac{1}{2} s^{\T} D s + (b + (W+D)z_{t})^{\T} s \right). \nonumber
\end{equation}
We also let $C$ be the normalization constant such that
\begin{equation}
    C = \sum\limits_{s\in \mathcal{S}} \exp(f(s)) \int \exp(-\frac{1}{2}v^{\T}(W+D)v)
dv. \nonumber
\end{equation}
Using the constants \( C_1 \) and \( C \) above, we express both the numerator and denominator of the ratio inside the acceptance probability explicitly as
    \begin{align}
        &\pi(s_{t}, v_{t+1/2})Q_{\phi}(s^{*}, v^{*}|s_{t}, v_{t+1/2}) \nonumber \\
        &= \frac{1}{C}\exp(\frac{1}{2}s_{t}^{\T}Ws_{t}+b^{\T}s_{t}-\frac{1}{2}v_{t+1/2}^{\T}(W+D)v_{t+1/2}) \nonumber \\
        &\cdot \frac{1}{C_1} \exp(-\frac{1}{2}s^{*\T}Ds^{*} + (b+(W+D)z_{t})^{\T}s^{*}) \nonumber \\
        & = \frac{1}{C_1C} \exp(-\frac{1}{2}s_{t}^{\T}Ds_{t}-\frac{1}{2}s^{*\T}Ds^{*} + b^{\T}(s_{t}+s^{*})-\frac{1}{2}z_{t}^{\T}(W+D)z_{t} +z_{t}^{\T}(W+D)(s_{t}+s^{*})),
        \label{eqn:prop6_proof_denom}
    \end{align}
    \begin{align}
        &\pi(s^{*}, -v^{*})Q_{\phi}(s_{t}, -v_{t+1/2}|s^{*}, -v^{*}) \nonumber \\
        &= \frac{1}{C}\exp(\frac{1}{2}s^{*\T}Ws^{*}+b^{\T}s^{*}-\frac{1}{2}v^{*\T}(W+D)v^{*}) \nonumber\\
        &\cdot \frac{1}{C_1} \exp(-\frac{1}{2}s_{t}^{\T}Ds_{t} + (b+(W+D)z_{t})^{\T}s_{t}) \nonumber \\
        & = \frac{1}{C_1C} \exp(-\frac{1}{2}s_{t}^{\T}Ds_{t}-\frac{1}{2}s^{*\T}Ds^{*} + b^{\T}(s_{t}+s^{*})-\frac{1}{2}z_{t}^{\T}(W+D)z_{t} +z_{t}^{\T}(W+D)(s_{t}+s^{*})).     \label{eqn:prop6_proof_numer}
    \end{align}
Comparing the formula of the denominator \eqref{eqn:prop6_proof_denom} with the formula of the numerator \eqref{eqn:prop6_proof_numer}, we observe that they are identical. Consequently, the acceptance probability \eqref{eqn:p-mod-acc} is always $1$ for a quadratic potential, confirming the rejection-free property.

\subsection{V-PDHAMS and PAVG}\label{sec:pham_to_pavg}

We show that V-PDHAMS reduces to PAVG when $\epsilon=0$ and $\phi=0$. If $\epsilon = 0$, the intermediate momentum $v_{t+1/2}$ and auxiliary variable $z_t$ can be written as
\begin{align}
    v_{t+1/2} &= -(L^{\T})^{-1}Z, \quad Z \sim \mathcal{N}(0,I),\label{eqn:hams_pavg_v12} \\
    z_t &= s_t + (L^{\T})^{-1}Z. \label{eqn:hams_pavg_z}
\end{align}
The construction of $z_t$ is the same as the construction of auxiliary variable in \eqref{eqn:PAVG_var_z} in PAVG. Substituting both \eqref{eqn:hams_pavg_v12} and \eqref{eqn:hams_pavg_z} into the proposal \eqref{eqn:phams_negation-s}, we have
\begin{align}
    s^{*}&\sim Q(s|z_t = s_t -v_{t+1/2};s_t) \nonumber \\
    &\propto \prod\limits_{i=1}^{d}\Softmax(-\frac{1}{2}d_{i}s_{i}^{2} + [\nabla f(s_{t})_{i} -(Ws_{t})_{i} + ((W+D)z_t)_{i}]s_{i}), \nonumber
\end{align}
which is also the same proposal of $s^*$ in \eqref{eqn:PAVG_var_s} in PAVG. We notice that when $\phi=0$, $z_t = s_t - v_{t+1/2} = s^*+v^*$, and the forward and backward proposals are \eqref{eqn:prop6_forward} and \eqref{eqn:prop6_backward}. We also notice that when $\phi=0$, the normalization constant $C_1$ \eqref{eqn:HAMSC1}, $C_2$\eqref{eqn:HAMSC2} is equivalent to the normalization constants of PAVG, \eqref{eqn:C1} and \eqref{eqn:C2}. Then the ratio inside acceptance probability  \eqref{eqn:HAMS_exact_acc} for Vanilla Discrete-HAMS can be simplified as
\begin{align}
    &\frac{\pi(s^{*}, -v^{*}) Q_{\phi}(s_{t}, -v_{t+1/2}|s^{*}, -v^{*})}{\pi(s_{t}, v_{t+1/2}) Q_{\phi} (s^{*}, v^{*}|s_{t}, v_{t+1/2})} \nonumber\\
    &=\frac{C_1\exp(f(s^*)-\frac{1}{2}\|L^{\T}v^*\|_2^2-\frac{1}{2}s_t^{\T}Ds_t +(\nabla f(s^{*})-Ws^{*} +(W+D)(s^*+v^*))^{\T}s_t)}{C_2 \exp(f(s_t)-\frac{1}{2}\|L^{\T}v_{t+1/2}\|_2^2-\frac{1}{2}s^{*\T}Ds^* +(\nabla f(s_t)-Ws_t +(W+D)(s_t-v_{t+1/2}))^{\T}s^*)} \nonumber \\
    &= \frac{C_1\exp(f(s^*)-\frac{1}{2}\|L^{\T}(z_t-s^*)\|_2^2-\frac{1}{2}s_t^{\T}Ds_t +(\nabla f(s^{*})-Ws^{*} +(W+D)z_t)^{\T}s_t)}{C_2 \exp(f(s_t)-\frac{1}{2}\|L^{\T}(z_t-s_t)\|_2^2-\frac{1}{2}s^{*\T}Ds^* +(\nabla f(s_t)-Ws_t +(W+D)z_t)^{\T}s^*)}. \label{eqn:acc_HAMS_PAVG}
\end{align}
The formula \eqref{eqn:acc_HAMS_PAVG} is also the same as the ratio in acceptance probability \eqref{eqn:PAVG_exact_acc} for PAVG. Therefore, the sampling scheme of Vanilla Discrete-HAMS becomes the same as PAVG when $\epsilon=\phi=0$.

\section{Properties of O-PDHAMS}
\subsection{Proof of Rejection-free Property}\label{sec:O-PHAMS_reject_free}
First, as mentioned in Supplement Section~\ref{sec:PHAMS_reject_free}, the gradient correction term in \eqref{eqn:p-mod1} vanishes when the target distribution has a quadratic potential, and we obtain \eqref{eqn:prop_hams_free_1} about $z_t$. From \eqref{eqn:over-phams-acc}, the forward transition $\tilde Q( s^* | z_t=s_t-v_{t+1/2};s_t)$ has reference distribution $Q(s|z_t=s_t-v_{t+1/2};s_t)$ in \eqref{eqn:phams_negation-s}, and the backward transition $\tilde Q (s_t |z_t=s^*+v^*;s^*)$ has reference distribution $Q(s|z_t=s^*+ v^*;s^*)$. The two reference distributions are:
    \begin{align}
        Q(s|z_t = s_{t}- v_{t+1/2};s_t) & \propto \exp(-\frac{1}{2}s^{\T}Ds + (\nabla f(s_{t})-Ws_{t}+(W+D)(s_{t}-v_{t+1/2}))^{\T}s) \nonumber \\
        & \propto \exp(-\frac{1}{2}s^{\T}Ds+ (b+(W+D)z_{t})^{\T}s),
            \label{eqn:reference_forward}
    \end{align}
    \begin{align}
        Q(s|z_t=s^{*}+v^{*};s^*) & \propto \exp(-\frac{1}{2}s^{\T}Ds + (\nabla f(s^{*})-Ws^{*}+(W+D)(s^{*}+v^{*}))^{\T}s) \nonumber \\
        & \propto \exp(-\frac{1}{2}s^{\T}Ds + (b+(W+D)z_{t})^{\T}s).
            \label{eqn:reference_backward}
    \end{align}
From \eqref{eqn:reference_forward} and \eqref{eqn:reference_backward}, the reference distribution for both forward and backward transition are the same, which we denote as $p(s)$. The common reference distribution suggests that $\tilde Q( s^* |z_t= s_t- v_{t+1/2};s_t) = p(s^*|s_t)$ and $\tilde Q (s_t |z_t=s^*+v^*;s^*) = p(s_t|s^*)$. From \eqref{eqn:prop_hams_free_1}, we also have
\begin{align}
    \pi(s_t, -v_{t+1/2}) & \propto \exp(\frac{1}{2}s_t^{\T}Ws_t + b^{\T}s_t -\frac{1}{2}v_{t+1/2}^{\T}(W+D)v_{t+1/2} ) \nonumber \\
    & \propto \exp(\frac{1}{2}s_t^{\T}Ws_t + b^{\T}s_t -\frac{1}{2}(z_t-s_t)^{\T}(W+D)(z_t-s_t) ) \nonumber \\
    &\propto \exp(-\frac{1}{2}s_t^{\T}Ds_t+ (b+(W+D)z_{t})^{\T}s_t) \nonumber \\
    & \propto p(s_t), \label{eqn:opdhams_reject_1}
\end{align}
\begin{align}
    \pi(s^*, v^*) & \propto \exp(\frac{1}{2}s^{*\T}Ws^* + b^{\T}s^* -\frac{1}{2}v^{*\T}(W+D)v^* ) \nonumber \\
    & \propto \exp(\frac{1}{2}s^{*\T}Ws^* + b^{\T}s^* -\frac{1}{2}(z_t-s^*)^{\T}(W+D)(z_t-s^*) ) \nonumber \\
    &\propto \exp(-\frac{1}{2}s^{*\T}Ds^*+ (b+(W+D)z_{t})^{\T}s^*) \nonumber\\
    & \propto p(s^*).\label{eqn:opdhams_reject_2}
\end{align}
By substituting \eqref{eqn:opdhams_reject_1} and \eqref{eqn:opdhams_reject_2} into \eqref{eqn:over-phams-acc}, the acceptance probability of O-DHAMS can be written as
\begin{align}
& \min\{1, \frac{\pi(s^{*}, -v^{*})\tilde{Q}_{\phi}(s_{t}, -v_{t+1/2}| s^{*}, -v^{*})}{\pi(s_{t}, v_{t+1/2})\tilde{Q}_{\phi}(s^{*}, v^{*}|s_{t}, v_{t+1/2})} \} \nonumber \\
& = \min\{1, \frac{p(s^{*})p(s_{t}| s^{*})}{p(s_{t})p(s^{*}| s_{t})}\} \nonumber \\
& = \min\{1, \frac{p(s^{*}, s_{t})}{p(s_{t}, s^{*})}\},
\label{eqn:overrelaxation-acc}
\end{align}
where the notation $p(\cdot, \cdot)$ represents the joint distribution of two variables in the over-relaxation Algorithm \ref{algo:over-relaxation}.
By Section 4.1 in DHAMS paper \citep{Zhou2025Dhams}, the joint distribution $p(\cdot, \cdot)$ obtained is symmetric, so that $p(s^{*}, s_{t})=p(s_{t}, s^{*})$ in \eqref{eqn:overrelaxation-acc}. Therefore, the acceptance probability is always $1$ for O-PDHAMS, confirming the rejection-free property.

\section{Model Calibration}
\subsection{Calibration of \textit{W} in the First Approach}\label{sec:W_estimate1}
For the first optimization problem in \eqref{eqn:W_calibration1}, define $D_f \in \mathbb{R}^{T \times d}$ such that the $t$-th row of $D_f$ is $(\nabla f(s_{t+1}) - \nabla f(s_t))^{\T}$, and similarly define $D_s \in \mathbb{R}^{T \times d}$ such that its $t$-th row is $(s_{t+1} - s_t)^{\T}$. Using these definitions, we rewrite the optimization problem as
\begin{equation}
\hat{W} = \argmin\limits_{W =W^{\T}} \sum\limits_{t=1}^{T} \|\nabla f(s_{t+1}) -\nabla f(s_{t})- W(s_{t+1}-s_{t})\|_{2}^{2} = \argmin\limits_{W=W^{\T}}\|D_{f}-D_{s}W\|_{F}^{2}.
\label{eqn:W1_target}
\end{equation}
The optimization problem above \eqref{eqn:W1_target} is convex in $W$, which enables us to solve it setting its first order derivatives with respect to $W$ as zero,
\begin{equation}
\frac{\partial \|D_{f}-D_{s}W\|_{F}^{2}}{\partial W} = (D_{s}^{\T}D_{s})W+W(D_{s}^{\T}D_{s}) - D_{s}^{\T}D_{f}-D_{f}^{\T}D_{s} =0.
\label{eqn:W1_grad}
\end{equation}
The equation \eqref{eqn:W1_grad} can be solved using the Bartels–Stewart algorithm \citepSupp{Bartel1972Lyapunov} for continuous Lyapunov equations \citepSupp{lyapunov1992stability}, available in standard computational packages (SciPy, MATLAB, etc.). When $D_{s}^{\T}D_{s}$ is positive definite, the Hurwitz stability condition is satisfied \citepSupp{Boyd1994Linear}, and the solution to \eqref{eqn:W1_target} is unique. Furthermore, when the dimension $d$ is relatively low, the equation can be solved directly by vectorizing the matrices involved,
\begin{equation}
\begin{aligned}
\vec(D_{s}^{\T}D_{f} +D_{f}^{\T}D_{s} ) & =\vec((D_{s}^{\T}D_{s})W+W(D_{s}^{\T}D_{s})) \\
&= [I \otimes (D_{s}^{\T}D_{s}) + (D_{s}^{\T}D_{s}) \otimes I] \vec(W). \nonumber
\end{aligned}
\end{equation}
We observe that when $f(s) = \frac{1}{2}s^{\top}Ws + b^{\top}s$, the matrix $W$ is a solution to \eqref{eqn:W1_target}. If the solution is unique, then the true matrix $W$ is guaranteed to be recovered.

\subsection{Calibration of \textit{W} in the Second Approach}\label{sec:W_estimate2}
By defining $a_t = f(s_{t+1}) - f(s_t) - \nabla f(s_t)^{\top}(s_{t+1} - s_t)$, $\beta_t = s_{t+1} - s_t$, and $B_t = \beta_t \beta_t^{\top}$, the optimization problem in \eqref{eqn:W_calibration2} can be rewritten as
    \begin{align}
         & \sum\limits_{t=1}^{T} [f(s_{t+1})-f(s_{t}) -\nabla f(s_{t})^{\T}(s_{t+1}-s_{t}) -\frac{1}{2}(s_{t+1}-s_{t})^{\T}W(s_{t+1}-s_{t}) ]^{2} \nonumber\\
         & = \sum\limits_{t=1}^{T} (a_{t}-\frac{1}{2}\beta_{t}^{\T}W\beta_{t})^{2} \nonumber\\
          & = \sum\limits_{t=1}^{T} (a_{t} -\frac{1}{2}\tr(WB_{t}))^{2} \nonumber \\
          & = \sum\limits_{t=1}^{T} [a_{t} -\frac{1}{2}(\sum\limits_{k}W_{kk}B_{t, kk} +2\sum\limits_{k<l}W_{kl}B_{t, kl})]^{2}.
          \label{eqn:W2_target}
    \end{align}
Given that $W$ is symmetric, let $\vech(W) \in \mathbb{R}^{\frac{d(d+1)}{2}}$ be the vectorization of its upper-triangular part. Define $B \in \mathbb{R}^{T \times \frac{d(d+1)}{2}}$ as the matrix whose $t$-th row contains the upper-triangular elements of $B_t$, and let $a \in \mathbb{R}^T$ be the vector consisting of the scalar terms $a_t$. The optimization problem in \eqref{eqn:W2_target} is convex in $\vech(W)$, and a solution $\hat{W}$ can be obtained as
\begin{align}
     \hat{W} & = \argmin\limits_{W =W^{\T}} \sum\limits_{t=1}^{T} [f(s_{t+1})-f(s_{t}) -\nabla f(s_{t})^{\T}(s_{t+1}-s_{t}) -\frac{1}{2}(s_{t+1}-s_{t})^{\T}W(s_{t+1}-s_{t}) ]^{2} \nonumber\\
     & = \argmin\limits_{\vech(W)} \|a -B \vech(W)\|_{2}^{2}.
     \label{eqn:W2_target_new}
\end{align}
The problem in \eqref{eqn:W2_target_new} is an ordinary least squares (OLS) problem for $\vech(W)$, which can be solved using various methods. The uniqueness of solution to \eqref{eqn:W2_target_new} is guaranteed when the matrix $B$ is of full column rank. We observe that when $f(s) = \frac{1}{2}s^{\top}Ws + b^{\top}s$, the matrix $W$ is a solution to \eqref{eqn:W2_target_new}. If the solution is also unique, then the true matrix $W$ is guaranteed to be recovered.

\subsection{Choice of Calibration Method for \textit{W}}\label{sec:W_choice}
When the dimension is high, solving the large linear system in \eqref{eqn:W2_target_new} can be numerically unstable. In contrast, the optimization problem \eqref{eqn:W1_target}, solved using the Bartels–Stewart algorithm, offers greater stability. Additionally, the first method—which directly relates gradient differences to the Hessian—serves as the foundation for many quasi-Newton methods (e.g., BFGS \cite{goldfarb1970family}).

On the other hand, in low-dimensional settings with sufficient samples, the OLS problem in \eqref{eqn:W2_target_new} is often easier to implement and computationally more efficient to solve.

\section{Scaling of State Variable} \label{sec:scaling}

For a discrete distribution $\pi(s) \propto \exp(f(s))$ supported on the lattice $\mathcal{S} = \mathcal{A}^{d}$, where $\mathcal{A} = \{a_1, \ldots, a_{K}\}$, consider the new variable $y = c s$ for a constant $c$. The induced distribution of $y$ is, by some abuse of notation, $\pi(y) \propto \exp(g(y))$, where $g(y) = f\left( \frac{1}{c} y \right)$, and has support $c \mathcal{S} = \{ c a_1, \ldots, c a_K \}^{d}$. The distributions $\pi(s)$ and $\pi(y)$ share the same probability mass function but are defined on lattices of different scales.

When applying PAVG or Preconditioned Discrete-HAMS to sample from the target $\pi(s)$, we need to calibrate a matrix $W_s$ according to Section~\ref{sec:W_calibration}. Similarly, if sampling from the target $\pi(y)$, we calibrate a corresponding matrix $W_y$. Denote the estimated matrix for $\pi(y)$ as $\hat{W}_y$, and for $\pi(s)$ as $\hat{W}_s$. From calibration method~\eqref{eqn:W_calibration1}, $\hat{W}_y$ is obtained as
\begin{align}
    \hat{W}_y
    &= \argmin\limits_{W = W^{\T}} \sum\limits_{t=1}^{T} \left\| \nabla g(y_{t+1}) - \nabla g(y_{t}) - W (y_{t+1} - y_{t}) \right\|_2^2 \nonumber \\
    &= \argmin\limits_{W = W^{\T}} \sum\limits_{t=1}^{T} \left\| \frac{1}{c} \nabla f\left( \frac{1}{c} y_{t+1} \right) - \frac{1}{c} \nabla f\left( \frac{1}{c} y_t \right) - c W (s_{t+1} - s_t) \right\|_2^2 \nonumber \\
    &= \argmin\limits_{W = W^{\T}} \sum\limits_{t=1}^{T} \left\| \nabla f(s_{t+1}) - \nabla f(s_{t}) - c^2 W (s_{t+1} - s_t) \right\|_2^2 \nonumber \\
    &= \frac{1}{c^2} \hat{W}_s. \label{eqn:W_calibration_1y}
\end{align}
Similarly, from calibration method~\eqref{eqn:W_calibration2}, $\hat{W}_y$ is obtained as
\begin{align}
    \hat{W}_y
    &= \argmin\limits_{W = W^{\T}} \sum\limits_{t=1}^{T} \left[ g(y_{t+1}) - g(y_{t}) - \nabla g(y_{t})^{\T} (y_{t+1} - y_{t}) - \frac{1}{2} (y_{t+1} - y_{t})^{\T} W (y_{t+1} - y_{t}) \right]^2 \nonumber \\
    &= \argmin\limits_{W = W^{\T}} \sum\limits_{t=1}^{T} \left[ f(s_{t+1}) - f(s_{t}) - \frac{1}{c} \nabla f\left( \frac{1}{c} y_{t} \right)^{\T} (c s_{t+1} - c s_t) - \frac{1}{2} (s_{t+1} - s_t)^{\T} (c^2 W) (s_{t+1} - s_t) \right]^2 \nonumber \\
    &= \argmin\limits_{W = W^{\T}} \sum\limits_{t=1}^{T} \left[ f(s_{t+1}) - f(s_{t}) - \nabla f(s_{t})^{\T} (s_{t+1} - s_t) - \frac{1}{2} (s_{t+1} - s_t)^{\T} (c^2 W) (s_{t+1} - s_t) \right]^2 \nonumber \\
    &= \frac{1}{c^2} \hat{W}_s. \label{eqn:W_calibration_2y}
\end{align}

In summary, from \eqref{eqn:W_calibration_1y} and~\eqref{eqn:W_calibration_2y}, we see that from either calibration method~\eqref{eqn:W_calibration1} or~\eqref{eqn:W_calibration2}, the estimated $W$ matrix satisfies $\hat{W}_y = \frac{1}{c^2} \hat{W}_s$.
As a result, when using the second-order expansion in~\eqref{eqn:PAVG_approx} with $\hat{W}_y$, the target distribution $\pi(y)$ is approximated by
\begin{align}
    \tilde{\pi}(y; y_t)
    &\propto \exp \left( \nabla g(y_t)^{\T} (y_{t+1} - y_t) + \frac{1}{2} (y_{t+1} - y_t)^{\T} \hat{W}_y (y_{t+1} - y_t) \right) \nonumber \\
    &\propto \exp \left( \nabla f(s_t)^{\T} (s_{t+1} - s_t) + \frac{1}{2} (c s_{t+1} - c s_t)^{\T} \left( \frac{1}{c^2} \hat{W}_s \right) (c s_{t+1} - c s_t) \right) \nonumber \\
    &\propto \exp \left( \nabla f(s_t)^{\T} (s_{t+1} - s_t) + \frac{1}{2} (s_{t+1} - s_t)^{\T} \hat{W}_s (s_{t+1} - s_t) \right). \nonumber
\end{align}
The distribution above, $\tilde{\pi}(y; y_t)$, coincides with $\pi(s; s_t)$ in~\eqref{eqn:PAVG_approx}. This equivalence implies that, after calibrating $\hat{W}_y$ using the methods from Section~\ref{sec:W_calibration}, the approximation to the target distribution $\pi(y)$ is the same as that to $\pi(s)$. Consequently, the sampling process from PAVG and Preconditioned Discrete-HAMS for $\pi(y)$ remains the same as for $\pi(s)$.

\section{Tuning Procedures}\label{sec:precond_tuning}

The PAVG sampler requires tuning of a single parameter $\delta$, from the calibration of matrix $D$ in Section~\ref{sec:D_calibration}. In contrast, the V-PDHAMS and O-PDHAMS samplers require tuning multiple parameters: the auto-regression parameter $\epsilon$, the stepsize $\delta$ as in PAVG, the gradient correction parameter $\phi$, and the over-relaxation parameter $\beta$ (for O-PDHAMS). For distributions with a quadratic potential, Supplement Sections~\ref{sec:PAVG_reject_free}, \ref{sec:PHAMS_reject_free}, and \ref{sec:O-PHAMS_reject_free} show that all preconditioned samplers are rejection-free, with acceptance rates equal to 1. As discussed in Section~\ref{sec:D_calibration}, $1/\delta$ serves as a stepsize parameter for the preconditioned samplers. We observe that for distributions with non-quadratic potentials, the acceptance rate increases as $\delta$ increases. For distributions with quadratic potentials, we directly run a grid search for $\delta$. For distributions with non-quadratic potentials, we first apply Algorithm~\ref{algo:target_acc_precond} to identify an approximate range of $\delta$ values that achieve an acceptance rate between $0.5$ and $0.9$.

\begin{algorithm}[tbp]
\caption{Stepsize tuning to target a certain acceptance rate}
\label{algo:target_acc_precond}
\begin{algorithmic}
    \State \textbf{Require:} Initial stepsize $\delta_0$, initial acceptance rate $\alpha_0$, target acceptance rate $\alpha$,
    \For{$m = 0$ to $M$}
        \State Obtain acceptance rate $\alpha_m$ with stepsize $\delta_m$
        \If{$\alpha_m < \alpha$}
            \State $\delta_{m+1} \gets \delta_m \exp((1 + m)^{-a})$
        \ElsIf{$\alpha_m > \alpha$}
            \State $\delta_{m+1} \gets \delta_m \exp(-(1 + m)^{-a})$
        \EndIf
    \EndFor,
    \State \Return $\delta_m$ with smallest $|\alpha_m -\alpha|$.
\end{algorithmic}

\end{algorithm}

For PAVG, we conduct a grid search in the approximate range of $\delta$ values, and then select the best parameter based on the highest ESS of $f(s)$. For Preconditioned Discrete-HAMS, when $\phi$ has a reasonable value between $0$ and $1$, the ESS of $f(s)$ is often higher. Guided by these observations, we implement the tuning procedure for Preconditioned Discrete-HAMS samplers as shown in Algorithm \ref{algo:precond-tuning}.

\begin{algorithm}[tbp]
\begin{enumerate}
    \item Fix the auto-regression parameter $\epsilon$ close to $1$, typically around $0.9$,
    \item Fix the over-relaxation parameter $\beta$ to achieve the desired proposal behavior: $\beta = 1$ for a random-walk proposal or $\beta \approx 0$ for an over-relaxed proposal,
    \item Perform a grid search over $\delta$ in $D$ (with $\phi=0$) to identify the optimal stepsize based on the highest ESS of $f(s)$,
    \item With $\delta$ fixed from the previous step, select $\phi$ from a set of candidate values between $0$ and $1$ based on the highest ESS of $f(s)$.
\end{enumerate}
\caption{Tuning procedure for Preconditioned Discrete-HAMS}
\label{algo:precond-tuning}
\end{algorithm}

\section{Additional Experiment Results}
\subsection{Additional Results for Discrete Gaussian}\label{sec:precond_gaussian_results}
\begin{figure}[tbp]
 \begin{subfigure}[b]{0.32\textwidth}
        \centering
        \includegraphics[width=0.8\linewidth]{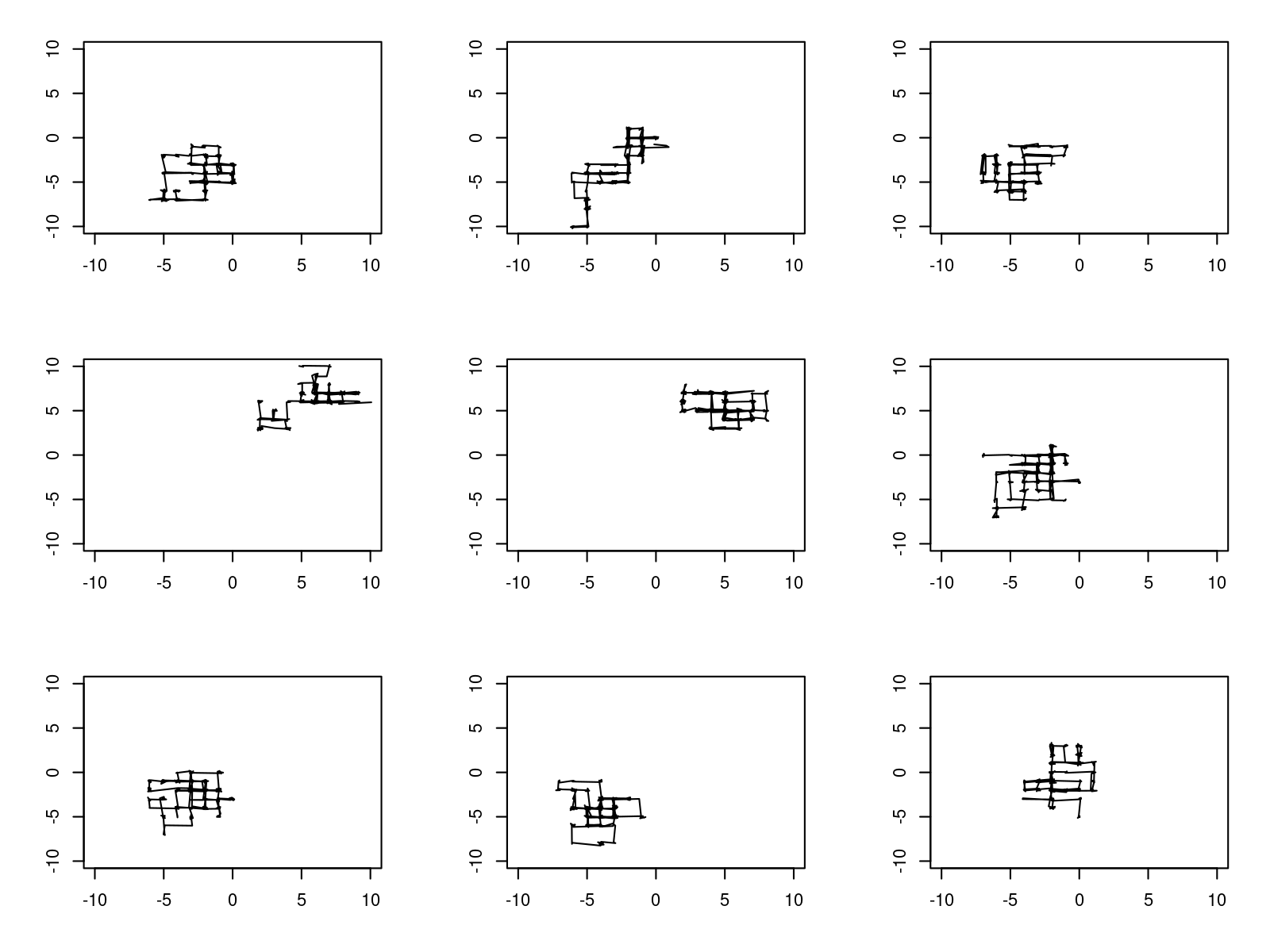}
        \caption{Trace plots from Metropolis}
        \label{fig:trace_precond_Metropolis}
    \end{subfigure}
     \begin{subfigure}[b]{0.32\textwidth}
        \centering
        \includegraphics[width=0.8\linewidth]{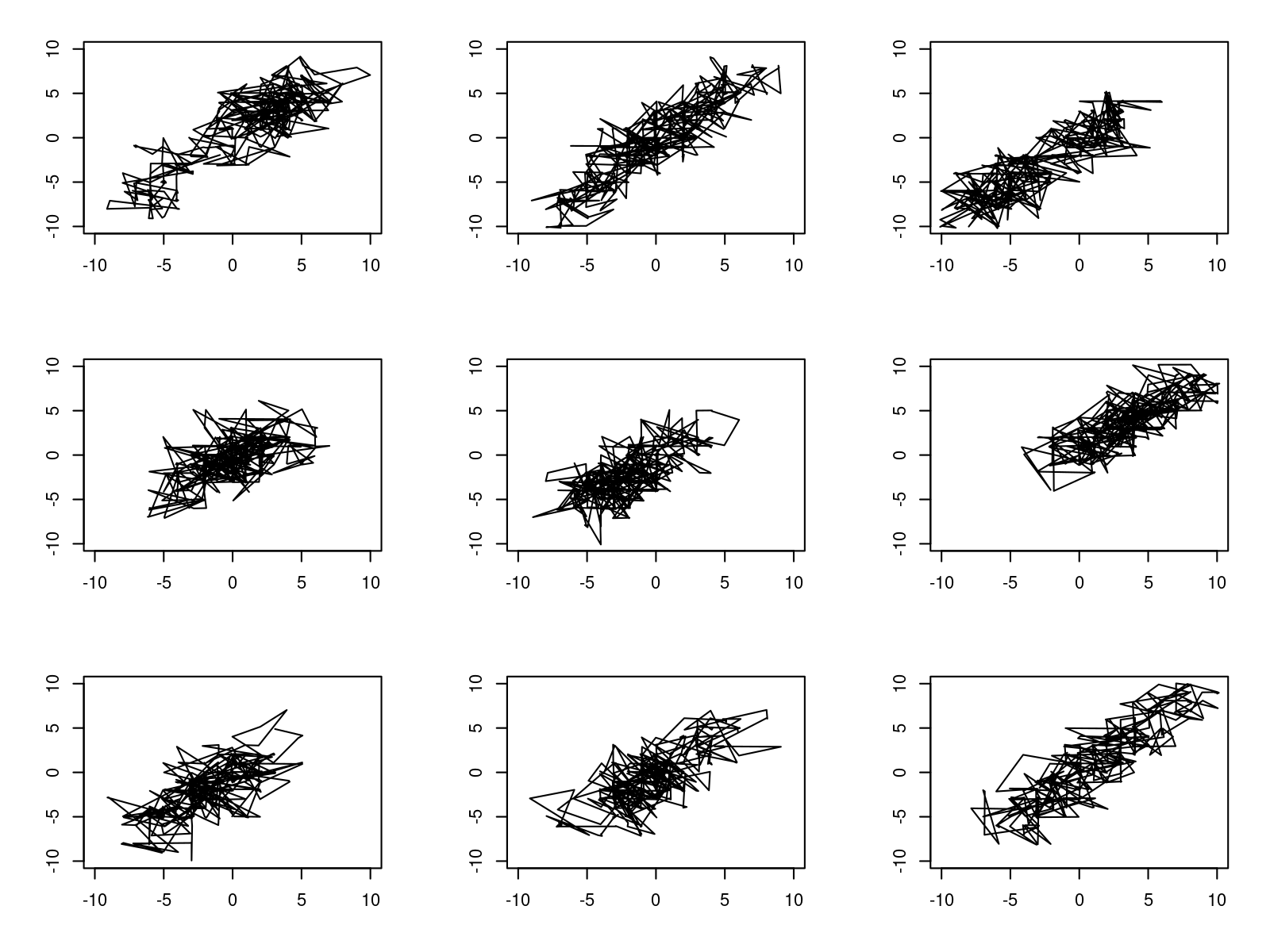}
        \caption{Trace plots from NCG}
        \label{fig:trace_precond_NCG}
    \end{subfigure}
     \begin{subfigure}[b]{0.32\textwidth}
        \centering
        \includegraphics[width=0.8\linewidth]{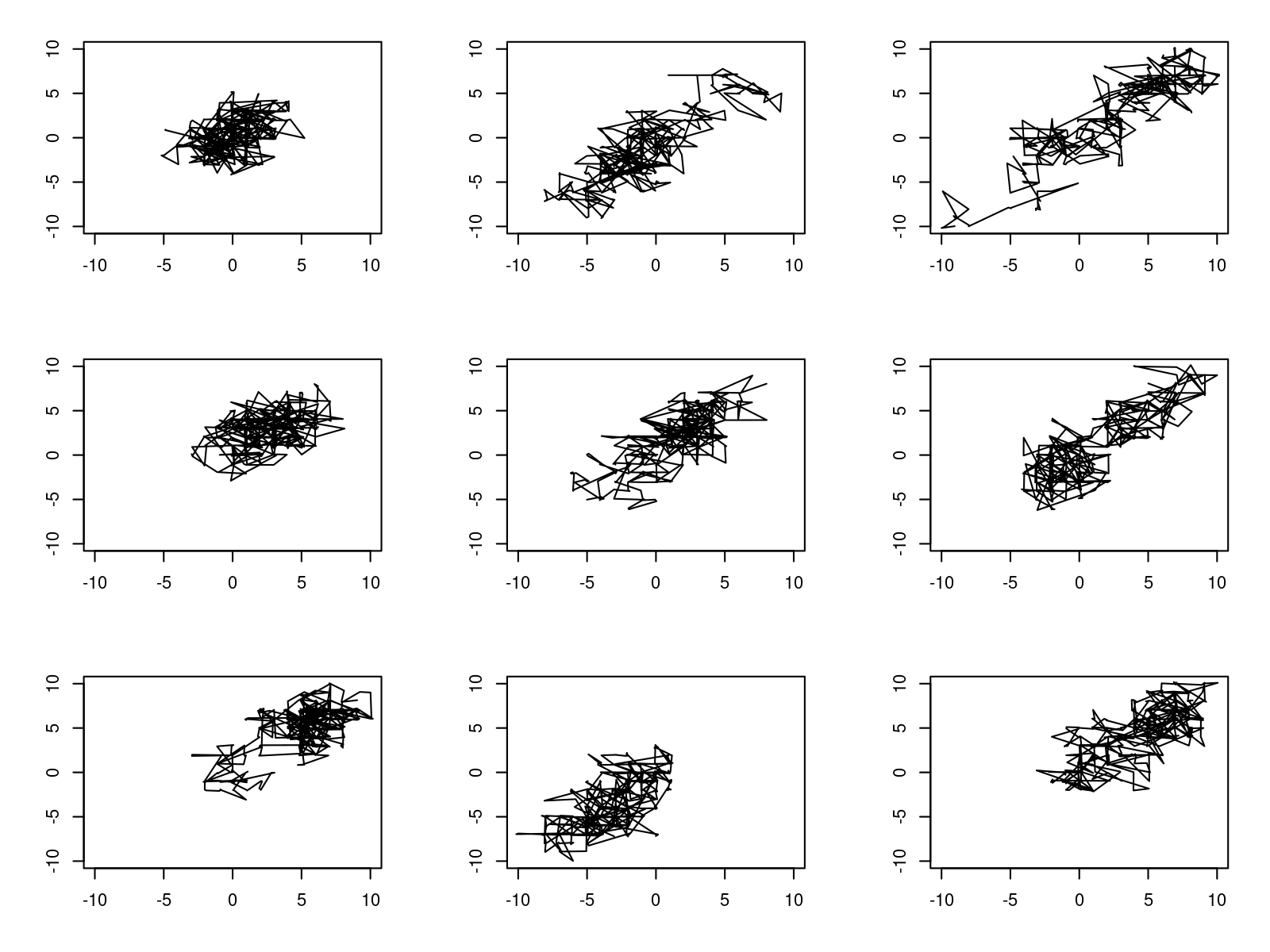}
        \caption{Trace plots from AVG}
        \label{fig:trace_precond_AVG}
    \end{subfigure}
     \begin{subfigure}[b]{0.32\textwidth}
        \centering
        \includegraphics[width=0.8\linewidth]{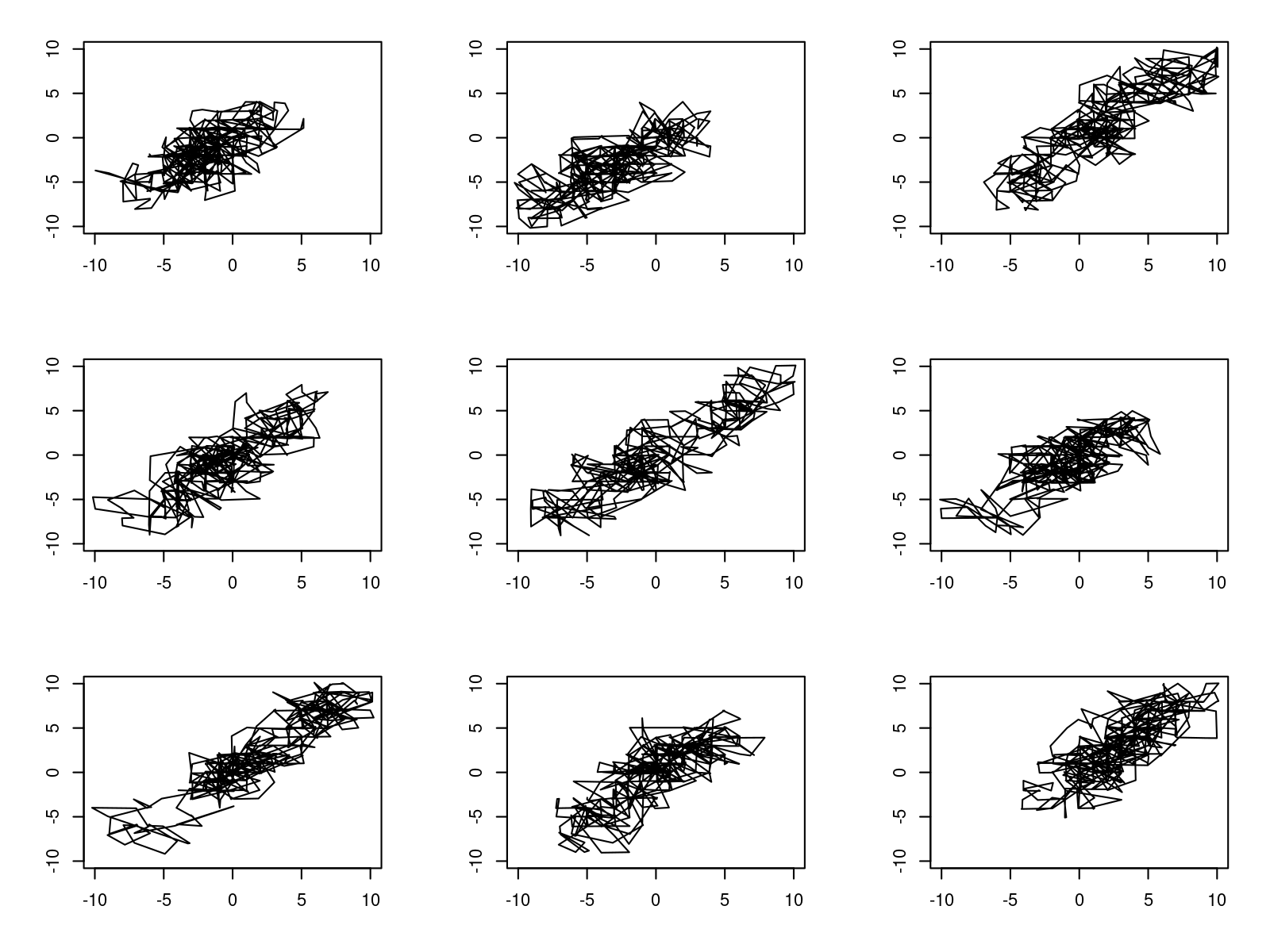}
        \caption{Trace plots from V-DHAMS}
        \label{fig:trace_precond_Hams}
    \end{subfigure}
     \begin{subfigure}[b]{0.32\textwidth}
        \centering
        \includegraphics[width=0.8\linewidth]{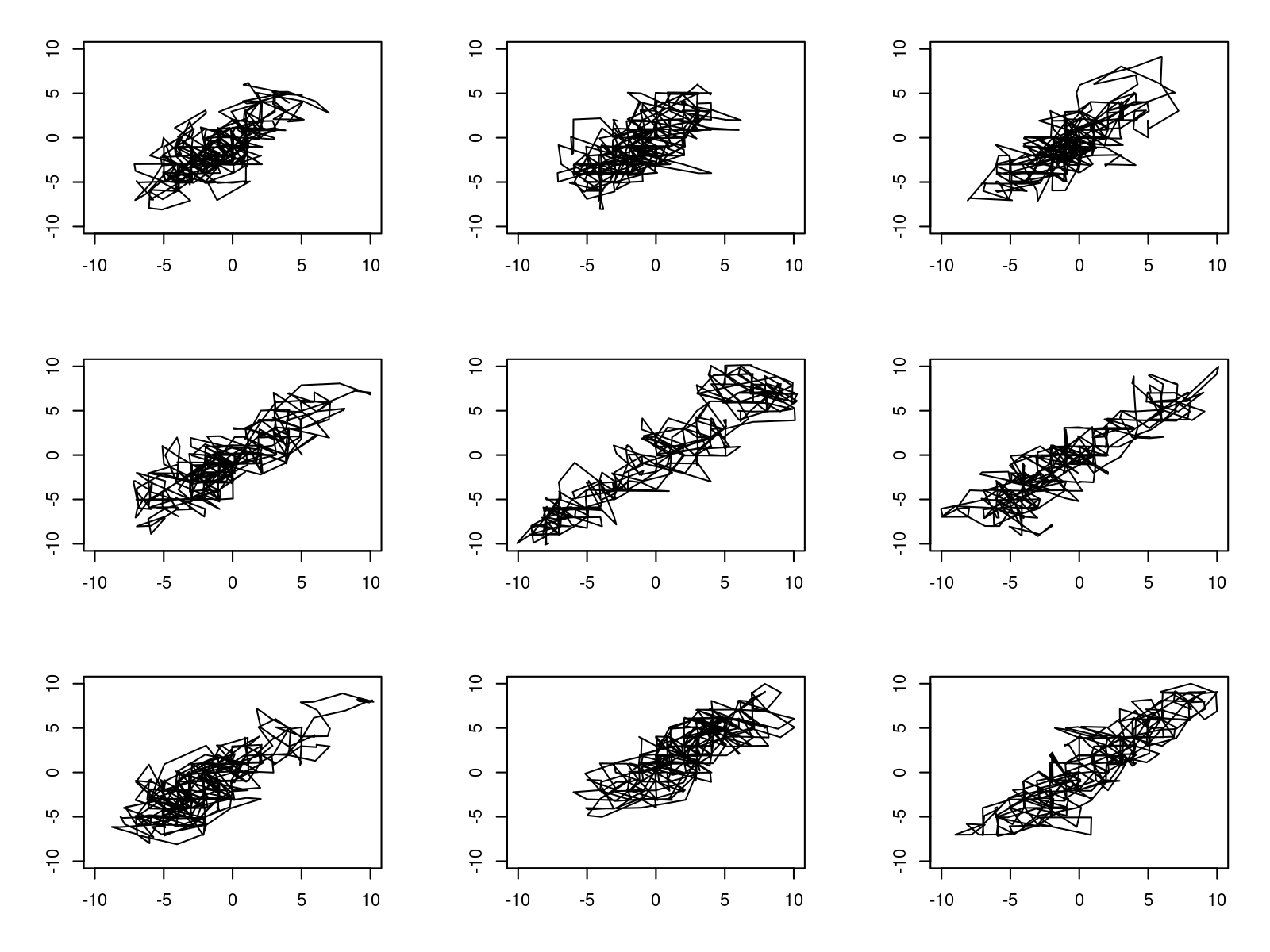}
        \caption{Trace plots from O-DHAMS}
        \label{fig:trace_precond_Overhams}
    \end{subfigure}
    \begin{subfigure}[b]{0.32\textwidth}
        \centering
        \includegraphics[width=0.8\linewidth]{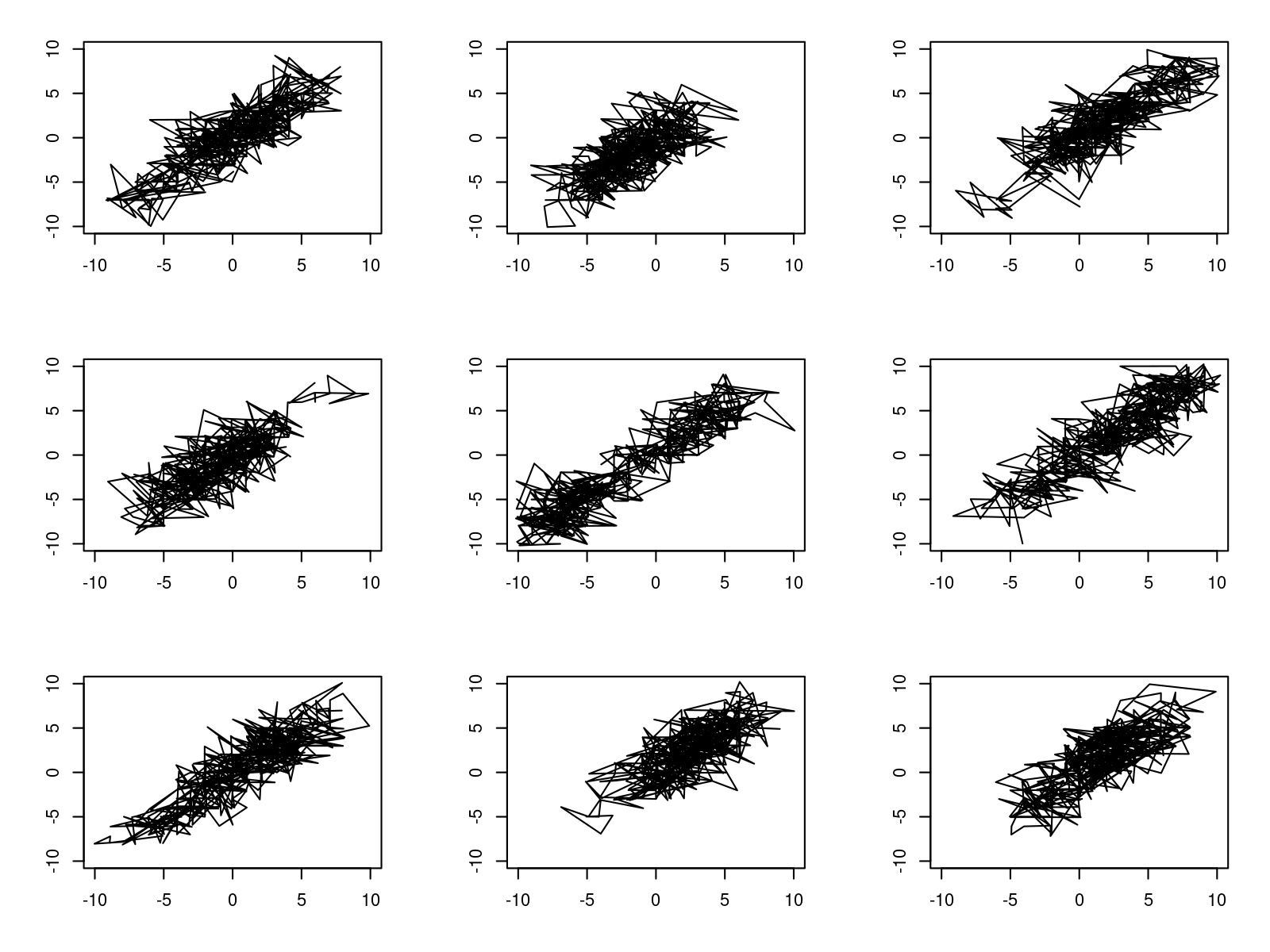}
        \caption{Trace plots from PAVG}
        \label{fig:trace_precond_PAVG}
    \end{subfigure}
    \begin{subfigure}[b]{0.32\textwidth}
        \centering
        \includegraphics[width=0.8\linewidth]{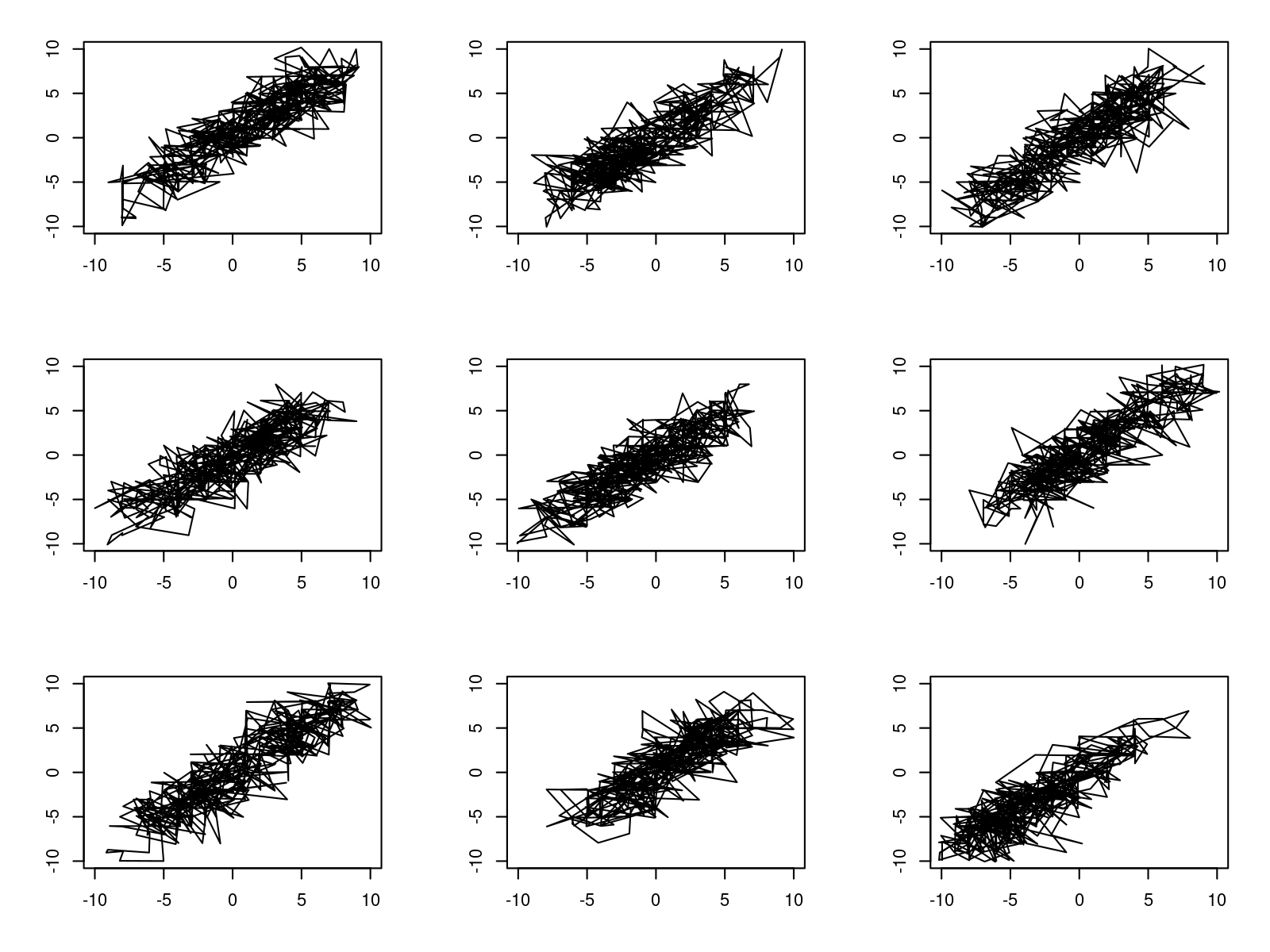}
        \caption{Trace plots from V-PDHAMS}
        \label{fig:trace_precond_vpdhams}
    \end{subfigure}
    \begin{subfigure}[b]{0.32\textwidth}
        \centering
        \includegraphics[width=0.8\linewidth]{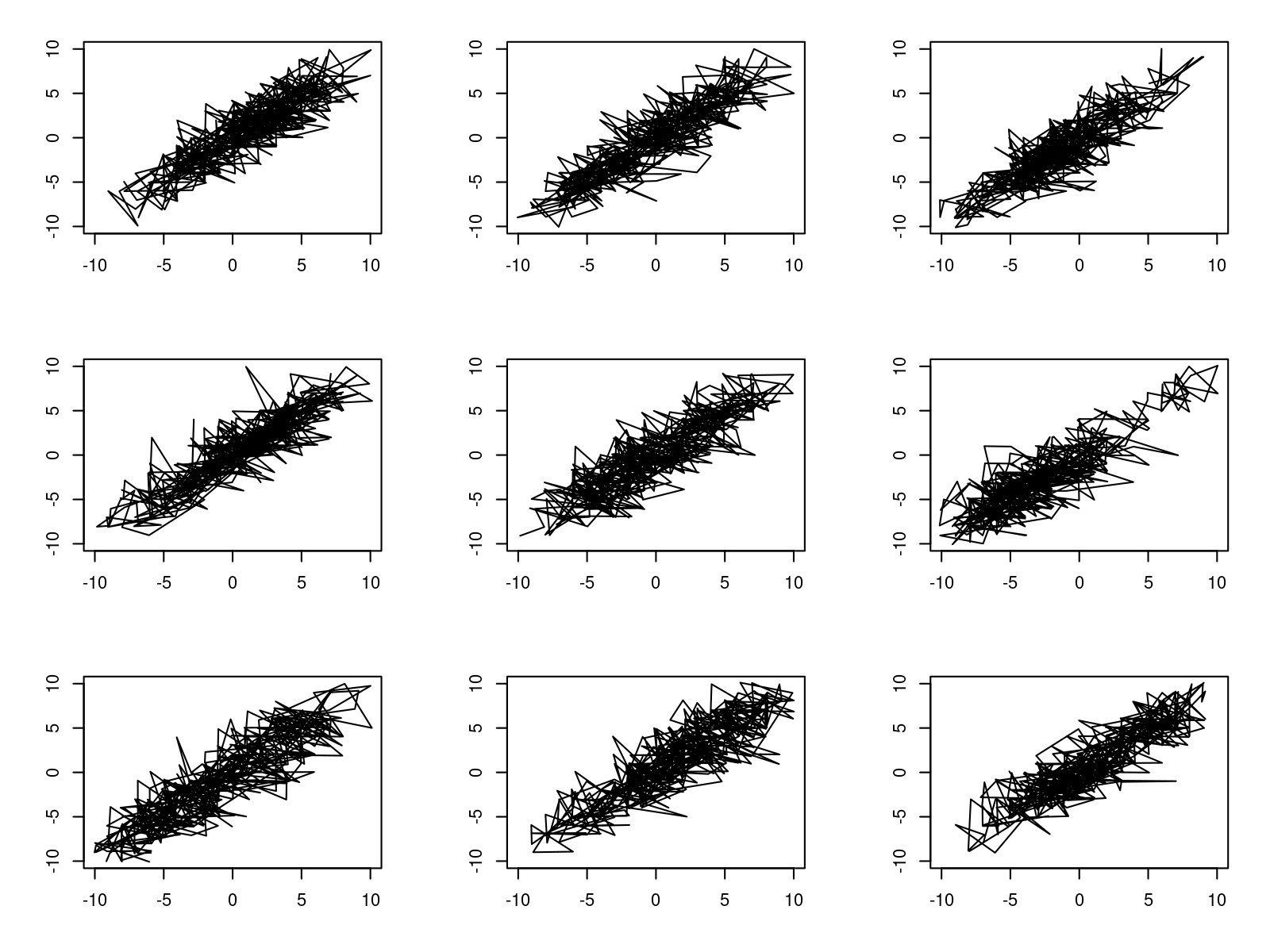}
        \caption{Trace plots from O-PDHAMS}
        \label{fig:trace_precond_opdhams}
    \end{subfigure}
\caption{Trace plots for discrete Gaussian distribution}
\label{fig:trace_precond_plots}
\end{figure}

The optimal parameters and associated acceptance rates for each method are presented in Table~\ref{tab:precond_parma_gaussian}. The smallest eigenvalue of preconditioning matrix $W$ is $-0.4$. We select 9 chains from the 100 parallel chains for each sampler and present their trace plots of the first two covariates for the first 450 draws after burn-in in Figure \ref{fig:trace_precond_plots}. Both V-PDHAMS and O-PDHAMS exhibit superior exploration capability in traversing the probability landscape.
\begin{table}[tbp]
\centering
\begin{tabular}{|c|c|c|c|c|}
\hline
    Sampler & Parameter & Acceptance Rate\\
    \hline
    Metropolis & $r=2$ & 0.73\\
    NCG & $\delta=3.5$ & 0.61\\
        AVG & $\delta=1.88$ & 0.58\\
   V-DHAMS & $\epsilon = 0.9, \delta=0.9, \phi=0.5$ & 0.86\\
    O-DHAMS & $\epsilon = 0.9, \delta=0.75, \phi=0.7, \beta=0.1$ & 0.79\\
    PAVG & $\delta =0.058$ & 1 \\
    V-PDHAMS & $\epsilon=0.9, \delta=0.058, \phi=0.0$ & 1 \\
    O-PDHAMS & $\epsilon=0.9, \delta=0.138, \phi=0.0, \beta=0.1$ & 1 \\
    \hline
\end{tabular}
\caption{Parameters for discrete Gaussian distribution}
\label{tab:precond_parma_gaussian}
\end{table}

\begin{figure}[tbp]
     \begin{subfigure}[b]{0.32\textwidth}
        \centering
        \includegraphics[width=0.8\linewidth]{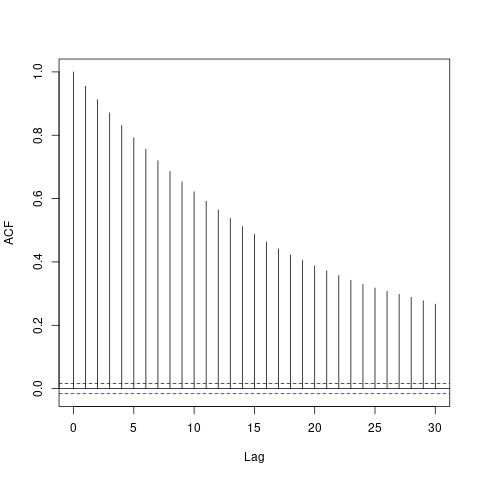}
        \caption{ACF from Metropolis}
        \label{fig:acf_precond_Metropolis}
    \end{subfigure}
     \begin{subfigure}[b]{0.32\textwidth}
        \centering
        \includegraphics[width=0.8\linewidth]{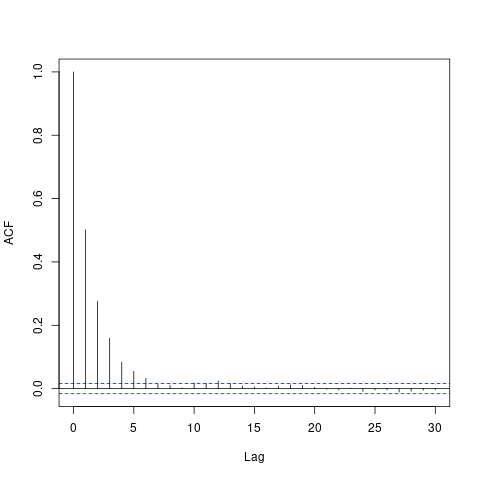}
        \caption{ACF from NCG}
        \label{fig:acf_precond_NCG}
    \end{subfigure}
     \begin{subfigure}[b]{0.32\textwidth}
        \centering
        \includegraphics[width=0.8\linewidth]{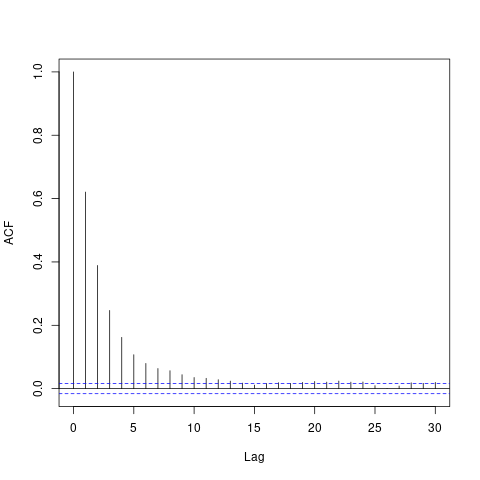}
        \caption{ACF from AVG}
        \label{fig:acf_precond_avg}
    \end{subfigure}
     \begin{subfigure}[b]{0.32\textwidth}
        \centering
        \includegraphics[width=0.8\linewidth]{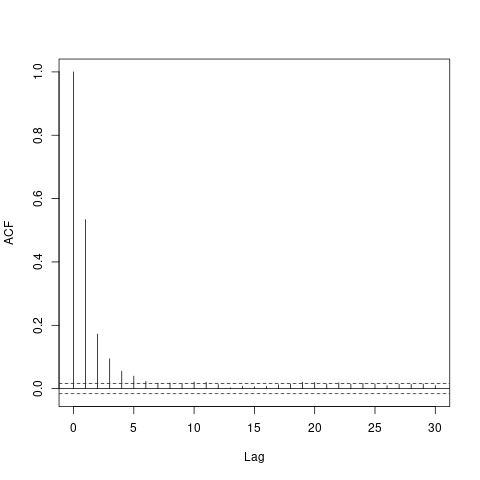}
        \caption{ACF from V-DHAMS}
        \label{fig:acf_precond_Hams}
    \end{subfigure}
     \begin{subfigure}[b]{0.32\textwidth}
        \centering
        \includegraphics[width=0.8\linewidth]{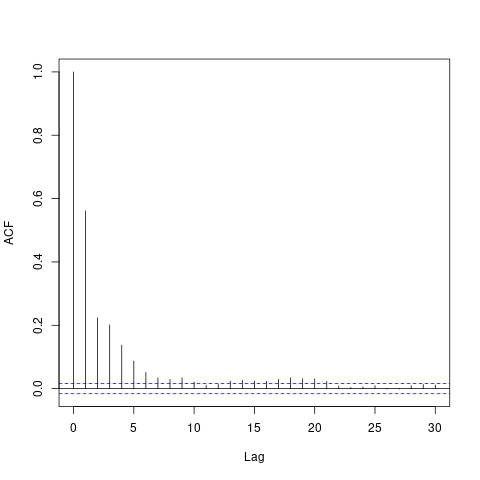}
        \caption{ACF from O-DHAMS}
        \label{fig:acf_precond_overhams}
    \end{subfigure}
     \begin{subfigure}[b]{0.32\textwidth}
        \centering
        \includegraphics[width=0.8\linewidth]{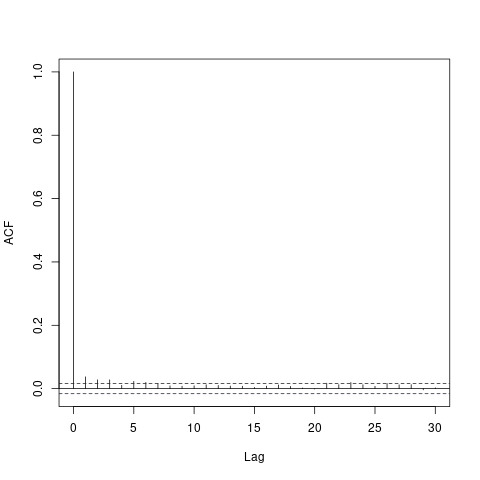}
        \caption{ACF from PAVG}
        \label{fig:acf_precond_PAVG}
    \end{subfigure}
     \begin{subfigure}[b]{0.32\textwidth}
        \centering
        \includegraphics[width=0.8\linewidth]{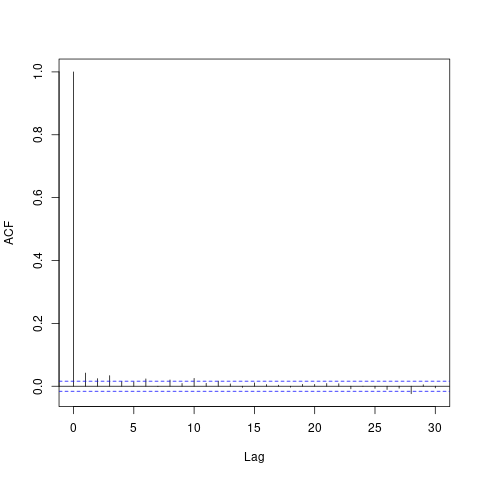}
        \caption{ACF from V-PDHAMS}
        \label{fig:acf_precond_vpdhams}
    \end{subfigure}
     \begin{subfigure}[b]{0.32\textwidth}
        \centering
        \includegraphics[width=0.8\linewidth]{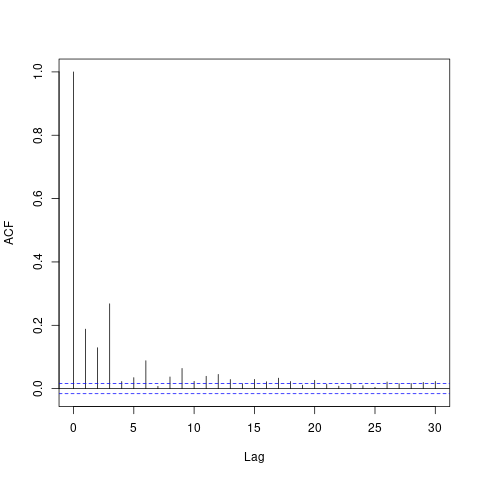}
        \caption{ACF from O-PDHAMS}
        \label{fig:acf_precond_opdhams}
    \end{subfigure}
\caption{ACF plots for discrete Gaussian distribution}
\label{fig:acf_precond_plots}
\end{figure}
\begin{figure}[tbp]
    \centering
    \includegraphics[width=0.4\linewidth]{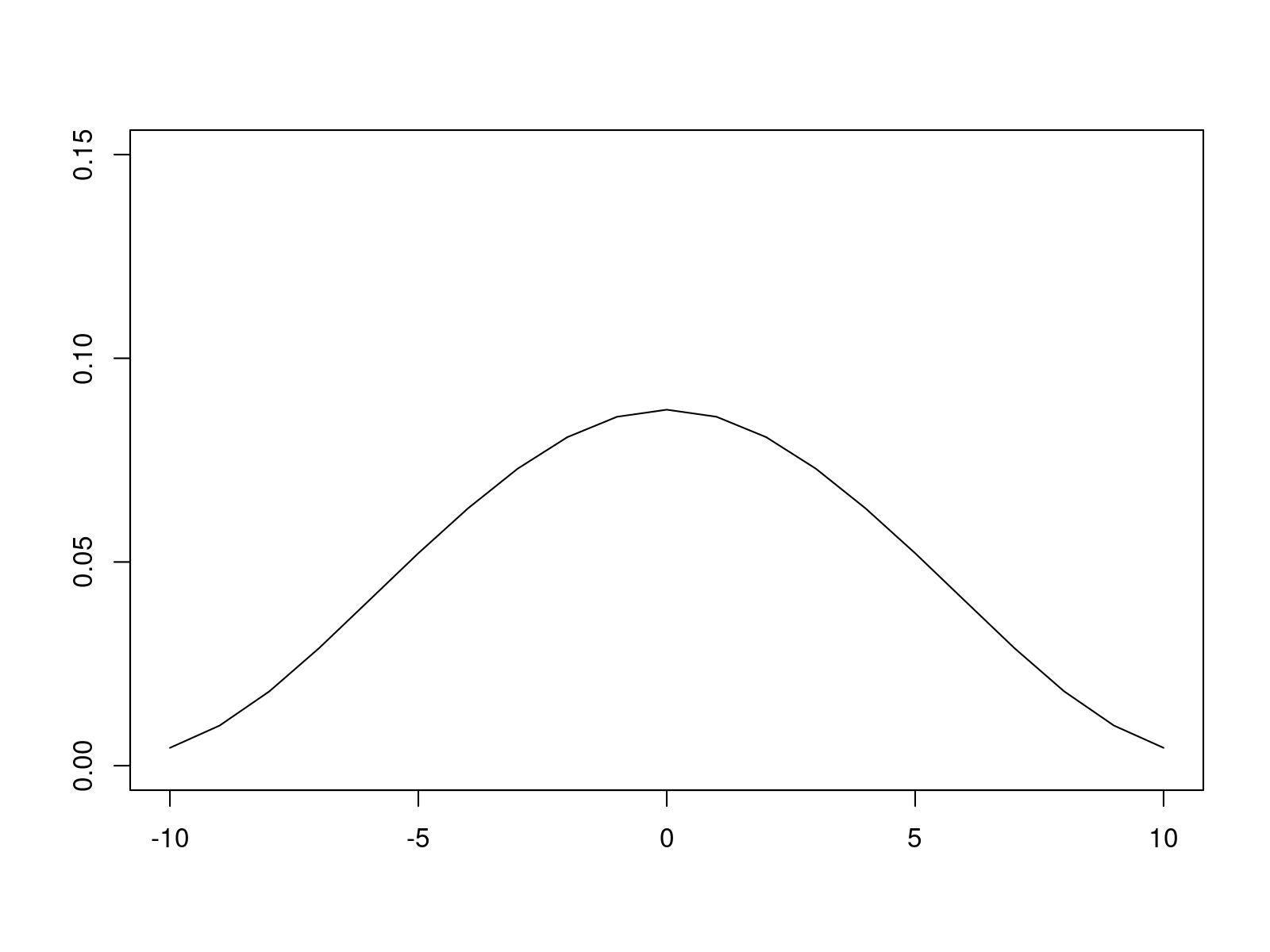}
    \caption{True marginal distribution of the first coordinate in discrete Gaussian distribution}
    \label{fig:freq_precond_truth}
\end{figure}
The plots of the auto-correlation functions (ACF) of the negative potential function $f(s)$ from a single chain are presented in Figure~\ref{fig:acf_precond_plots}. Preconditioned samplers show much lower auto-correlations, indicating reduced dependencies among draws and improved mixing efficiency. Notably, PAVG and V-PDHAMS demonstrate almost zero auto-correlations.

\begin{figure}[H]
     \begin{subfigure}[b]{0.32\textwidth}
        \centering
        \includegraphics[width=0.8\linewidth]{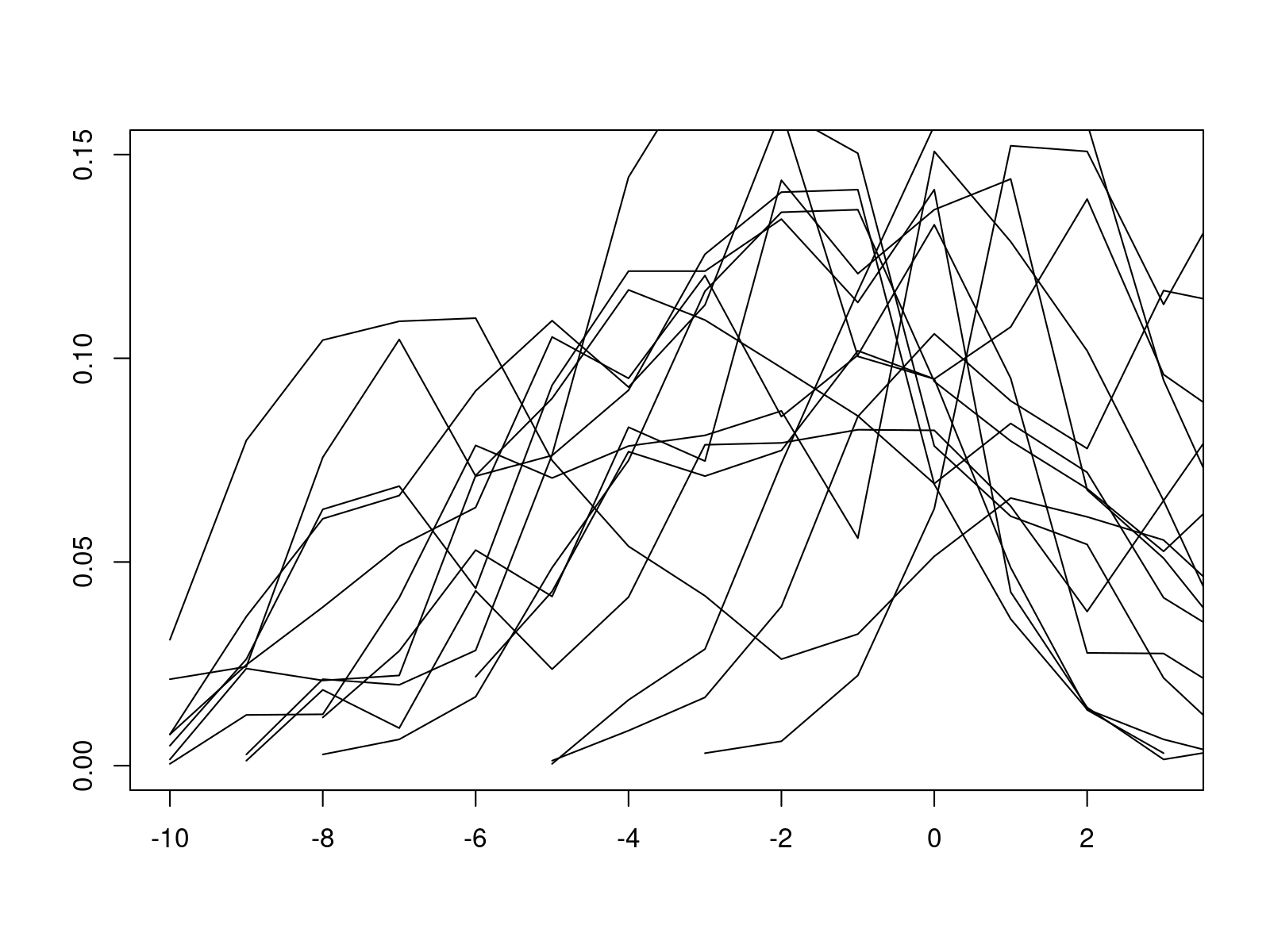}
        \caption{Metropolis}
        \label{fig:freq_precond_Metropolis}
    \end{subfigure}
     \begin{subfigure}[b]{0.32\textwidth}
        \centering
        \includegraphics[width=0.8\linewidth]{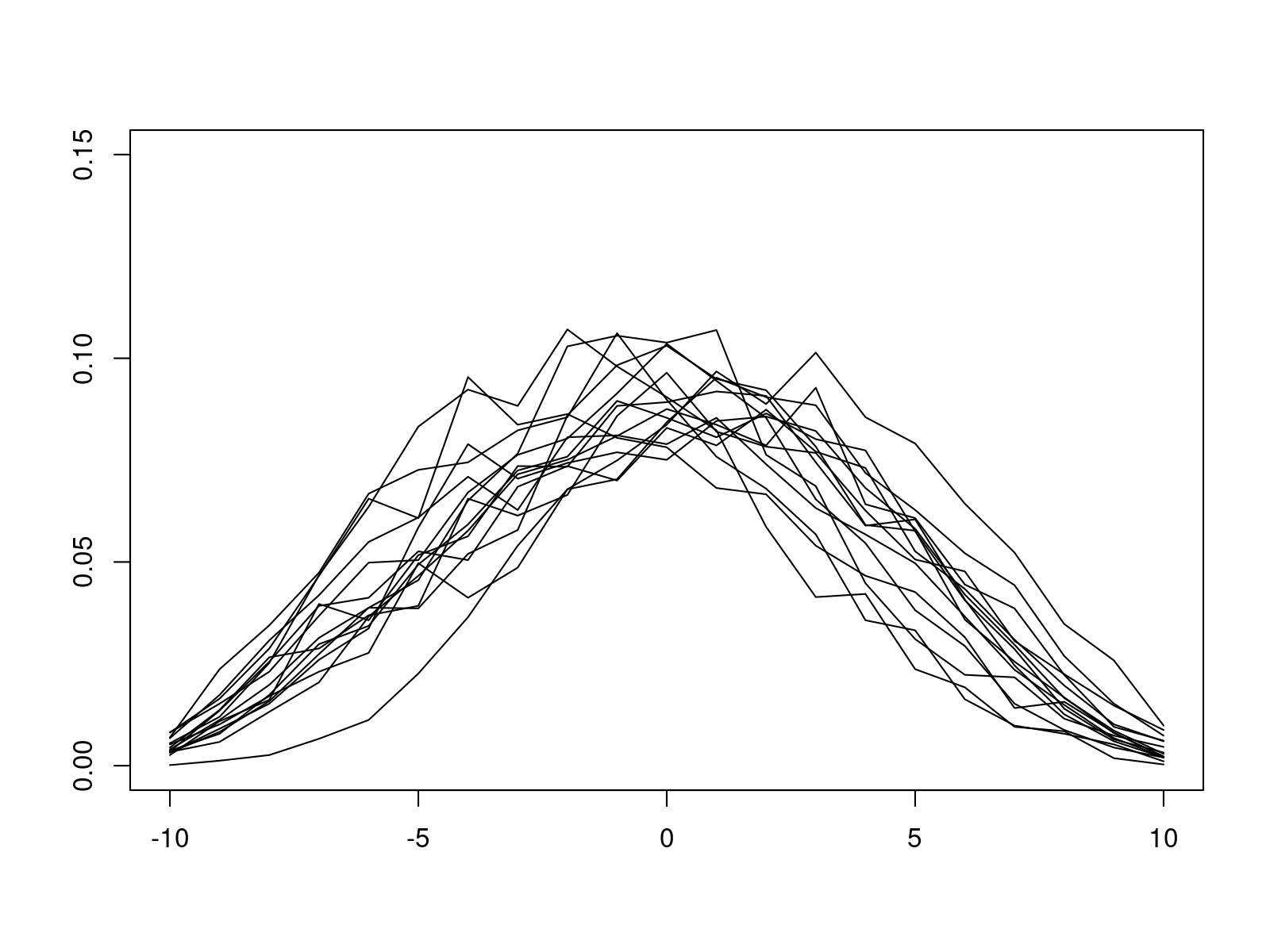}
        \caption{ NCG}
        \label{fig:freq_precond_NCG}
    \end{subfigure}
     \begin{subfigure}[b]{0.32\textwidth}
        \centering
        \includegraphics[width=0.8\linewidth]{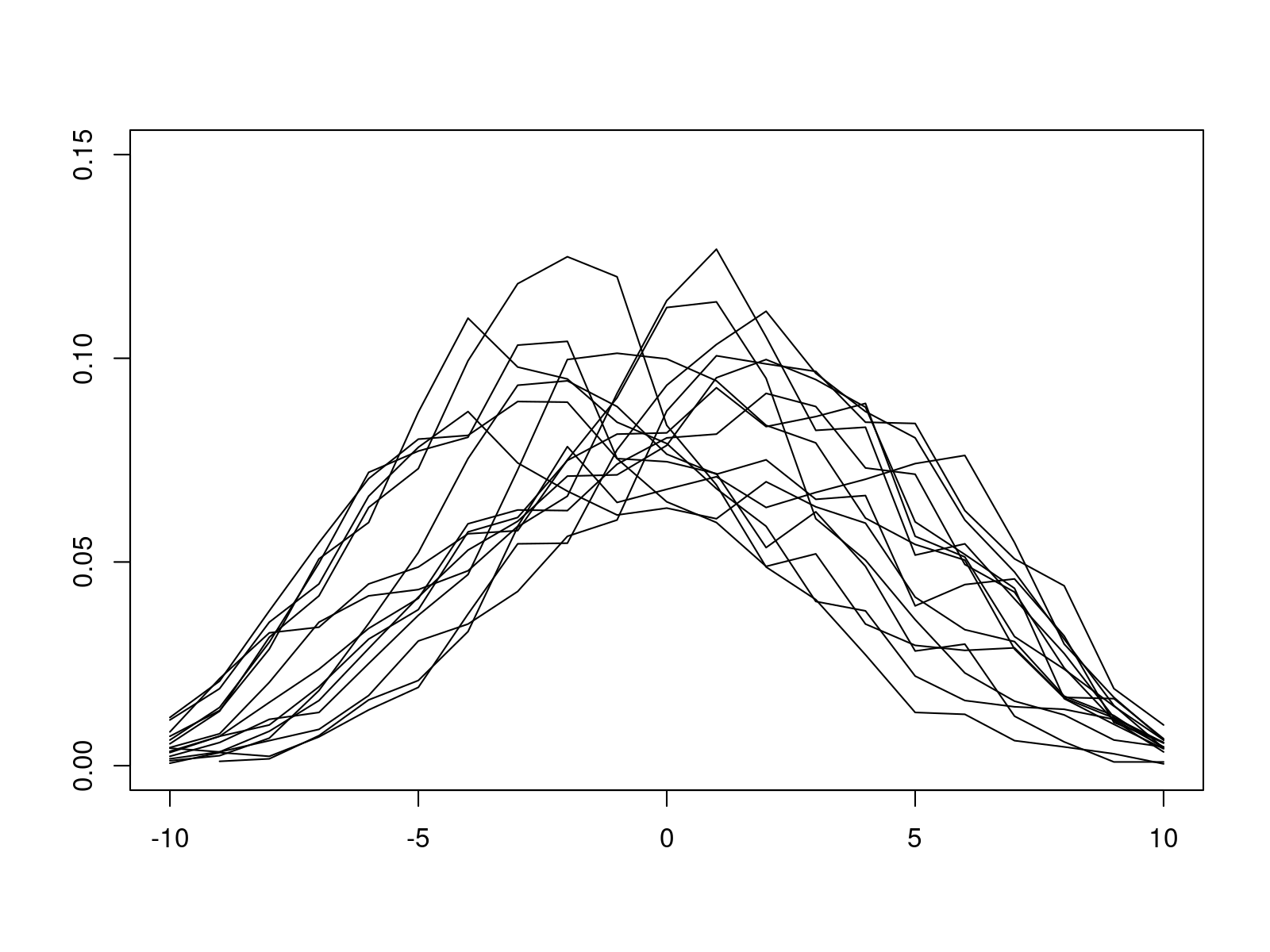}
        \caption{ AVG}
        \label{fig:freq_precond_avg}
    \end{subfigure}
     \begin{subfigure}[b]{0.32\textwidth}
        \centering
        \includegraphics[width=0.8\linewidth]{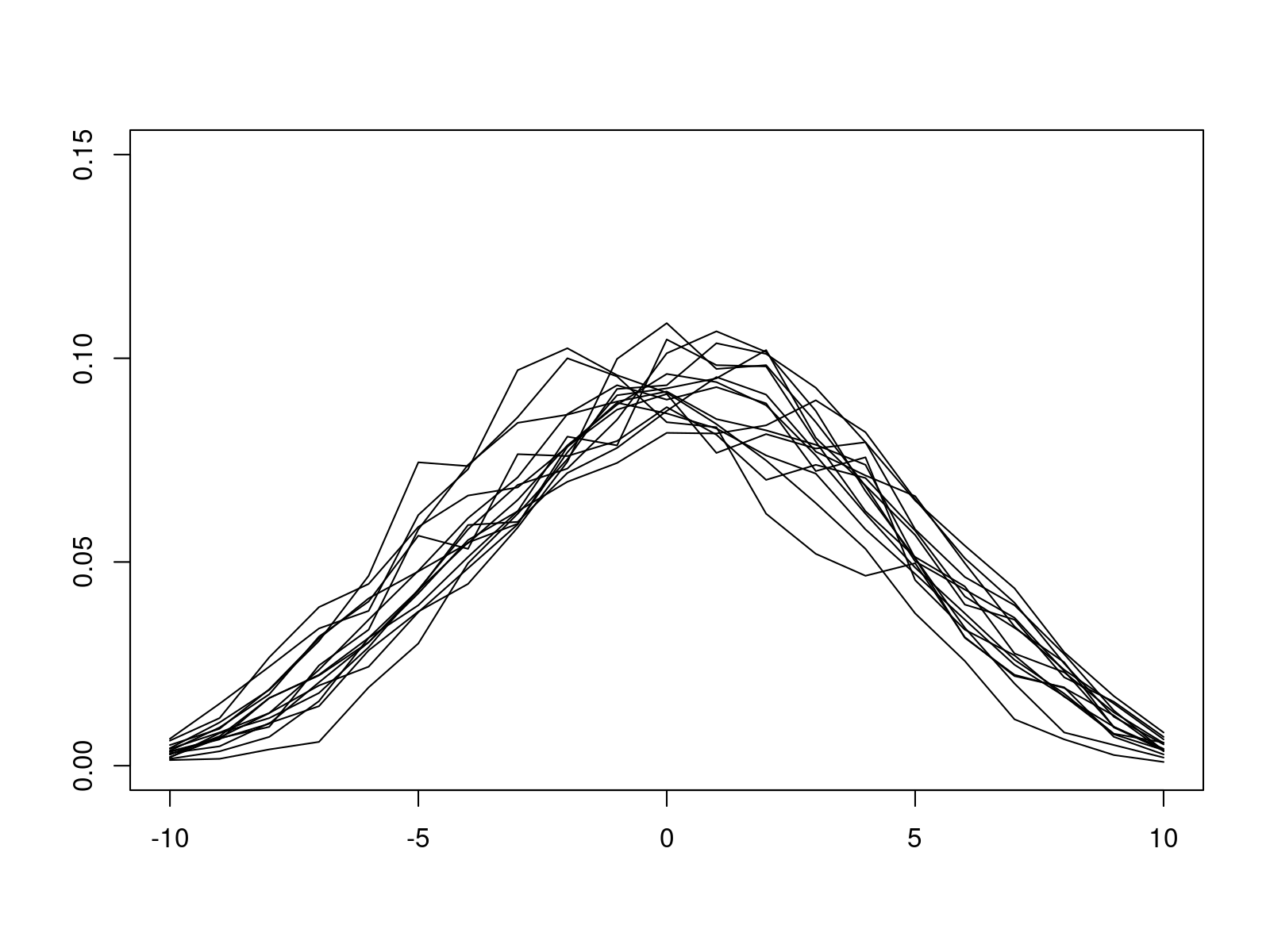}
        \caption{ V-DHAMS}
        \label{fig:freq_precond_Hams}
    \end{subfigure}
     \begin{subfigure}[b]{0.32\textwidth}
        \centering
        \includegraphics[width=0.8\linewidth]{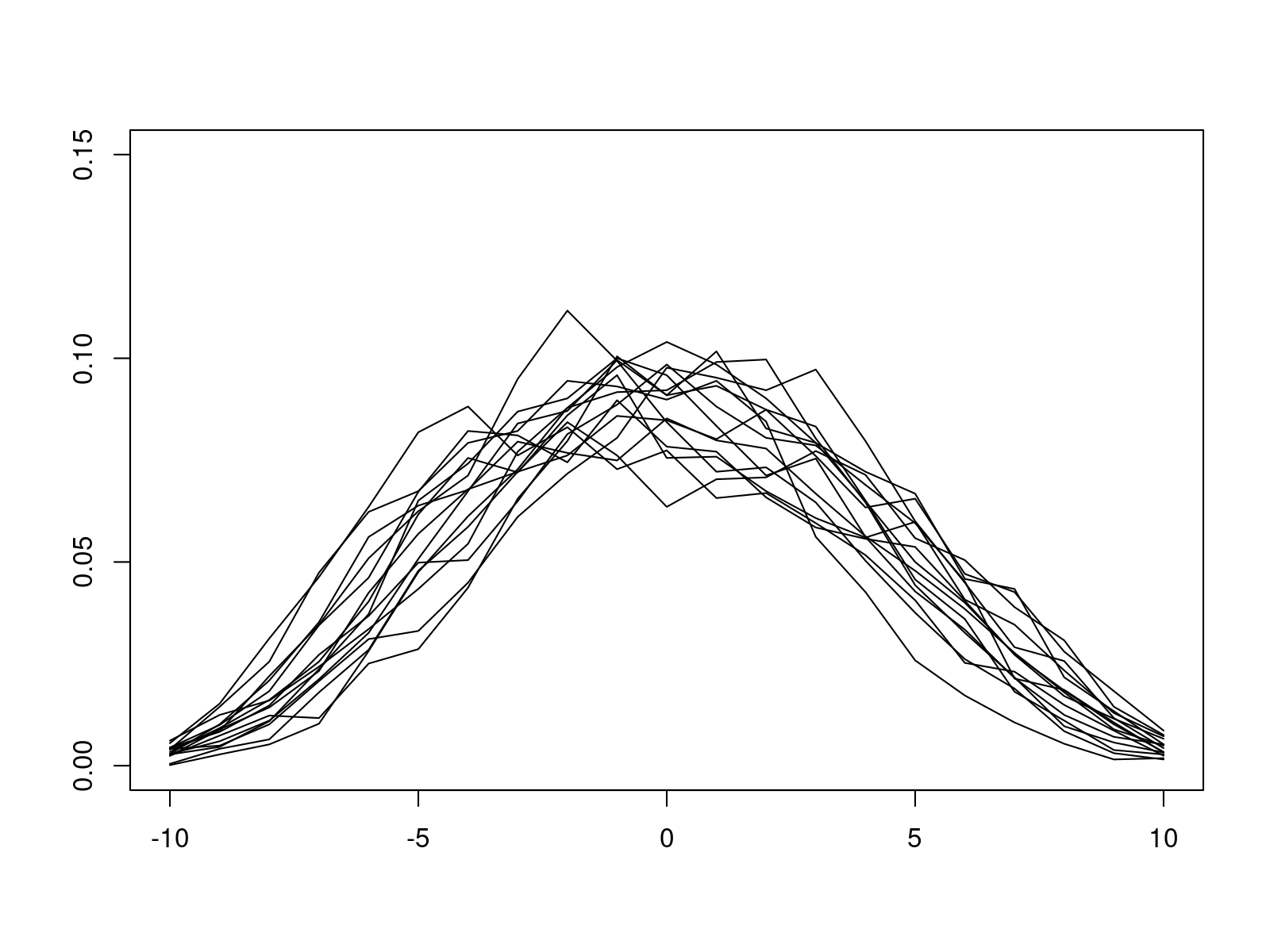}
        \caption{O-DHAMS}
        \label{fig:freq_precond_overhams}
    \end{subfigure}
    \begin{subfigure}[b]{0.32\textwidth}
        \centering
        \includegraphics[width=0.8\linewidth]{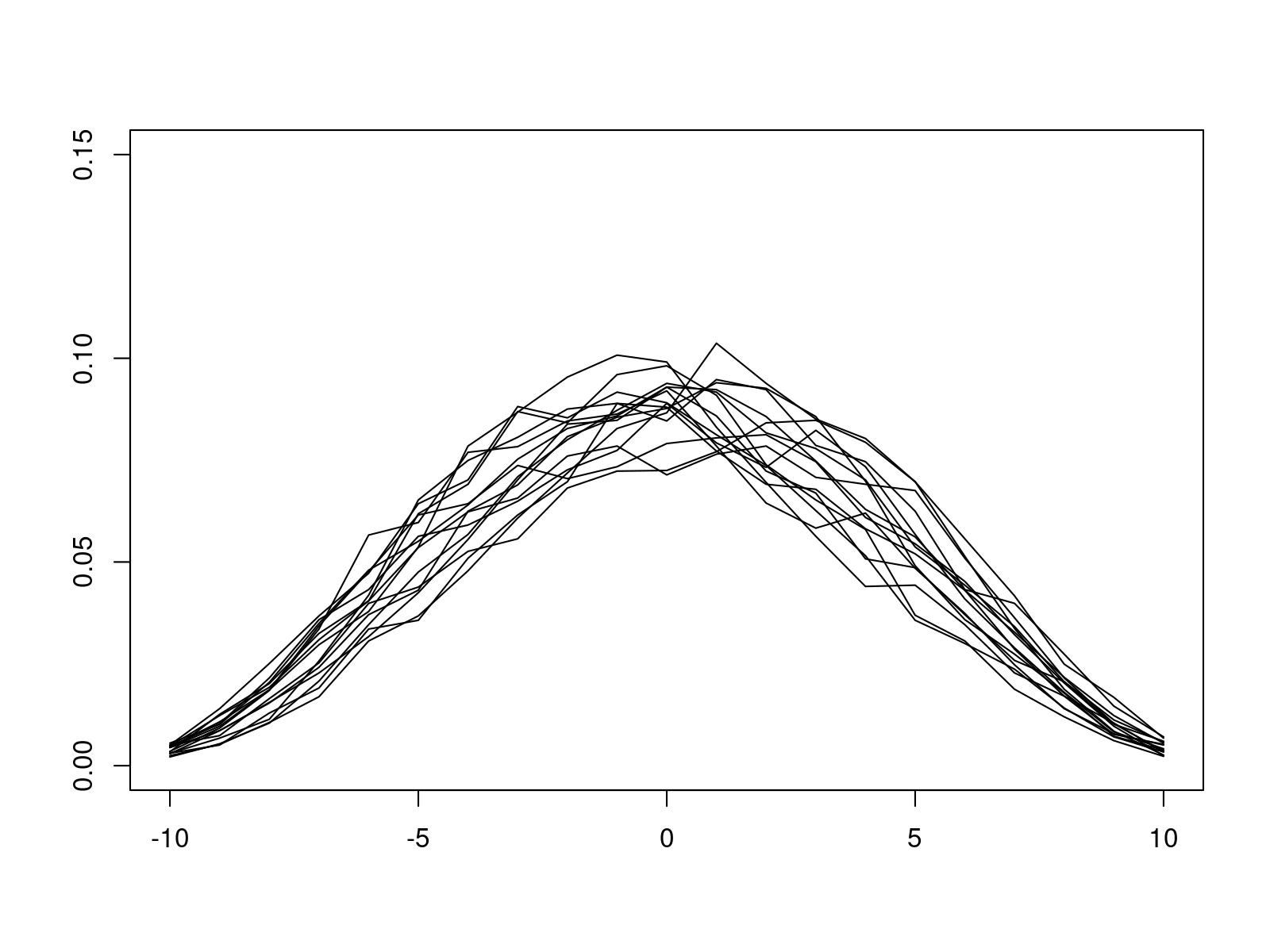}
        \caption{PAVG}
        \label{fig:freq_precond_pavg}
    \end{subfigure}
    \begin{subfigure}[b]{0.32\textwidth}
        \centering
        \includegraphics[width=0.8\linewidth]{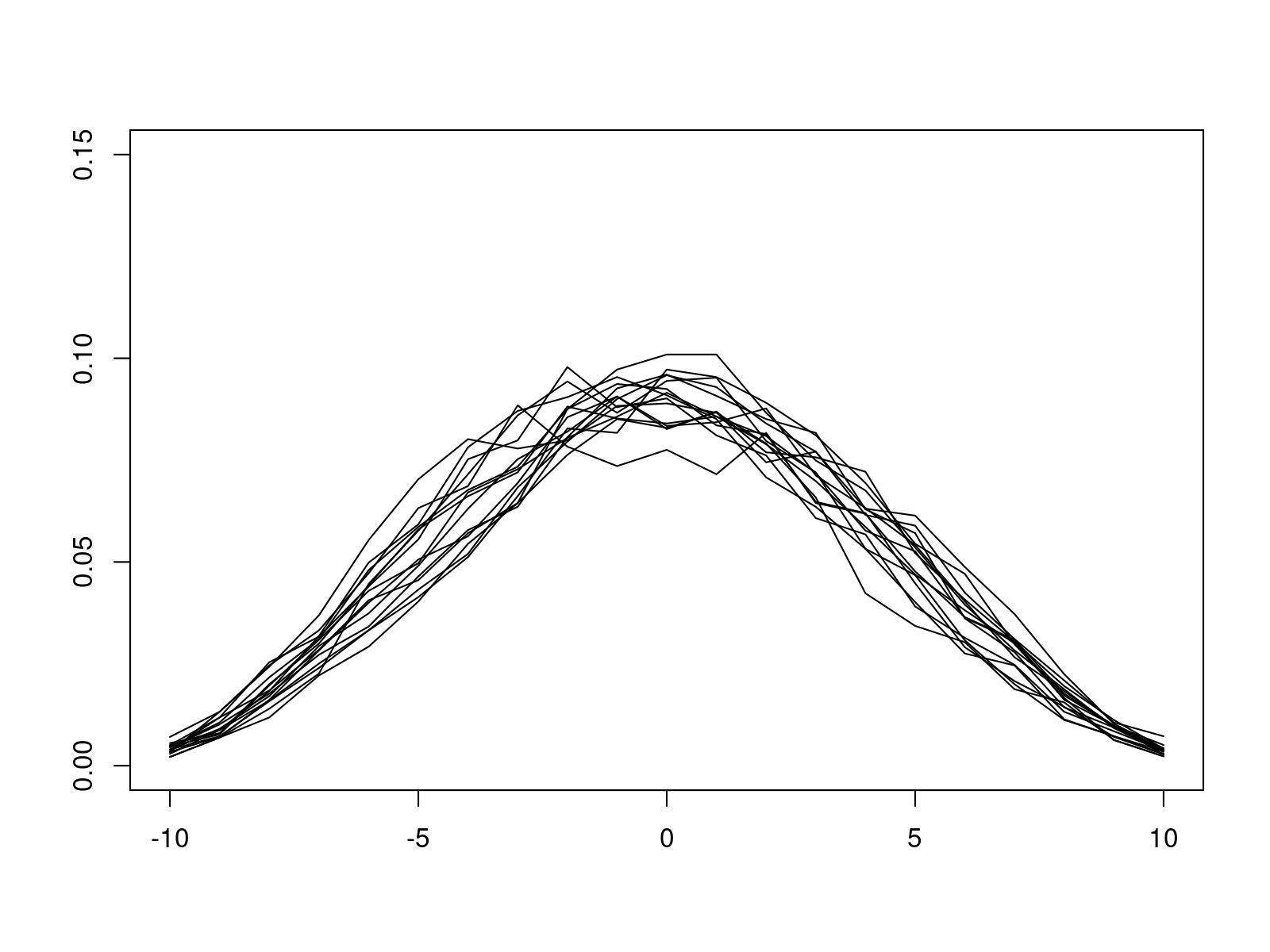}
        \caption{V-PDHAMS}
        \label{fig:freq_precond_vpdhams}
    \end{subfigure}
    \begin{subfigure}[b]{0.32\textwidth}
        \centering
        \includegraphics[width=0.8\linewidth]{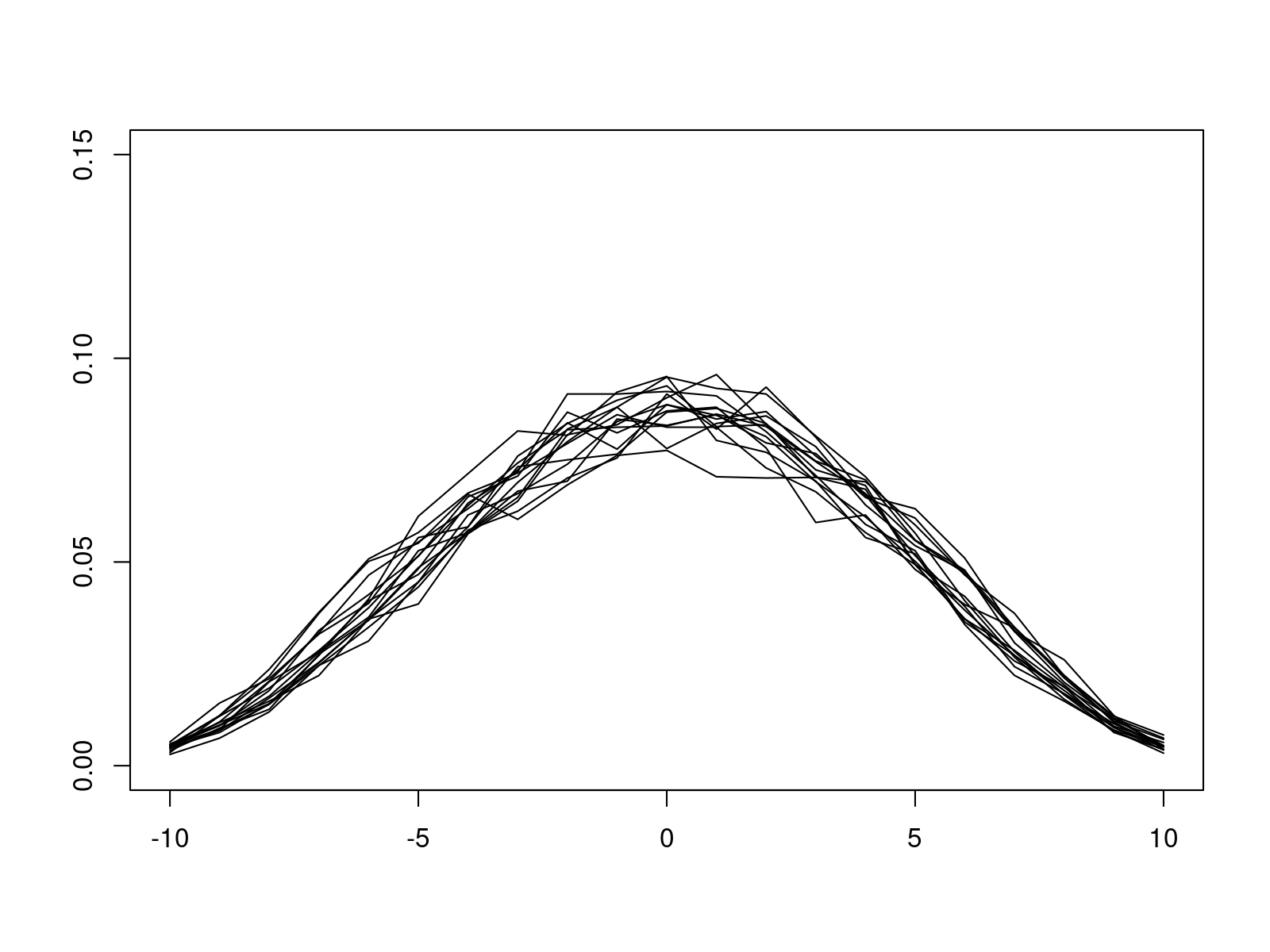}
        \caption{O-PDHAMS}
        \label{fig:freq_precond_opdhams}
    \end{subfigure}
\caption{Frequency plots of the first coordinate in discrete Gaussian distribution}
\label{fig:freq_precond_plots}
\end{figure}

For each sampler, we present frequency plots of the first coordinate across 10 independent chains, each based on 6, draws, as shown in Figure~\ref{fig:freq_precond_plots}.
The corresponding true marginal distribution is displayed in Figure~\ref{fig:freq_precond_truth}. Among all methods, the frequency plots generated by V-PDHAMS and O-PDHAMS are the closest to the ground truth, indicating more accurate sampling performance.

\subsection{Additional Results for Quadratic Mixture}\label{sec:precond_mixture_results}
\begin{figure}[tbp]
 \begin{subfigure}[b]{0.32\textwidth}
        \centering
        \includegraphics[width=0.8\linewidth]{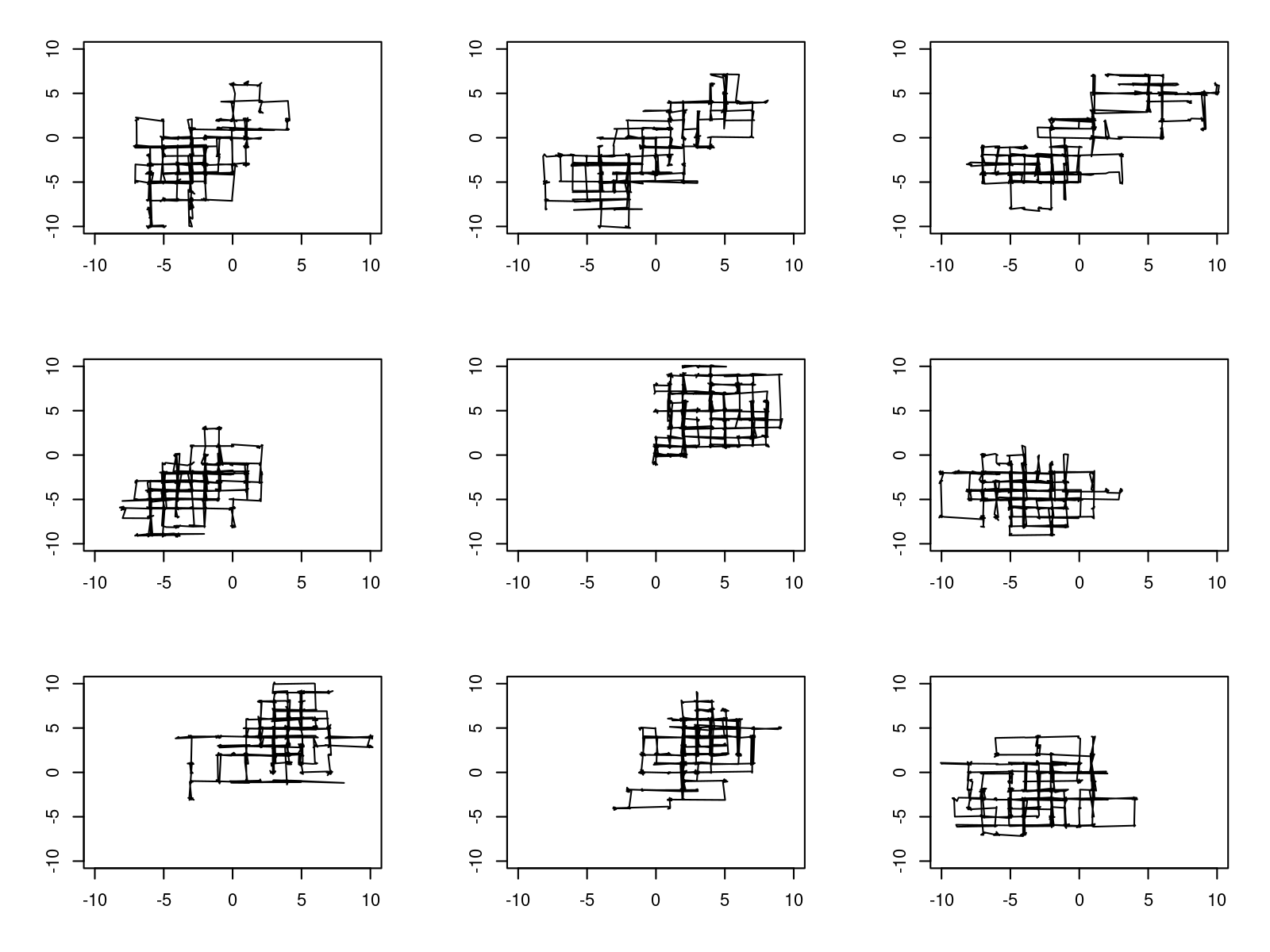}
        \caption{Trace plots from Metropolis}
        \label{fig:poly_less_precond_trace_Metropolis}
    \end{subfigure}
     \begin{subfigure}[b]{0.32\textwidth}
        \centering
        \includegraphics[width=0.8\linewidth]{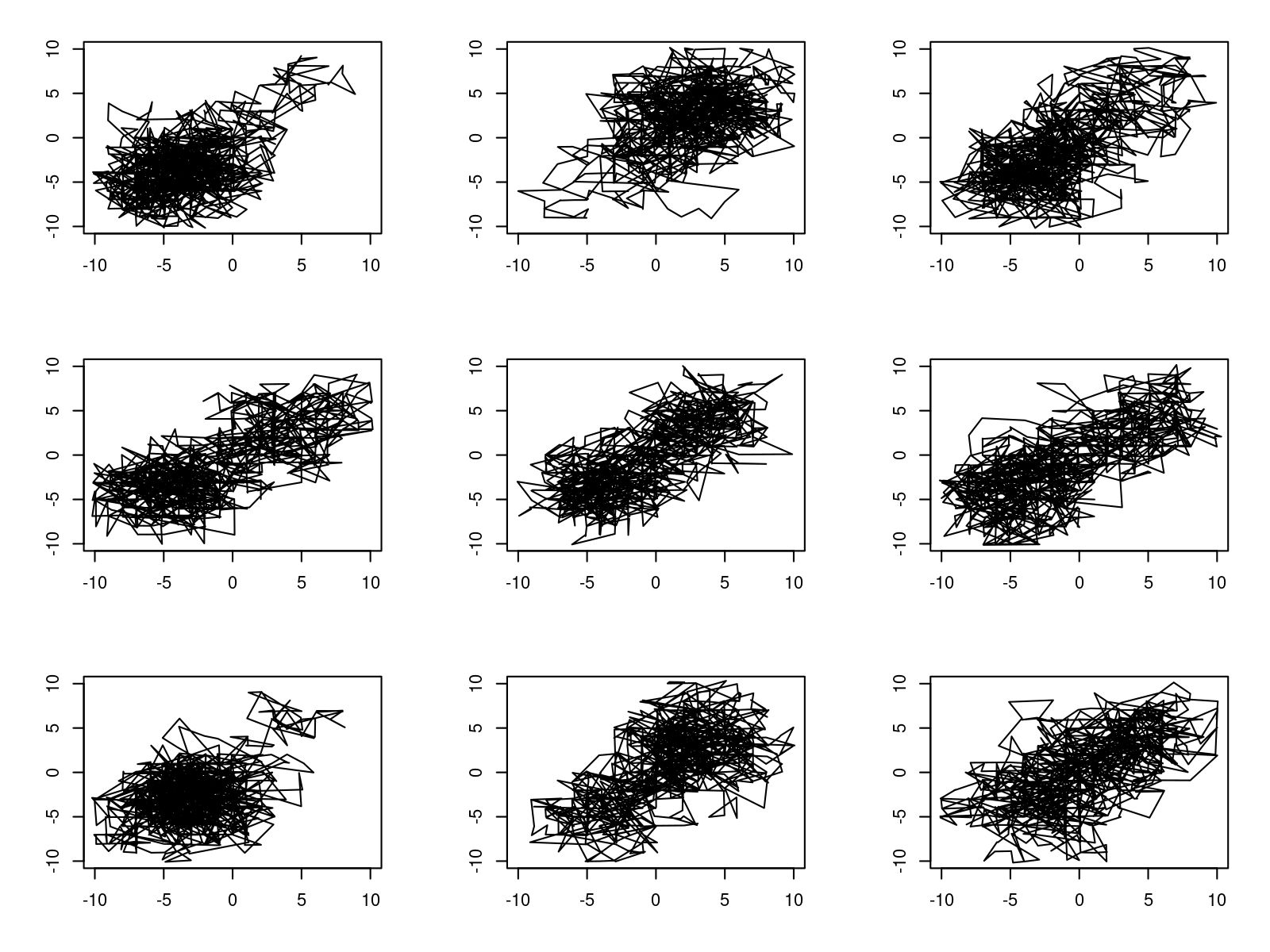}
        \caption{Trace plots from NCG}
        \label{fig:poly_less_precond_trace_NCG}
    \end{subfigure}
     \begin{subfigure}[b]{0.32\textwidth}
        \centering
        \includegraphics[width=0.8\linewidth]{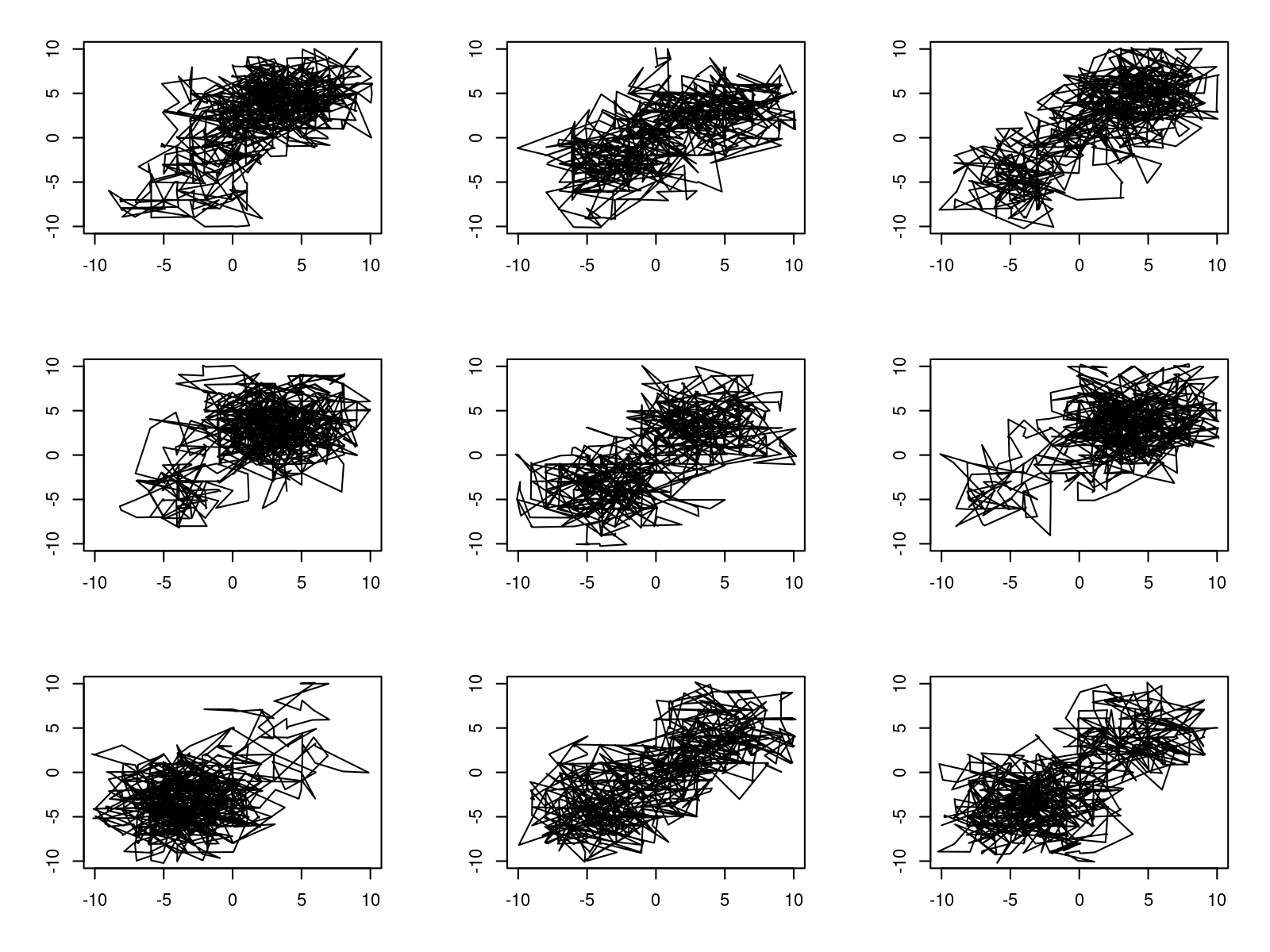}
        \caption{Trace plots from AVG}
        \label{fig:poly_less_precond_trace_AVG}
    \end{subfigure}
     \begin{subfigure}[b]{0.32\textwidth}
        \centering
        \includegraphics[width=0.8\linewidth]{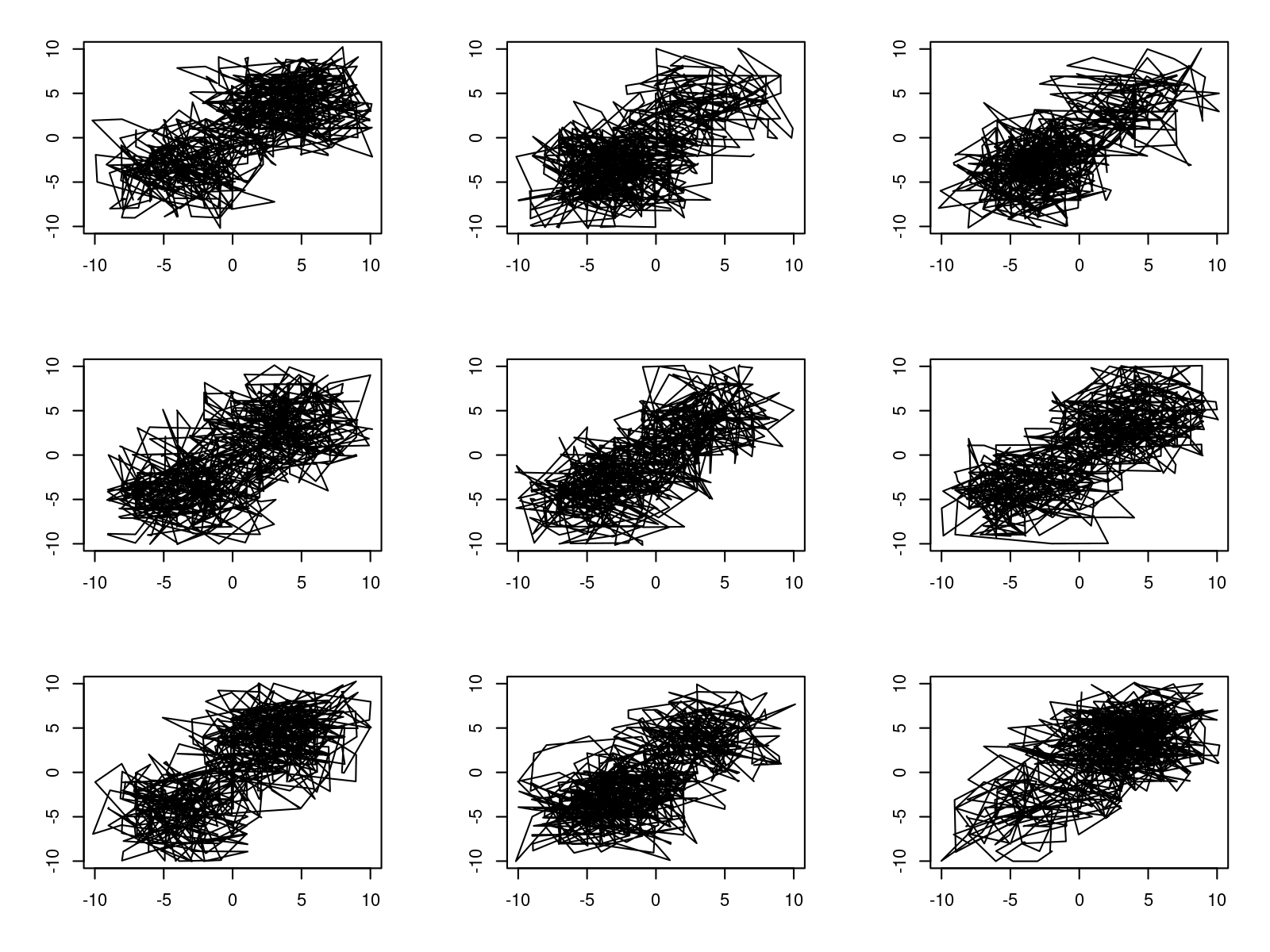}
        \caption{Trace plots from V-DHAMS}
        \label{fig:poly_less_precond_trace_Hams}
    \end{subfigure}
     \begin{subfigure}[b]{0.32\textwidth}
        \centering
        \includegraphics[width=0.8\linewidth]{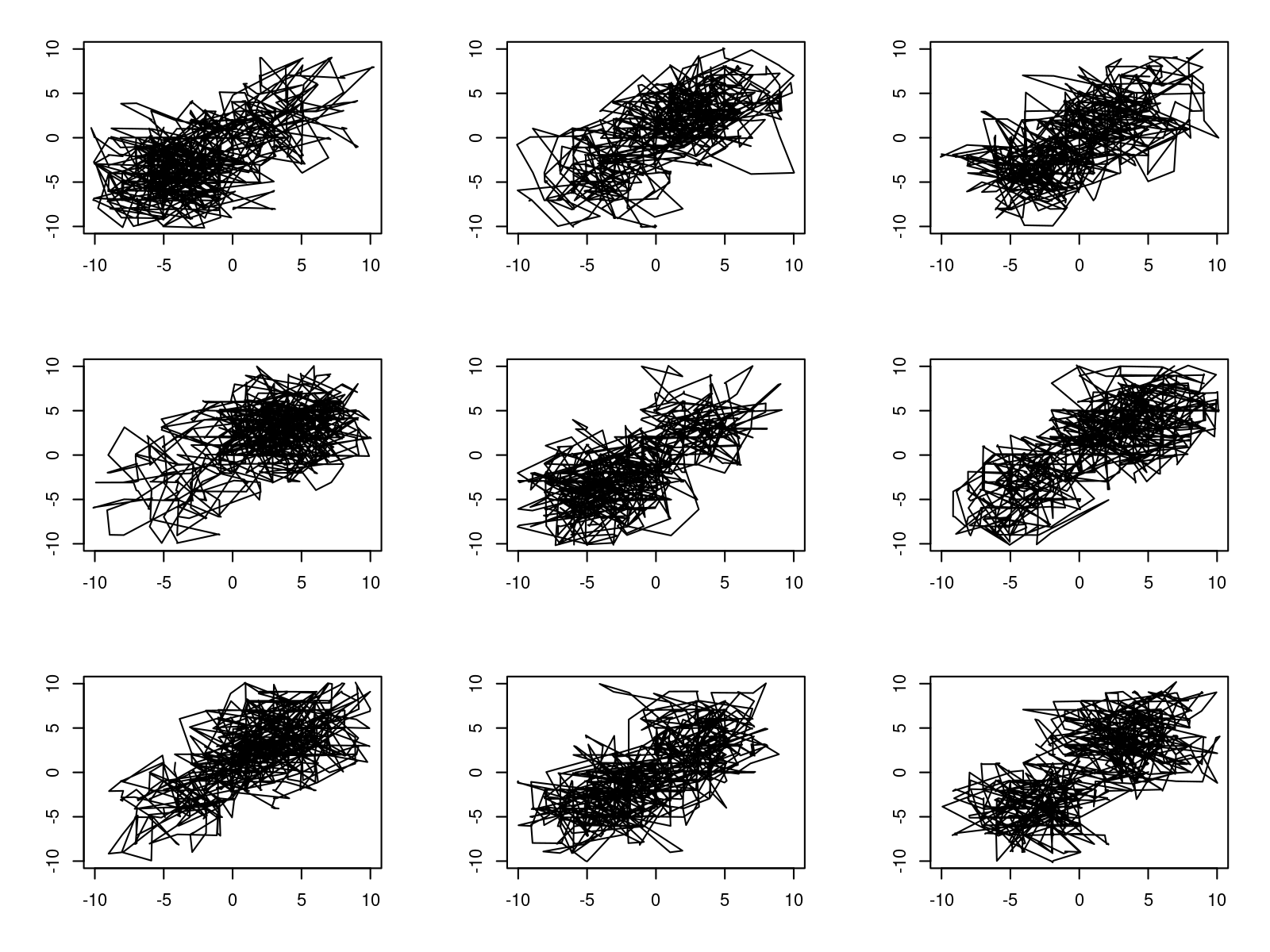}
        \caption{Trace plots from O-DHAMS}
        \label{fig:poly_less_precond_trace_Overhams}
    \end{subfigure}
    \begin{subfigure}[b]{0.32\textwidth}
        \centering
        \includegraphics[width=0.8\linewidth]{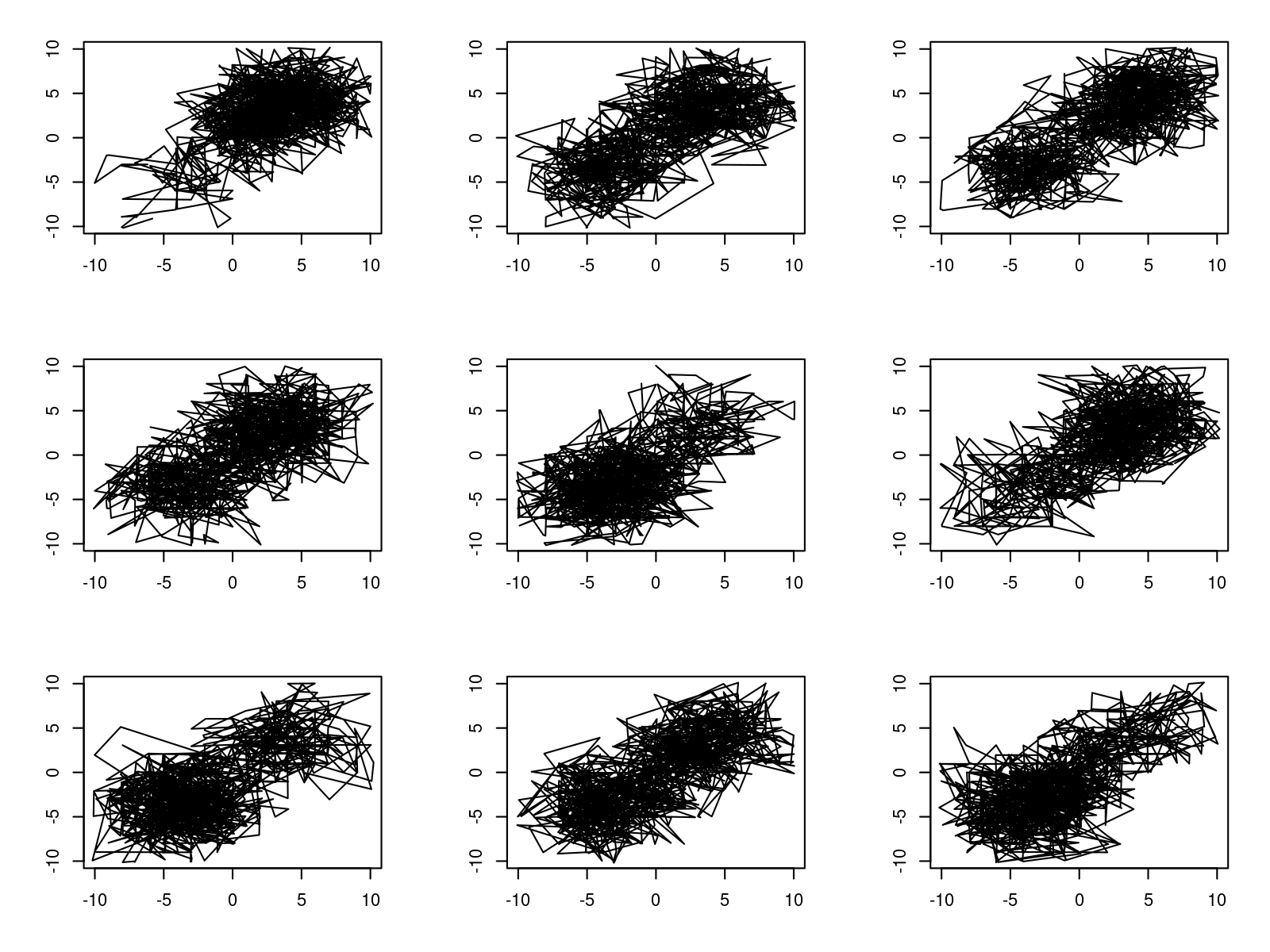}
        \caption{Trace plots from PAVG}
        \label{fig:poly_less_precond_trace_pavg}
    \end{subfigure}
    \begin{subfigure}[b]{0.32\textwidth}
        \centering
        \includegraphics[width=0.8\linewidth]{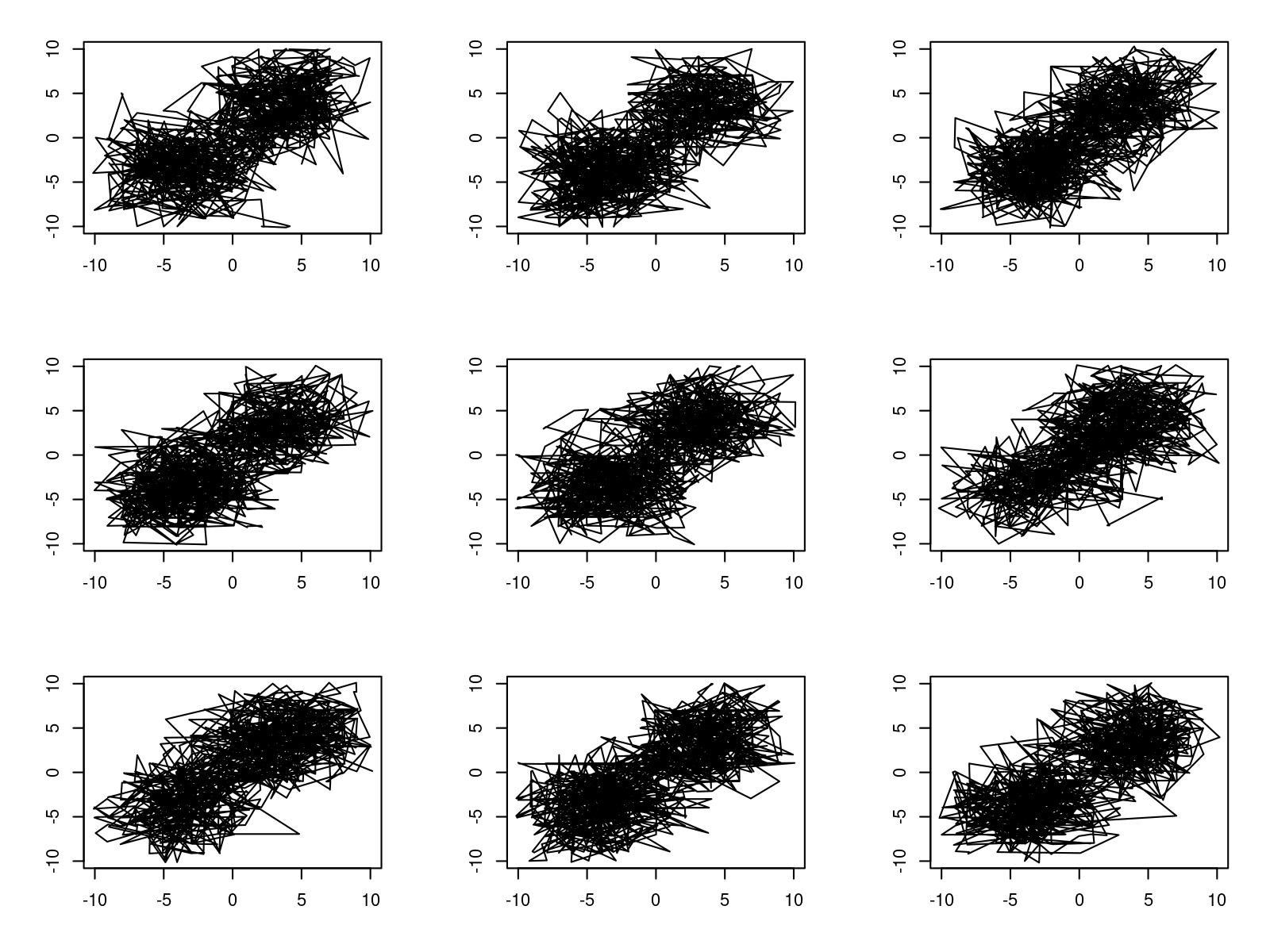}
        \caption{Trace plots from V-PDHAMS}
        \label{fig:poly_less_precond_trace_vpdhams}
    \end{subfigure}
    \begin{subfigure}[b]{0.32\textwidth}
        \centering
        \includegraphics[width=0.8\linewidth]{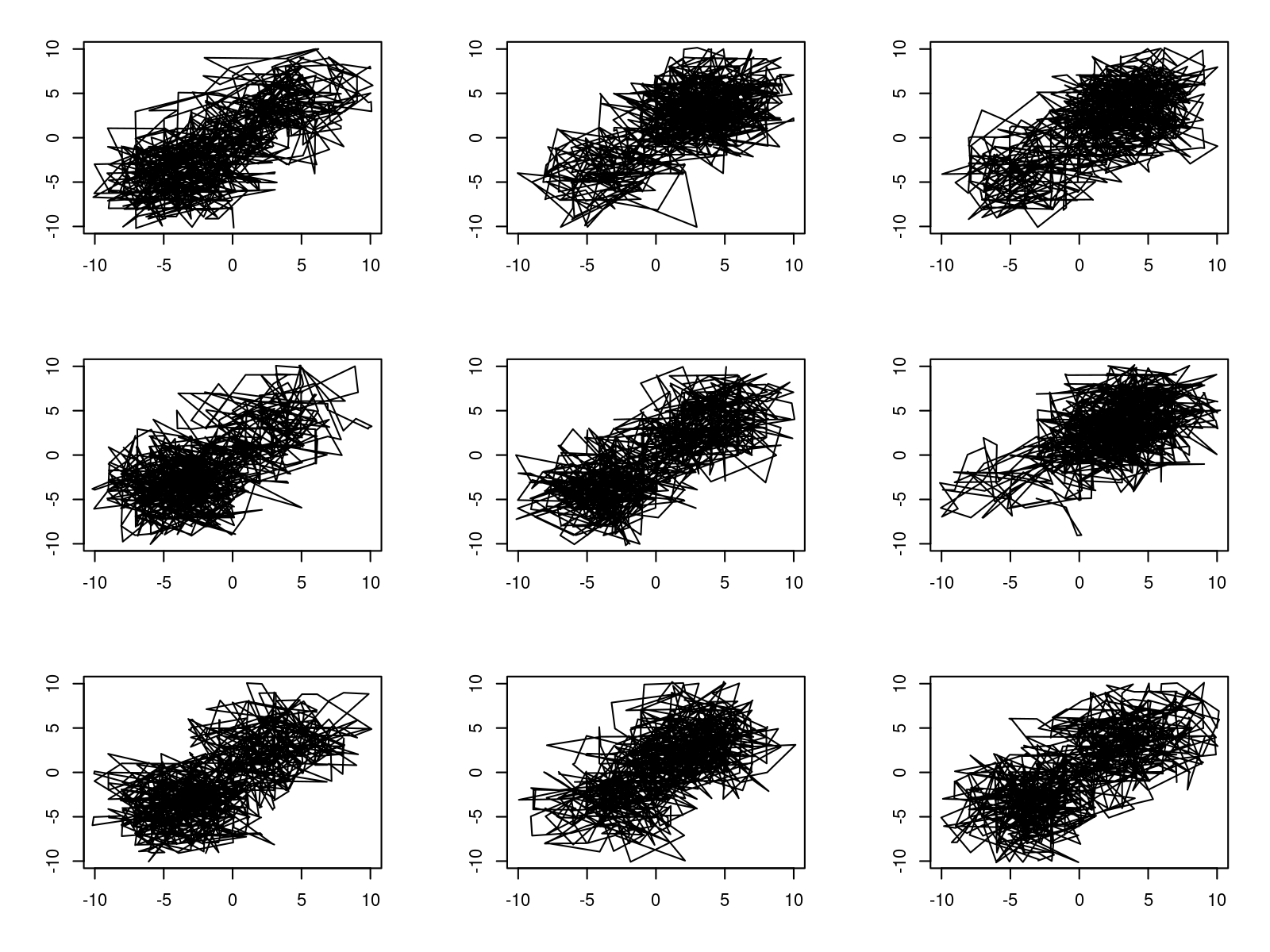}
        \caption{Trace plots from O-PDHAMS}
        \label{fig:poly_less_precond_trace_opdhams}
    \end{subfigure}
\caption{Trace plots for quadratic mixture distribution}
\label{fig:trace_precond_plots_poly}
\end{figure}

The optimal parameters and associated acceptance rates for each method are presented in Table \ref{tab:precond_parma_poly}. The smallest eigenvalue of preconditioning matrix $W$ is $-0.16$. We select 9 chains from the 100 parallel chains for each sampler and present their trace plots of the first two covariates for the first 1,000 draws after burn-in in
Figure~\ref{fig:trace_precond_plots_poly}. V-PDHAMS and O-DHAMS exhibit significantly better exploration capability than the other samplers,
in traversing the probability landscape.

\begin{table}[tbp]

\centering
\begin{tabular}{|c|c|c|c|c|}
\hline
    Sampler & Parameter & Acceptance Rate \\
    \hline
    Metropolis & $r = 4$ & 0.70\\
    NCG & $\delta=2.6$ & 0.94\\
        AVG & $\delta=3.5$ & 0.69\\
   V-DHAMS & $\epsilon = 0.85, \delta=1.65, \phi=0.5$& 0.73\\
    O-DHAMS & $\epsilon = 0.85, \delta=1.5, \phi=0.3, \beta=-0.9$ & 0.64\\
    PAVG & $\delta=0.16$ & 0.92\\
    V-PDHAMS & $\epsilon=0.85, \delta=0.12, \phi=0.125$ & 0.91\\
    O-PDHAMS & $\epsilon=0.85, \delta=0.14, \phi=0.2, \beta=-0.9$ & 0.86\\
    \hline
\end{tabular}
\caption{Parameters for quadratic mixture distribution}
\label{tab:precond_parma_poly}
\end{table}

\begin{figure}[tbp]
\begin{subfigure}[b]{0.32\textwidth}
        \centering
        \includegraphics[width=0.8\linewidth]{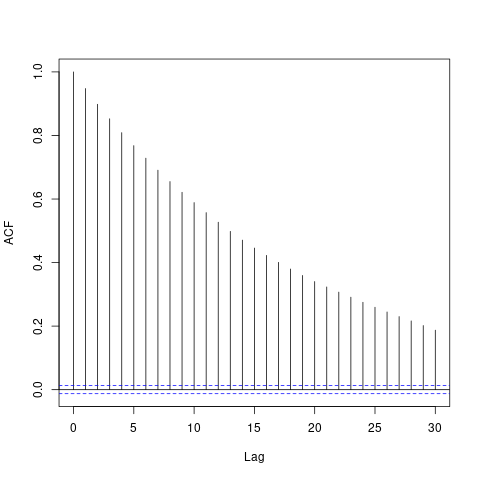}
        \caption{ACF from Metropolis}
        \label{fig:acf_precond_Metropolis_poly}
    \end{subfigure}
     \begin{subfigure}[b]{0.32\textwidth}
        \centering
        \includegraphics[width=0.8\linewidth]{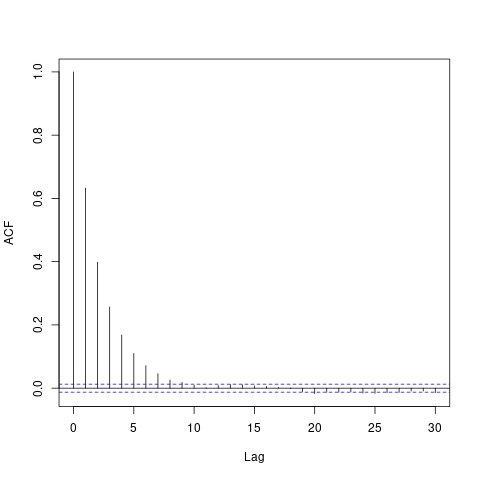}
        \caption{ACF from NCG}
        \label{fig:acf_precond_NCG_poly}
    \end{subfigure}
     \begin{subfigure}[b]{0.32\textwidth}
        \centering
        \includegraphics[width=0.8\linewidth]{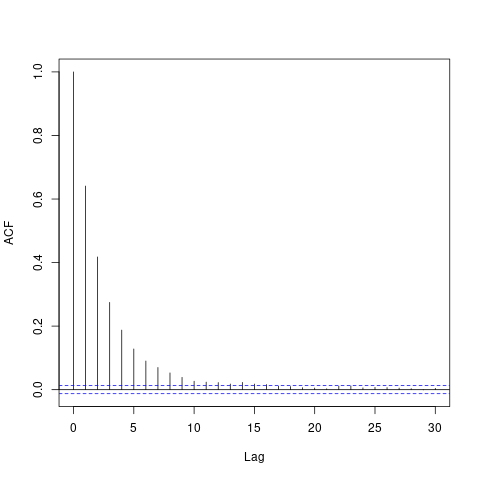}
        \caption{ACF from AVG}
        \label{fig:acf_precond_AVG_poly}
    \end{subfigure}
     \begin{subfigure}[b]{0.32\textwidth}
        \centering
        \includegraphics[width=0.8\linewidth]{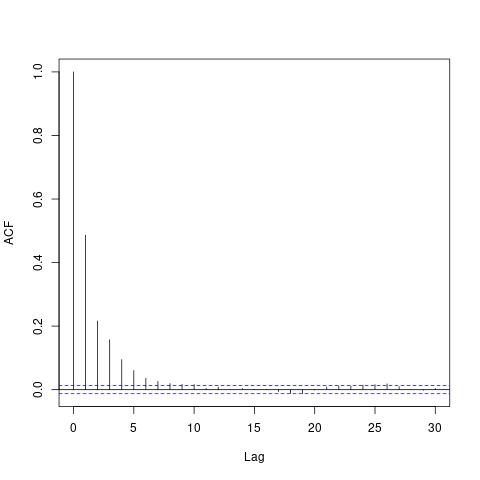}
        \caption{ACF from V-DHAMS}
        \label{fig:acf_precond_Hams_poly}
    \end{subfigure}
     \begin{subfigure}[b]{0.32\textwidth}
        \centering
        \includegraphics[width=0.8\linewidth]{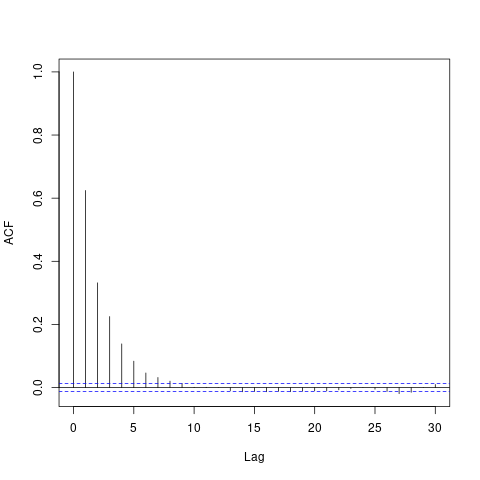}
        \caption{ACF from O-DHAMS}
        \label{fig:acf_precond_overhams_poly}
    \end{subfigure}
     \begin{subfigure}[b]{0.32\textwidth}
        \centering
        \includegraphics[width=0.8\linewidth]{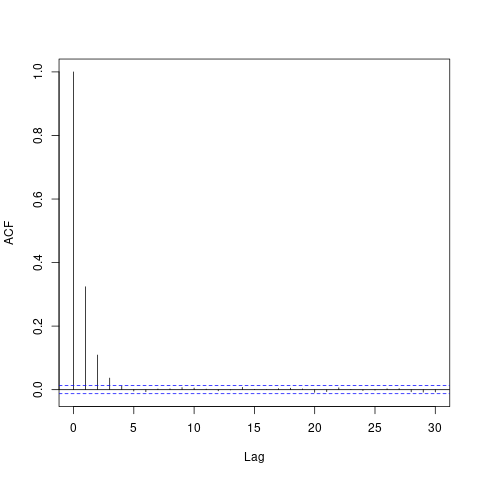}
        \caption{ACF from PAVG}
        \label{fig:acf_precond_pavg_poly}
    \end{subfigure}
     \begin{subfigure}[b]{0.32\textwidth}
        \centering
        \includegraphics[width=0.8\linewidth]{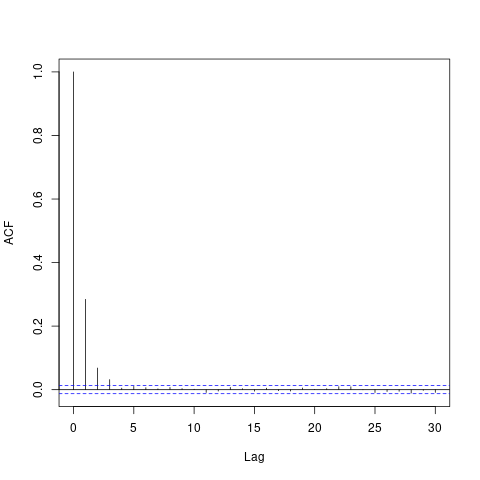}
        \caption{ACF from V-PDHAMS}
        \label{fig:acf_precond_vpdhams_poly}
    \end{subfigure}
     \begin{subfigure}[b]{0.32\textwidth}
        \centering
        \includegraphics[width=0.8\linewidth]{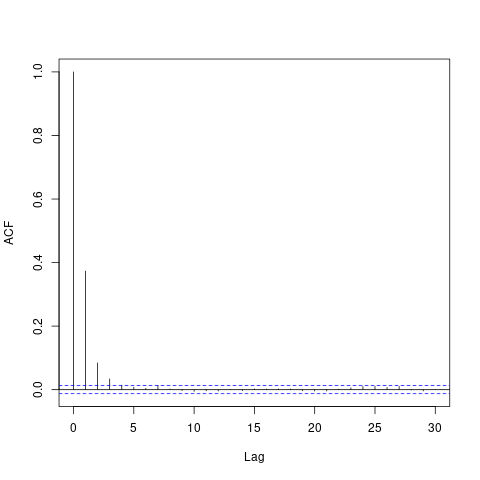}
        \caption{ACF from O-PDHAMS}
        \label{fig:acf_precond_opdhams_poly}
    \end{subfigure}
\caption{ACF plots for quadratic mixture distribution}
\label{fig:acf_precond_plots_poly}
\end{figure}

\begin{figure}[tbp]
    \centering
    \includegraphics[width=0.4\linewidth]{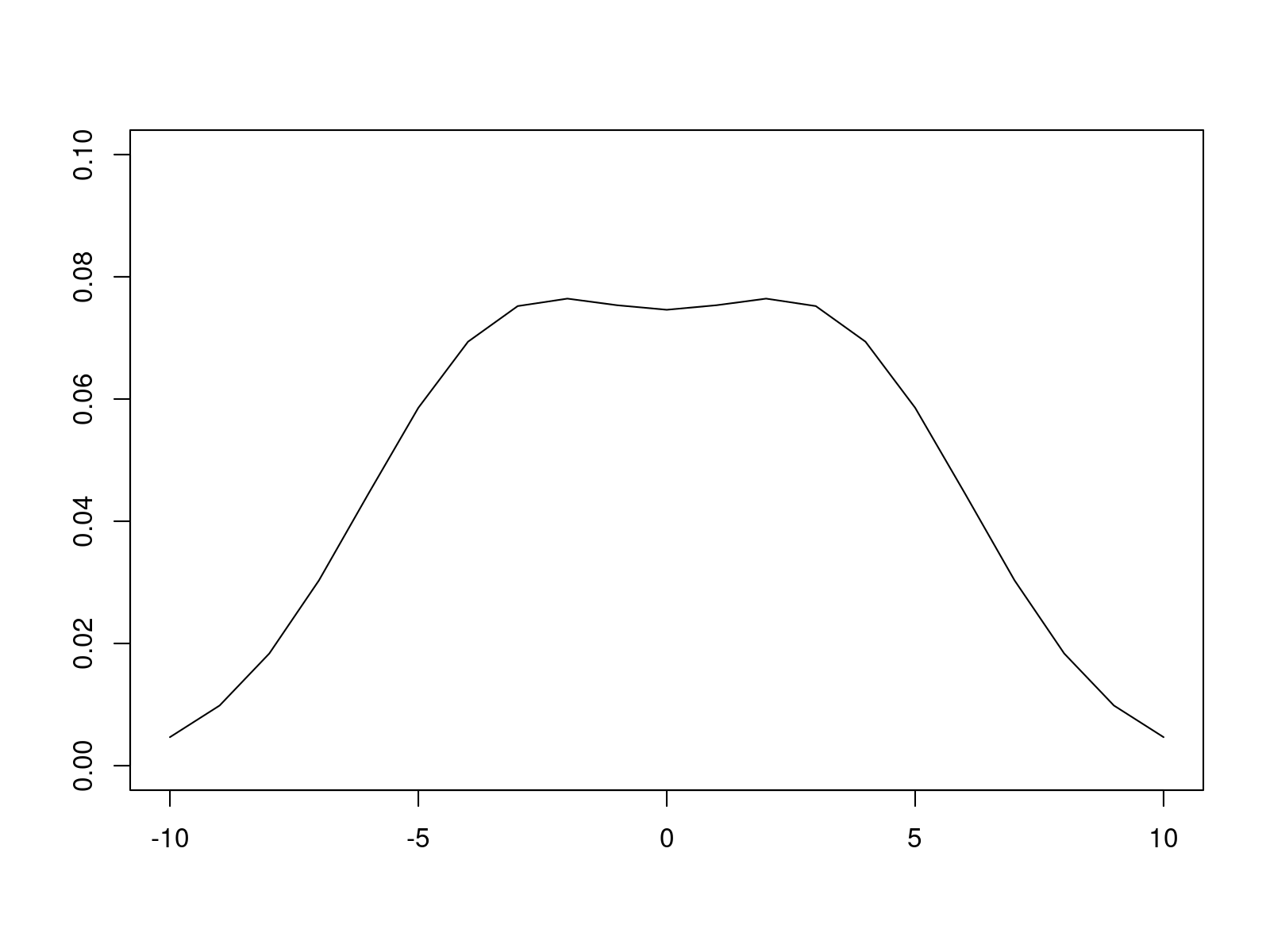}
    \caption{True marginal distribution of the first coordinate in quadratic mixture distribution}
    \label{fig:freq_precond_truth_poly}
\end{figure}
The plots of the auto-correlation functions (ACF) of $f(s)$ from a single chain are presented in Figure~\ref{fig:acf_precond_plots_poly}. PAVG, V-DHAMS and O-DHAMS exhibit auto-correlations lower than other samplers, indicating reduced dependencies among draws.

\begin{figure}[H]
\begin{subfigure}[b]{0.32\textwidth}
        \centering
        \includegraphics[width=0.8\linewidth]{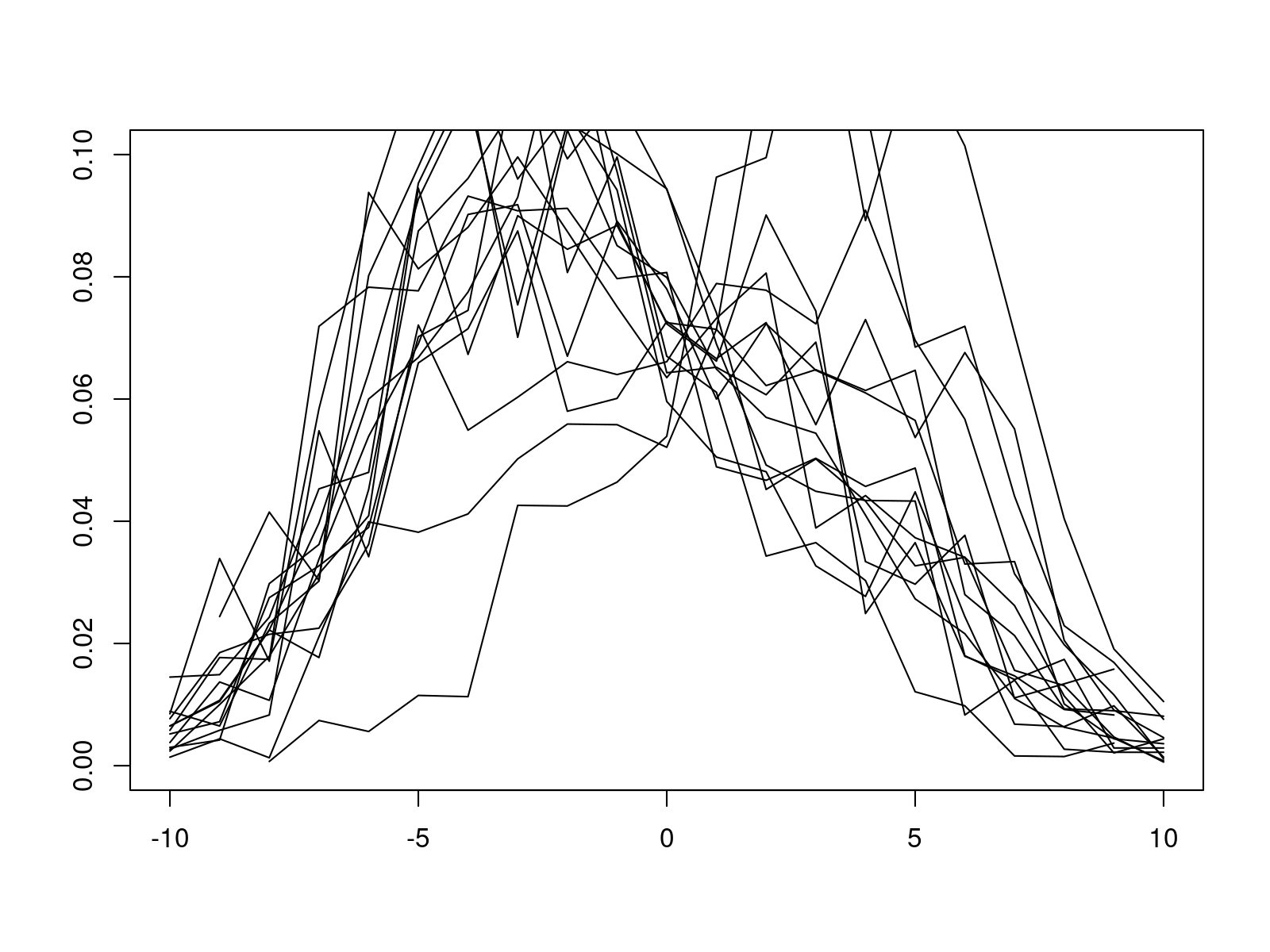}
        \caption{Metropolis}
        \label{fig:freq_precond_Metropolis_poly}
    \end{subfigure}
     \begin{subfigure}[b]{0.32\textwidth}
        \centering
        \includegraphics[width=0.8\linewidth]{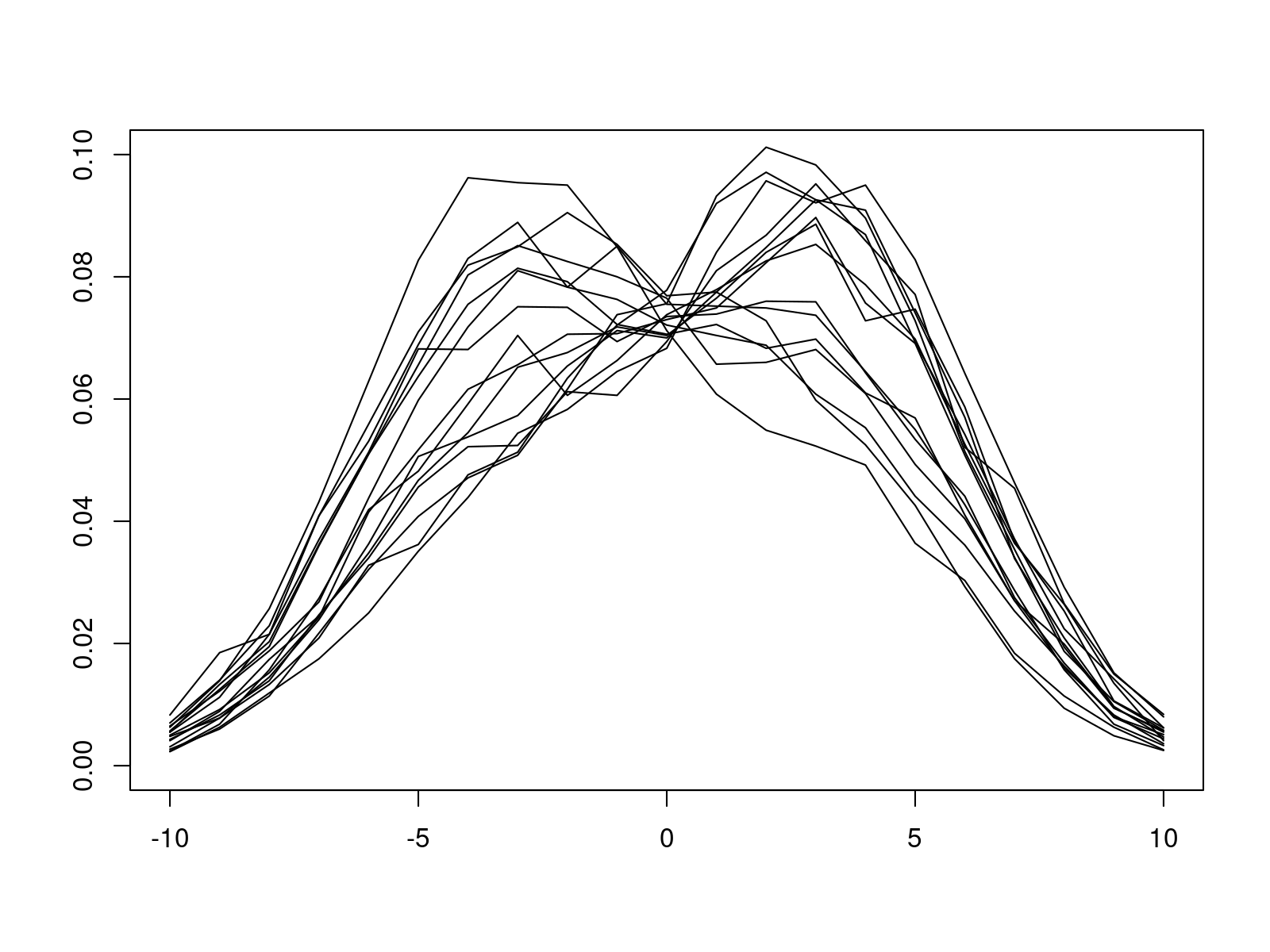}
        \caption{NCG}
        \label{fig:freq_precond_NCG_poly}
    \end{subfigure}
     \begin{subfigure}[b]{0.32\textwidth}
        \centering
        \includegraphics[width=0.8\linewidth]{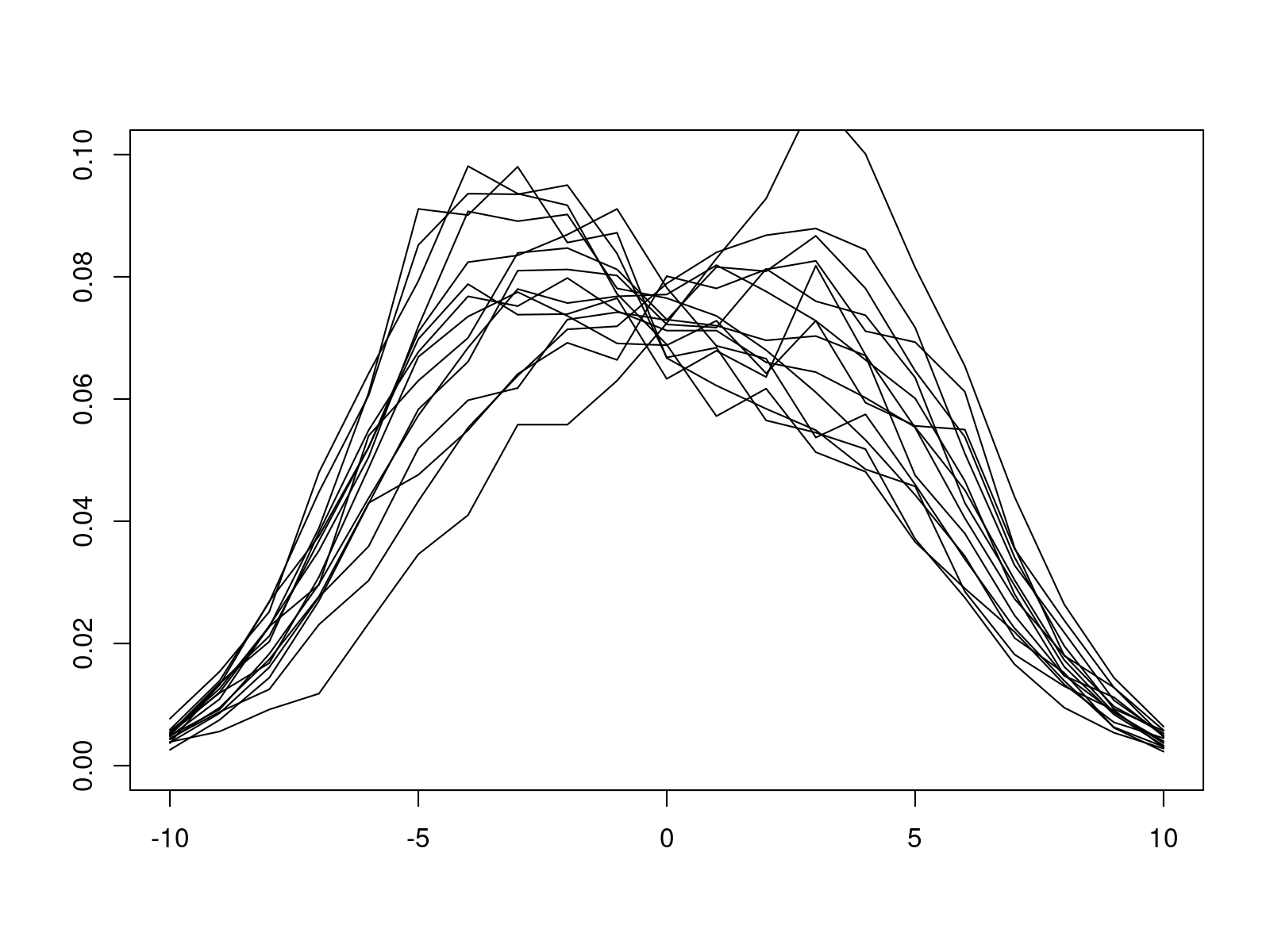}
        \caption{AVG}
        \label{fig:freq_precond_avg_poly}
    \end{subfigure}
     \begin{subfigure}[b]{0.32\textwidth}
        \centering
        \includegraphics[width=0.8\linewidth]{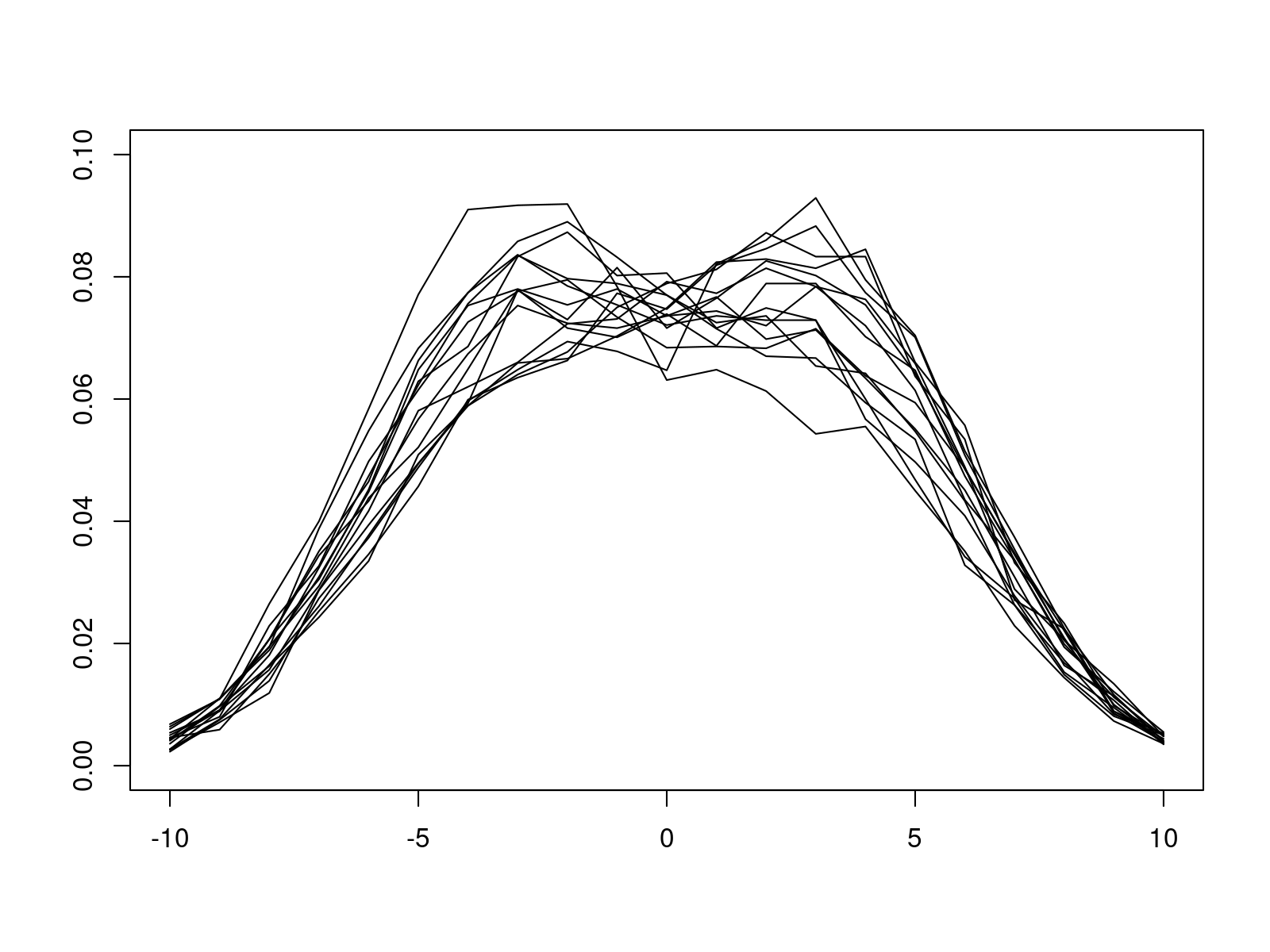}
        \caption{V-DHAMS}
        \label{fig:freq_precond_Hams_poly}
    \end{subfigure}
     \begin{subfigure}[b]{0.32\textwidth}
        \centering
        \includegraphics[width=0.8\linewidth]{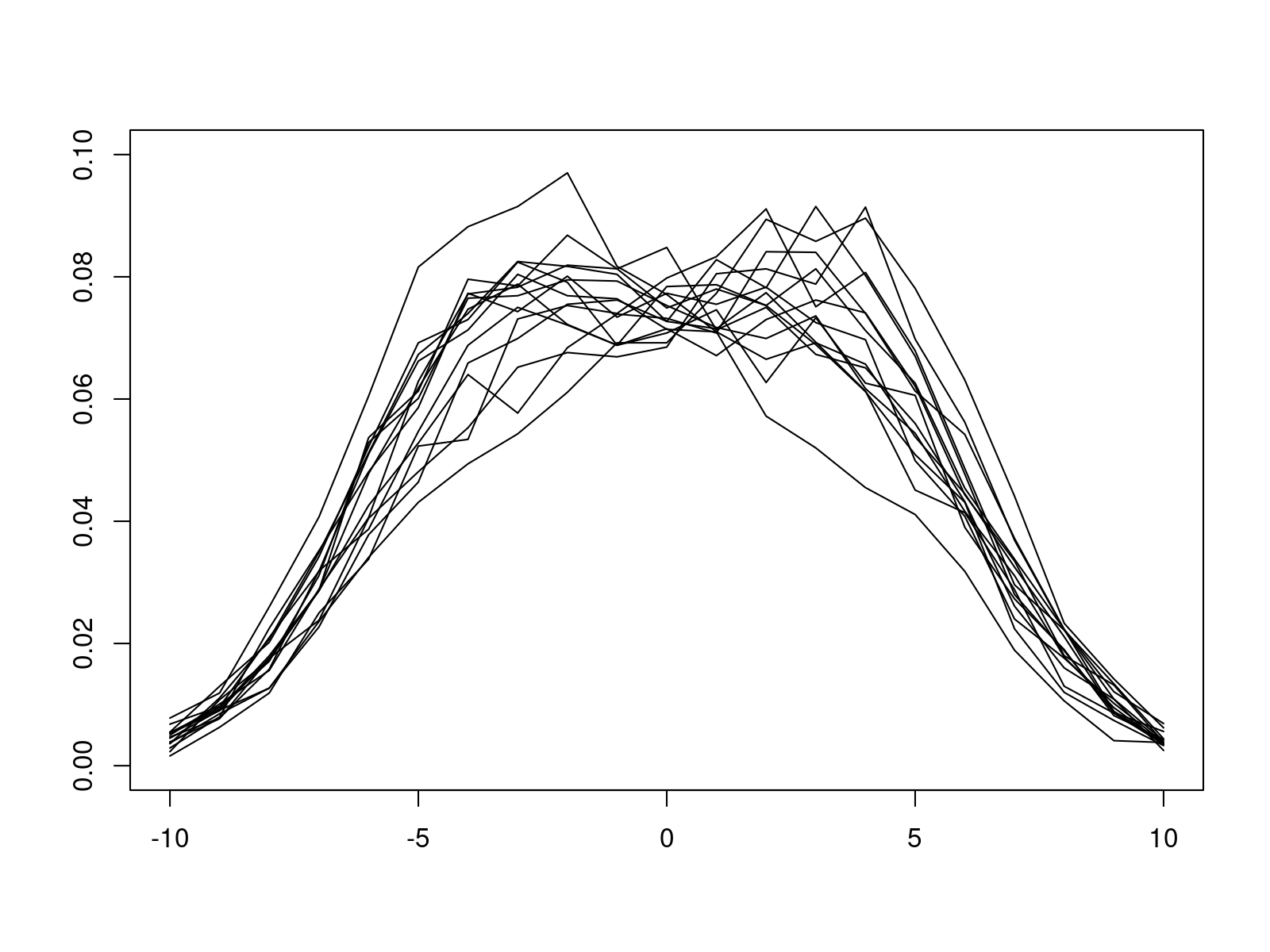}
        \caption{O-DHAMS}
        \label{fig:freq_precond_overhams_poly}
    \end{subfigure}
    \begin{subfigure}[b]{0.32\textwidth}
        \centering
        \includegraphics[width=0.8\linewidth]{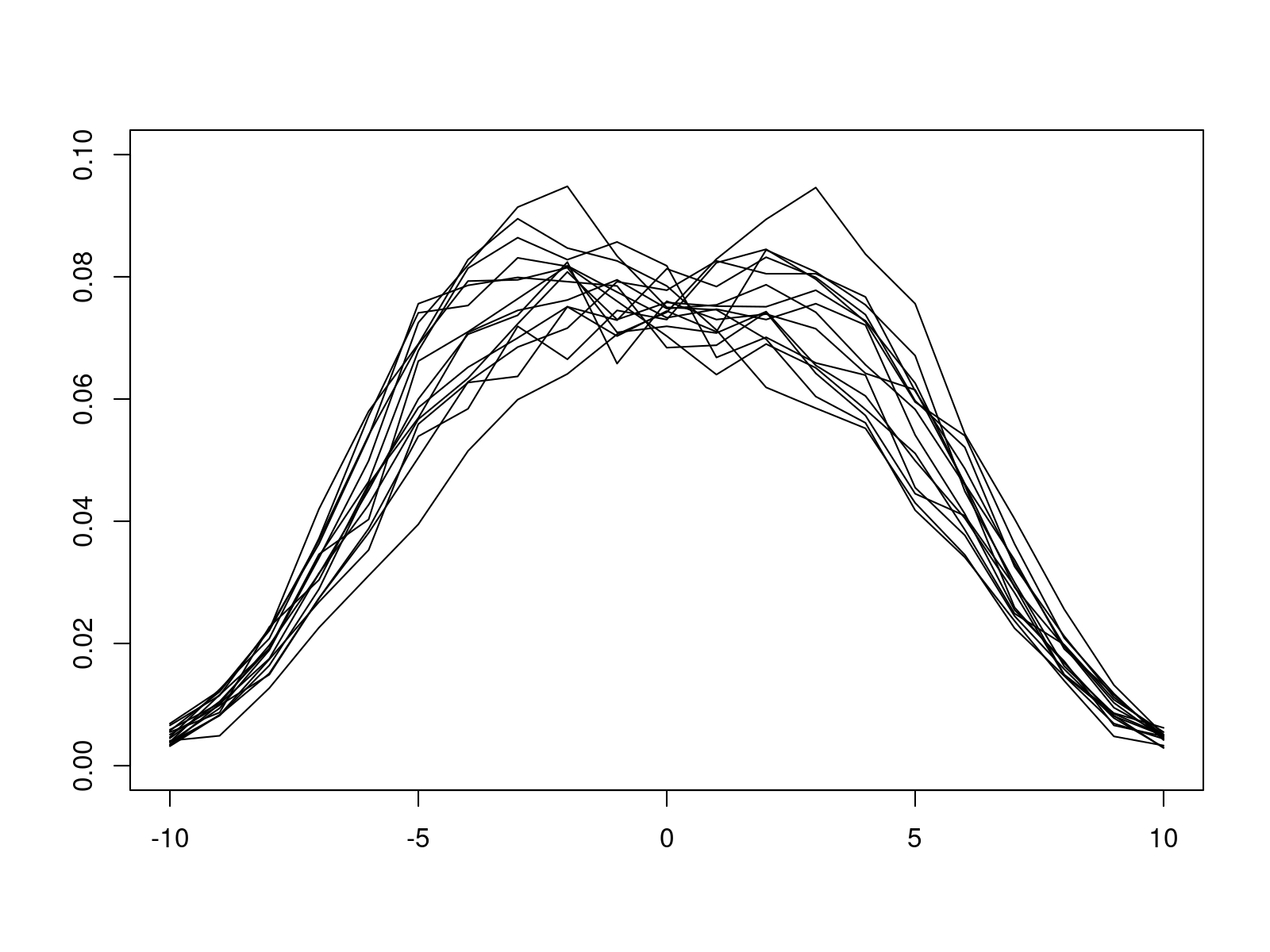}
        \caption{PAVG}
        \label{fig:freq_precond_pavg_poly}
    \end{subfigure}
    \begin{subfigure}[b]{0.32\textwidth}
        \centering
        \includegraphics[width=0.8\linewidth]{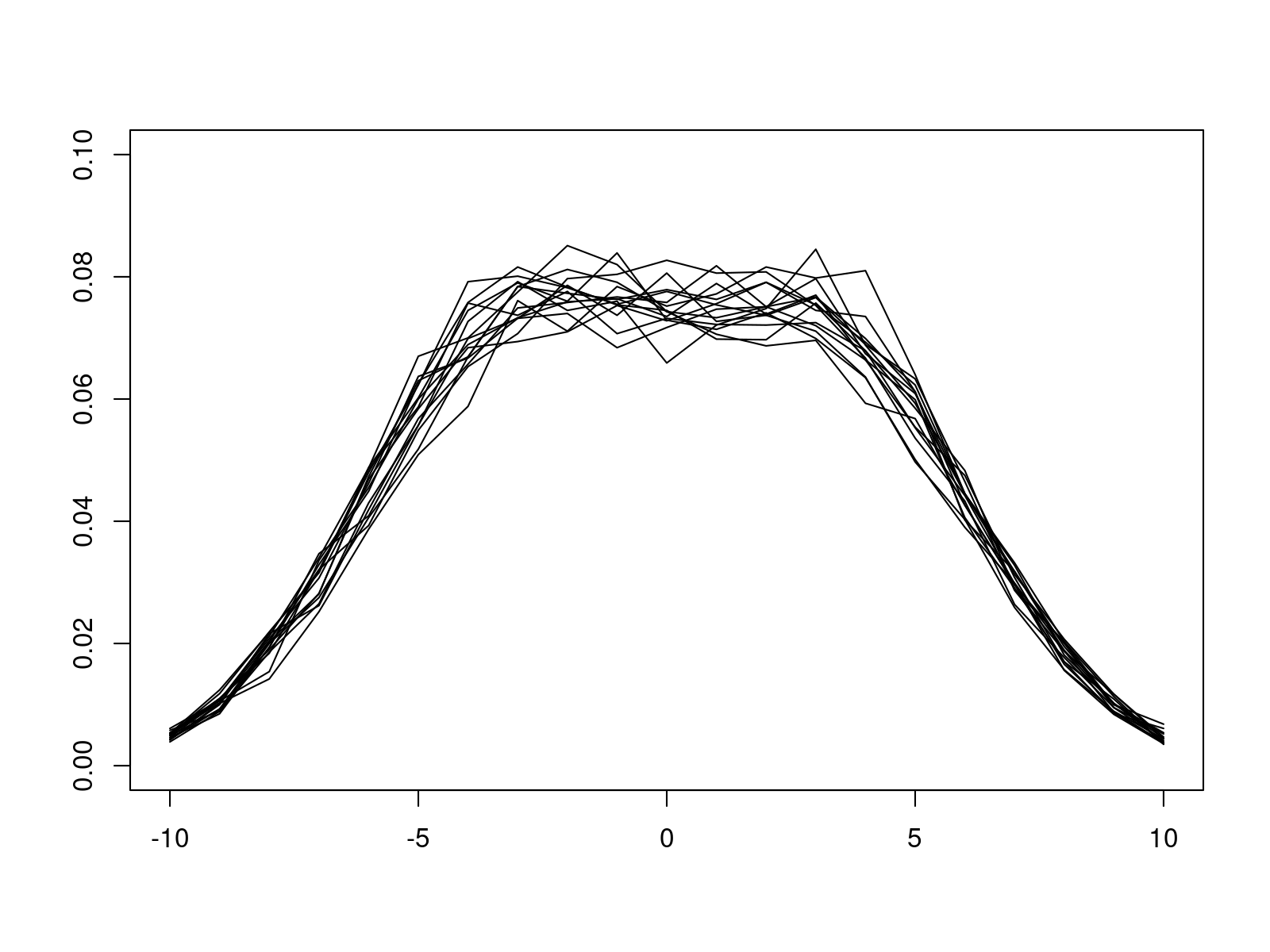}
        \caption{V-PDHAMS}
        \label{fig:freq_precond_vpdhams_poly}
    \end{subfigure}
    \begin{subfigure}[b]{0.32\textwidth}
        \centering
        \includegraphics[width=0.8\linewidth]{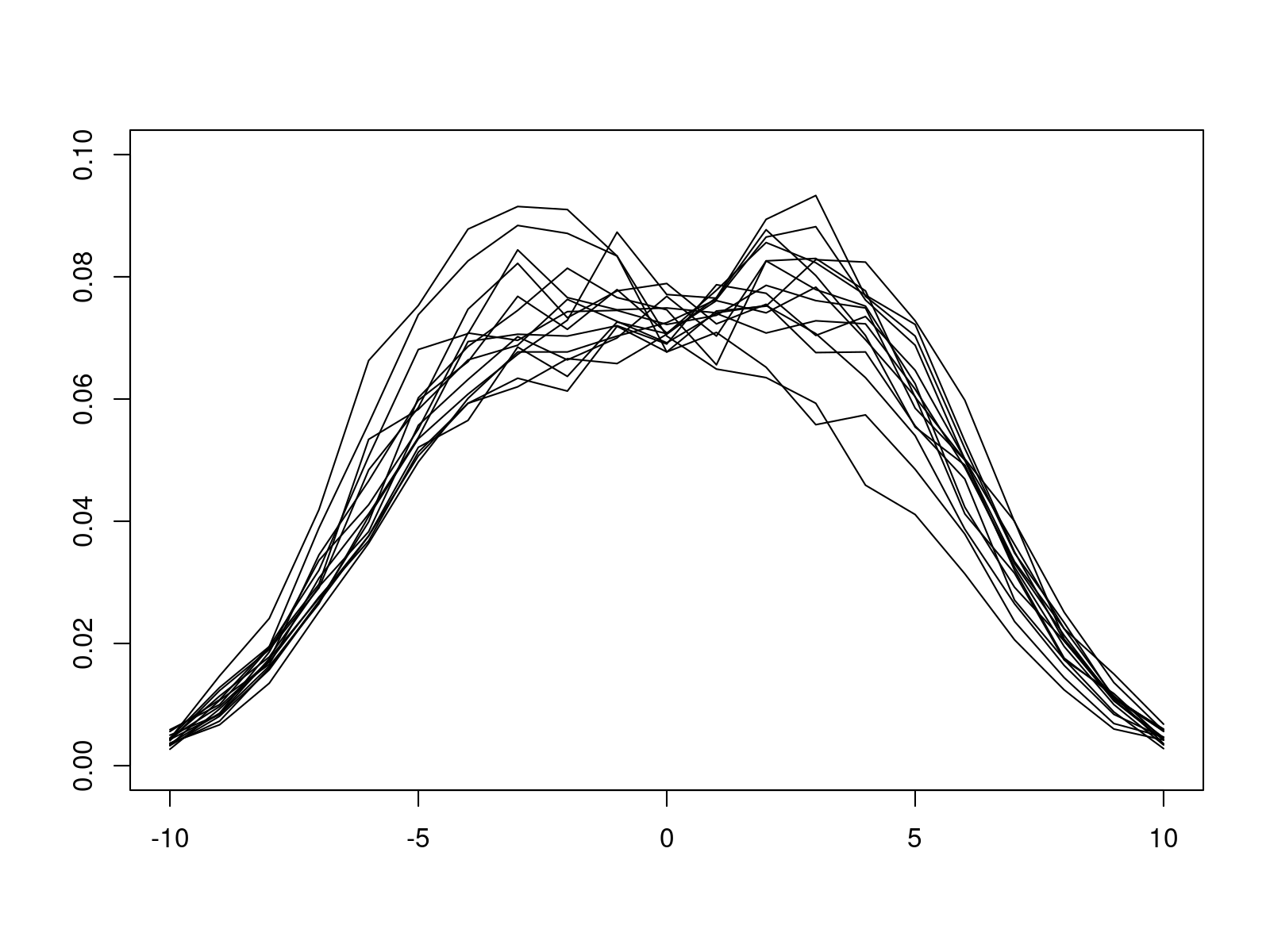}
        \caption{O-PDHAMS}
        \label{fig:freq_precond_opdhams_poly}
    \end{subfigure}
\caption{Frequency plots of the first coordinate in quadratic mixture distribution}
\label{fig:freq_precond_plots_poly}
\end{figure}

For each sampler, we present frequency plots of the first coordinate across 10 independent chains, each based on 10,000 draws, as shown in Figure~\ref{fig:freq_precond_plots_poly}. The corresponding true marginal distribution is depicted in Figure~\ref{fig:freq_precond_truth_poly}. Due to the multi-modal nature of the distribution, Figure~\ref{fig:freq_precond_truth_poly} appears relatively flat in the middle region, where the probability mass is highest. Among all methods, the frequency plots produced by V-PDHAMS most closely resembles the ground truth, indicating superior exploration of mixture components.

\subsection{Additional Results for Clock Potts Model}\label{sec:precond_potts_results}
\begin{figure}[h]
\begin{subfigure}[b]{0.32\textwidth}
        \centering
        \includegraphics[width=0.8\linewidth]{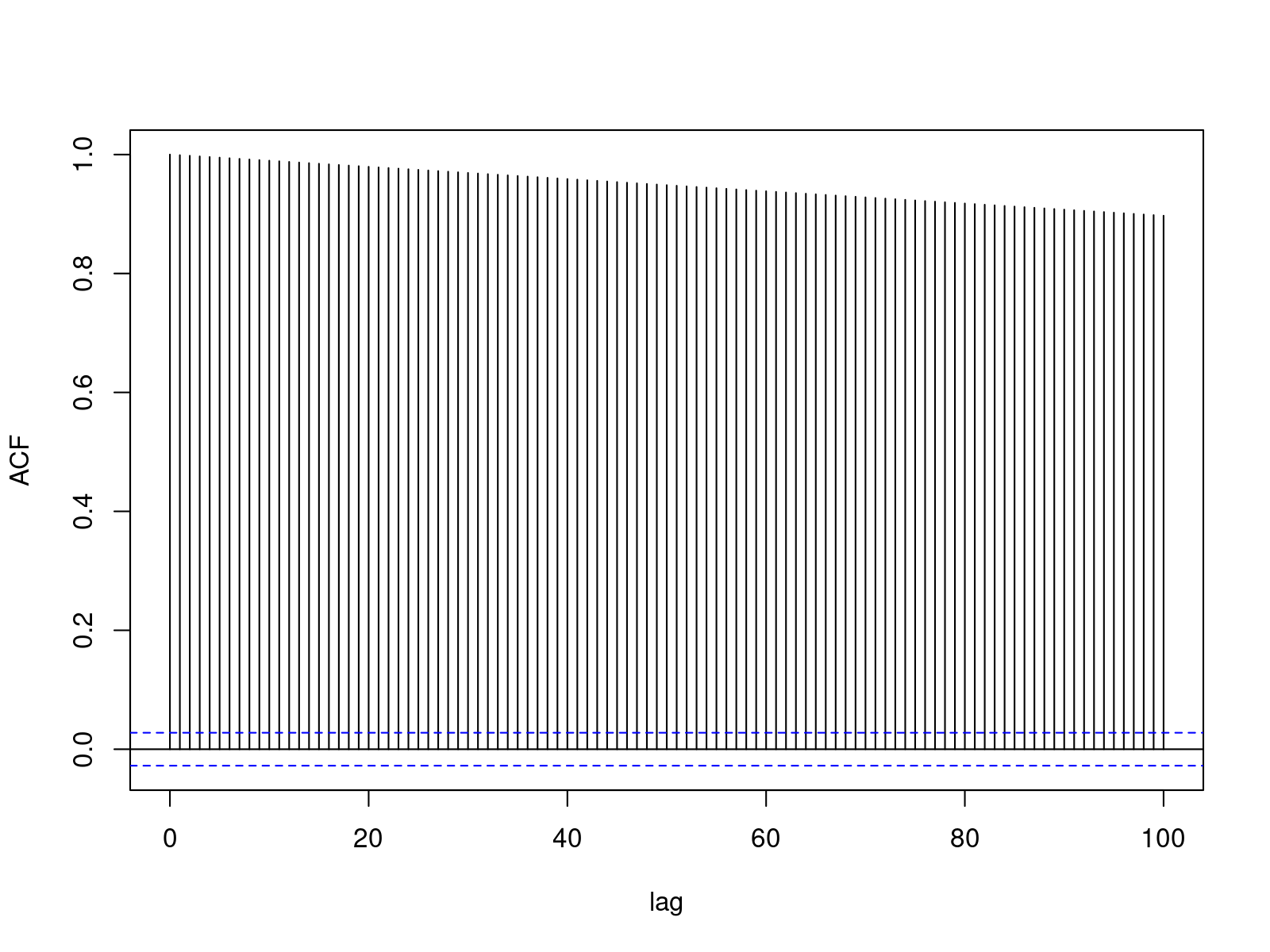}
        \caption{Metropolis}
        \label{fig:acf_precond_Metropolis_potts_ferro}
    \end{subfigure}
     \begin{subfigure}[b]{0.32\textwidth}
        \centering
        \includegraphics[width=0.8\linewidth]{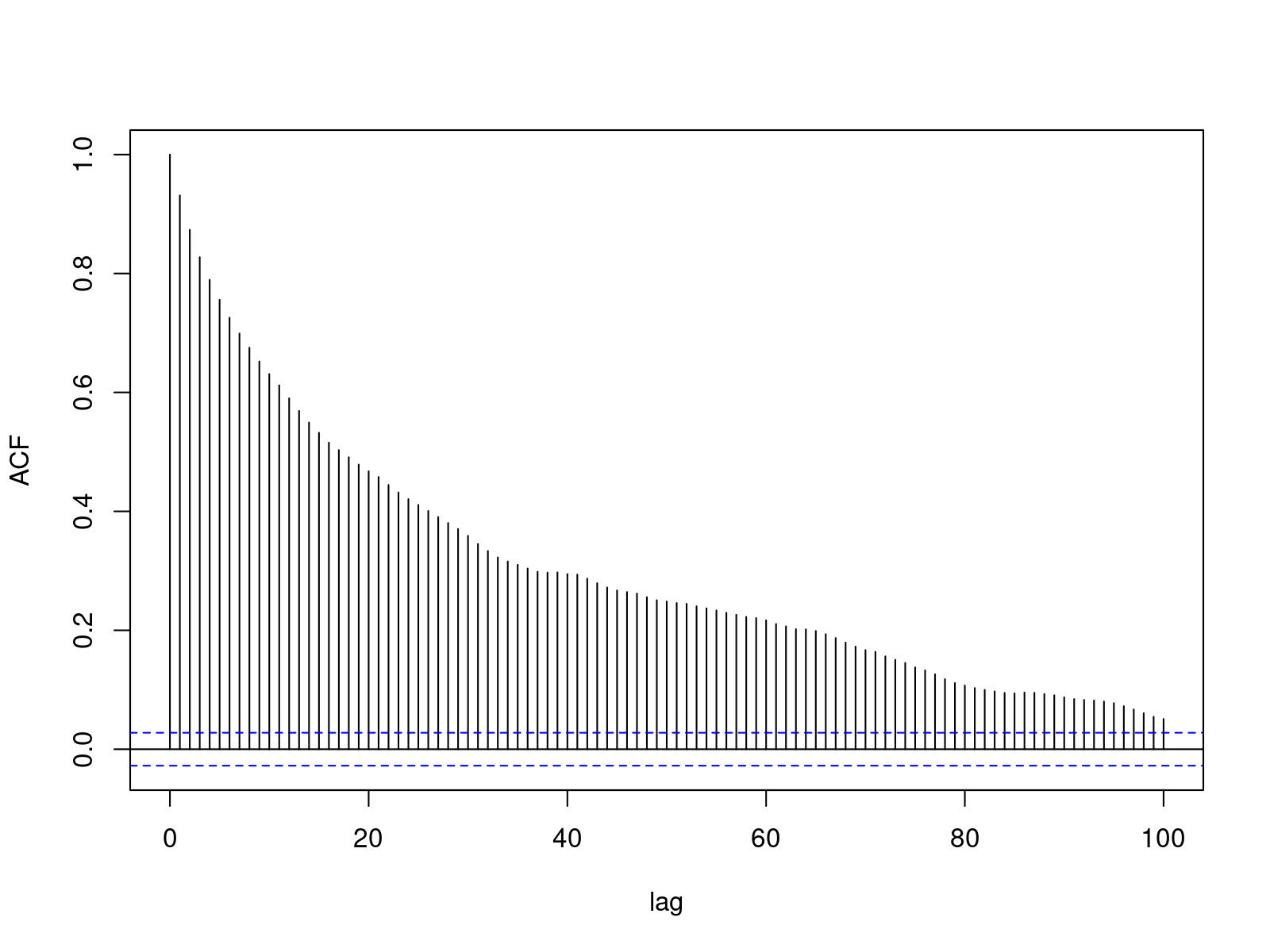}
        \caption{NCG}
        \label{fig:acf_precond_NCG_potts_ferro}
    \end{subfigure}
     \begin{subfigure}[b]{0.32\textwidth}
        \centering
        \includegraphics[width=0.8\linewidth]{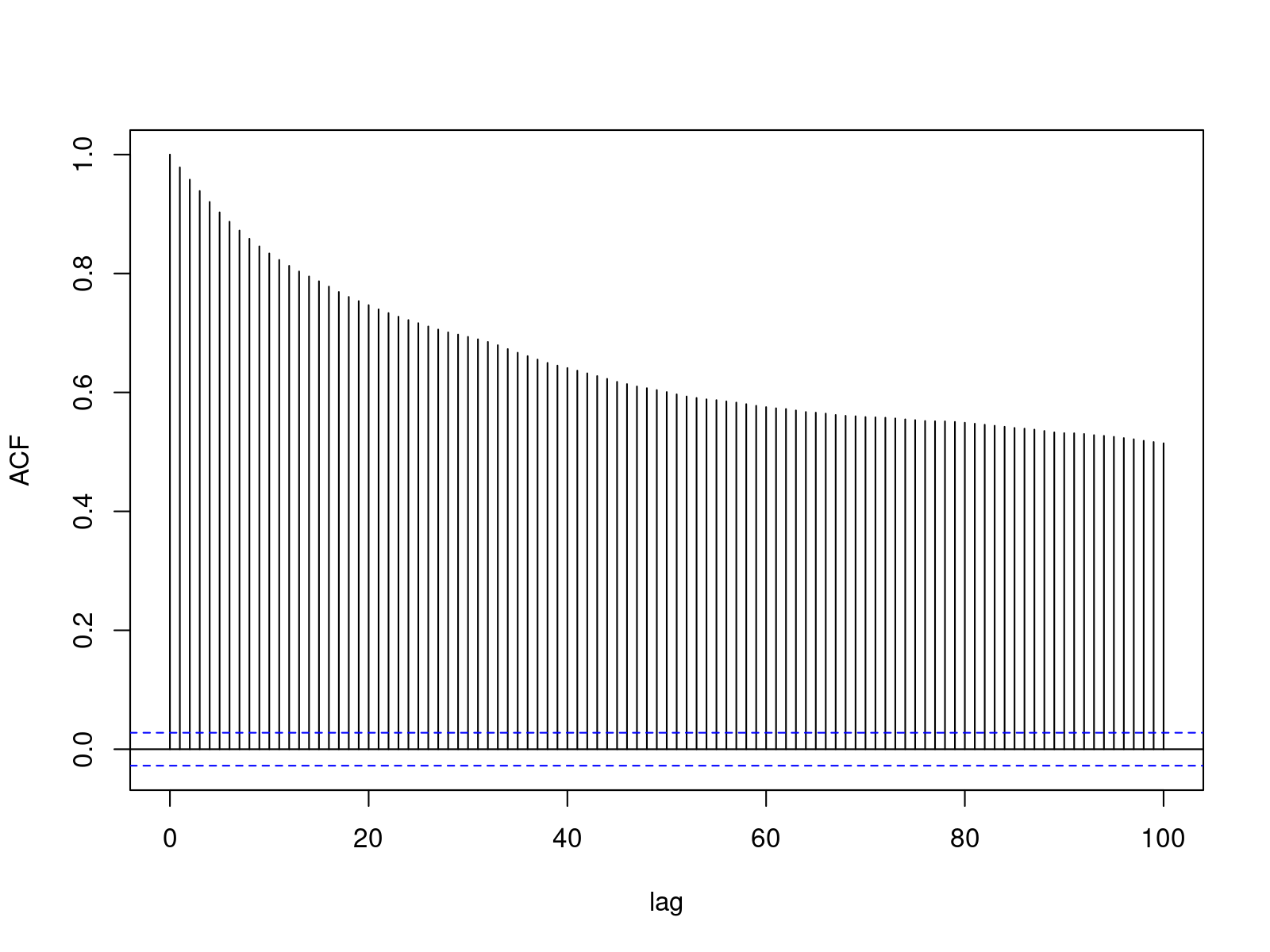}
        \caption{AVG}
        \label{fig:acf_precond_avg_potts_ferro}
    \end{subfigure}
     \begin{subfigure}[b]{0.32\textwidth}
        \centering
        \includegraphics[width=0.8\linewidth]{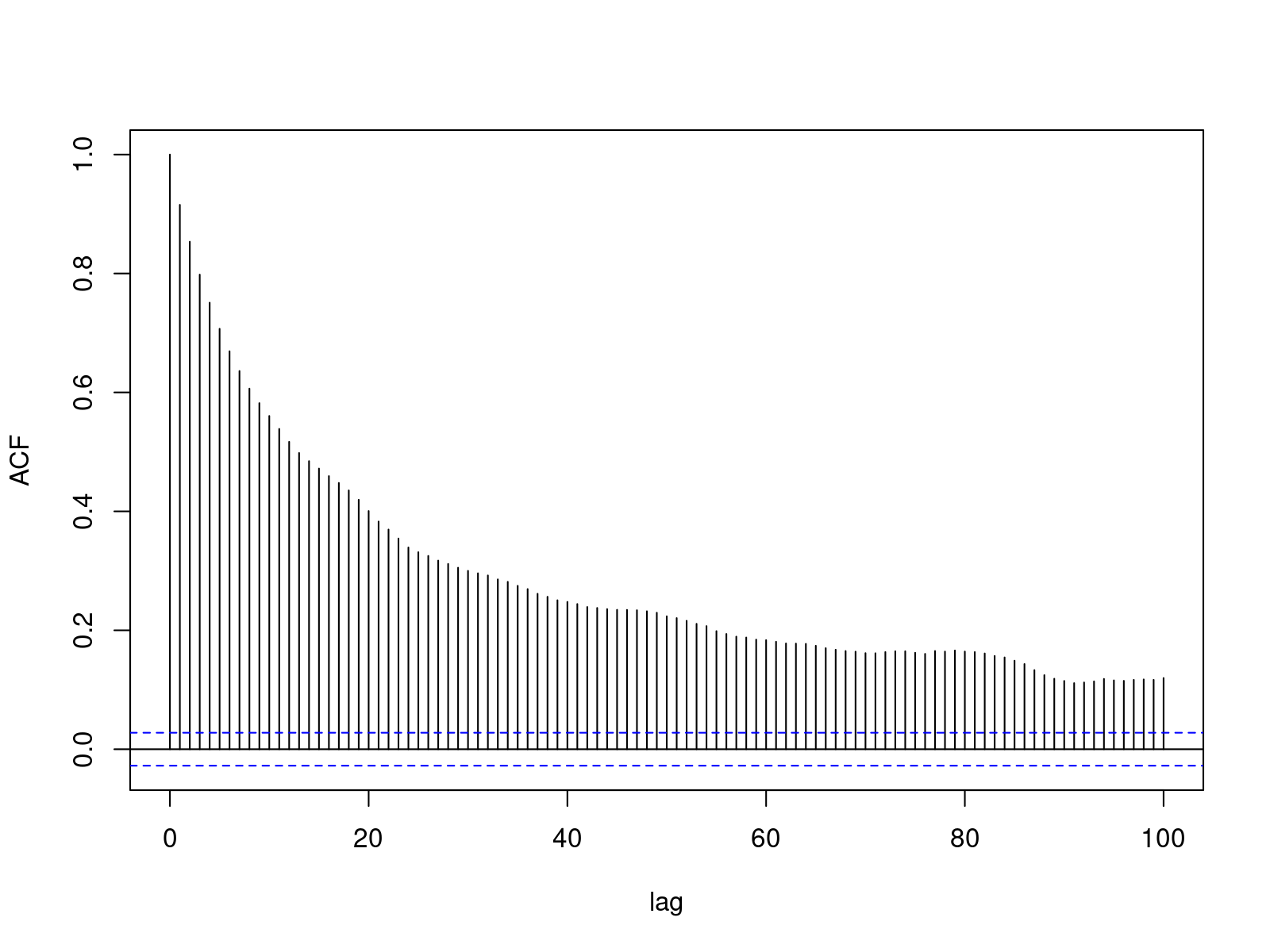}
        \caption{V-DHAMS}
        \label{fig:acf_precond_Hams_potts_ferro}
    \end{subfigure}
     \begin{subfigure}[b]{0.32\textwidth}
        \centering
        \includegraphics[width=0.8\linewidth]{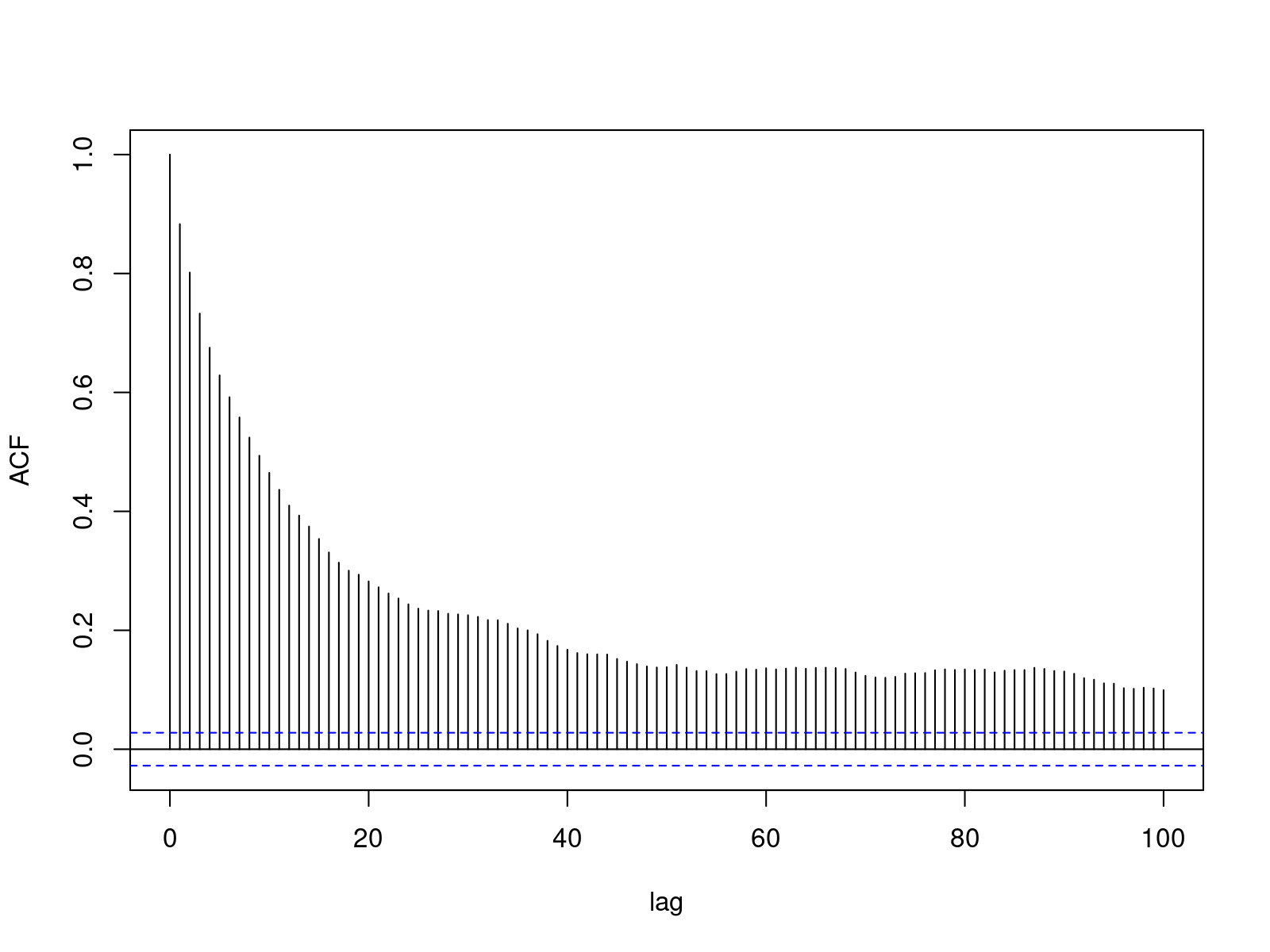}
        \caption{O-DHAMS}
        \label{fig:acf_precond_overhams_potts_ferro}
    \end{subfigure}
    \begin{subfigure}[b]{0.32\textwidth}
        \centering
        \includegraphics[width=0.8\linewidth]{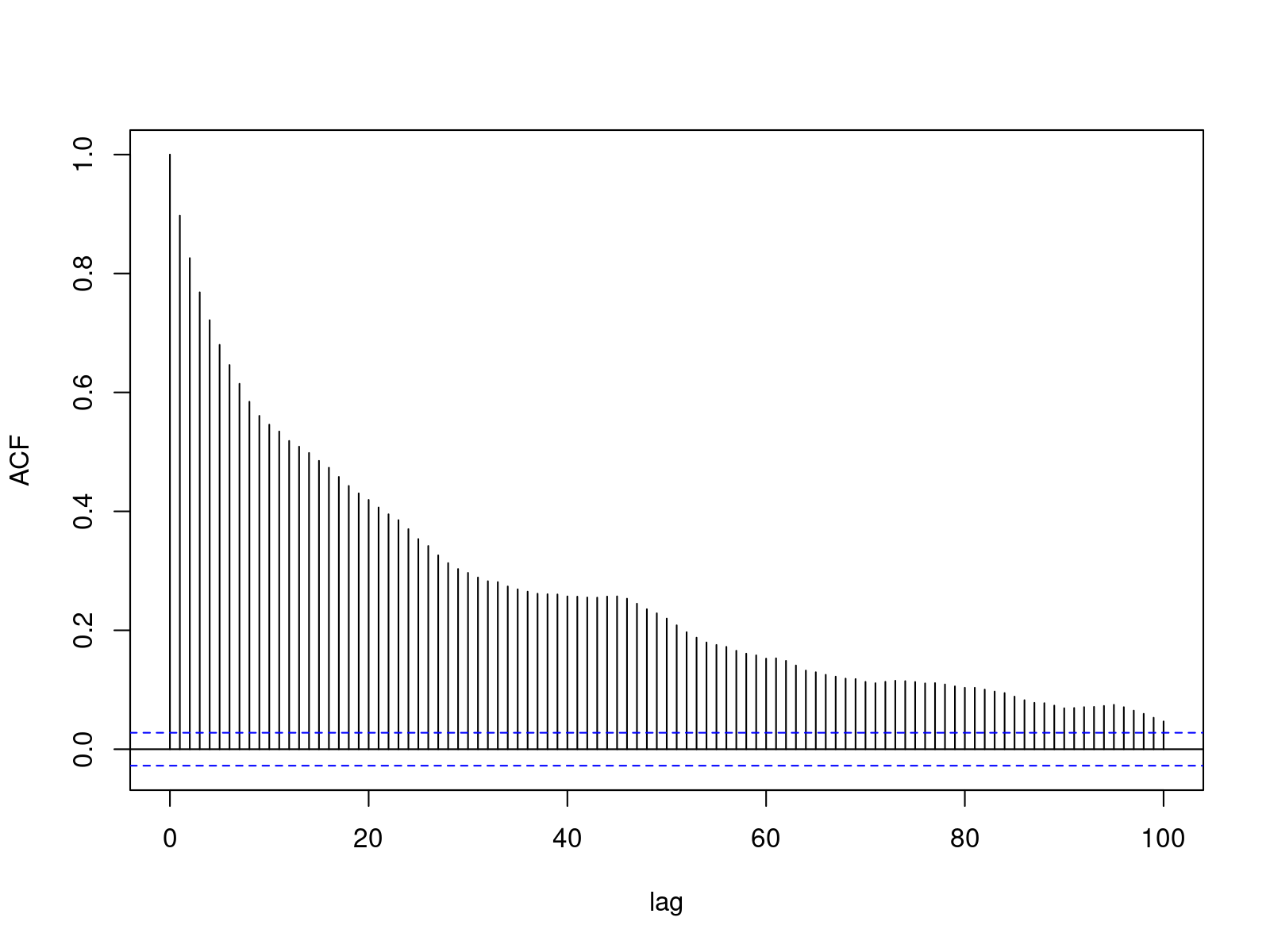}
        \caption{PAVG}
        \label{fig:acf_precond_pavg_potts_ferro}
    \end{subfigure}
    \begin{subfigure}[b]{0.32\textwidth}
        \centering
        \includegraphics[width=0.8\linewidth]{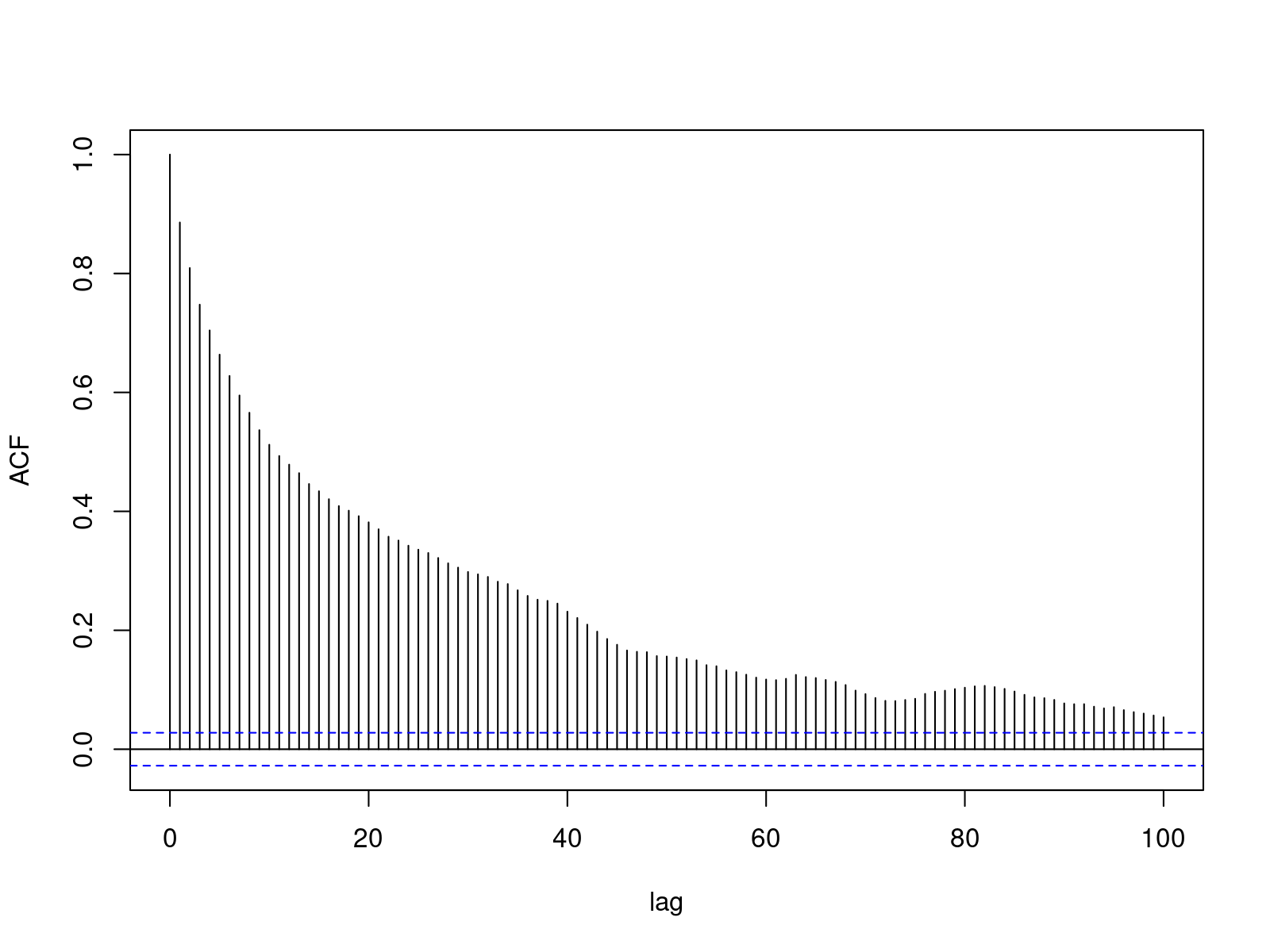}
        \caption{V-PDHAMS}
        \label{fig:acf_precond_vpdhams_potts_ferro}
    \end{subfigure}
    \begin{subfigure}[b]{0.32\textwidth}
        \centering
        \includegraphics[width=0.8\linewidth]{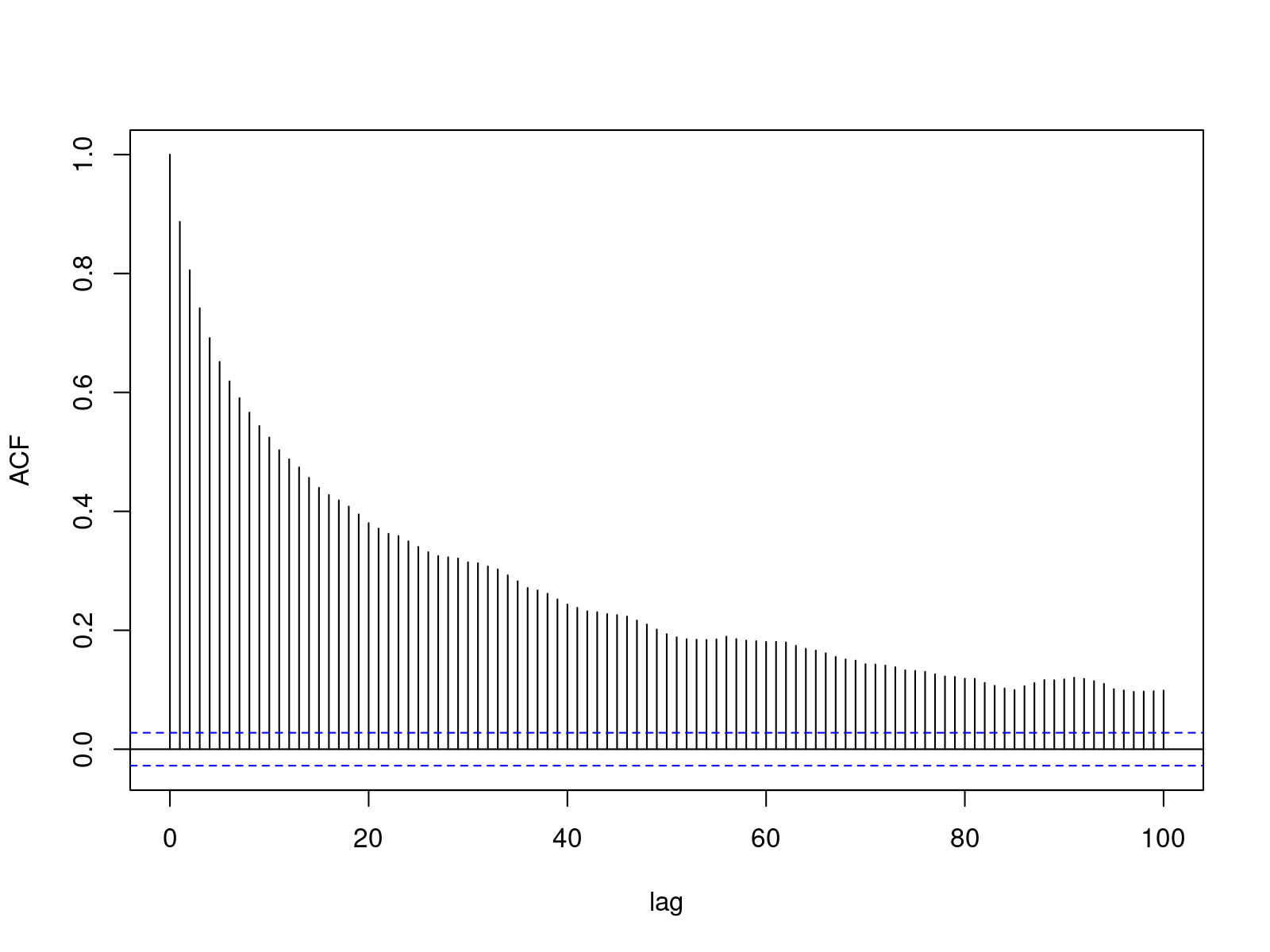}
        \caption{O-PDHAMS}
        \label{fig:acf_precond_opdhams_potts_ferro}
    \end{subfigure}
\caption{ACF plots of negative potential for clock Potts model (Ferromagnetic model)}
\label{fig:acf_precond_plots_potts_ferro}
\end{figure}

\begin{figure}[h]
\begin{subfigure}[b]{0.32\textwidth}
        \centering
        \includegraphics[width=0.8\linewidth]{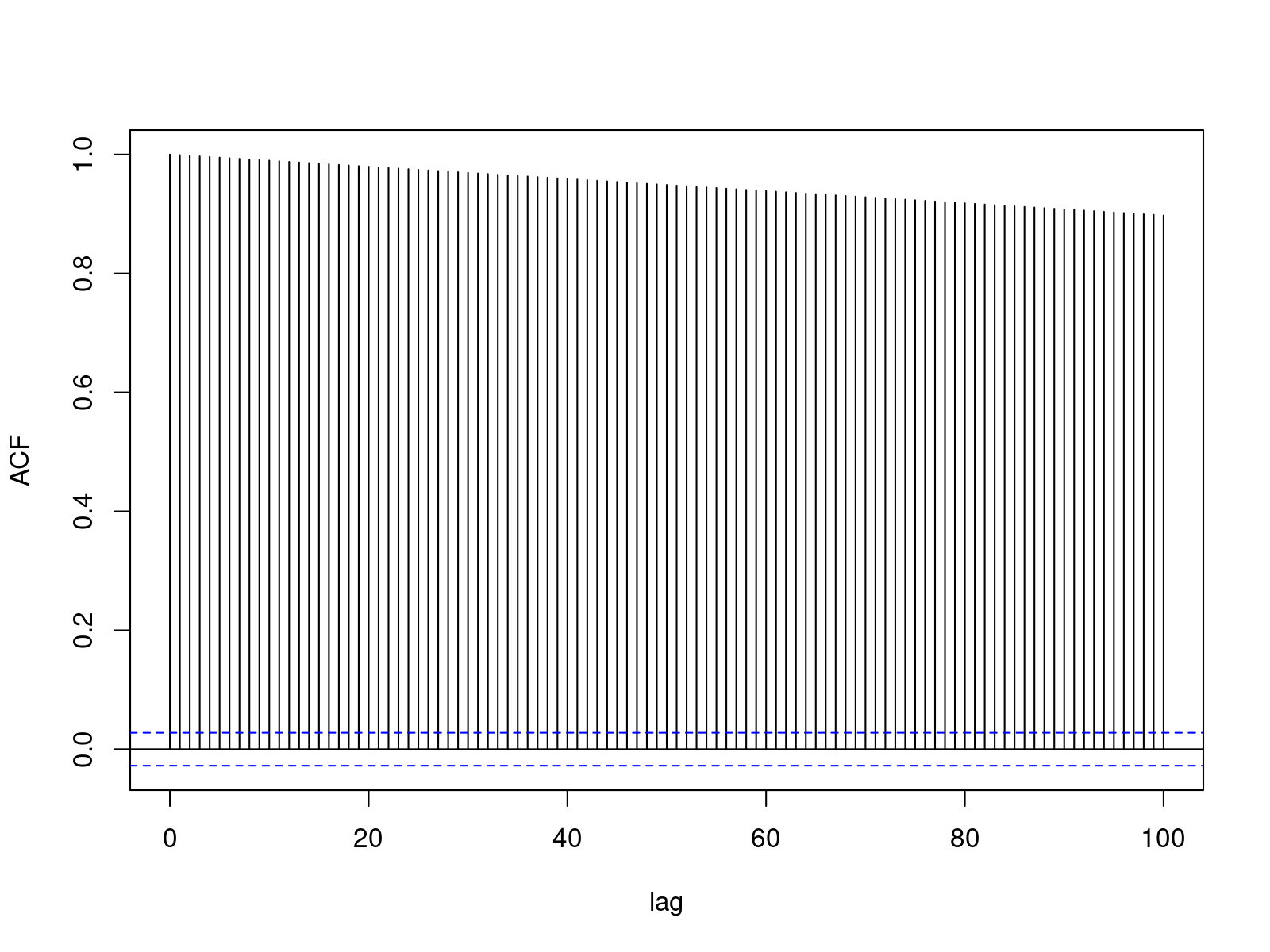}
        \caption{Metropolis}
        \label{fig:acf_precond_Metropolis_potts}
    \end{subfigure}
     \begin{subfigure}[b]{0.32\textwidth}
        \centering
        \includegraphics[width=0.8\linewidth]{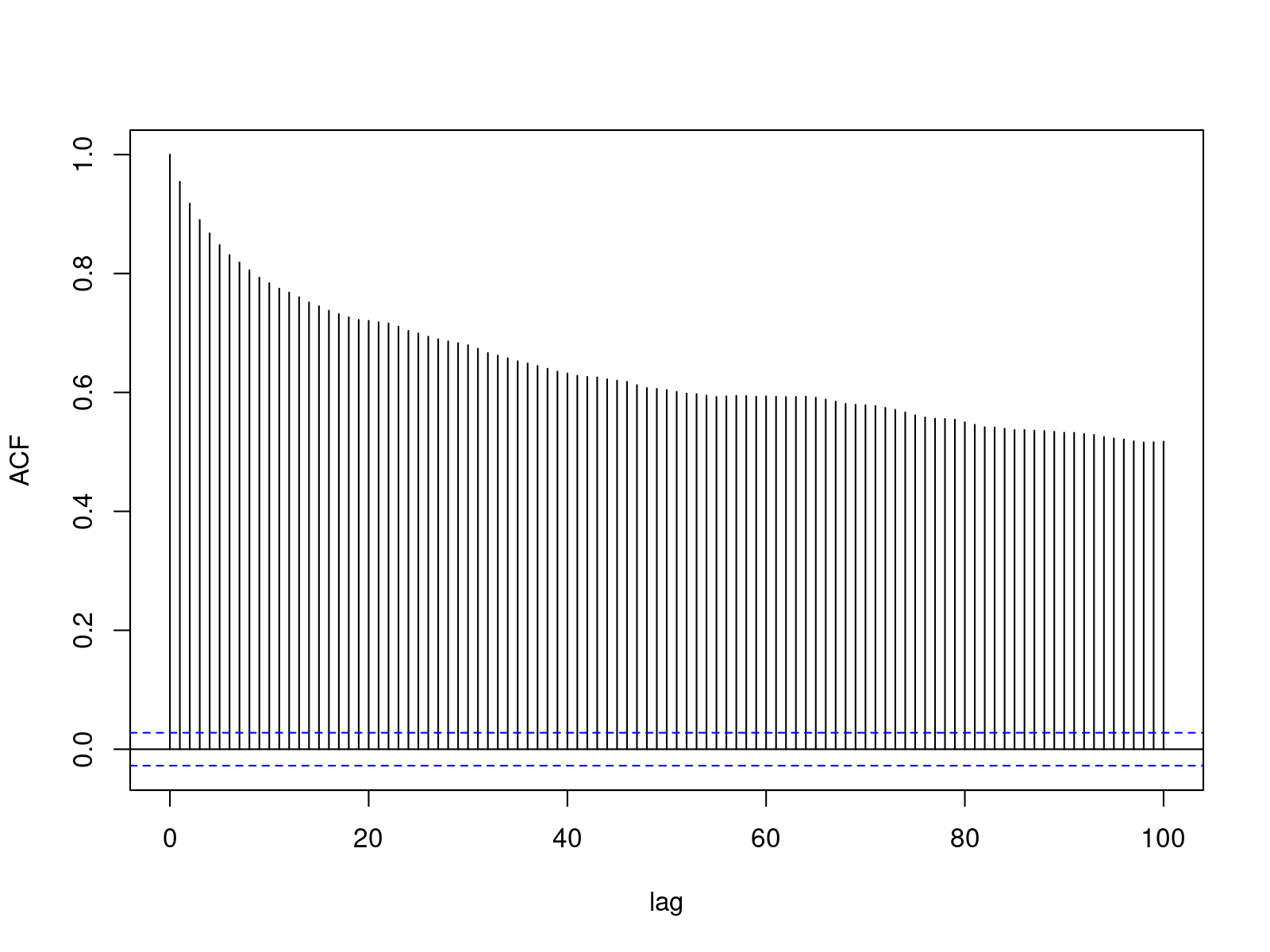}
        \caption{NCG}
        \label{fig:acf_precond_NCG_potts}
    \end{subfigure}
     \begin{subfigure}[b]{0.32\textwidth}
        \centering
        \includegraphics[width=0.8\linewidth]{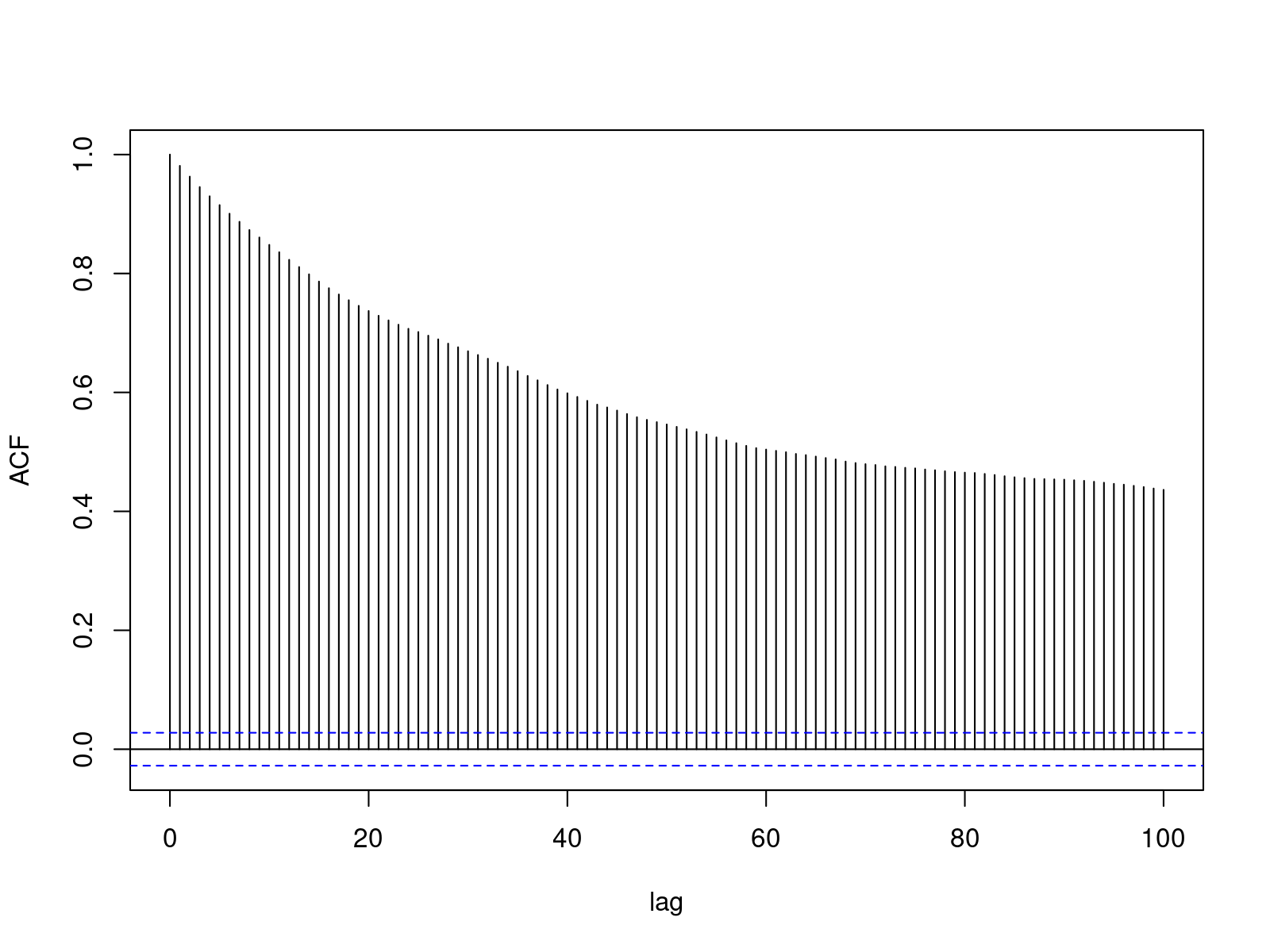}
        \caption{AVG}
        \label{fig:acf_precond_avg_potts}
    \end{subfigure}
     \begin{subfigure}[b]{0.32\textwidth}
        \centering
        \includegraphics[width=0.8\linewidth]{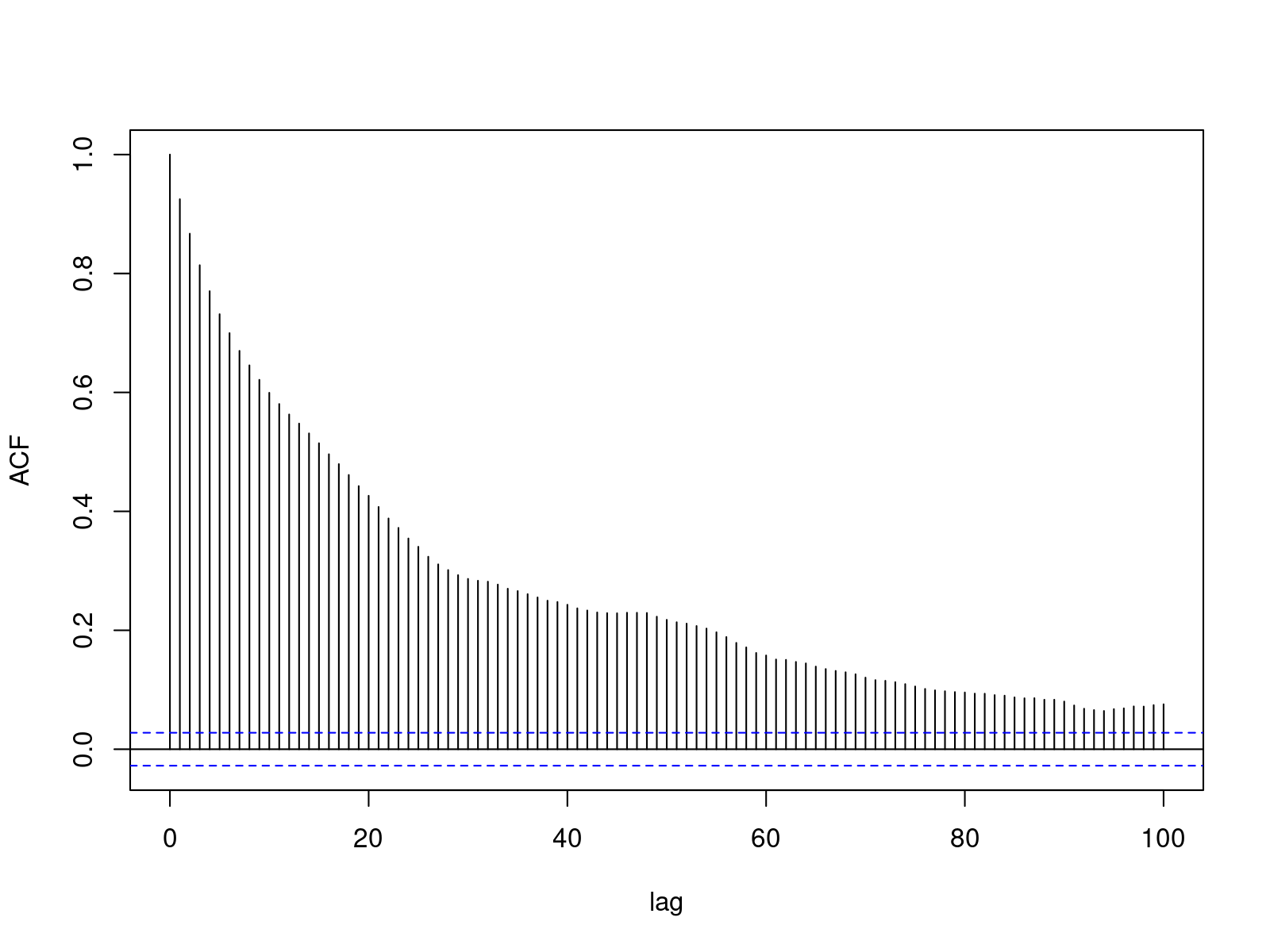}
        \caption{V-DHAMS}
        \label{fig:acf_precond_Hams_potts}
    \end{subfigure}
     \begin{subfigure}[b]{0.32\textwidth}
        \centering
        \includegraphics[width=0.8\linewidth]{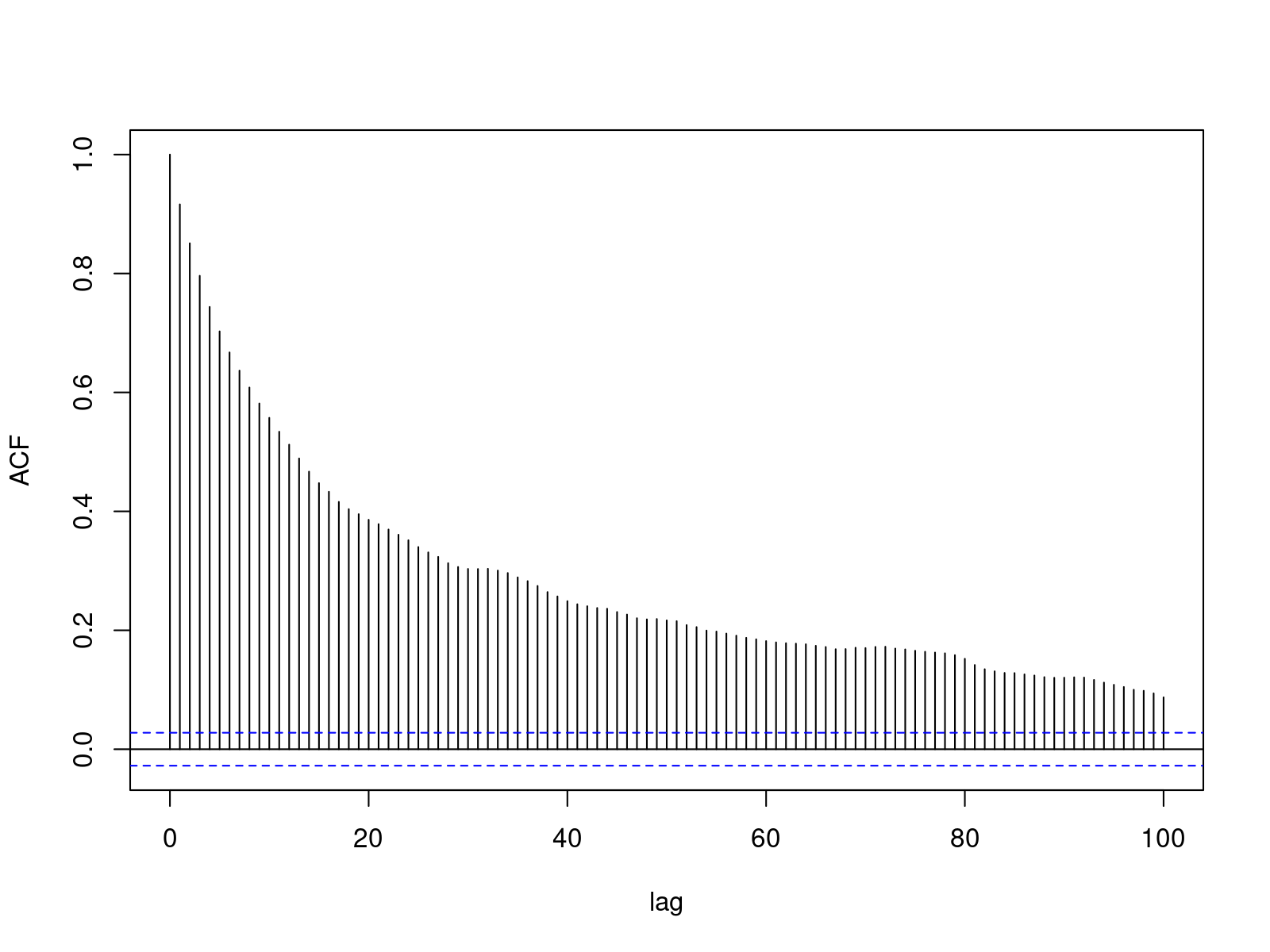}
        \caption{O-DHAMS}
        \label{fig:acf_precond_overhams_potts}
    \end{subfigure}
    \begin{subfigure}[b]{0.32\textwidth}
        \centering
        \includegraphics[width=0.8\linewidth]{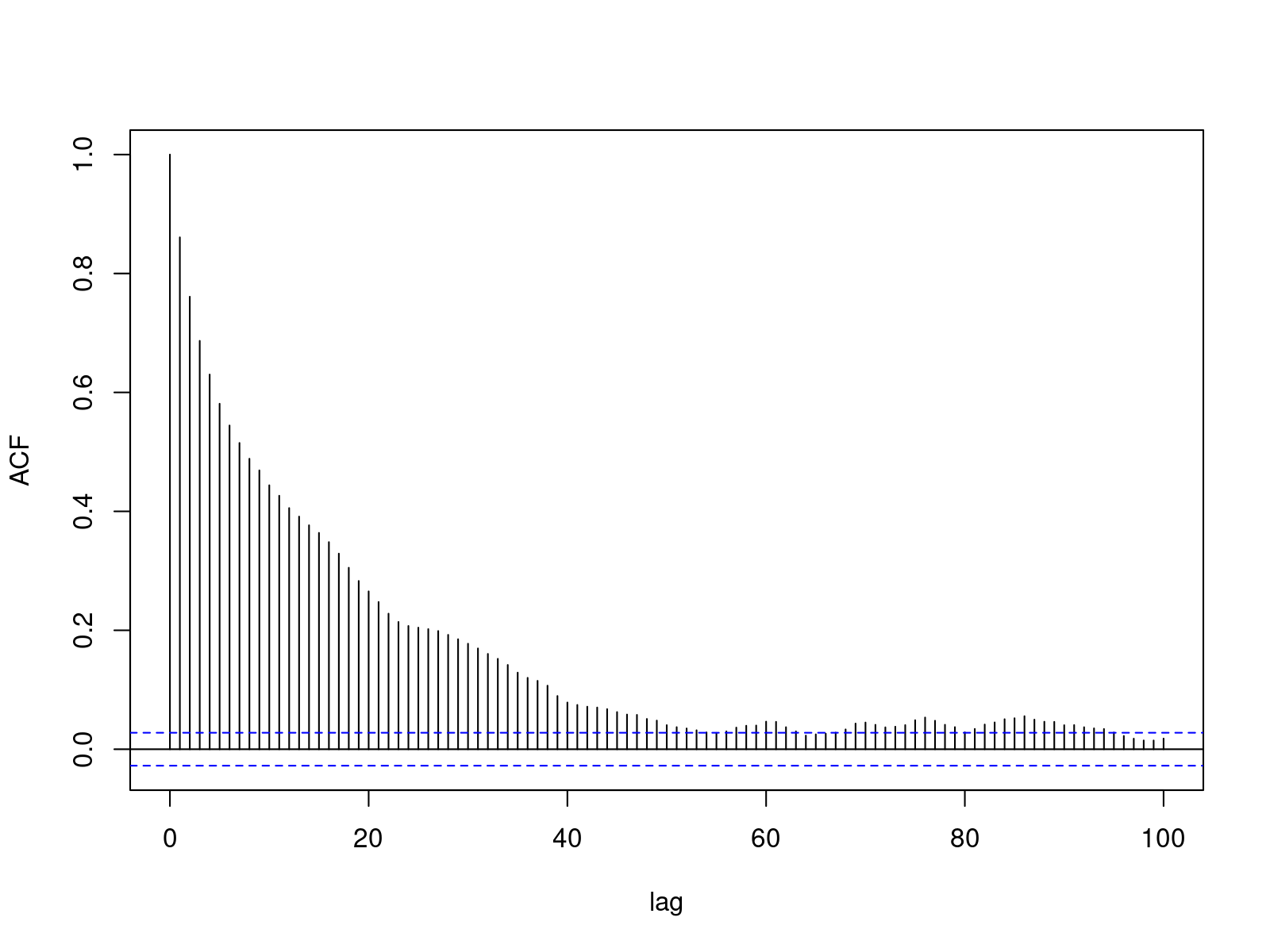}
        \caption{PAVG}
        \label{fig:acf_precond_pavg_potts}
    \end{subfigure}
    \begin{subfigure}[b]{0.32\textwidth}
        \centering
        \includegraphics[width=0.8\linewidth]{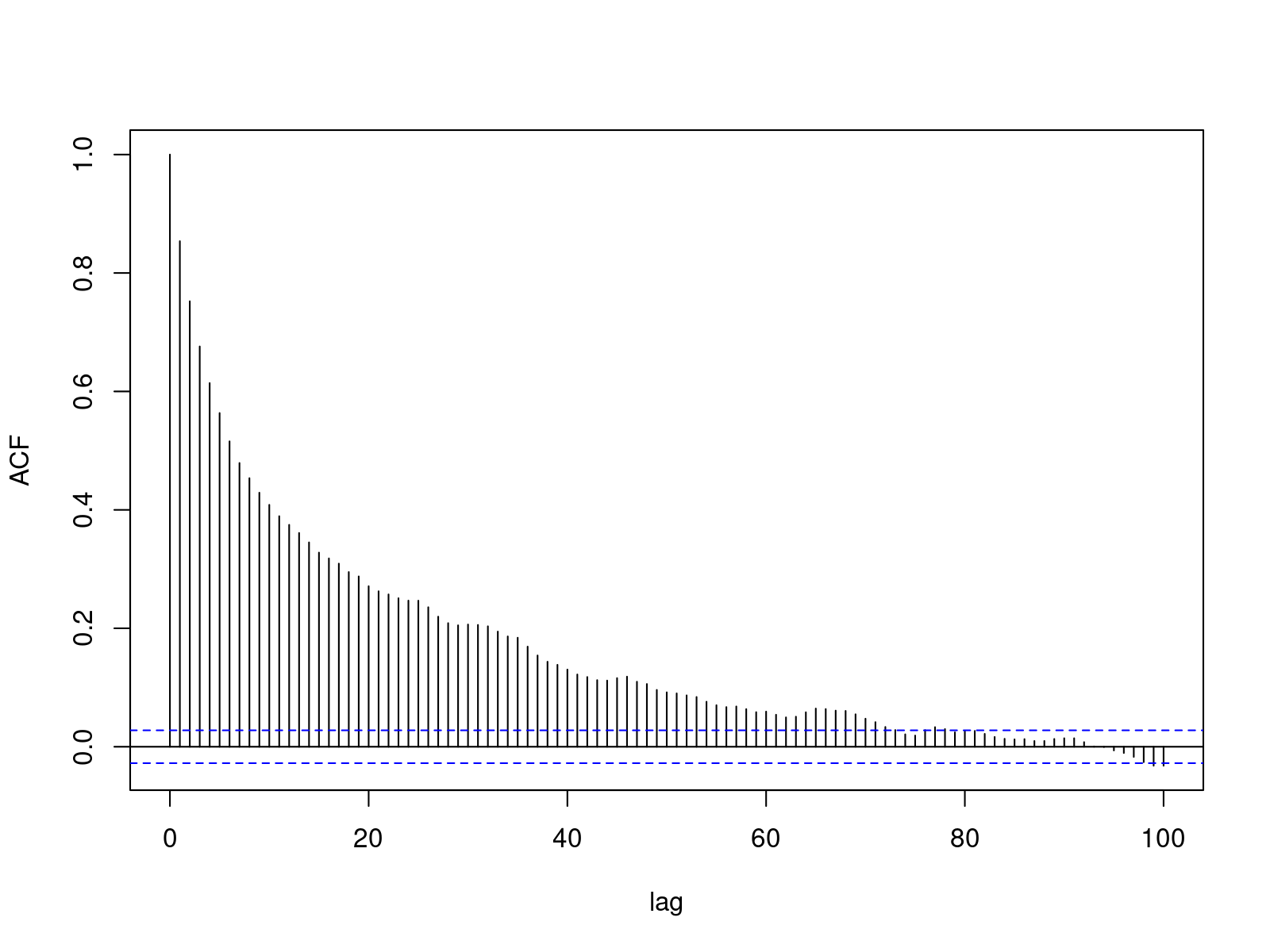}
        \caption{V-PDHAMS}
        \label{fig:acf_precond_vpdhams_potts}
    \end{subfigure}
    \begin{subfigure}[b]{0.32\textwidth}
        \centering
        \includegraphics[width=0.8\linewidth]{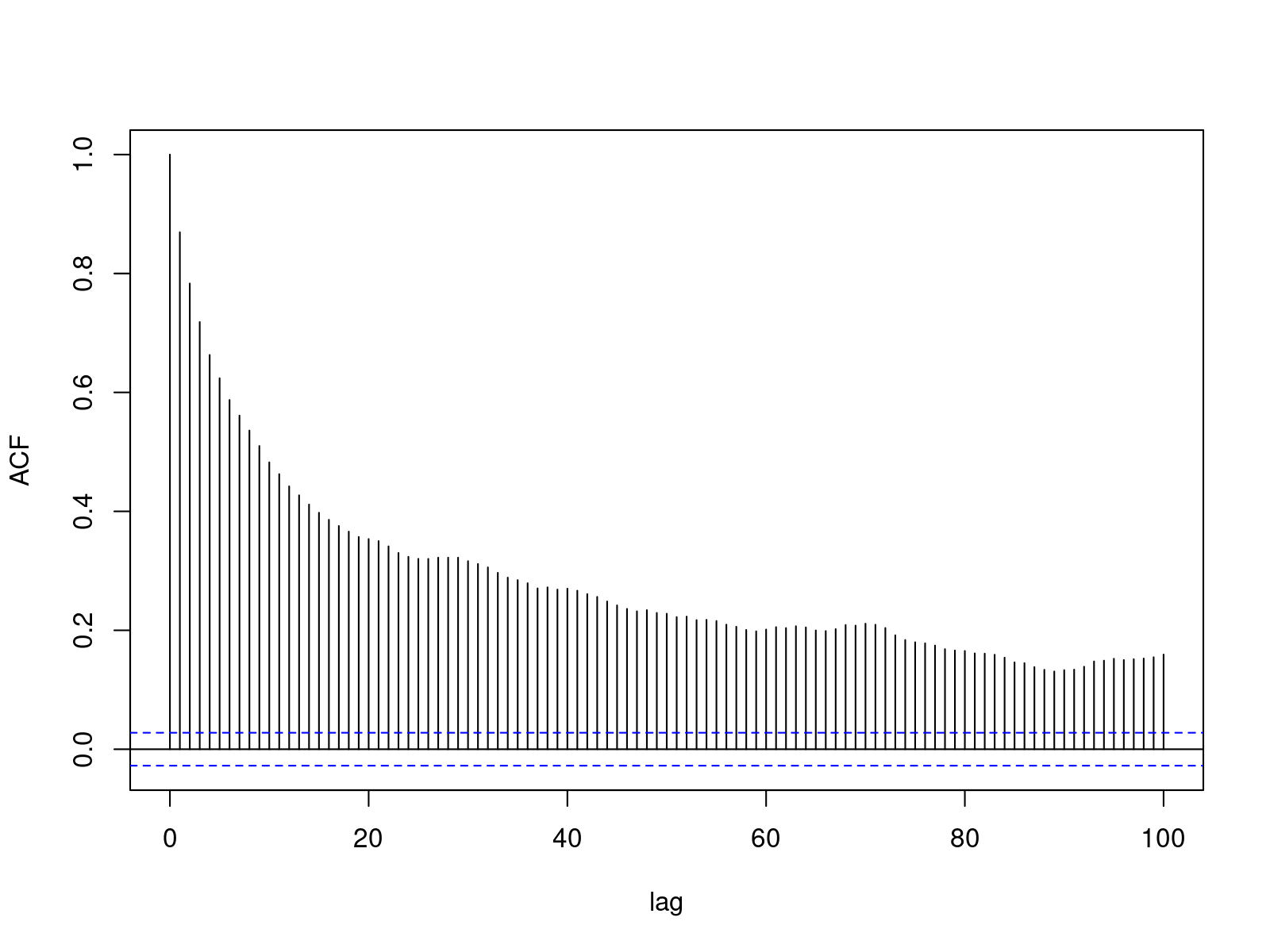}
        \caption{O-PDHAMS}
        \label{fig:acf_precond_opdhams_potts}
    \end{subfigure}
\caption{ACF plots of negative potential for clock Potts model (Anti-ferromagnetic model)}
\label{fig:acf_precond_plots_potts}
\end{figure}

\begin{table}[tbp]
\centering
\begin{tabular}{|l|cc|cc|}
\hline
   {Sampler} & \multicolumn{2}{c|}{Parameter} & \multicolumn{2}{c|}{Acceptance Rate} \\
 & FM & AFM & FM & AFM \\
    \hline
    Metropolis & r= 2  & r= 2 & 0.01 & 0.01\\
    NCG & $\delta=0.1$ & $\delta=0.105$  & 0.80 &0.73 \\
        AVG & $\delta=0.065$ & $\delta=0.055$ & 0.59 & 0.69\\
   V-DHAMS & $\epsilon = 0.85, \delta=0.21, $ & $\epsilon = 0.85, \delta=0.2,$ & 0.64 & 0.65\\
    & $\phi=0.12$ & $\phi=0.15$ & & \\
    O-DHAMS & $\epsilon = 0.85, \delta=0.22, $ & $\epsilon = 0.85, \delta=0.5, $ & 0.47 & 0.55\\
     & $\phi=0.15, \beta=0.1$ & $\phi=0.15, \beta=0.1$ & & \\
    PAVG & $\delta=16$  & $\delta=14$ & 0.58 &  0.46\\
    V-PDHAMS & $\epsilon=0.85, \delta=15.5,$ & $\epsilon=0.85, \delta=13.5,$ & 0.66 & 0.56\\
     & $\phi=0.04$ & $\phi=0.04$ & & \\
    O-PDHAMS & $\epsilon=0.85, \delta=15,$ & $\epsilon=0.85, \delta=12.5,$ & 0.60 & 0.44\\
    & $\phi=0.04, \beta=0.1$ & $\phi=0.04, \beta=0.1$ & & \\
    \hline
\end{tabular}
\caption{Parameters for clock Potts model}
\label{tab:precond_parma_potts}
\end{table}
The optimal parameters and associated acceptance rates for each method are presented in Table~\ref{tab:precond_parma_potts}. The smallest eigenvalue of the preconditioning matrix $W$ is $-4.56$ for the Ferromagnetic model and $-4.49$ for the Anti-ferromagnetic model. The autocorrelation function (ACF) plots of $f(s)$ from a single chain are shown in Figure~\ref{fig:acf_precond_plots_potts_ferro} for the ferromagnetic model, and in Figure~\ref{fig:acf_precond_plots_potts} for the anti-ferromagnetic model.

PAVG, V-DHAMS, and O-DHAMS exhibit lower auto-correlations compared to other samplers for both ferromagnetic and anti-ferromagnetic model, suggesting weaker dependencies between successive draws. 

\clearpage

\end{document}